\documentclass{article}
\usepackage[margin=1in]{geometry}
\usepackage{graphicx} 
\usepackage{subcaption}
\usepackage{color}
\usepackage{verbatim}
\usepackage{url}
\usepackage{amsmath}
\usepackage{amsfonts}
\usepackage{authblk}
\usepackage{float}
\usepackage[normalem]{ulem} 
\usepackage[numbers,sort&compress]{natbib}
\usepackage{tabularray}
\usepackage{float}
\usepackage{graphicx}
\usepackage{codehigh}
\usepackage{authblk}
\usepackage{bibunits}
\UseTblrLibrary{booktabs}
\UseTblrLibrary{siunitx}

\NewTableCommand{\tinytableDefineColor}[3]{\definecolor{#1}{#2}{#3}}
\defaultbibliographystyle{unsrtnat}

\title{Introducing AI to an Online Petition Platform Changed Outputs but not Outcomes}

\author[1,2]{Isabel Corpus\thanks{Corresponding author. Contact at isc36@cornell.edu}}
\author[3]{Eric Gilbert}
\author[1,2]{Allison Koenecke}
\author[1,2]{Mor Naaman}

\affil[1]{Department of Information Science, Cornell University, Ithaca, NY}
\affil[2]{Cornell Tech, New York, NY}
\affil[3]{School of Information, University of Michigan, Ann Arbor, MI}

\date{\today}

\begin{document}
\begin{bibunit}

\maketitle

\begin{abstract}
The rapid integration of AI writing tools into online platforms raises critical questions about their impact on content production and outcomes.
We leverage a unique natural experiment on Change.org, a leading social advocacy platform, to causally investigate the effects of an in-platform ``write with AI'' tool.
To understand the impact of the AI integration, we collected 1.5 million petitions and employed a difference-in-differences analysis. Here we show that in-platform AI access significantly altered the lexical features of petitions and increased petition homogeneity, but did not improve petition outcomes.
We confirmed the results in a separate analysis of repeat petition writers who wrote petitions before and after introduction of the AI tool. 
The results suggest that while AI writing tools can profoundly reshape online content, their practical utility for improving desired outcomes may be less beneficial than anticipated, and introduce unintended consequences like content homogenization.

\end{abstract}

The rapid integration of AI writing tools into online platforms raises critical questions about their impact on content production and outcomes.
We leverage a unique natural experiment on Change.org, a leading social advocacy platform, to causally investigate the effects of an in-platform ``write with AI'' tool.
To understand the impact of the AI integration, we collected 1.5 million petitions and employed a difference-in-differences analysis. Here we show that in-platform AI access significantly altered the lexical features of petitions and increased petition homogeneity, but did not improve petition outcomes.
We confirmed the results in a separate analysis of repeat petition writers who wrote petitions before and after introduction of the AI tool. 
The results suggest that while AI writing tools can profoundly reshape online content, their practical utility for improving desired outcomes may be less beneficial than anticipated, and introduce unintended consequences like content homogenization.

How does the integration of AI writing tools change content production, participation, and outcomes on online platforms? 
New AI writing tools and assistants based on Large Language Models (LLMs) are capable of generating high-quality text that is more articulate, persuasive, and creative than if the users had written the text themselves~\cite{Noy2023-sf, doshiHauser2024, matz2024potential, argyle2025testing, kumar2025human}.
These tools and assistants are increasingly introduced into online platform user flows~\cite{experienceAmazonSellersCan2024, WriteAIFacebook}.
However, the effect of in-platform AI writing tools on content and its outcomes on online platforms remains uncertain, with conflicting evidence regarding the contexts in which AI can augment human performance~\cite{vaccaroWhenCombinationsHumans2024}.
While generally helpful, the effect of AI assistance on text performance could be task-specific; AI might not be as effective in tasks that are more nuanced or require personal context. 
Furthermore, AI writing tools might reduce the overall diversity of text produced across writers~\cite{doshiHauser2024, wu2024generativemonoculturelargelanguage, haoArtificialIntelligenceTools2026,agarwal2025ai,anderson2024homogenization}, perhaps leading to less compelling, individualized text.
To study the potential trade-offs of AI writing tools on a real-world online platform, we explore AI-assisted petition-writing on Change.org.

Change.org is a social advocacy platform where individuals around the world can write, read, and sign petitions aiming to effect change in their communities. 
Change.org aims to facilitate democratic deliberation and advocacy by enabling individuals to enact grassroots change and connect with their community around shared issues.
By gathering community support in the form of signatures and comments, petitions are able to gain media attention and persuade decision-makers.
In 2023, Change.org integrated an AI writing tool built on an OpenAI-provided LLM into the petition writing workflow.
The site's home page encouraged visitors to ``create a compelling petition in minutes'' with the help of AI.
Petition writers can use this tool to generate petition drafts from only a short, user-entered description of a cause.
Petition outcomes, and thereby the outcomes of the causes they champion, may be influenced by AI tools that impact the petitions' content. 
Conceivably, integration of AI could bolster the effectiveness of the Change.org platform, as well as individual petitions, by lowering the barrier to participation and enabling users to produce high quality content with ease. 
As suggested by prior research, it is possible that this lower barrier for participation could increase petition production and completion rates on the platform, while improving the quality of content produced by low-skill writers~\cite{Noy2023-sf, brynjolfsson2025generativeAIatwork}.

In this work, we leveraged a unique natural experiment to estimate the causal impact of access to the Change.org AI~writing tool on published petitions in terms of lexical features and petition outcomes. 
Historically, the user experience on Change.org had been similar in the United States, Great Britain, Canada and Australia.
However, this newly introduced AI tool was first rolled out in the United States, Great Britain, and Canada, with a delayed release in Australia.
We conducted a difference-in-differences analysis that leverages this staggered roll-out in AI tool access to measure the causal impact. 
Unlike off-platform AI tools, the Change.org in-platform AI tool is uniquely tailored to the task and thereby channels AI usage through a single model, a single backend prompt, and a standard interface.
This in-platform AI tool and its staggered roll-out allowed us to avoid concerns that arise in many observational studies of generative AI use in the wild, where AI use can be difficult to detect and driven by idiosyncratic user flows.

Our research questions are twofold: (a) does access to the AI-generated draft tool result in changes to petition lexical features, and (b) does access to the tool result in changes to petition outcomes? 
To study these questions, we collected 1.5 million petitions written on Change.org between 2022 and 2024.
We then estimated the impact of AI tool access using trends in petitions produced during the eleven weeks of differential access between the full launch of the AI tool in the treated countries and introduction of the AI tool in Australia.
We find that access to the AI-generated draft tool significantly changed petition lexical features. 
Specifically, the lexical features of petitions written with access to AI had greater resemblance to those of pre-AI petitions that achieved strong outcomes. 
We also show that AI access resulted in increased homogeneity: petitions became more similar to each other.
At the same time, access to the AI tool failed to improve petition outcome metrics such as hitting comment and signature thresholds. 
The shift in features and lack of improvement to petition outcomes is not driven by changes to platform participation; user participation remains constant after access to AI. 
In fact, we show that the observed trends of lexical features and outcomes are replicated when looking at repeat petition writers who wrote petitions with and without the AI tool access.
Ultimately, our analyses show that integration of AI does not bolster the effectiveness of Change.org, despite significantly altering the characteristics of the petition text. 
\section*{Results} 

To estimate the causal impact of access to the Change.org in-platform AI tool, we conducted a series of difference-in-differences analyses~\cite{callaway2021difference}
that leveraged a natural experiment in AI tool access on Change.org driven by a staggered roll-out by geographic region: the United States, Great Britain, and Canada received full access eleven weeks prior to Australia. 
We confirmed country-specific launch dates for the AI-generated draft tool with members of the Change.org team.
We could thus compare country-level trends in petition lexical style and outcomes during the eleven-week period (October 2, 2023 - December 15, 2023) where access differed between geographic regions.

The hypothesis underlying our analyses is that since all countries followed similar trends in petition style and outcomes prior to AI tool release, we can attribute differences in trends during the eleven-week period to the introduction of the AI tool.
We verify the existence of similar trends through a parallel trends analysis (shown in SI Appendix Figure 10) in which we show that before the release of AI, trends in style and outcomes of petitions produced in the control country (Australia) were parallel to those in treated countries (the United States, Great Britain, Canada).  
We also qualitatively observe that time series of petition style and outcome are again parallel after the period of differential access, when all countries have AI access, as shown in SI Appendix Figure 23.
Our comparison of trends follows a difference-in-differences quasi-experimental approach, 
in which we measure the change in trends between similar groups after some groups received an intervention.
We focus on the eleven weeks of differential access between the launch of the AI tool in treated countries and the launch of the AI tool in Australia. 
We leverage two types of specifications for the difference-in-differences approach: static and dynamic. 
In the static difference-in-differences specification, we measure the overall effect across countries and weeks of AI access for treated countries. 
In the dynamic difference-in-differences specification, we measure the effect of AI access for each week in which treated countries have AI access~\cite{callaway2021difference}. 
We also confirm our results through a series of robustness tests (see SI Appendix Figures 11-20, 30, 31) and a synthetic controls analysis (see SI Appendix Figures 34-36), which reaffirm our difference-in-differences results. 

We defined the intervention in our causal analysis as AI access, rather than the actual use of AI in the petition. 
While we know when users had access to the in-platform AI tool, we cannot robustly tell when and how they used it, as AI detection models may be biased in their ability to identify AI-generated content.
Users may also use off-platform generic AI tools; however, we expect off-platform usage to be similar in both treated and control countries and thereby to not influence our estimates. 
We did develop an AI classifier specific to the Change.org AI tool to understand if the tool was utilized when users had access.
With this method, we estimate that during the eleven-week period of differential access, over half of petitions that were written with access to the in-platform AI tool were AI-generated (see SI Appendix Figure 2). 
We also find further support for parallel trends: Australia and treated countries had similar detected levels of (general, off-platform) AI use prior to the release of the in-platform AI tool. 

In our main analysis we use Australia as the control country to estimate causal effects.
We confirm that our estimates are robust to multiple geographic regions with a synthetic control analysis that leverages additional control countries (New Zealand and India) in the SI Appendix, Figures 34-36.
The results of the synthetic control analysis are consistent with the difference-in-differences analysis, demonstrating that our causal estimates are consistent across regions and outside of Australia. 

We preregistered our difference-in-differences analysis with AsPredicted (\url{https://aspredicted.org/xtcd-tvqk.pdf}), and note below the few instances where we expanded from the preregistration. 

\subsection*{Lexical Features}
We started by estimating the effect of in-platform AI tool access on petition lexical style through three features: lexical diversity, readability, and text length.
We selected these lexical features because of their strong predictive relationship to petition outcomes prior to the release of the AI tool (see SI Appendix Figure 32).
Prior work has used lexical diversity and readability to assess writing competency in a variety of contexts such as language proficiency and clarity~\cite{mesgar2018neural, bizzoni2023good, cowgill2026}, and sufficient length is necessary for a petition to convey a compelling message. 
We operationalized lexical diversity through Moving Average Type-Token Ratio (MATTR), a measure of lexical diversity calculated as a moving average of the ratio of unique words to total words across windows of the text. 
Higher MATTR scores denote a more varied vocabulary. 
We operationalized readability using the Flesch-Kincaid Grade Level (FKGL), a common measure of readability calculated with the number of words per sentence and syllables per word. 
Higher FKGLs denote more complex text that requires a higher grade level to understand. 
For text length we used a simple word count of the number of words in the petition text and title combined.

Figure 1a demonstrates that access to the AI tool had a significant and consistent effect on lexical features: petitions were longer, with more complicated and varied vocabularies.
Figure 1a shows the results of a dynamic difference-in-differences estimation, in which we estimate the average treatment effect in treated countries by week (x-axis), compared to the baseline (week -1). 
The figure shows average treatment effects for treated countries in weeks before any AI tool access (orange, left of week 0), during AI tool A/B testing (gray, weeks 0-25), and after the full launch of the tool (blue, right of week 26).  
We measure the shift (y-axis) via average group-time treatment effect estimates on the treated users, relative to the baseline week (week -1). We exclude the A/B period from our analysis (but show it in the figure) as we do not know which users or what share of users accessed the tool during this period. 
The spikes in estimated effect on petition style during the A/B test period reflect experimentation with prompt engineering that informed the in-platform AI tool. After full launch, the model and prompt behind the AI tool remain consistent during the eleven-week period for which we estimate the effect of AI tool access.
For example, the Word Count graph in Figure 1a shows that in weeks after full launch (blue, right), AI access resulted in an estimated addition of 50-70 words to the median petition word count, relative to the baseline week.
Similarly, access to the AI tool increased median MATTR values and FKGLs in weeks with AI access. 

Figure 1b shows the result of a static difference-in-differences estimation, by which the average treatment effect is estimated across the post-AI (after week 26) period, relative to the pre-AI period (before week 0).
We estimate that over all weeks in the post-AI period, the MATTR (top row of Figure 1b) increases by 0.049 units (95\% CI = [0.047, 0.050], t(17) = 61.48, p $<$ 0.001), median petition grade level (second row of Figure 1b) increases by 1.49 grades (95\% CI = [1.14, 1.83], t(17) = 9.16, p $<$ 0.001), and median word count (bottom row of 1b) increases by 53.63 words (95\% CI = [47.08, 60.18], t(17) = 17.27, p $<$ 0.001).
An estimated increase of 53.63 words during this eleven-week period corresponds to a 49\% increase in median petition length, from 110 words pre-AI.
Similarly, the median FKGL increases by 19\%, from grade 8 pre-AI, and the MATTR increases by 6\%, from 0.8 units pre-AI.
A descriptive analysis shows that petition vocabularies qualitatively changed as well: prior to AI access the most frequently used verbs in petition titles were mono-syllabic Old English verbs such as ``stop'' and ``let''.
Following AI access, multi-syllabic verbs like ``implement'' and ``establish'' jumped in frequency (see time series of verb frequency in the SI Appendix Figures 21 and 22).

\begin{figure}[htbp]
    \centering
    \includegraphics[width=0.8\textwidth]{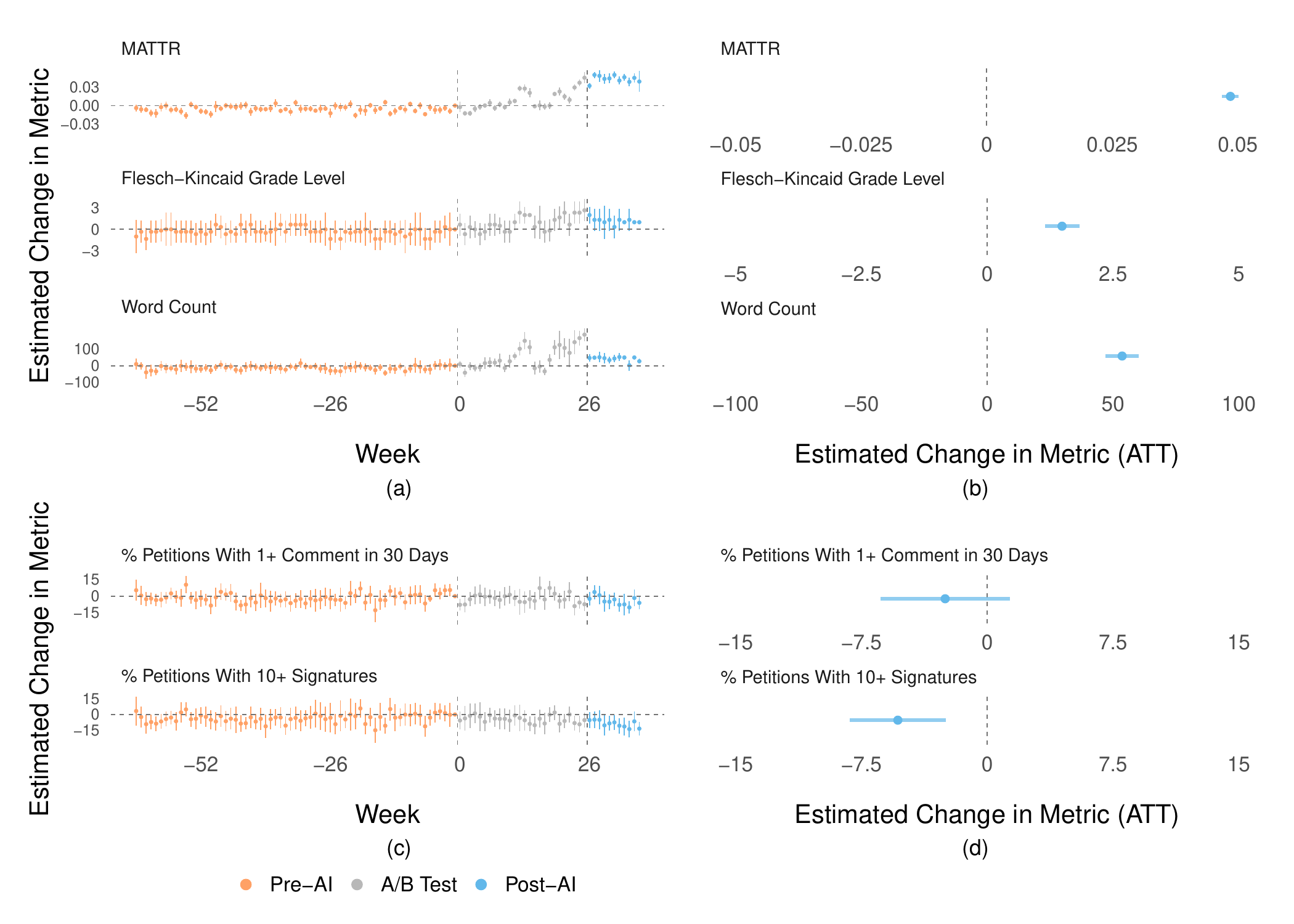}
    \caption[]{Platform trends in lexical features (panels a, b) changed significantly with AI access while trends in outcomes (panels c, d) did not improve, or worsened, based on a difference-in-differences analysis of platform trends before and after Change.org AI tool release.
Panels (a, c) show dynamic difference-in-differences, with points reflecting estimated average treatment effects on treated units (ATT) by week, relative to the baseline week (Week -1), on lexical features (a) and outcomes (c). Simultaneous 95\% confidence bands are shown.
Panels (b, d) show static difference-in-differences, with points reflecting the aggregated estimated average treatment effects (ATT), relative to pre-AI period, on lexical features (b) and outcomes (d). 
The figure includes 95\% CI's.
Lexical feature effects (a, b) reflect changes in weekly median values, while outcome effects (c, d) reflect changes in percentage points.
The A/B testing period, shown in gray between the dashed lines, is excluded from the aggregate estimates in panels (b, d).
The Pre-AI period was 65 weeks long (N = 237,020 petitions), the A/B test period was 26 weeks (N = 82,504 petitions), and the Post-AI period was 11 weeks (N = 42,111 petitions).}
\label{fig:figure_1}
\end{figure}

\begin{figure}[htbp]
    \centering
    \includegraphics[width=0.6\textwidth]{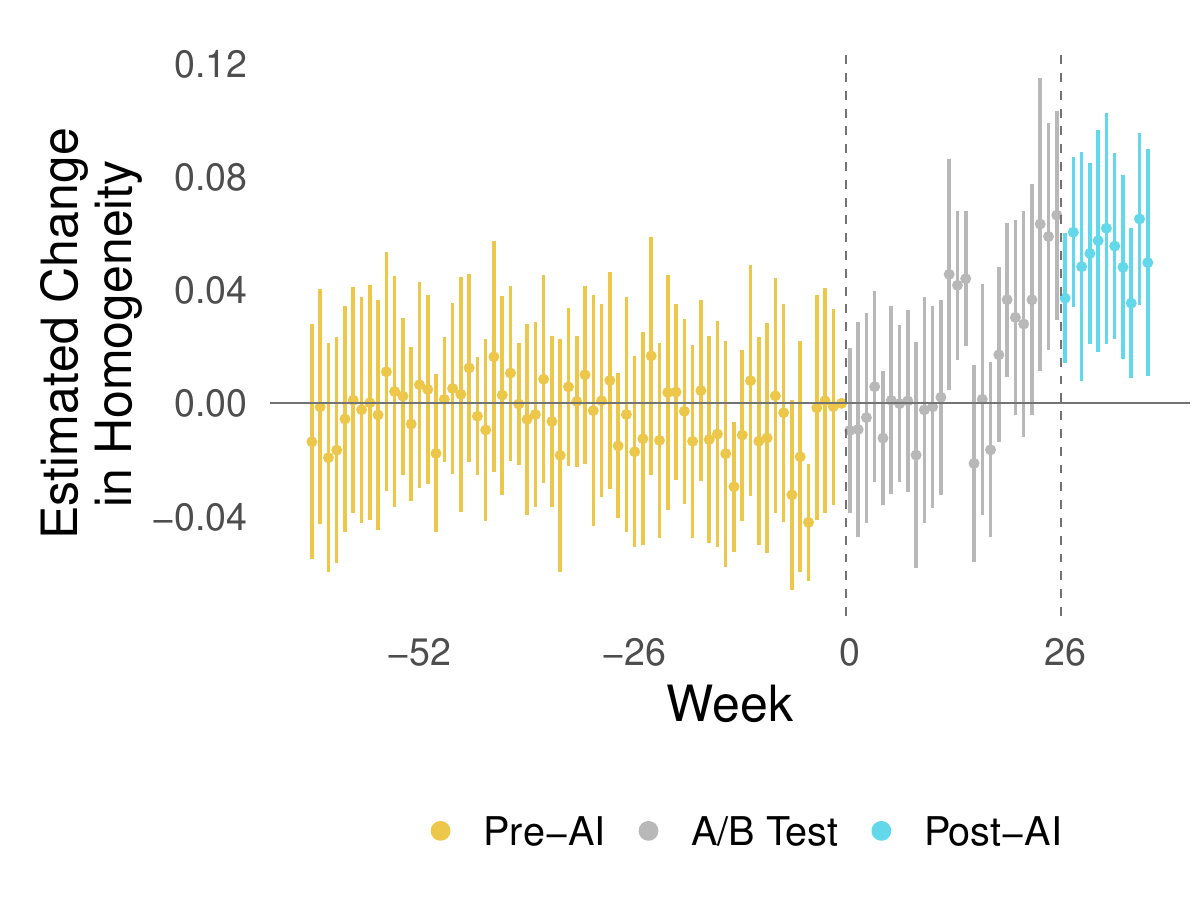}
    \caption[]{Dynamic difference-in-differences estimates for homogeneity shows that with access to the AI tool, platform homogeneity increases.
Data are presented as estimated average treatment effects on treated units (ATT) by week, with simultaneous 95\% confidence bands.
Estimates are compared to the baseline of the week prior to A/B testing in treated countries (Week -1). 
Pre-AI period was 65 weeks (N = 237,020 petitions), A/B test period was 26 weeks (N = 82,504 petitions), and the Post-AI period was 11 weeks (N = 42,111 petitions).}
    \label{fig:figure_2}
\end{figure}

\subsection*{Platform Homogeneity} 
We augmented our preregistered analysis with a difference-in-differences study of the impact of AI access on platform homogeneity.
The results show an increase in inter-petition homogeneity, as measured through the mean pairwise similarity between petitions~\cite{wu2024generativemonoculturelargelanguage, padmakumar2023does}.
To measure homogeneity, we first created 3,072-dimension embeddings with the OpenAI text-embedding-3-large model (\url{https://platform.openai.com/docs/models/text-embedding-3-large})
for all petitions.
We then calculated the cosine similarity between all pairs of petitions posted in a given week and country, and took the mean of these pairwise scores to obtain the mean pairwise similarity for that week and country.
We used these weekly country-level time series of homogeneity to estimate the effect of in-platform AI access on treated countries through a dynamic difference-in-differences estimation.

The dynamic difference-in-differences results shown in Figure 2 demonstrate that access to the AI tool caused petitions to be more similar to one another than if they had been written without access to the AI tool.  
In weeks after week 26 (shown in blue), following the introduction of AI, AI access had a consistent effect on petition homogeneity, increasing mean pairwise similarity by about 0.05~units (on a scale of~-1 to~1) for treated units in each period. 
We also conducted a static difference-in-differences estimation (shown in the SI Appendix Table 14), finding that overall, the post-AI period had a~0.056 (95\% CI = [0.049,~0.063], t(17) = 17.83, p $<$ 0.001) increase in pairwise similarity, which is a 23\% increase from the average pre-AI weekly mean similarity value of 0.240.

\subsection*{Outcomes}
We find that petitions' outcomes did not improve---and in some cases even worsened---with access to the AI tool. 
We operationalized petition outcomes through two metrics: the share of petitions that reach 1 comment in 30 days, and the share that reach 10 signatures at the time of data collection.
We preregistered our primary outcome as the presence of a comment posted within 30 days after the petition was posted, since the time of comments (but not signatures) is available in our data.
As each petition was posted at least one year prior to data collection, we expect recency was not a factor in whether petitions reached these minimal thresholds.
For example, 99\% of petitions receive their first comment in the first month after posting, if they receive a comment at all. 
We include additional analysis with different thresholds for comments and signatures in SI Appendix Figure 11, showing that results remain consistent.
We use these minimal thresholds
instead of raw engagement counts because engagement is sparse (during the period of our main analysis, 39\% of petitions reach 10 signatures and 35\% ever gain a single comment). 
Furthermore, petitions on Change.org concern a wide variety of communities, from hyper-local neighborhood issues that persuade decision-makers with tens of supporters, to global issues that gain hundreds of thousands of signatures. 
On the other hand, the minimum thresholds for petitions to ``get off the ground'' remain the same across time and context.
By using minimum thresholds we ensure our analysis is not biased towards issues that affect more populous communities while still reflecting a meaningful indicator of a petition's outcomes.

We do not find evidence that AI tool access positively impacted petition outcomes.
Instead, we observe that with AI access, outcomes are either unchanged, with no significant differences observed post-treatment, or worsened.
Figure 1c shows the results of a dynamic difference-in-differences estimation, in which we find that the average estimated change in treated countries' comments and signatures tends to be worse than the baseline week (Week -1) during the weeks post-AI access. 
For example, the top graph in Figure 1c shows that in the weeks after the full launch (first row, blue, right of week 26), the share of petitions that received at least ten signatures is consistently lower than the baseline week. 
During most weeks with AI access, the percent of petitions that reached one comment in 30 days declined by about 5 percentage points relative to the baseline week. 
Figure 1d shows that after AI tool release, the share of petitions that receive one comment in 30 days decreased by -2.51 percentage points relative to the share of petitions pre-AI tool release, although that difference was not statistically significant (95\% CI = [-6.36, 1.34], t(17) = -1.37, p = 0.187).
However, the share of petitions that reached 10 signatures when written with access to AI decreased by -5.33 percentage points, which was statistically significant (95\% CI = [-8.21, -2.45], t(17) = -3.91, p = .001).
In sum, we do not find evidence that petition outcomes are improved when users have access to AI. 

\subsection*{Petition Text Quality}
We find that petitions written with access to the AI tool have, on average, higher writing quality and persuasiveness.
One possible explanation for the lack of improvement to platform-level outcomes could be that the AI tool is simply not very effective: petitions produced with the tool could have lower quality text or be less persuasive than petitions written without the tool. 
To understand the quality of petitions written with and without AI access, we conducted a labeling task in which independent raters labeled petitions for writing quality and persuasiveness.
Our descriptive results (full details in the SI Appendix Section Petition Writing Quality and Persuasiveness, and Figures 4-7) demonstrate with two-sided t-tests that average petition writing quality is significantly higher among petitions written with in-platform AI access, relative to petitions written without AI access (\(\mu_{post-pre}=0.70\), t(412.46) = 9.76, \( p<\) 0.001, Cohen's d = 0.96, 95\% CI = [0.75, 1.16]).
Average petition persuasiveness scores are similarly significantly higher among petitions written with in-platform AI access, relative to petitions written without AI access (\(\mu_{post-pre}=0.66\), t(414.56) = 8.75, \(p<\) 0.001, Cohen's d = 0.86, 95\% CI = [0.65, 1.06]).
In the SI Appendix Figures 5-7, we show that prior to AI access, writing quality and persuasiveness are highly associated with lexical features and predictive of petition outcomes. 
However, after AI access, writing quality and persuasiveness are weaker, or negative, predictors of petition outcomes. 
Overall, as writing quality and petition persuasiveness are significantly higher among petitions written with access to in-platform AI, it does not seem that changes to the quality of petition text explain the unimproved platform-level outcomes that we observe. 
We consider potential explanations and ramifications in the Discussion section. 
Next, we consider participation as another potential explanation for unimproved platform-level outcomes.

\subsection*{Participation}
Participation on Change.org did not change after users gained access to the in-platform AI tool. 
While the AI tool did not improve petition outcomes, one possible explanation could be that the tool's launch attracted a different population.
For example, the launch could have resulted in an influx of low-skill writers, thereby reducing the overall skill of writers on the platform. 
In this hypothetical case, platform-level trends might not improve, even if the tool improved the outcomes of individual petitions.
However, we did not find evidence that the effect of AI access on platform trends was driven by changed participation patterns.
Instead, the volume of user participation by week remained consistently in the same range as user participation prior to AI tool access.
We were unable to conduct a difference-in-differences analysis of petition participation because pre-AI participation and volume trends varied between treated and control countries, violating the test requirements.
A time-series analysis of participation as measured by the volume of users writing petitions on the platform showed that participation did not increase with AI access. 

To test the consistency of platform participation, we used pre-AI user participation data from treated and untreated countries to fit a Bayesian Structural Time Series (BSTS) model (see details in SI Appendix Figures 37, 38 and Section Bayesian Structural Time Series)~\cite{brodersen2015causalimpact}.
This BSTS model uses pre-AI long-term trends and seasonal fluctuations to predict a synthetic control for counterfactual participation in treated countries had there been no AI tool launch. 
This forecast for platform participation closely reflected actual user participation, with actual participation falling within a 95\% credible interval of forecast participation.
Through comparing the synthetic counterfactual participation against actual platform participation, we do not find evidence of a significant effect of AI access on user participation, in either direction (\(\Delta_{treated-control}\) = -354 users (s.d. = 901), $p$ = 0.343, 95\% credible interval [-2167, 1263]). 

\begin{figure}[htbp]
    \centering
    \includegraphics[width=0.8\textwidth]{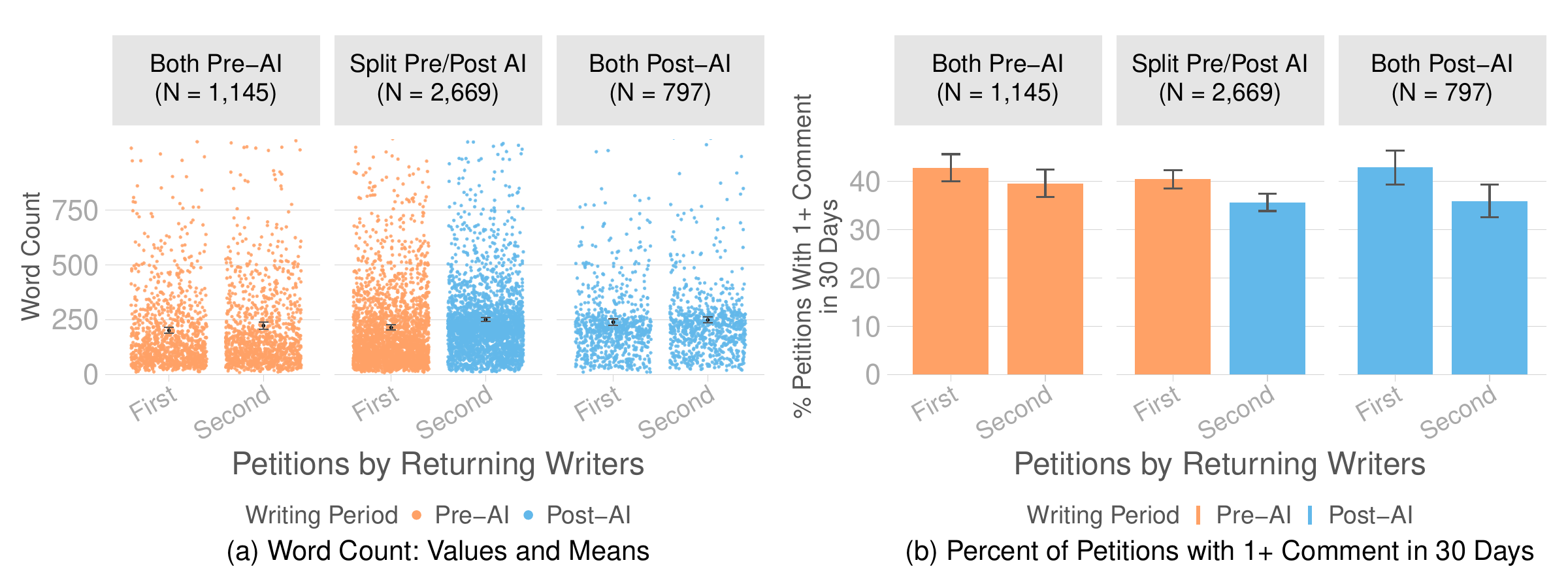}
    \caption[]{Returning petition writers (N=4,611 users) from treated countries, on average, write longer petitions with worse outcomes when their second petition is written with access to AI. 
Panel (a) shows petition length is higher in petition writers' second petitions, in all conditions.
The greatest increase in word count occurs for writers in the Pre/Post split cohort.
We calculate the average word count using all petitions within each cohort of writers. For legible visualization, we do not display outliers in the figure, as defined as petitions with a word count greater than 3 standard deviations from the mean (N=149 petitions, 1.6\% of petitions from repeat writers). 
The black point reflects average word count with 95\% CI's (bootstrapped, 1,000 iterations).
Panel (b) shows petition outcomes are worse in writers' second petitions, in all conditions. 
The greatest drop in outcomes for users' second petitions occurs when the second petition is written with access to AI. 
The data are presented as the percentage of petitions that reach 1 comment in 30 days (height of bar) with 95\% CI's.
Returning petition writers who return within 6 months are excluded, to enforce the same minimum time gap between petitions across cohorts.
}
    \label{fig:figure_3}
\end{figure}

\subsection*{Repeat Petition Writers} 
We further demonstrate that changes to platform composition were not responsible for changes to platform trends by showing that the platform-level trends---shifting petition features and unimproved outcomes---are consistent even when examining trends within repeat users. 
We focus our analysis here on one lexical feature (word count) and one outcome (comment engagement). 
The results for the other variables were similar, and are shown in the SI Appendix Figure 33. 
We do not conduct a panel difference-in-differences analysis because of the rarity of repeat petition writing; only 6.38\% of users in our dataset wrote multiple petitions, which amounted to too few untreated users for a reliable balanced panel.
Instead, we model the predictive relationship between AI access and word count or comment engagement among repeat writers from treated countries (the United States, Great Britain, and Canada) over the full period for which we collected data (through November 30, 2024).

We find that word count increases significantly between users' first and second petitions when their first petition is written without AI tool access and their second petition is written with access to the AI tool (Pre/Post cohort); there is no significant change, however, when users have access to AI for neither of their petitions (Pre/Pre cohort) or both of their petitions (Post/Post cohort).
Figure 3 shows word count (3a) and comment engagement (3b) in petitions written by three cohorts of returning writers: users who wrote both petitions without access to the AI tool (left facet of each figure), users who wrote their first petition without access and their second petition with access to the AI tool (middle), and users who wrote both petitions with access to the AI tool (right). 
Figure 3a shows that across cohorts, on average, returning writers' second petitions were longer than their first petitions. 
For example, in the Pre/Post cohort (middle facet), writers' first petitions were written without access to AI and had an average word count of 213.70 words (95\% CI = [202.44, 227.68]) (left column in orange).
The same users' second petitions were written with AI access, and had an average word count of 250.27 words (95\% CI = [241.49, 259.20]) (right column in blue).
We fit a linear mixed-effects regression model to predict log-transformed word count with fixed effects for cohort (Pre/Pre, Pre/Post, Post/Post) and petition order, and random effects for user ID.
We then used pairwise contrasts with Bonferroni correction~\cite{scottpairwise1980} for multiple comparisons to test the contrasts between first and second petitions written within each cohort. 
We use pairwise contrasts within the Pre/Post group to estimate that users' second petitions are 39\% longer than their first petitions (log scale, ratio = 1.394, z = 20.353, p \(<\) .0001, Cohen's d = 0.399, ratio 95\% CI = [1.340, 1.450]). 
The estimated increase in first to second petition word count was smaller in the Pre/Pre group (log scale, ratio = 1.061, z = 2.370, p = 0.0534, Cohen's d = 0.072, 95\% CI = [1.000, 1.127]) and Post/Post group (log scale, ratio = 1.066, z = 2.140, p = 0.0950, Cohen's d = 0.077, 95\% CI = [0.993, 1.145]).
Overall, within repeat petition writers, the positive relationship between AI access and word count reaffirms the global trends we observed in the platform difference-in-differences analysis.

We also found that across cohorts, returning writers' second petitions had worse outcomes than their first petitions, and that this decline was greater when users had AI access for their second petition.
We use our preregistered primary outcome metric, the presence of at least one comment in 30 days threshold, as an indicator of petition outcomes.
We saw similar results for the outcome of reaching at least 10 signatures, included along with model summaries in SI Appendix Table 20.
In the Pre/Post cohort (middle facet of Figure 3b), 40.50\% (95\% CI = [38.67, 42.38]) of writers' first petitions (first bar, orange) and 35.71\% (95\% CI = [33.87, 37.32]) of writers' second petitions (second bar, blue) received at least one comment in 30 days. 
We estimated the predictive relationship of AI access on comment engagement by fitting a mixed-effects logistic regression model.
Through Bonferroni-adjusted pairwise contrasts, we find that Pre/Post users' second petitions, which were written with access to AI, have 25\% lower odds of a positive outcome than their first petitions, with this difference being significant (OR=0.754, z=-4.236, \(p =  0.0001\), 95\% CI = [0.642, 0.884]).
On the other hand, the first and second petitions written by users in the Pre/Pre cohort, for whom both petitions were written without AI access, have no significant difference in outcomes (OR=0.825, z=-1.911, p=0.168, 95\% CI = [0.648, 1.050]).
In the Post/Post cohort we observe a similar pattern to the Pre/Post cohort, with a significant difference in odds of first and second petition positive outcome (OR=0.662, z=-3.382, \(p=0.0022\), 95\% CI = [0.495, 0.887]).
The decline in petition outcomes for petition writers' second petitions across cohorts reflects a predictable regression to the mean.
However, we also see that among repeat petition writers, AI access is associated with significantly worse petition outcomes.
The negative relationship between AI access and petition outcomes among repeat-petition writers extends our findings from the platform-level difference-in-differences analysis, in which we see no evidence that AI access improved petition outcomes.

Finally, our repeat-writer data allowed us to see whether AI access was associated with improved outcomes for returning users whose first petitions had poor outcomes. 
Previous research found that AI has the greatest benefit for low-performing users~\cite{dellacquaNavigatingJaggedTechnological2023, Noy2023-sf}. 
We thus tested the relationship between cohort (Pre/Pre, Pre/Post, Post/Post) and second-petition outcomes among users who had unsuccessful first petitions. 
We again focus on petition outcomes through an indicator of whether petition writers receive at least one comment in 30 days.
Of users who had unsuccessful first petitions (no comments within 30 days), we found no evidence that cohorts with AI access had better second-petition outcomes than expected compared to the other cohorts (\(\chi^2(2) = 1.179, p = 0.555, N = 2,697, w = 0.021\)).
In summary, AI access was not associated with improved outcomes for low-performing users.

\section*{Discussion}

Our work provides a causal analysis of the impact of AI integration into a real-world online platform. 
Our findings extend prior work on the impact of AI on online platforms at large: while online platforms might be influenced by both off-platform AI and in-platform AI tools, past studies have focused on general-use off-platform AI.
As online platforms increasingly integrate AI, it is important to understand the potential impact on platform dynamics.
In our case, the staggered roll-out of the Change.org in-platform AI tool allowed us to circumvent the challenges of noisy AI detection, and instead directly measure the impact of in-platform AI access through causal methods.
We find that in-platform AI on Change.org shifts petition lexical features, likely improves quality,
and increases site-wide homogeneity. 
At the same time, the in-platform AI tool did not improve platform outcomes, nor did it seem to affect platform participation.

Our results suggest that, on average, with AI, writers produce higher-quality petition text with lower effort than if they had written without AI.
Prior literature, as well as our results on petition writing quality and persuasiveness, suggests that the average quality of petition text and arguments would improve with AI assistance~\cite{HAGEN2016783, chen2019multi, schoenegger2025large, Noy2023-sf}.

So, why doesn't AI access improve platform-level outcomes on Change.org? 
We have three hypotheses for mechanisms that might drive the lack of improvement in outcomes: first, AI might not improve the substance of petitions; second, AI might increase readers' AI suspicion; and third, AI-assisted writing might reduce petition-writer ownership and effort. 

First, even with higher text quality, petition outcomes might not improve if the petitions generated by the AI tool do not have improved substance.
AI tools are not always guaranteed to augment human performance~\cite{vaccaroWhenCombinationsHumans2024}. 
While prior literature has explained human-AI collaboration through a ``jagged technological frontier'' in which AI either enhances human productivity and performance (inside the frontier) or performs worse than humans alone (outside the frontier)~\cite{dellacquaNavigatingJaggedTechnological2023}, it may be the case that the technological capability of AI does not drive petition-writing success.
In particular, the technological capabilities of AI to produce believable, high-quality text seem to be insufficient to explain the outcomes of content produced on this type of advocacy platform, where evaluations are determined by readers' dynamic, subjective, and social perceptions. 
Specifically, it is possible that petitions, with their real-world, context-specific missions, fall outside of AI's capabilities for augmentation because the task requires personal and community-specific knowledge and engagement~\cite{wojtowicz2025undermining}.
For example, we show in Figure 9 of the SI Appendix that petitions written with access to AI tended to use less specific language in petition titles.
In sum, AI on Change.org might contribute to what others have called ``semantic garbage'' or ``AI slop''~\cite{feuerriegel2023research, floridi2020gpt, moller2025impact, tullis2025sifting}: although AI shapes petitions to resemble previous articulate, well-researched petitions, the changes may be superficial.

Our second proposed mechanism is AI suspicion: readers' perceptions of petitions, and the Change.org platform, might deteriorate if they suspect the use of AI.
Prior work has shown that AI suspicion has been attributed to social penalties since the release of ChatGPT in 2022~\cite{jakesch2023human, reif2025socialpenalty}.
While people use biased heuristics to detect AI~\cite{jakesch2023human}, when people suspect AI use, they may perceive the content as less trustworthy and assign lower evaluations to the writers~\cite{kadoma2025generative, reif2025socialpenalty, jakesch2019ai, hohenstein2023artificial}. 
If the AI-written petitions induce higher reader suspicion, outcomes are indeed likely to be affected.  
This hypothesis is difficult to test with our data since human perceptions of AI assistance may have shifted considerably in the year since our study's petitions were posted~\cite{duan2024understanding, matatov2024examining}. 
On the other hand, there are reasons to discount the ``AI suspicion'' hypothesis in this case. 
Petition readers were not primed to suspect AI use on Change.org; during the eleven-week period for which we estimate the causal impact of AI, Change.org provided no information to readers that petitions could have been written with AI. 
Despite the absence of platform-provided signals of AI use, readers may have heightened AI suspicion driven by the lexical style, paragraph structure, and vocabulary of petitions written with AI access.
If AI suspicion is driving petition outcomes, this problem might worsen as Change.org increases their advertising, and consequently readers' suspicion, of the AI tool. 

Our third hypothesis is author commitment: when writing with AI, authors may feel and act differently toward their own petitions.
Petition writers may feel diminished ownership over petitions they write with AI, and therefore be less likely to take actions that would attract signatures and comments, such as sharing the petition on social media.
Previous literature on ownership in AI systems supports this explanation, showing that writers do not consider themselves to be the owners of text that is written with substantial AI support, especially when they use AI tools that require minimal user involvement~\cite{draxler2024ai, kadoma2024inclusion, xu2024what, he2025which}. 
As perceptions of ownership contribute to how people construct their personal and collective identities~\cite{pierce2003state}, and identity is central to sharing behaviors on social media~\cite{zhao2013faces}, we would expect the diminished ownership to result in writers who are less likely to promote their petition in online communities. 
As off-platform promotion is critical to petition success~\cite{prosser2011DoSocialNetworking,proskurnia2017PredictingtheSuccess, dumas2015ExaminingPoliticalMobilization, Carlson2019PleaseSignHere, Harrison2022ExploringEpetitioning}, with less promotion petitions would likely get fewer comments and signatures.
One way to address this concern could be through adding friction to the AI-assisted petition writing process, such as by requiring a lengthy user input or prompting the user to edit the draft produced by the AI tool.
Research in platform design has demonstrated that friction can improve decision-making in collaborative human-AI systems by forcing users to have cognitive engagement~\cite{buccinca2021trust}. 
Such mechanisms could potentially improve authors' perceptions of ownership and their likelihood of sharing petitions.

While petition outcomes remain the same, or potentially worsen, our data also indicates that Change.org participation is unchanged by AI access.
Previous work has shown that general-purpose AI can cause participation to significantly grow on content sharing platforms~\cite{tullis2025sifting}.
In particular, researchers, journalists, and community moderators have warned that the lowered cost of content production will enable the mass production of AI-generated misinformation and AI slop~\cite{feuerriegel2023research, floridi2020gpt, moller2025impact, tullis2025sifting, lloyd2023there}.
However, if users only enter the petition-writing flow with strong intent to write a petition, the marginal increase in production could be limited, like we have seen in this case. 
Furthermore, since Change.org, unlike other platforms, does not offer financial or social incentives for mass production of content, users may have little incentive to create superfluous petitions. 
Overall, the participation trends offer a mixed bag for Change.org: on one hand, participation did not grow, despite prior research suggesting that AI assistance would
broaden access to the platform. 
On the other hand, Change.org so far has likely avoided the fate of other platforms which are subjected to excessive AI spam and slop~\cite{tullis2025sifting}. 

It is especially important to consider the impact of AI integration on Change.org, where AI integration can influence which causes succeed and shift how users connect to their own communities.
We found that AI has already caused a dramatic shift to petition lexical features on Change.org and driven platform-level homogenization.
If AI continues to homogenize the language of protest, it has the capacity to undermine signals of petition authenticity and legitimacy~\cite{wojtowicz2025undermining}.
As we show in the SI Appendix Figures 5, 6 and 32, the predictive strength of features of petition text (i.e., lexical features, writing quality, persuasiveness) on petition outcomes has weakened, or reversed, with platform-wide shifts that correspond with in-platform AI access. 
Regardless of lexical style, the introduction of in-platform AI can impact readers' evaluations and writers' engagement with their communities. 
Furthermore, with AI minimizing the cognitive load of petition-writing, petition writers might not only feel diminished ownership but also less engagement with the petition's cause and their community at large. 
In sum, AI integration in petition-writing might significantly change the dynamics of civic engagement in ways that are difficult to predict and measure. 
Despite potential harms, Change.org has deepened the integration of the AI tool into the petition writing flow. 
As of early 2026, some users visiting the Change.org home page were no longer offered the choice to start a petition without AI support; instead, they were met with a text-box that automatically launches the AI tool.
Overall, our study serves as a cautionary tale for platforms that seek to integrate AI into their platforms: while user-generated content may change, AI tools may not yield dividends in participation or effectiveness, while potentially introducing additional harms. 

Our research has several limitations. Using only publicly accessible data, we could not observe or control for various ways in which petitions might be amplified by in-platform features (e.g., being highlighted on the Change.org home page) which may impact outcomes.  
Similarly, we could not account for off-platform amplification (e.g., how petitions are ranked by a Facebook newsfeed) that may be impacted by AI use in writing petitions.
However, some platform amplifications, like the Change.org home page, only feature petitions that are already victorious or ``trending'' with hundreds to thousands of signatures.
As such, we would not expect the home page to influence whether petitions are able to reach minimal thresholds for petition outcomes, which are the focus of our study.

Furthermore, this work does not directly address the impact of AI access on individual users who take advantage of it, but instead shows the impact of AI access on the dynamics of the platform at large. 
As online platforms involve deeply interconnected networks of users, causal estimates of interventions on individual-level outcomes may suffer from bias and require internal platform access~\cite{bak2025moving}. 
Instead, we estimate here the causal impact of the intervention (in-platform AI access) to the system (platform-level trends in petition style and outcomes).
While our analysis does not identify the causal mechanisms that drive the results we observe, future work could use field or lab experiments to test different mechanisms, such as AI suspicion, or the effect of AI writing support on writers' downstream investment in their petitions' success. 

\section*{Materials and Methods}
Our research complies with the ethical regulations established by the Cornell Institutional Review Board (FWA00004513). 
We consulted with the Cornell IRB office, which deemed our main analysis --- including scraping petitions and using causal inference methods for analysis --- ``not human subjects research.''
Our labeling task was deemed exempt from formal ethics review by the Cornell IRB (protocol \#IRB0150383).

\subsection*{Data}
Our data consisted of 1,517,433 petitions posted to Change.org between January 1, 2022 and December 31, 2024.
Our difference-in-differences analyses focus on petitions posted prior to December 15, 2023, to capture the period of differential access. 
However, we also conduct descriptive analyses that include petitions posted through 2024, as in the analyses of repeat-petition writers (Figure 3), sentiment (SI Appendix Figure 8), title concreteness (SI Appendix Figure 9), vocabulary shifts (SI Appendix Figures 21-22), longitudinal trends of lexical style and outcome metrics (SI Appendix Figure 23), English language variety (SI Appendix Figure 24).
Change.org is the largest online petition service, providing a platform for individuals to create their own petitions or support others' petitions. 
When users visit a Change.org petition, the platform encourages them to sign the petition by sharing their name and zipcode. 
Their signature then counts toward the total count of signatures displayed on the petition's page. 
The platform then encourages signees to support the petition by posting a comment. 
Comments are therefore only posted by petition supporters who already signed a petition, and can be considered an exceptionally strong vote of support for the petition. 
Petition writers can declare their petitions to be a victory at any point --- however, victory status is a scarce and ambiguous signal.
Some petitions are declared victories without ever accruing any signatures, and of the petitions we collected, only 1.38\% were ever declared victories.

Prior research into online petitions has demonstrated that e-petitions are dependent on off-platform promotion through social networks, news outlets, or direct shares to gain signatures~\cite{prosser2011DoSocialNetworking,proskurnia2017PredictingtheSuccess, dumas2015ExaminingPoliticalMobilization, Carlson2019PleaseSignHere, Harrison2022ExploringEpetitioning}.
Prior work has found crowdfunding platforms, which operate similarly to Change.org, to display ``friendfunding'', by which campaigns tend to be funded by friends and family, unless shared broadly on external social media platforms~\cite{lee2022NewDigitalSafetyNet, borst2018FromFriendfunding}. 
Based on this extensive prior work, we hypothesize that few readers browse Change.org in pursuit of petitions to support, but instead find petitions off-platform through social networks, news coverage, or direct shares from friends. 

To collect the petitions posted to Change.org we used a web-scraping library (\url{https://docs.scrapy.org/en/latest/}) to extract the text, comments, and metadata of all petitions posted during the time period. 
This dataset contains all petitions that had not been deleted or taken down as of the time of collection (January 9, 2025).
Our data includes full petition text, title, and metadata about the author and the petition. 
Petition metadata includes user ID, engagement (signatures, comments), and labels such as language tag.
Since each petition is connected to the author's user ID, we could track all petitions written by the same user during the period for which we collected data.
The main causal analysis of our study uses a subset (N = 361,635) of petitions that are English language, posted by users in the United States, Canada, Great Britain, and Australia between January 3, 2022 and December 15, 2023, and met minimum length requirements (at least 2 sentences and 5 words). 
We include only English-language petitions in all analyses (main and supplementary) to allow for consistent interpretation of our lexical metrics.
We consider petitions posted prior to December 15, 2023 for the period of differential access, as Australia received in-platform AI access in late December. 
For robustness we repeat our main analysis with full month of December 2023, in SI Appendix Figure 14.
Since we were interested in the lexical attributes of the text, this subset only included petitions with text that included at least two sentences and five or more words.
We conducted a robustness check to estimate difference-in-differences results with the full text, including petitions with fewer than five words or two sentences.
The SI Appendix contains details on summary statistics (Tables 1-3, Figure 1) as well as additional robustness checks (Figures 11-20, 30, 31).

\subsection*{Petition Lexical Features}
Our lexical features characterize petition text through length, lexical diversity, and reading grade level. 
We calculate these metrics on the combined text of the petition title and full text of the petition, with the title treated as a first sentence. Word count is measured as the number of tokens (separated by a space) in the title and full body of the petition. 

Lexical diversity is calculated through the moving average type-token ratio (MATTR), which is resistant to varied text length~\cite{ZENKER2021100505}. 
The MATTR is calculated (\url{https://github.com/kristopherkyle/lexical_diversity})
by first taking the ratio of unique token types to total tokens for every window of a fixed size over the full text and title. 
In our analysis, each word (separated by a space) is a token, and unique token types are unique words.
We use a window size of 50, as is typical in prior studies~\cite{ZENKER2021100505}.
The MATTR of each petition is then the average of these ratios over all windows.
Formally, the petition \(p\) MATTR score is as follows, with \(W_p\) as the set of the petition's full text words, \(N_{win}\) as the window size, and \(U_{p,i}\) as the set of the petition's unique words within each window of \(N_{win}\) words starting with word \(i\).

\begin{align}
\text{MATTR}(p) = \frac{1}{|W_p| - N_{win} + 1} \sum_{i=0}^{|W_p| - N_{win}} \frac{|U_{p,i}|}{N_{win}} 
\end{align}

Readability is calculated through the Flesch-Kincaid grade level, for which higher grade levels correspond to more complicated text~\cite{kincaid1975derivation}.
The Flesch-Kincaid grade level is a common metric to evaluate text complexity in a variety of high-impact contexts, such as education and public health~\cite{tanprasert2021flesch, jindal2017assessing}.
The Flesch-Kincaid grade level metric seeks to capture complexity at both the sentence level, by considering average number of words per sentence, and word level, by considering average number of syllables per word.
With the petition's \(p\) sets of words \(W_p\), sentences \(Se_p\), and syllables \(Sy_p\), The Flesch-Kincaid grade level (FKGL) is calculated on the full text and title as follows (via \url{https://github.com/cdimascio/py-readability-metrics}):
\begin{align}
\text{FKGL}(p) = 0.39 \times \frac{|W_p|}{|Se_p|} + 11.8 \times \frac{|Sy_p|}{|W_p|}
- 15.59
\end{align}

\subsection*{Petition Outcomes}
On Change.org, signatures and comments are the core outcomes that drive a petition's success. 
It is unlikely for a petition to have a positive outcome without broad support, as is demonstrated by signatures and comments.
These indicators are the primary way by which petitions gain legitimacy, media attention, and ultimately convince decision-makers to enact change. 
We consider minimal thresholds of petition outcomes: that a petition must receive 10 signatures and have 1 comment in 30 days. 
It would be unlikely that any petition, regardless of scale or motivation, is able to reach any strong outcomes without achieving these minimum thresholds. 

Comments are a signal of especially strong support from a signee, and demonstrate a higher level of commitment for the petition. 
We report the number of comments written within the first 30 days after a petition is posted. 
We exclude comments written by the user who posted the petition.
Even though most engagement occurs soon after a petition was posted, by limiting the time window to 30 days, we ensure that all petitions written during our period of analysis have the same opportunity for receiving a comment, and avoid biasing outcomes in favor of petitions that have been posted longer.
While this was the primary outcome metric that we preregistered, our analysis shows that the time-boxing likely made no difference compared to using the full window.
Of the 416,914 petitions that ever gain a comment and were posted at least 30 days before data collection, only 0.26\% of petitions received their first comment outside of the 30-day window.

Signatures are the primary capital by which petitions gain legitimacy and are able to persuade decision-makers.
The Change.org platform even states, ``Petitions with 1,000+ supporters are 5x more likely to win!''
We report petition signatures at the threshold of 10 total signatures. 
We do not have data on when petitions gain signatures.
However, petitions included in our difference-in-differences analysis had between 1 and 3 years to gain signatures. 
Provided that less than 1\% of petitions gained their first comment after the first 30 days, we expect that most petitions also reach 10 signatures within the first 30 days, if at all.

\subsection*{Difference-in-Differences Design}
Change.org conducted a staggered roll-out of its AI writing tool by geographic region. 
The United States, Great Britain, and Canada were among the countries that participated in A/B testing during April-September 2023 and then received a full launch of the tool in October, 2023. 
However, there were several countries that had delayed access to the AI tool: Australia, New Zealand, and India. 
We used Australia as the control for our difference-in-differences analysis, as it showed adequately similar pre-AI trends to trends in lexical features and outcomes in the United States, Great Britain, and Canada.
While New Zealand and India failed to demonstrate adequately similar pre-AI trends, we include petitions written in these countries in supplementary synthetic control analyses, included in the SI Appendix Figures 34-36.

We used static and dynamic difference-in-differences to measure the causal impact of access to the in-platform AI tool on petition lexical features and outcomes. 
For the static difference-in-differences approach, we use ordinary least squares (OLS) regression with country and week period fixed-effects, to control for differences in groups and over time.
We measure time period fixed effects at the week level, following previous work, and use countries as the groups.
We follow the difference-in-differences model specified by del Rio-Chanona et al. when studying the impact of AI on Stack Overflow participation~\cite{rio-chanonaAreLargeLanguage2024} where researchers measured production volume changes based on off-platform access to external LLMs such as ChatGPT. 
\begin{align}
Y_{c,t} = 
\alpha_c+ 
\lambda_t + 
\beta \cdot Treated_{c,t} + 
\boldsymbol{\eta}\cdot\mathbf{X_{c,t}} + 
\epsilon_{c,t}
\label{eq:did}
\end{align}
Here, the dependent variable \(Y_{c,t}\) reflects the metric for which we are estimating change.
When studying lexical features we use the median lexical feature (e.g. median word count) as the dependent variable. 
When studying outcomes, we use the share of petitions that achieved the outcome threshold (e.g. share of petitions reaching 10 signatures). 
\( \alpha_c\) are country fixed effects, \(\lambda_{t}\) are time-period (i.e. week) fixed effects,  and \(\epsilon_{c,t} \) is the error term. 
Pre-treatment covariates can be included through \(\mathbf{X_{c,t}}\) and coefficients of these covariates through \(\boldsymbol{\eta}\).
We do not include any covariates in our main analysis; the SI Appendix provides robustness checks with additional covariates, such as day of week (Figure 16). 
The $Treated_{c,t}$ term reflects the interaction between time period and country, thereby taking the value of 1 if a country had AI access during a given week, and 0 otherwise. 
The value \(\beta\) thus captures the average estimated treatment effect on treated units (ATT) of access to the in-platform AI tool in countries that received access. 
We cluster standard errors at the month level, again following previous difference-in-differences studies of online platform activity~\cite{rio-chanonaAreLargeLanguage2024}.

In total, 279,131 petitions are considered for the static difference-in-differences analysis, which includes 65 weeks pre-AI and 11 weeks post-AI, and excludes the AI tool's experimental A/B test period as noted above. 
We create our static difference-in-differences estimation using the \texttt{did} package in R to estimate average treatment effect on treated groups (ATT), weighted by group size.
To account for the small number of groups included in our month-clustered standard errors, rather than reporting the Wald test under an asymptotic normal distribution, we use a t-distribution with 17 degrees of freedom (one fewer than the 18 monthly clusters, which corresponds to the 65 pre-AI weeks and 11 post-AI weeks)~\cite{rio-chanonaAreLargeLanguage2024}.
We conducted additional robustness checks such as shortening the pre-AI period (SI Appendix Figure 13), using a longer post-AI period (SI Appendix Figure 14), including short-text petitions (SI Appendix Figure 15), clustering standard errors by country, rather than month (SI Appendix Figure 17), using a subset of petitions about civic engagement topics (SI Appendix Figure 30), and using a subset of petitions about non-civic engagement topics (SI Appendix Figure 31). 
In Figure 16 of the SI Appendix, we leverage a small set of covariates (for each country, the weekly share of petition-writers that have commented before; share of petition writers that are new to the platform; share of petitions with decision makers specified; and share of petitions posted on a weekend) to calculate doubly robust estimates of the ATT, as proposed by Sant'Anna and Zhao~\cite{SANTANNA2020101}.
With the doubly robust approach, we observe similar results.
With AI access, lexical features change significantly while outcomes do not improve.

In addition to calculating aggregate estimates for the effect of AI in treated countries, we conducted a dynamic difference-in-differences analysis to measure how lexical features and outcomes changed dynamically over time periods. 
In this setting we create difference-in-differences estimates that are grouped by time period (week) relative to a universal baseline (the week prior to A/B testing). 
We used the week prior to A/B testing (Week -1) as a reference point to estimate the effect of varying lengths of exposure to the treatment (access to in-platform AI).
To calculate the dynamic difference-in-differences estimates shown in Figures 1a and 1c, we used the method and software developed by Callaway and Sant'Anna~\cite{callaway_santanna_2025_did} 
to measure group-time average treatment effects that account for heterogeneity~\cite{callaway2021difference}. 
In particular, instead of calculating the overall average treatment effect on treated users (ATT) that is captured in equation (\ref{eq:did}), Callaway and Sant'Anna calculate ATT(g,t) values, which are specific to units in group $g$ that are treated at time $t$, as shown in equation (\ref{eq:event_study}). 
Here, $Y_t(g)$ reflects the outcome of group $g$ at time t, and $Y_t(0)$ reflects the potential outcome of group $g$ at time $t$ had it never been treated.
Since our causal analysis terminates before any of the delayed-release countries receive AI access, countries are all either untreated or treated across the full time period, and as such our ATT(g,t) estimates simplify to ATT(t) estimates for each time period.

\begin{align}
ATT(g,t) \;=\;
\mathbb{E}\!\left[\, Y_t(g) \;-\; Y_t(0) \;\middle|\; G = g \,\right]
\label{eq:event_study}
\end{align}

\subsection*{Repeat Petition Writers}
For our study of repeat petition writers, we sought to understand how petition lexical features and outcomes differed between users' first and second petitions, and how this difference varied by AI access. 
To prevent bias against petition-writers who posted their petitions during the last month of 2024 (less than a full month prior to our data collection), we subset to petitions posted between January 2022 and November 30, 2024.

We first fit a mixed-effects linear regression model to predict word count (natural log transformed due to the right skew) among repeat petition users. 
To fit this model we used the R \texttt{lme4} package, version 1.1.36.
In this model we have fixed effects for user cohort (Pre/Pre, Pre/Post, Post/Post), petition order (First, Second), and the interaction between cohort and petition order. 
We also include random effects for user ID, as we have repeated measures for each user.
The word count model is captured in equation (\ref{eq:mixed_effects_wc}). 
\begin{align}
\ln(\text{Word Count}_{ij}) = 
\beta_{c}\cdot\mathbf{1}\{\text{Cohort}_{ij} = c\} + 
\beta_{p}\cdot\mathbf{1}\{\text{Petition Order}_{ij} = p\} + \notag  \\ 
\beta_{c,p}\cdot(\mathbf{1}\{\text{Cohort}_{ij} = c\}\times\mathbf{1}\{\text{Petition Order}_{ij} = p\})  + 
\gamma_{i} + 
\epsilon_{ij} 
\label{eq:mixed_effects_wc}
\end{align}
In this model, $j$ reflects the observations for each user ID, $i$. 
\(\gamma_i\) reflects the random effects attributed to each user ID. 
We are primarily focused on the coefficient of the interaction between cohort and petition order, which is captured by \(\beta_{c,p}\).
We compare the levels of the interaction term \(\beta_{c,p}\) through pairwise contrasts, in which each contrast corresponds to a difference in expected log-word count between the first and second petitions for a given cohort. 
We then adjust the pairwise contrast estimates with a Bonferroni correction, to account for multiple hypothesis testing~\cite{scottpairwise1980}. 
Due to our sample size (9,222 petitions from repeat-writers), we use asymptotic estimation for the test statistics, as is default under the \texttt{lme4} package.
In the SI Appendix we use the same model specification to understand the predictive relationship between repeat-users cohort and our other lexical measures, MATTR and log-transformed Flesch-Kincaid grade level (SI Appendix Figure 33, Table 21). 

We fit a similar logistic mixed-effects model to predict differences in petition outcomes among repeat petition users. 
We use the same specification of random and fixed effects as in equation (\ref{eq:mixed_effects_wc}), but fit a logistic rather than linear regression model as our dependent variable, whether a petition received a comment in 30 days, is a binary value.
To fit our models we use the R \texttt{lme4} package (version 1.1.36) with the bound optimization by quadratic approximation (BOBYQA) optimizer with 200,000 maximum iterations for outcomes.
Again, we use asymptotic estimation for the test statistics.
The model specification is shown in equation (\ref{eq:mixed_effects_comments}). 
\begin{align}
\ln\left(\frac{P(\text{Outcome}_{ij} = 1)}{1 - P(\text{Outcome}_{ij} = 1)}\right) = 
\beta_{c}\cdot\mathbf{1}\{\text{Cohort}_{ij} = c\} + 
\beta_{p}\cdot\mathbf{1}\{\text{Petition Order}_{ij} = p\} + \notag  \\ 
\beta_{c,p}\cdot(\mathbf{1}\{\text{Cohort}_{ij} = c\}\times\mathbf{1}\{\text{Petition Order}_{ij} = p\}) + \gamma_{i}  + \epsilon_{ij}
\label{eq:mixed_effects_comments}
\end{align}
We again compare pairwise contrasts, adjusted for multiple hypothesis testing, between levels of the interaction term. 
In this model, we interpret contrasts as explaining the difference in log-odds of a positive outcome between first and second petition order, depending on cohort.
Through these pairwise contrasts, we can compare the association between petition order and outcome variable across user cohorts. We show the results for our other outcome metric, the share of petitions that reach a 10+ signature threshold, in SI Appendix Figure 33 and  Table 20.
\subsection*{Data Availability}
We make petition metadata available at the following OSF project link, included  here: \url{https://osf.io/q3hj5/?view_only=3fc5b89d05eb4741957af431b107cf2a}.
All analyses included in the main text and figures, as well as in the supplementary material analysis and figures, can be reproduced with the data shared here. Note, we do not make full petition text data or user names available, even though they are publicly available, for user agency, privacy, and copyright concerns. The full petition text of all non-deleted petitions can be retrieved from the Web using the petition IDs provided in the data. Published academic researchers will be able to request the full text dataset from the authors.

\subsection*{Code Availability}
We make code available at the same OSF project link, included here: \url{https://osf.io/q3hj5/?view_only=3fc5b89d05eb4741957af431b107cf2a}.
All analyses included in the main text and figures, as well as in the supplementary material analysis and figures, can be reproduced with the code shared here.

\subsection*{Acknowledgments}
We thank Maria DeCaro for her contributions to labeling petition topics described in the SI Appendix and Travis Lloyd for advising our data collection process. 
We thank Jessica Klein for pointing us to the Change.org implementation and for helpful conversations.
The authors received no specific funding for this work; no funder had a role in study design, data collection and analysis, decision to publish or preparation of the manuscript.
We thank the Cornell Bowers College of Computing and Information Science (CIS) for their support of the first author through a Deans’ Excellence and Hopper-Dean Fellowship.

\subsection*{Author Contributions}
IC: conceptualization, data curation, formal analysis, investigation, methodology, project administration, resources, software, supervision, validation, visualization, writing (original draft, review, editing). 
EG: conceptualization, methodology,
writing (review, editing). 
AK: conceptualization, methodology, project administration, supervision, writing (review, editing). 
MN: conceptualization, methodology, project administration, supervision, resources, writing (original draft, review, editing).

\subsection*{Competing Interests}
The authors declare no competing interests.
The authors do not have any funding, employment (recent, present, or anticipated), personal financial interests or non-financial interests from organizations that may gain or lose from this publication.

\putbib[references]
\end{bibunit}

\newpage
\section*{SI Appendix}
\setcounter{figure}{0} 
\begin{bibunit}
\thispagestyle{empty}
\listoffigures
\listoftables
\tableofcontents

\newpage

\begin{figure}[H]
  \centering
  \begin{subfigure}[b]{0.8\textwidth}
    \centering
    \includegraphics[width=\textwidth]{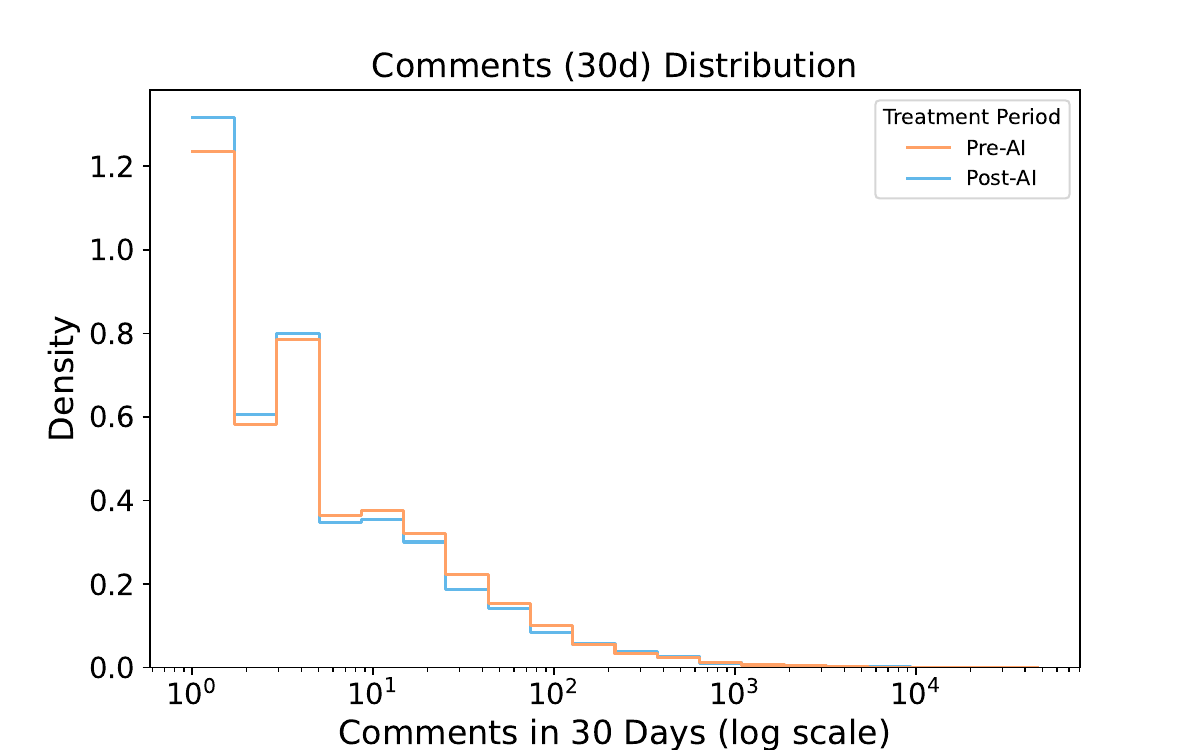}
    \label{fig:hist_comments}
    \caption{}
  \end{subfigure}
  \vfill
  \begin{subfigure}[b]{0.8\textwidth}
    \centering
    \includegraphics[width=\textwidth]{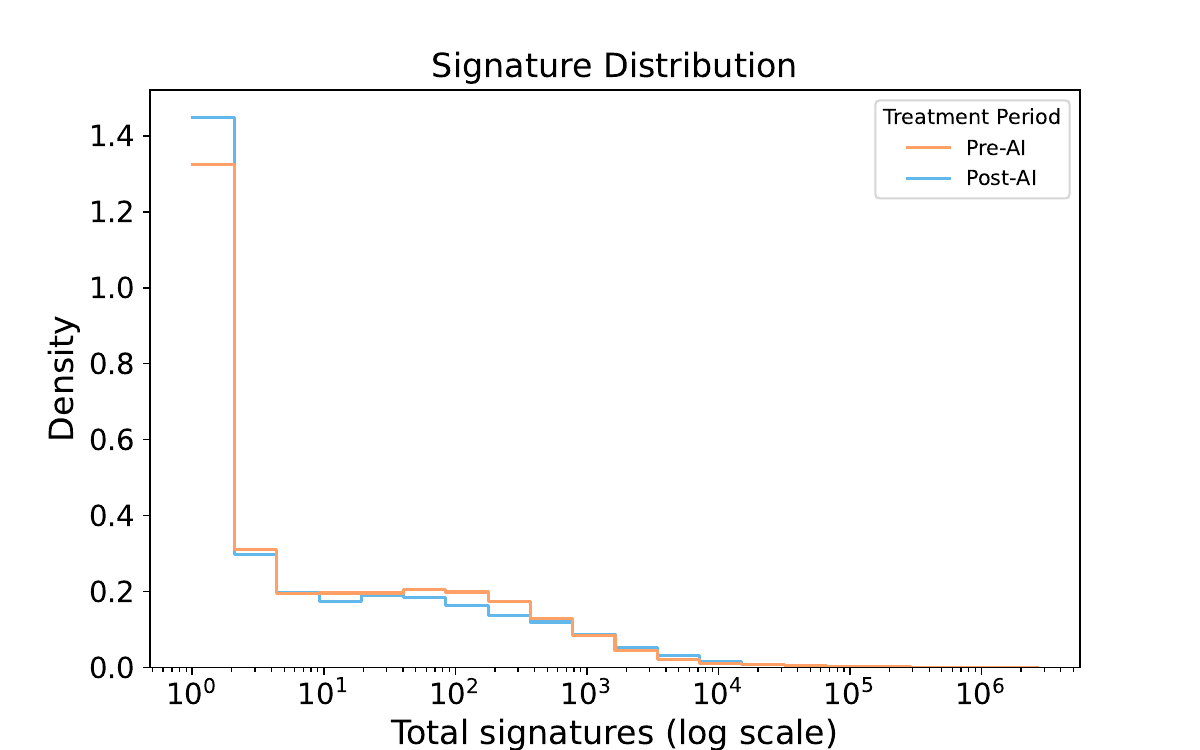}
    \label{fig:hist_signatures}
    \caption{}
  \end{subfigure}

  \caption[Empirical CDF of Outcomes]{Distribution of outcome measures (a) total comments in 30 days and (b) total signatures in pre- and post-AI periods (excluding A/B test period) during the period of study for the main analysis, January 2022 - December 2023, using 237,020 petitions written during the pre-AI period and 42,111 petitions written during the post-AI period. Post AI-the density of low-outcome petitions increases.}
  \label{fig:outcome_hists}
\end{figure}

\begin{figure}[H]
  \centering
  \includegraphics[width=\textwidth]{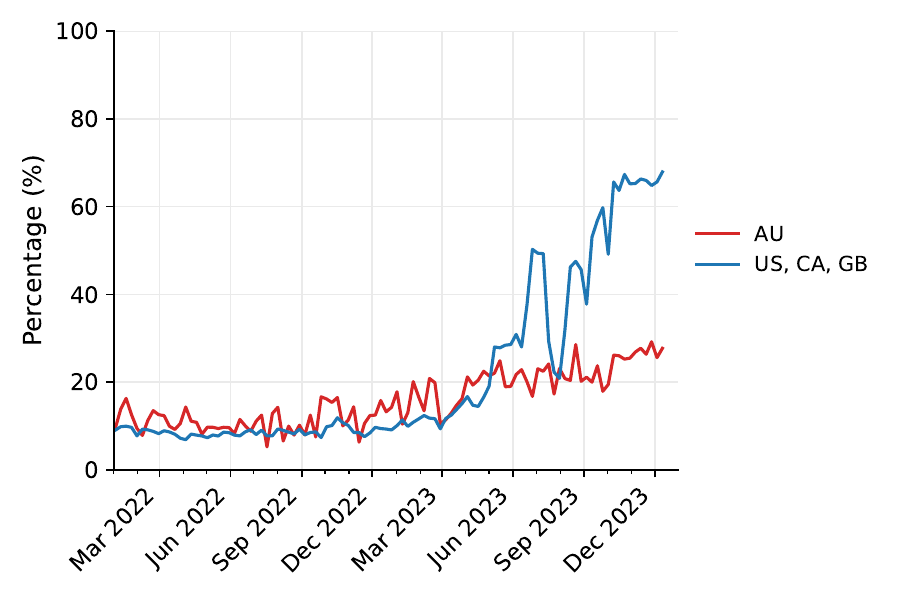}
  \caption[AI Detection Time Series]{Time series showing the share of petitions, by week, that our ensemble model classifies as AI-generated in English petitions from treated and control countries (N = 361,635 petitions). 
  The share of AI-generated petitions increases sharply after the introduction of the AI-generated draft feature in countries with access to the tool (US, GB, CA). As of the week of December 11, 2023 (the last week prior to AU gaining access to the in-platform AI tool), most (67.9\%) of petitions written with access to the AI-generated draft feature on Change.org were generated with AI, while only 27.7\% of petitions written in AU were labeled as AI-generated. 
  Positive AI detection prior to April, 2023 reflects false positives or off-platform AI usage.
  Our ensemble classifier achieved 99\% accuracy and 99\% precision on a test set of 674 annotated petitions.}
  \label{fig:ai_detection}
\end{figure}

\begin{figure}[H]
    \centering
    \includegraphics[width=0.8\linewidth]{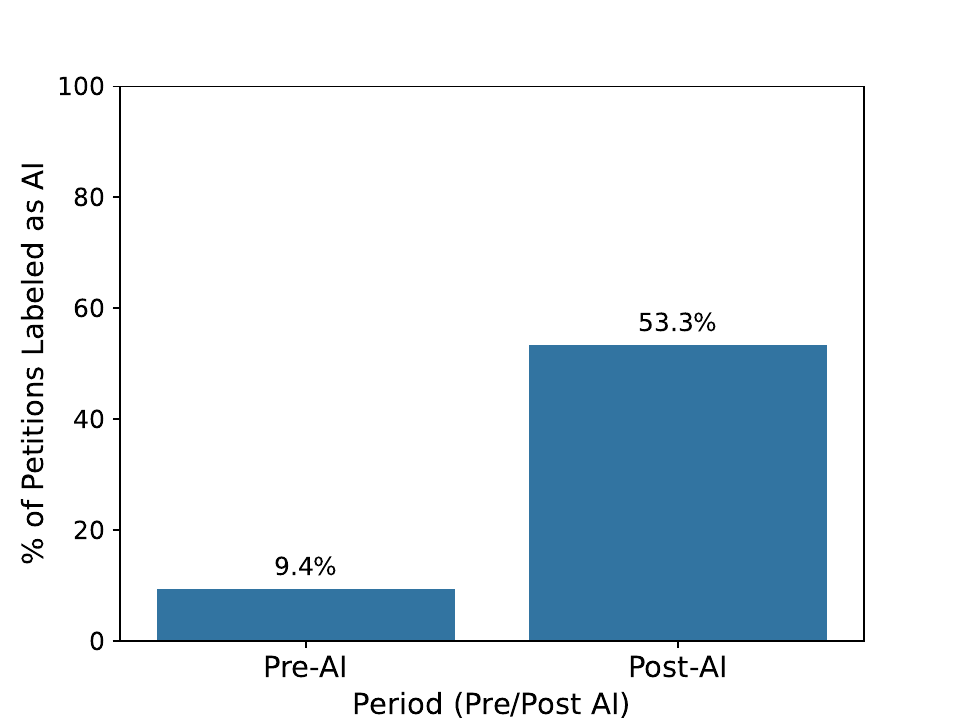}
    \caption[AI Detection Labeling Task]{We use our ensemble method for AI classification to predict if petitions in the sample used for labeling writing quality (N = 417 petitions) are written by the AI tool. 
    9.4\% of petitions in the pre-AI sample were flagged as AI-written, while 53.3\% of the post-AI petitions were flagged as AI-written.}
    \label{fig:survey ai detection}
\end{figure}

\begin{figure}[H]
  \centering
  \begin{subfigure}[b]{.8\textwidth}
    \centering
    \includegraphics[width=\textwidth]{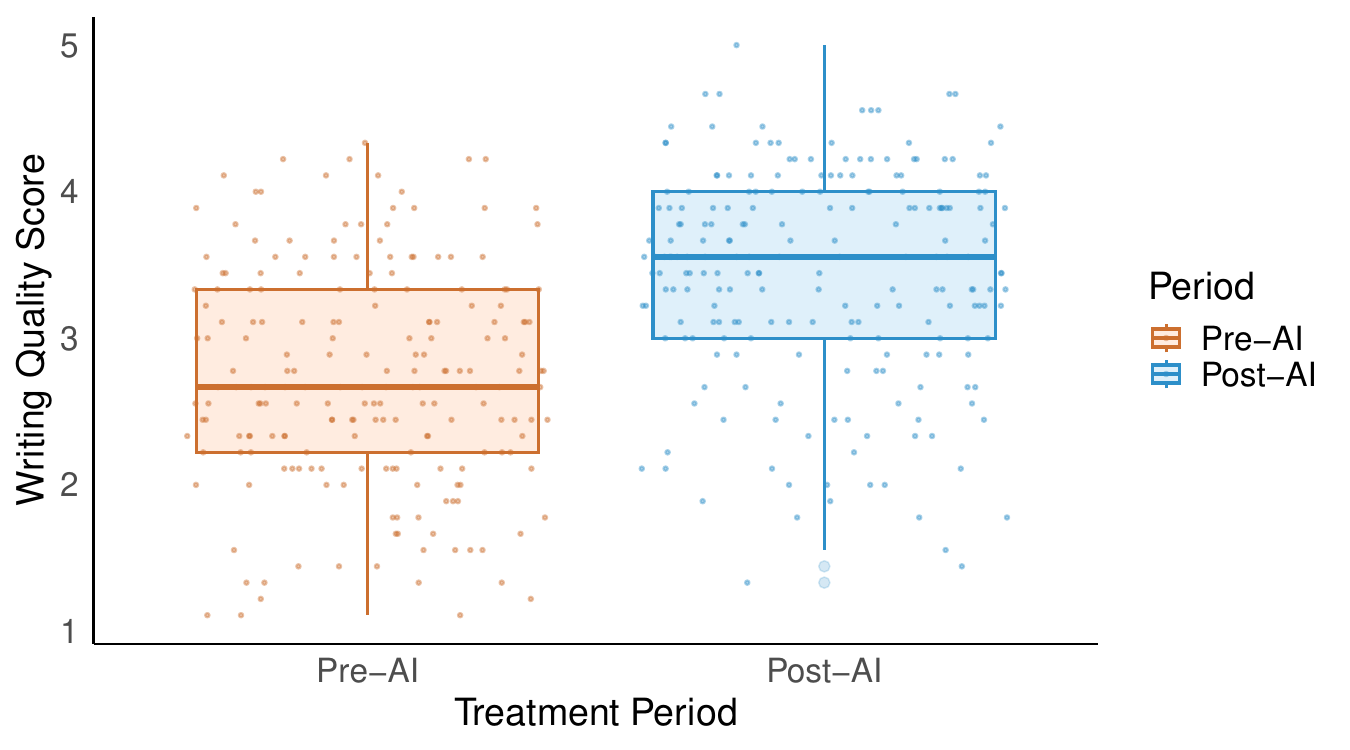}
    \caption{}
    \label{fig:writing_quality_boxplot}
  \end{subfigure}
  \vfill
  \begin{subfigure}[b]{.8\textwidth}
    \centering
    \includegraphics[width=\textwidth]{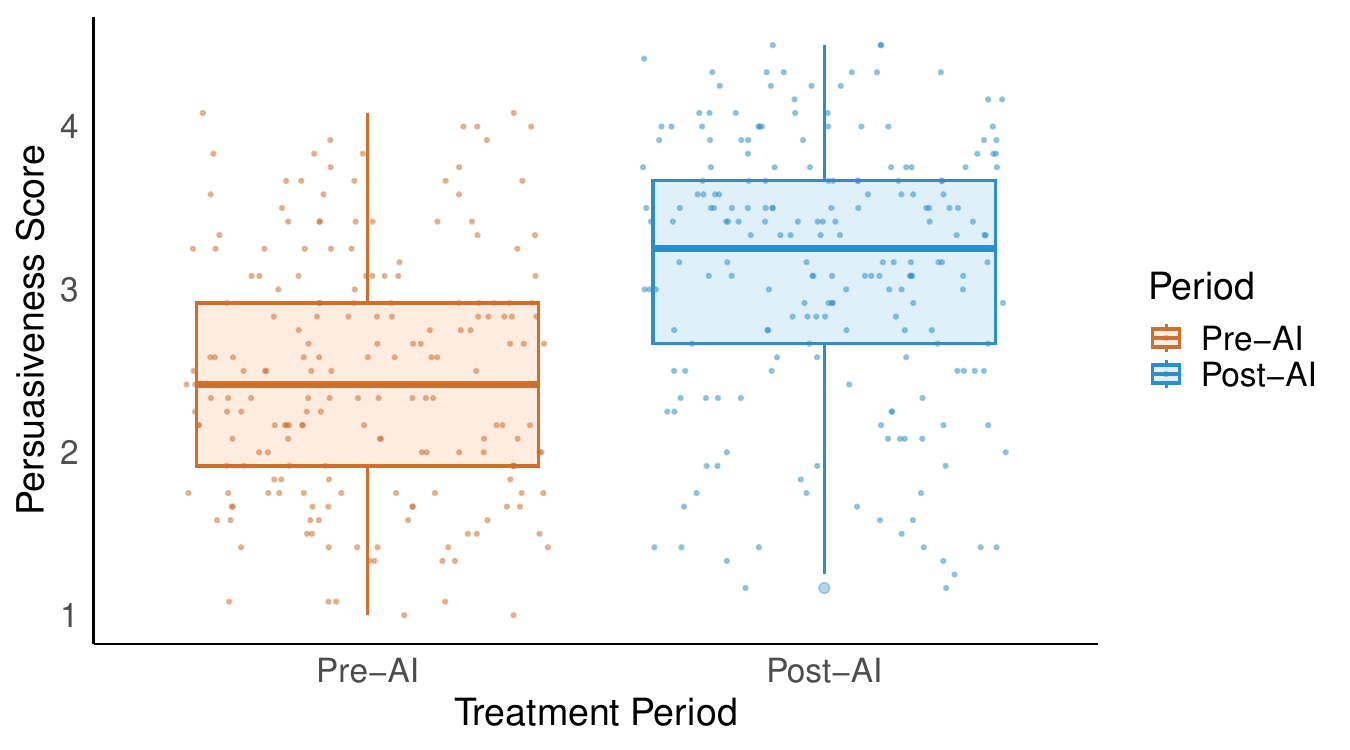}
    \caption{}
    \label{fig:persuasive_boxplot}
  \end{subfigure}
  \caption[Writing Quality and Persuasiveness Boxplots]{Boxplots showing the distribution of (a) writing quality score and (b) persuasiveness scores of 203 petitions written Pre-AI (before in-platform AI access) and 214 petitions written Post-AI (after in-platform AI access).
  The distribution of writing quality and persuasiveness scores shifts to be higher among petitions written with AI access.
  }
  \label{fig:survey_boxplots}
\end{figure}

\begin{figure}[H]
  \centering
  \begin{subfigure}[b]{.8\textwidth}
    \centering
    \includegraphics[width=\textwidth]{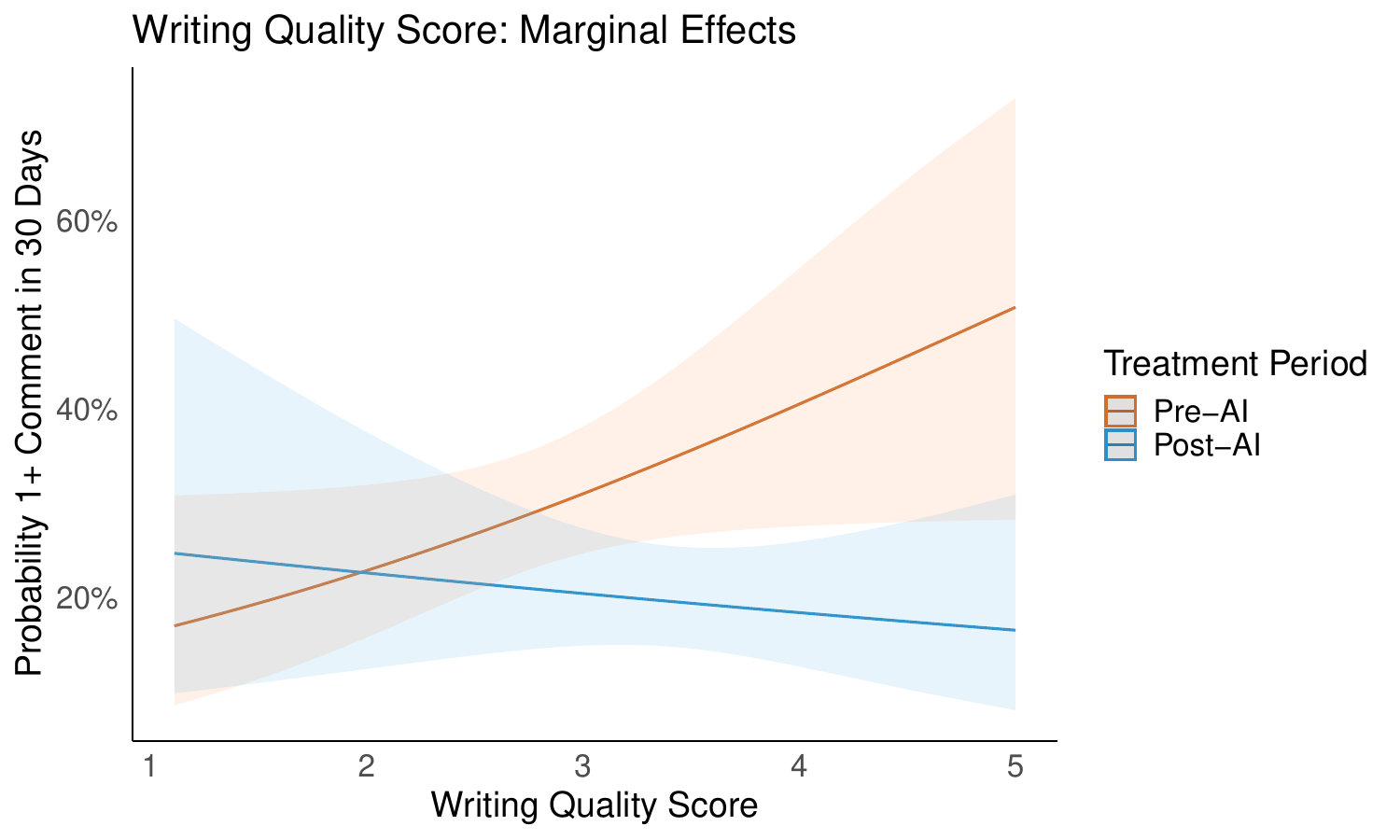}
    \caption{}
    \label{fig:writing_quality_marginal_effects}
  \end{subfigure}
  \vfill
  \begin{subfigure}[b]{.8\textwidth}
    \centering
    \includegraphics[width=\textwidth]{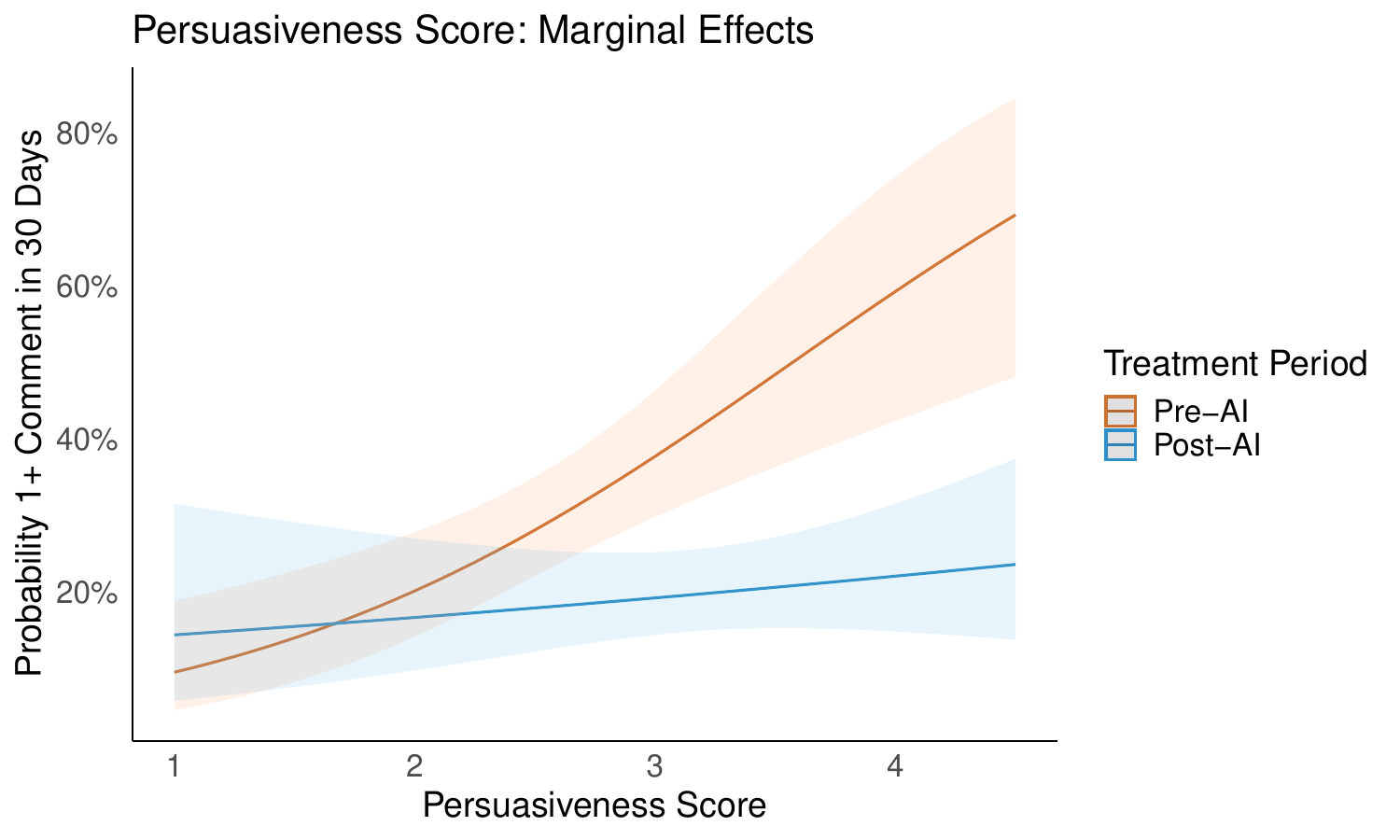}
    \caption{}
    \label{fig:persuasive_marginal_effects}
  \end{subfigure}
  \caption[Writing Quality and Persuasiveness Marginal Effect on 1+ Comment Outcome]{Marginal effect of (a) writing quality score and (b) persuasiveness score regressed on outcome metric (1+ comment in 30 days), by treatment period (Pre-AI, Post-AI).
  Model uses a sample of 417 petitions labeled for writing quality and persuasiveness.
  Pre-AI, increasing writing quality and persuasiveness scores were predictive of an increased probability of receiving a successful outcome (i.e., reaching 1+ comment in 30 days). 
  Post-AI, increasing persuasiveness and writing quality scores were not predictive of improved probability of receiving a successful outcome. 
  In the case of writing quality, higher writing quality scores were weakly negatively predictive of the outcome metric.
  Figure shows 95\% CI.
  }
  \label{fig:survey_marginal_effects}
\end{figure}

\begin{figure}[H]
  \centering
  \begin{subfigure}[b]{.8\textwidth}
    \centering
    \includegraphics[width=\textwidth]{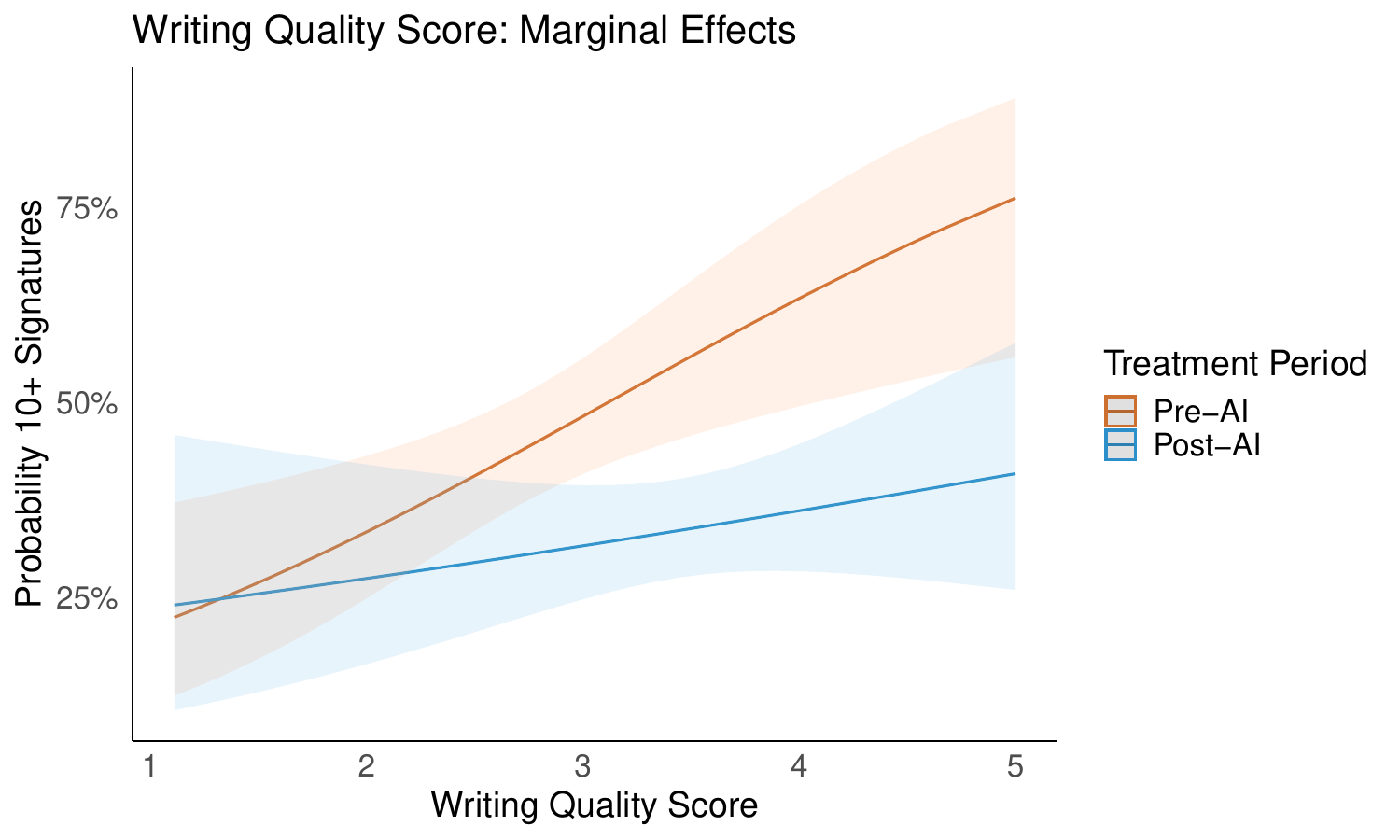}
    \caption{}
    \label{fig:writing_quality_marginal_effects_signatures}
  \end{subfigure}
  \vfill
  \begin{subfigure}[b]{.8\textwidth}
    \centering
    \includegraphics[width=\textwidth]{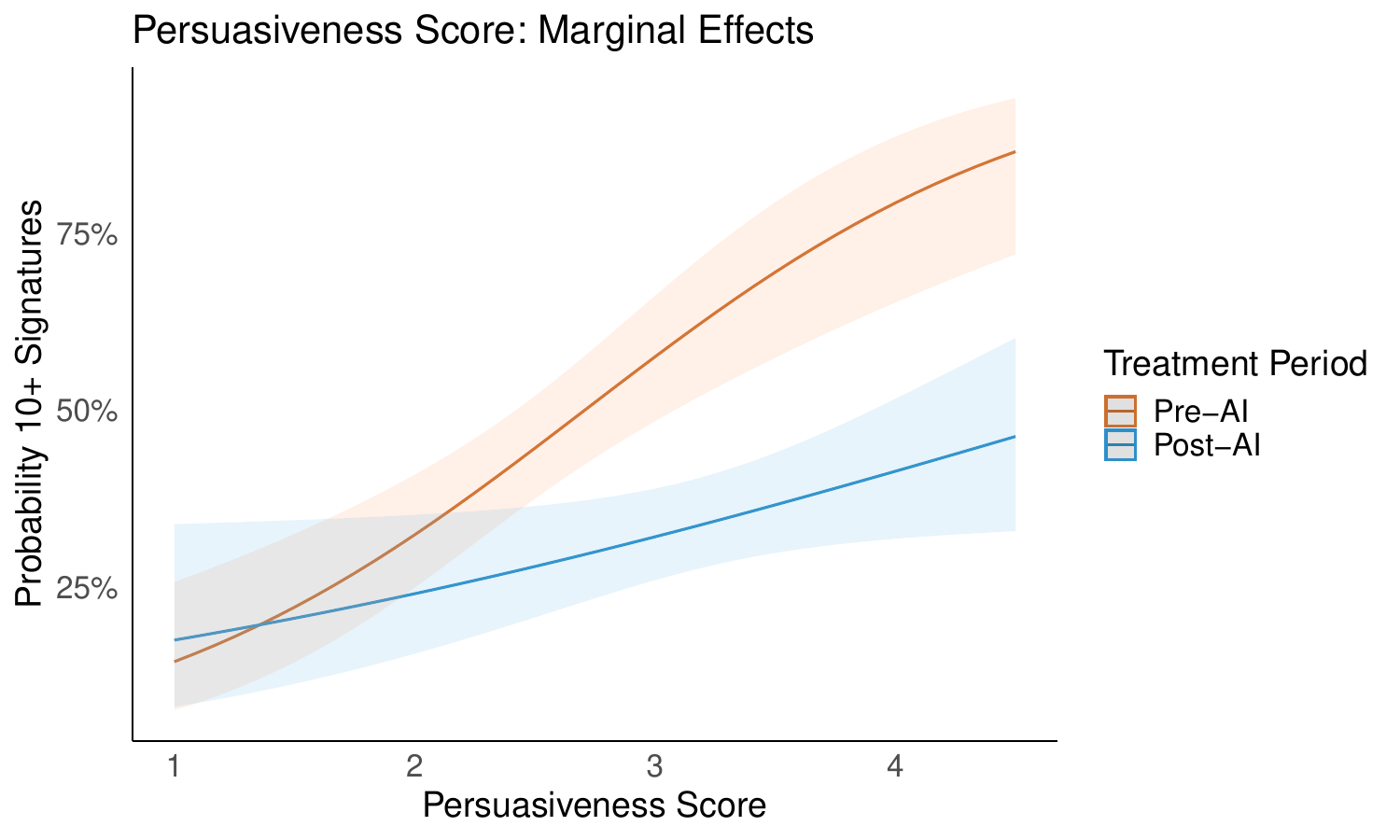}
    \caption{}
    \label{fig:persuasive_marginal_effects_signatures}
  \end{subfigure}
  \caption[Writing Quality and Persuasiveness Marginal Effect on 10+ Signature Outcome]{Marginal effects of (a) writing quality score and (b) persuasiveness score regressed on petition outcome (10+ signatures), by treatment period (Pre-AI, Post-AI).
  Model uses a sample of 417 petitions labeled for writing quality and persuasiveness.
  Pre-AI, higher writing quality and persuasiveness scores were predictive of a higher probability of a successful outcome. 
  Post-AI, higher persuasiveness and writing quality scores were attributed with a reduced increase in probability of a successful outcome. 
  Figure shows 95\% CI.
  }
  \label{fig:survey_marginal_effects_signatures}
\end{figure}

\begin{figure}[H]
  \centering
  \includegraphics[width=\textwidth]{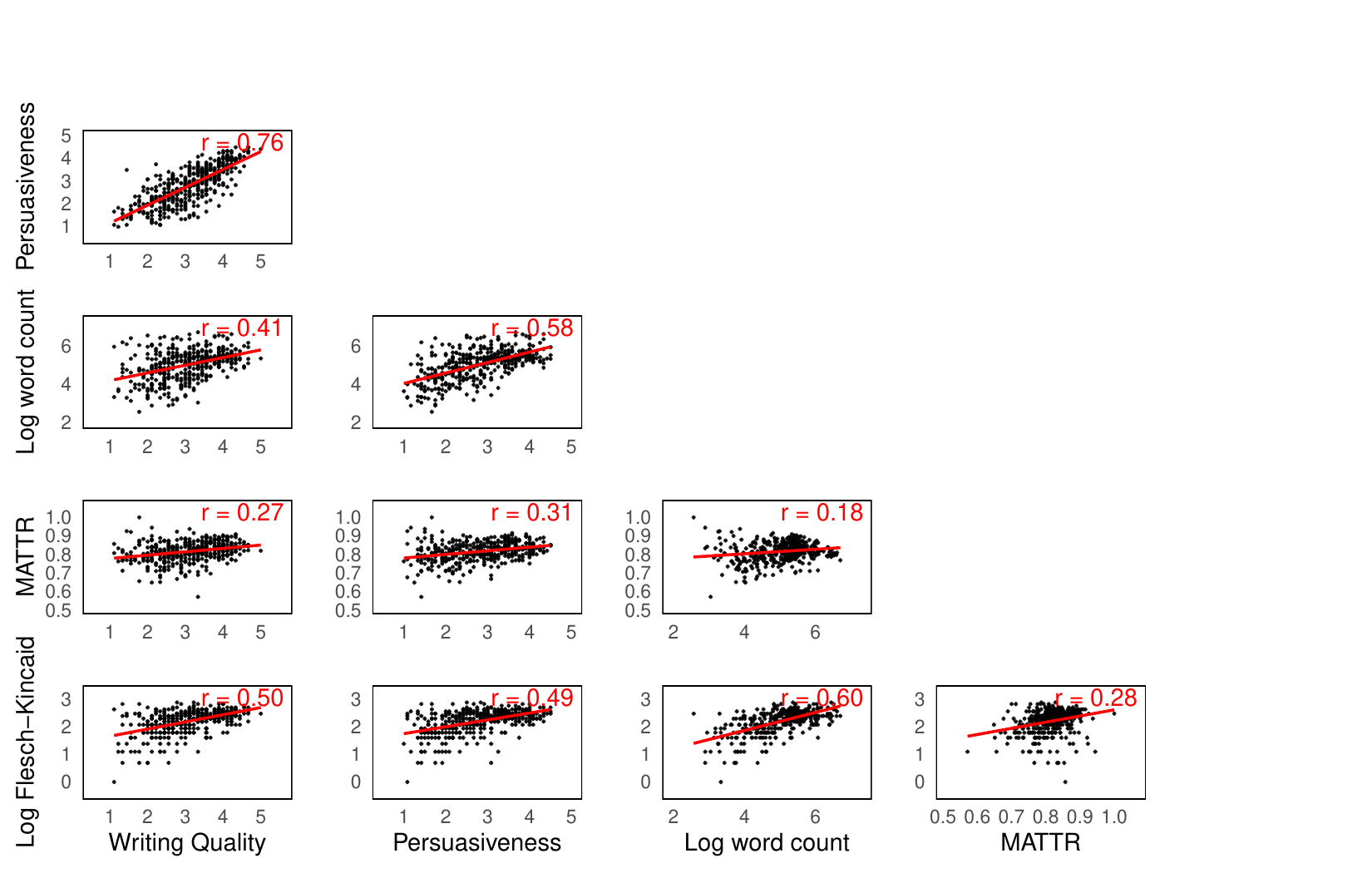}
  \caption[Correlation Between Lexical Features and Labeled Writing Quality/Persuasiveness]{Correlation between lexical measures (log-transformed word count, MATTR, and log-transformed Flesch-Kincaid grade level) and measures of petition text quality (writing quality, persuasiveness) for a random sample of pre- and post-AI annotated petitions.
  We then drop four petitions for which Flesch-Kincaid score is negative to permit log transformation (N = 413 petitions).
  We observe positive correlations between lexical features and quality measures. 
  Scatterplots include a linear fit line and the Pearson correlation coefficient in red text.}
  \label{fig:correlation plot lexical features and survey measures}
\end{figure}

\begin{figure}[H]
\centering

\begin{subfigure}[t]{0.5\textwidth}
  \centering
  \includegraphics[width=\linewidth]{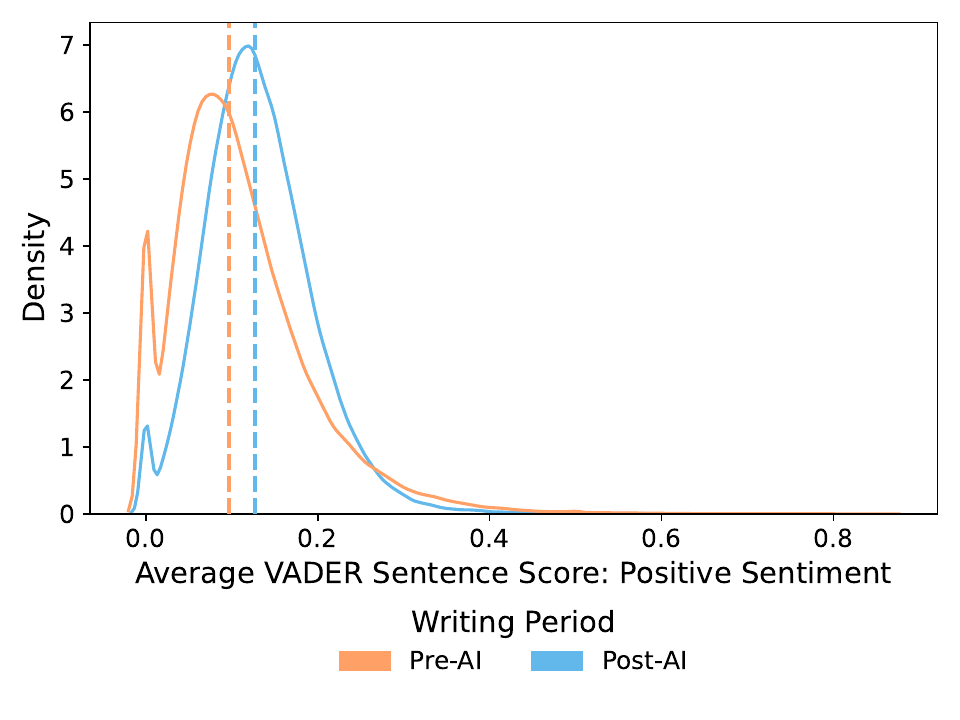}
  \caption{}
  \label{fig:positive}
\end{subfigure}\hfill
\begin{subfigure}[t]{0.5\textwidth}
  \centering
  \includegraphics[width=\linewidth]{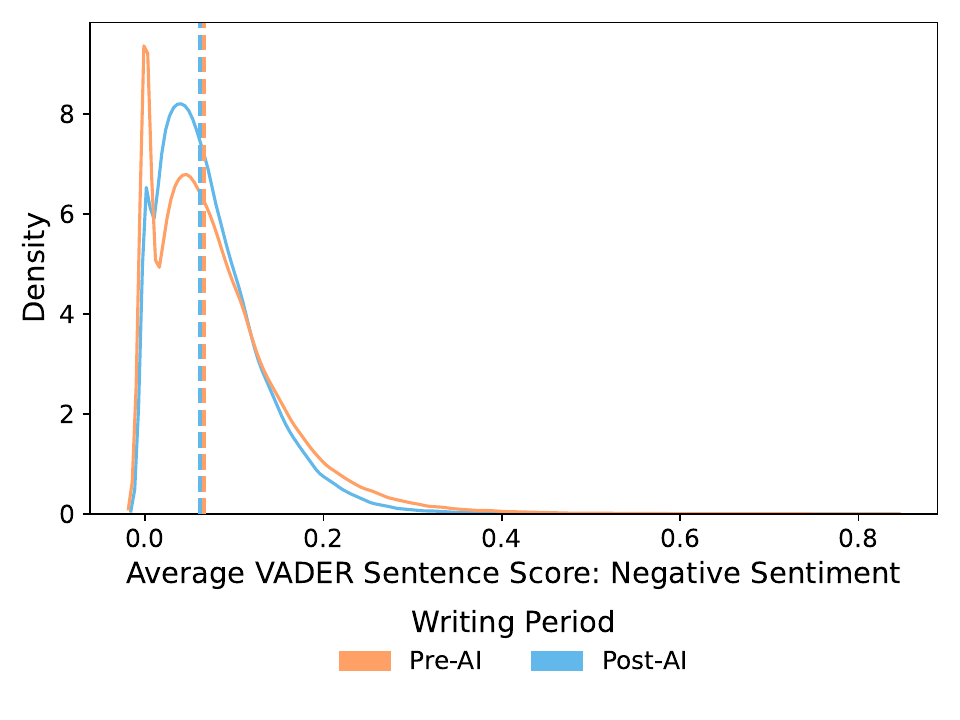}
  \caption{}
  \label{fig:negative}
\end{subfigure}

\vspace{0.5em}

\begin{subfigure}[t]{0.48\textwidth}
  \centering
  \includegraphics[width=\linewidth]{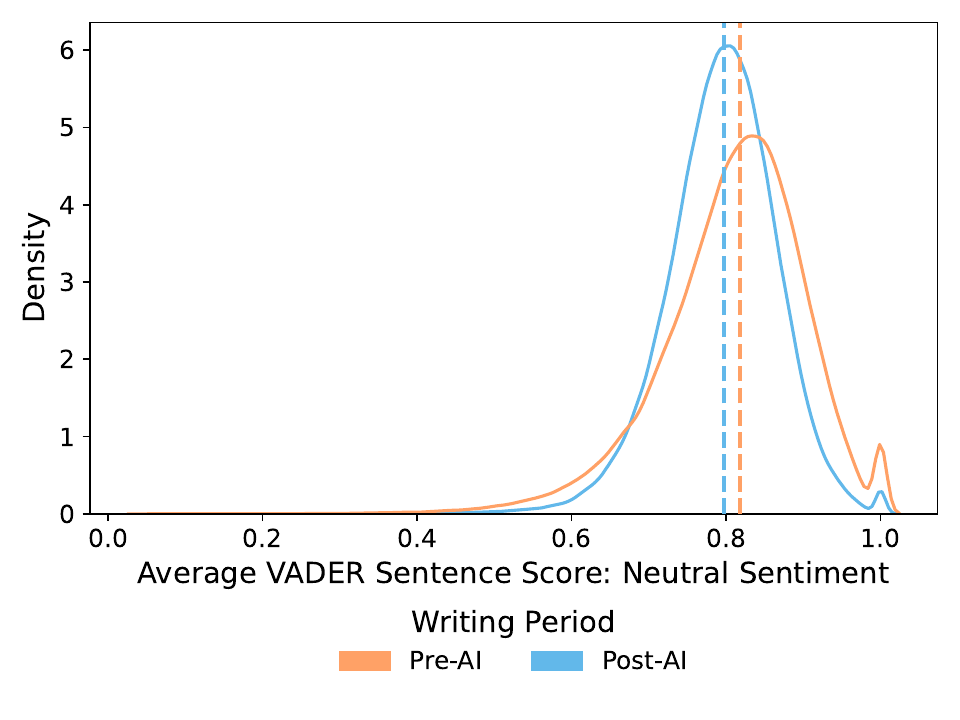}
  \caption{}
  \label{fig:neutral}
\end{subfigure}

\caption[Sentiment KDE]{Distribution of average sentence VADER sentiment scores for English language petitions (237,020 petitions pre-AI, 182,229 petitions post-AI) written in the US, GB, CA, AU between January 2022-December 2024 (excluding A/B test period and period of differential access) by (a) positive sentiment, (b) negative sentiment, and (c) neutral sentiment. 
Vertical lines reflect median values. 
We observe that post-AI, the distribution of positive sentiment shifts to the right, suggesting petitions written with access to AI tend to be more positive in valence. 
The distribution of negative sentiment also changed post-AI, with a greater density of petitions at the mode of 0.1 negative sentiment. Finally, we see a decrease in neutral sentiment, with the distribution shifting to the left.}
\label{fig:vader}
\end{figure}

\begin{figure}[H]
    \centering
    \includegraphics[width=\textwidth]{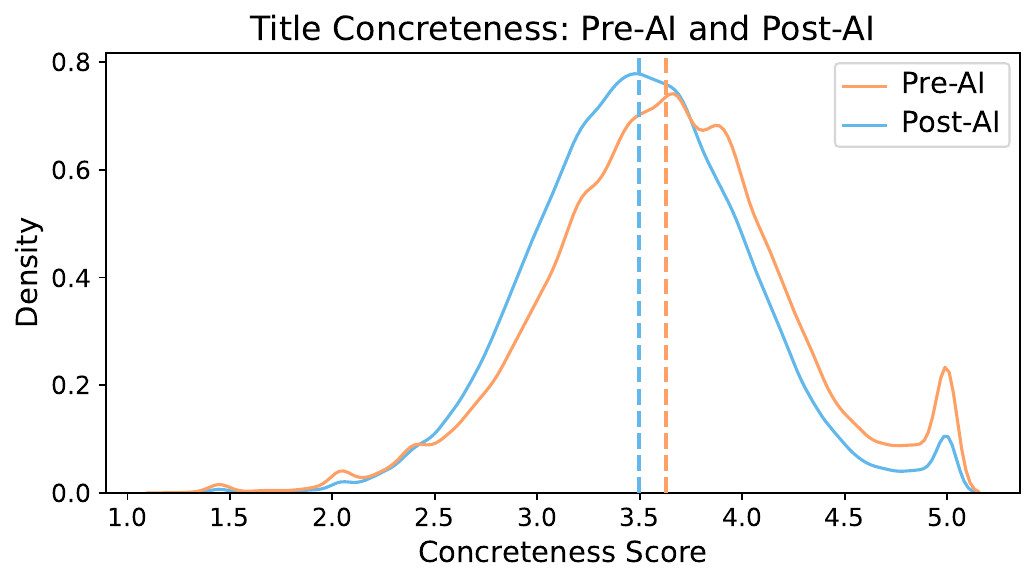}
    \caption[Title concreteness KDE]{Distribution of title concreteness for English language petitions (237,020 petitions pre-AI and 182,229 petitions post-AI) written in the US, GB, CA, AU between January 2022-December 2024 (excluding A/B test period and period of differential access).
    Vertical lines display median title concreteness scores pre- and post- AI.
    We observe that the distribution of title concreteness decreases with in-platform AI access, with the median value decreasing from 3.63 to 3.50 (Kruskal-Wallis test H(1)=5341.433, p \( <\) 0.001, \( \eta^2=0.011\)).}
    \label{fig:title_concreteness}
\end{figure}

\begin{figure}[H]
  \centering
  \begin{subfigure}[b]{0.45\textwidth}
    \centering
    \includegraphics[width=\textwidth]{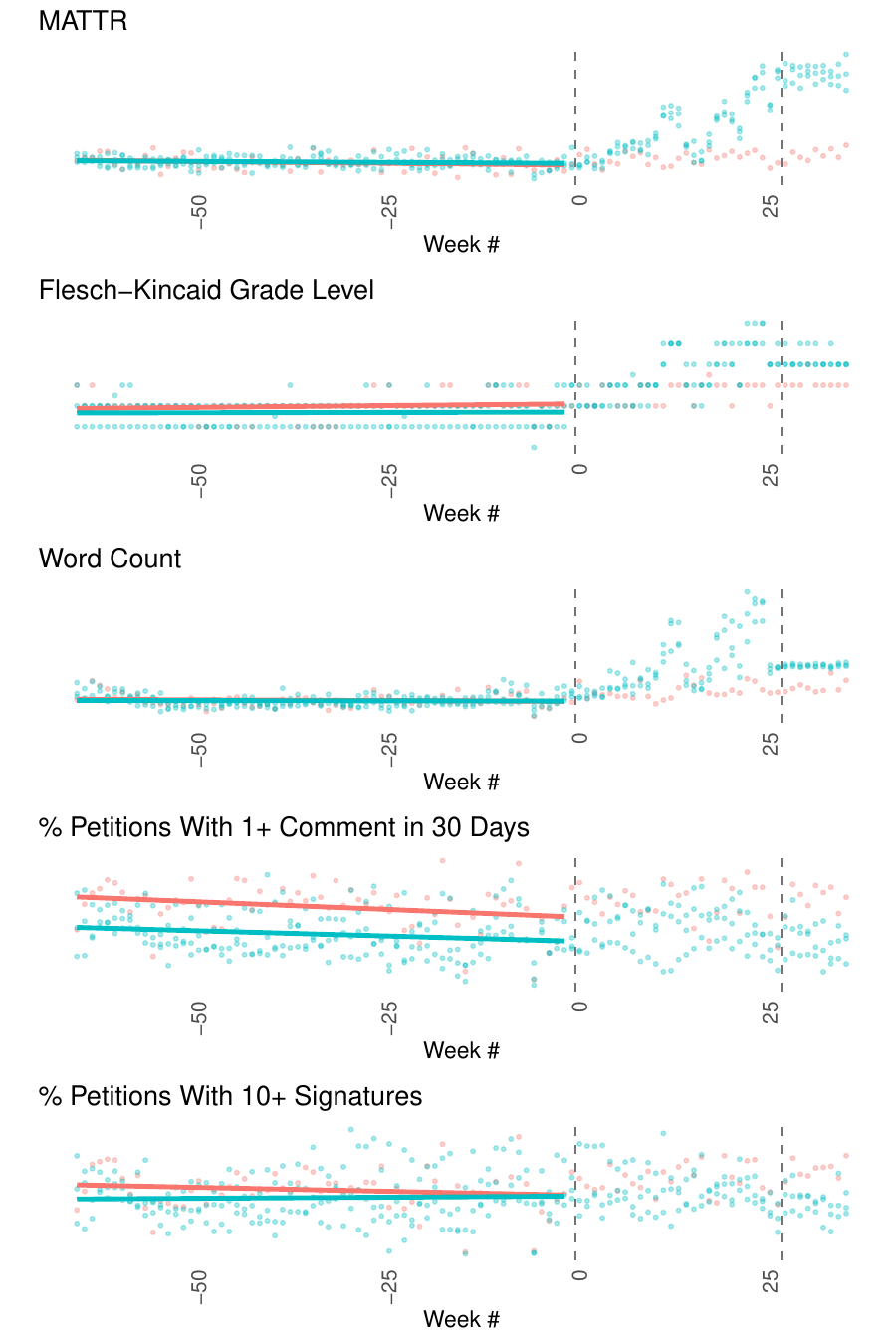}
    \caption{}
    \label{fig:parallel_trends_lr}
  \end{subfigure}
  \hfill
  \begin{subfigure}[b]{0.5\textwidth}
    \centering
    \includegraphics[width=\textwidth]{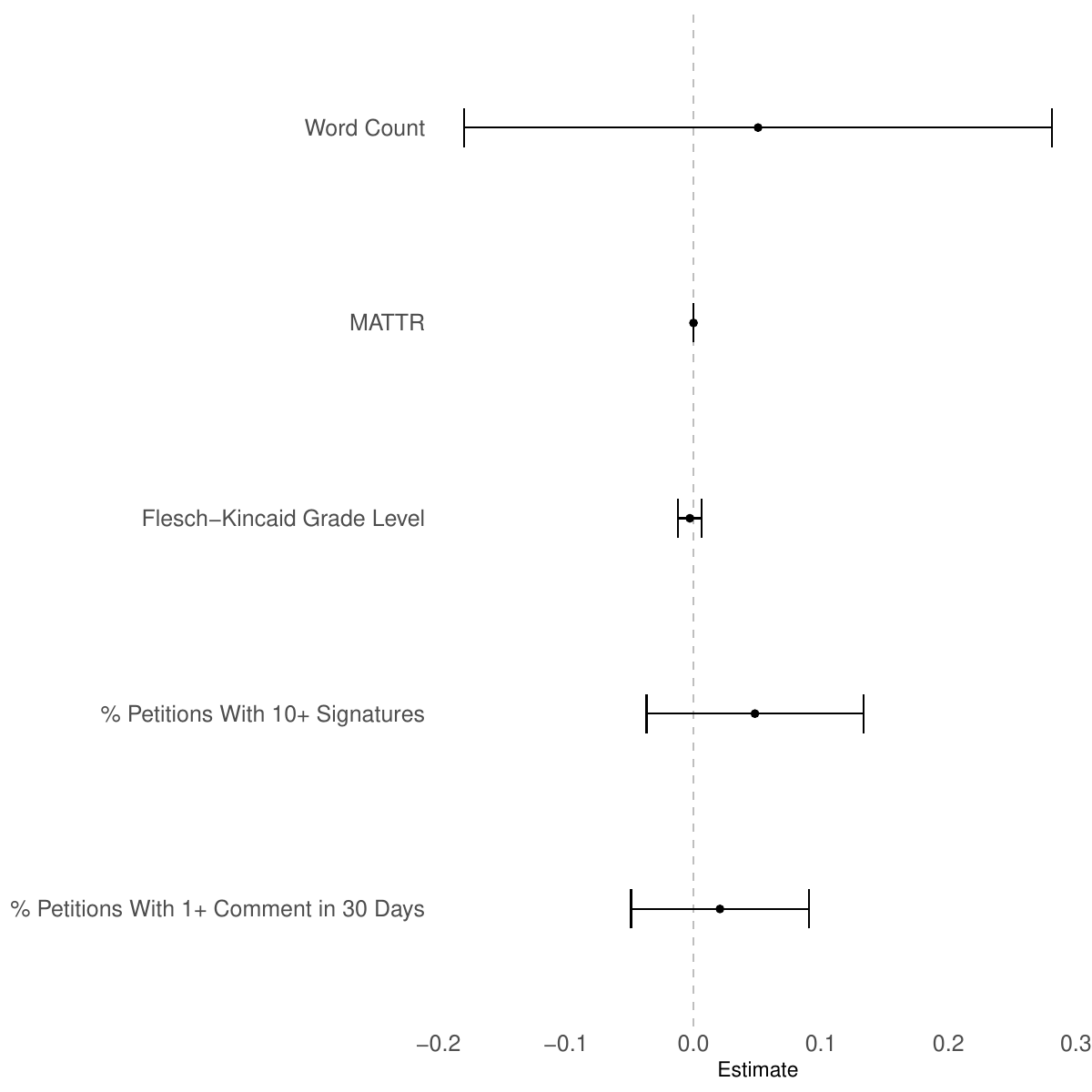}
    \caption{}
    \label{fig:parallel_trends_slope_CI}
  \end{subfigure}
  \caption[Parallel Trends]{To assess if our core lexical features and petition outcomes satisfy the assumption of parallel trends pre-intervention, we fit simple linear models to predict the metric by week for both treated and untreated groups. We used N = 237,020 petitions from the pre-treatment period.
  Panel (a) shows the fit regression line to control and treated units in the pre-AI period.
  Panel (b) shows 95\% CI on the coefficient of the interaction term by week and treatment group. 
  We find for each metric that the interaction is insignificant, as the confidence intervals all include 0.}
  \label{fig:parallel_trends}
\end{figure}

\begin{figure}[H]
\centering
\begin{subfigure}{0.48\textwidth}
    \includegraphics[width=\textwidth]{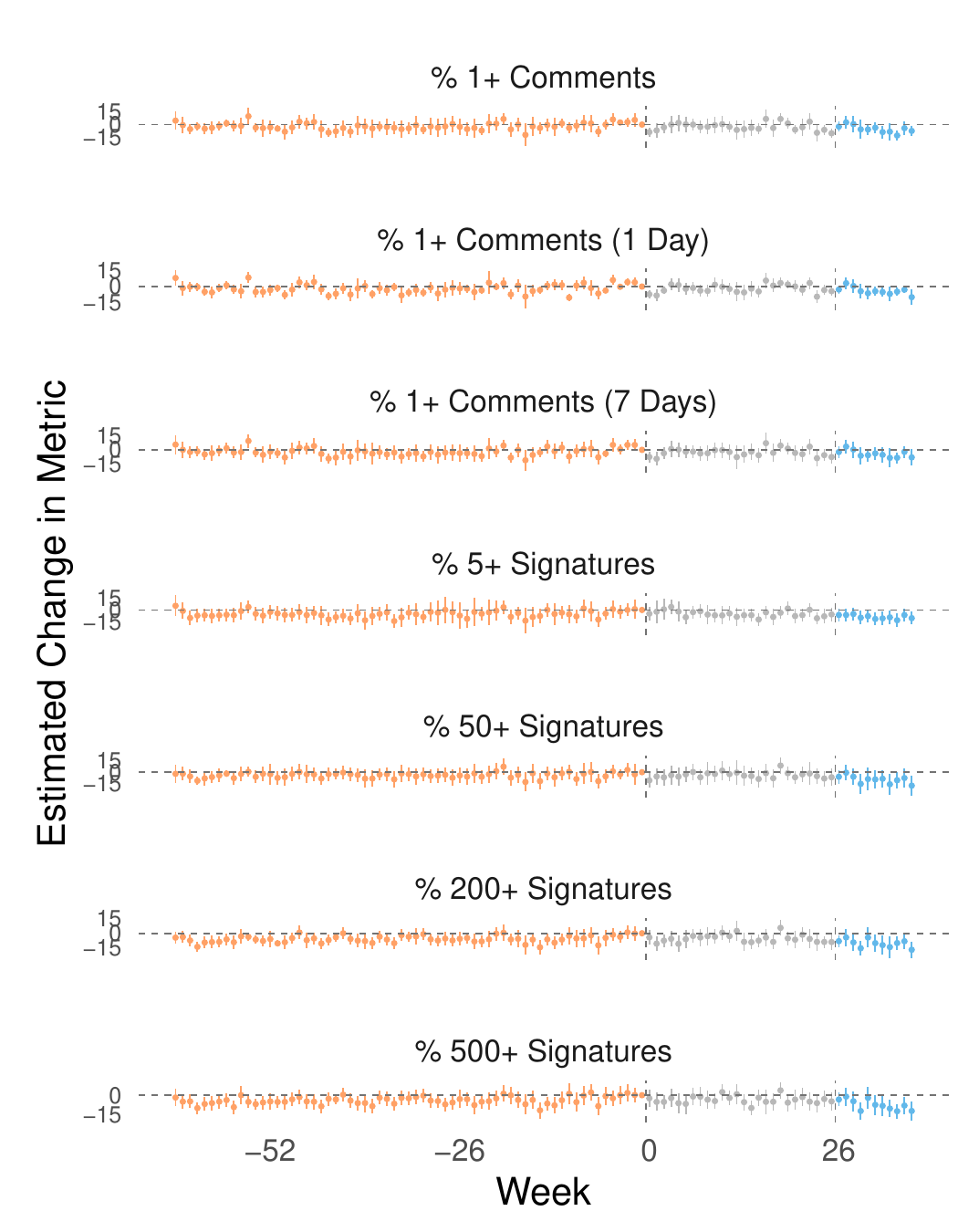}
    \caption{}
    \label{fig:additional_outcomes_thresholds_event_study}
\end{subfigure}
\hfill
\begin{subfigure}[b]{0.48\textwidth}
    \includegraphics[width=\textwidth]{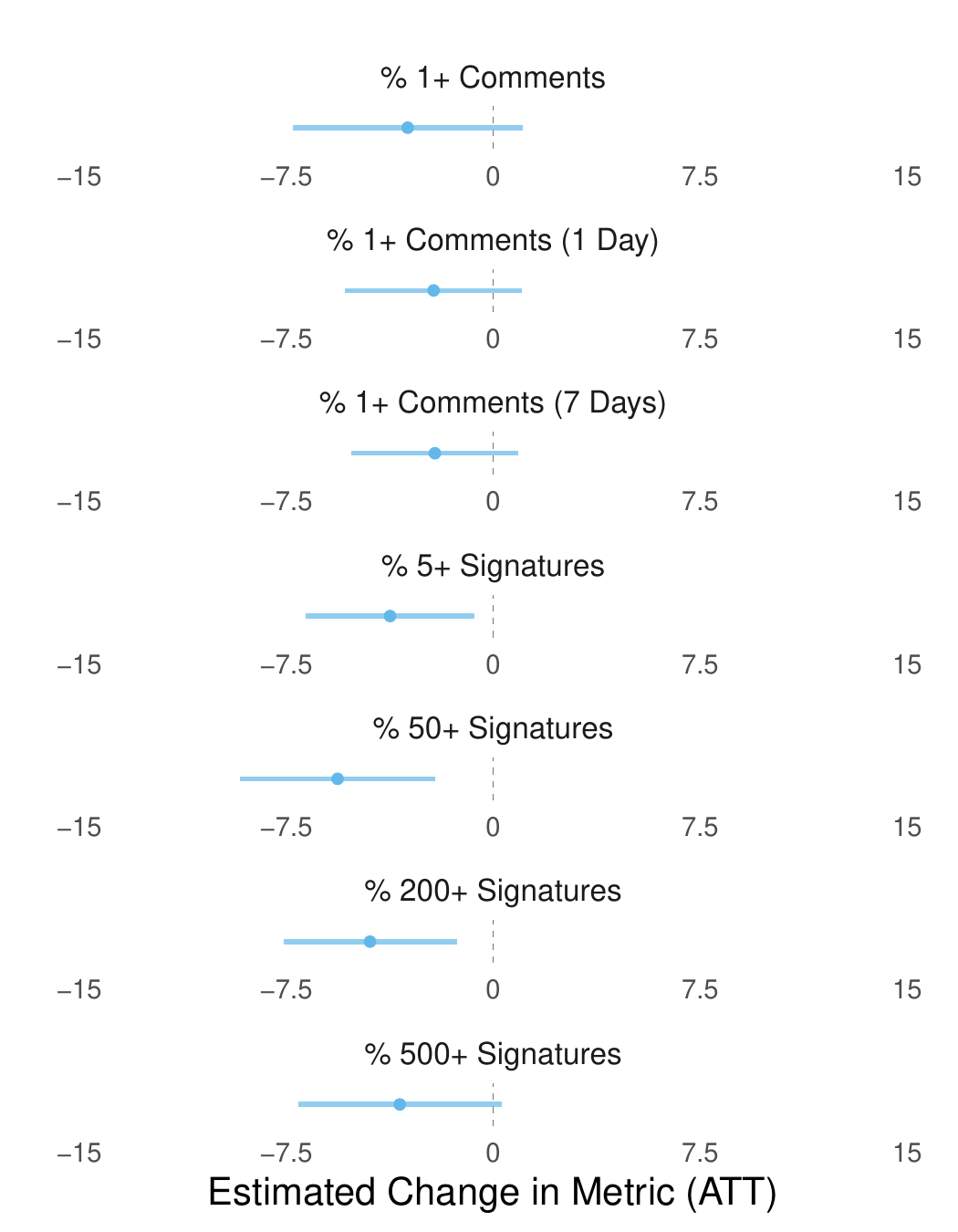}
    \caption{}
    \label{fig:additional_outcomes_thresholds_pre_post}
\end{subfigure}

\vspace{0.5em} 

\begin{subfigure}[b]{0.48\textwidth}
    \includegraphics[width=\textwidth]{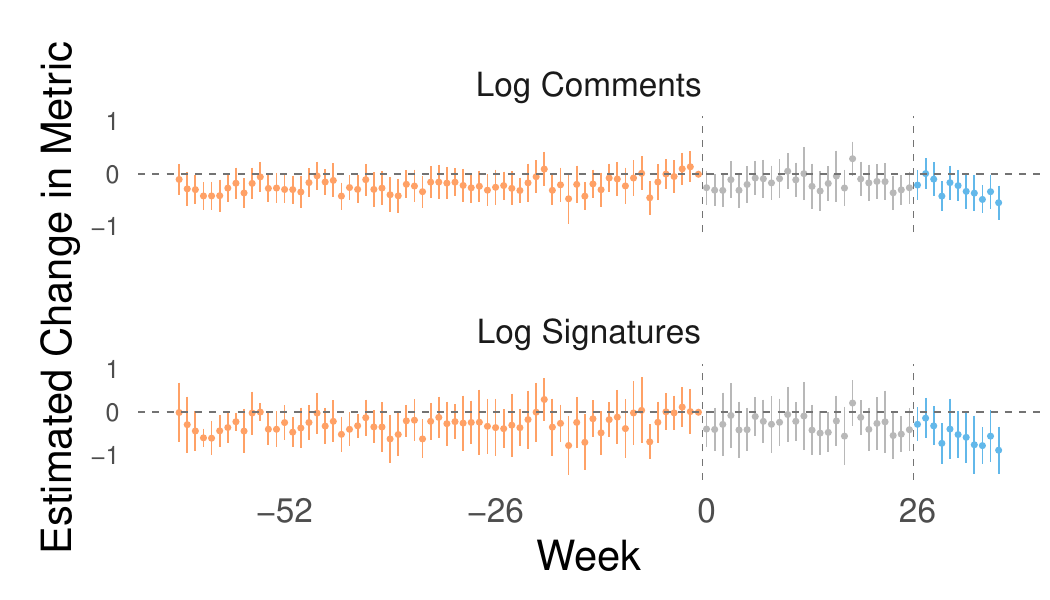}
    \caption{}
    \label{fig:additional_outcomes_log_event_study}
\end{subfigure}
\hfill
\begin{subfigure}[b]{0.48\textwidth}
    \includegraphics[width=\textwidth]{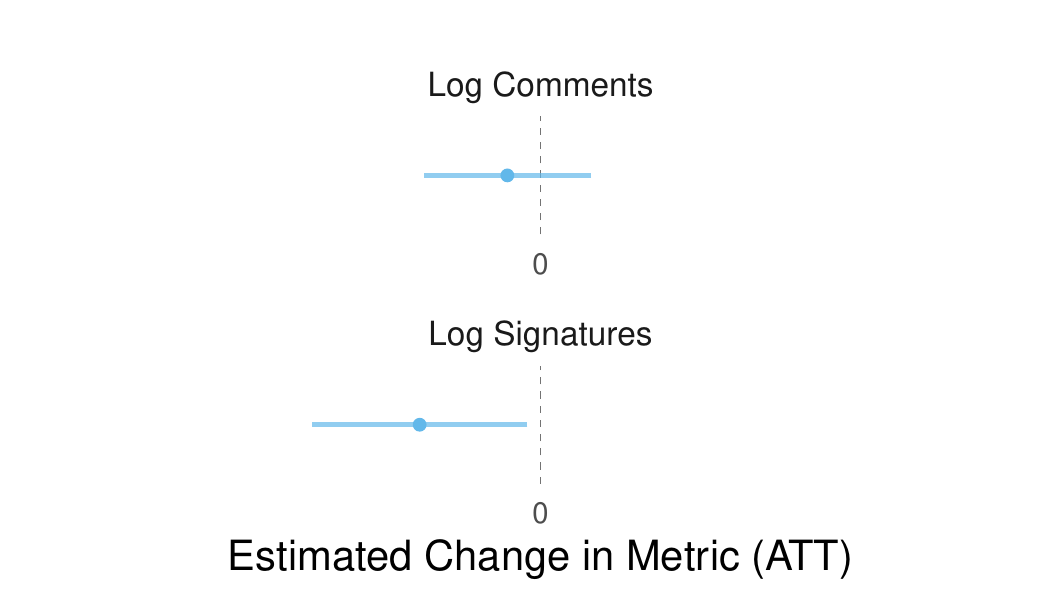}
    \caption{}
    \label{fig:additional_outcomes_log_pre_post}
\end{subfigure}

\caption[Difference-in-differences for Alternative Outcome Metrics]{
Static and dynamic difference-in-differences estimates for additional outcome measures.
In the main text we report one set of thresholds for petitions: 1 comment in 30 days and 10 signatures. 
Here we show that the results observed in the main text (insignificant changes, or significant negative effects) are consistent for a variety of alternative thresholds.
Panels (a) and (b) show the estimated change in percentage points for additional thresholds of comments and signatures. 
Panels (c) and (d) show the estimated change in log-transformed comment and signature counts, normalized by the number of petitions posted during the time period. 
We log-transform comments and signatures due to heavy positive skew.
We report 95\% CI bands and use the same model specification as outcomes in main analysis.
Pre-AI period was 65 weeks (N = 237,020 petitions), A/B test period was 26 weeks (N = 82,504 petitions), and the Post-AI period was 11 weeks (N = 42,111 petitions).}
\label{fig:additional_outcomes}
\end{figure}

\begin{figure}[H]
\centering
\begin{subfigure}[b]{0.48\textwidth}
    \includegraphics[width=\textwidth]{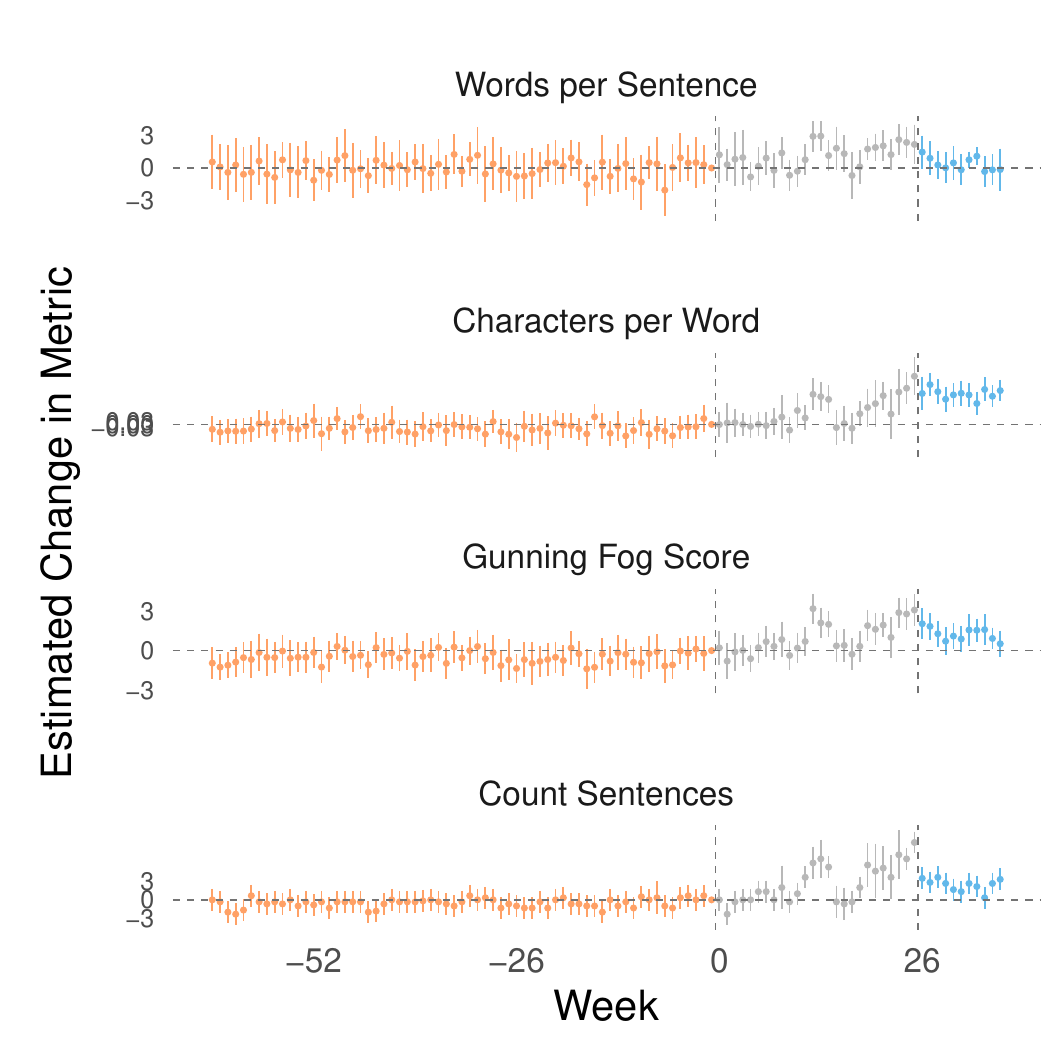}
    \caption{}
\end{subfigure}
\hfill
\begin{subfigure}[b]{0.48\textwidth}
    \includegraphics[width=\textwidth]{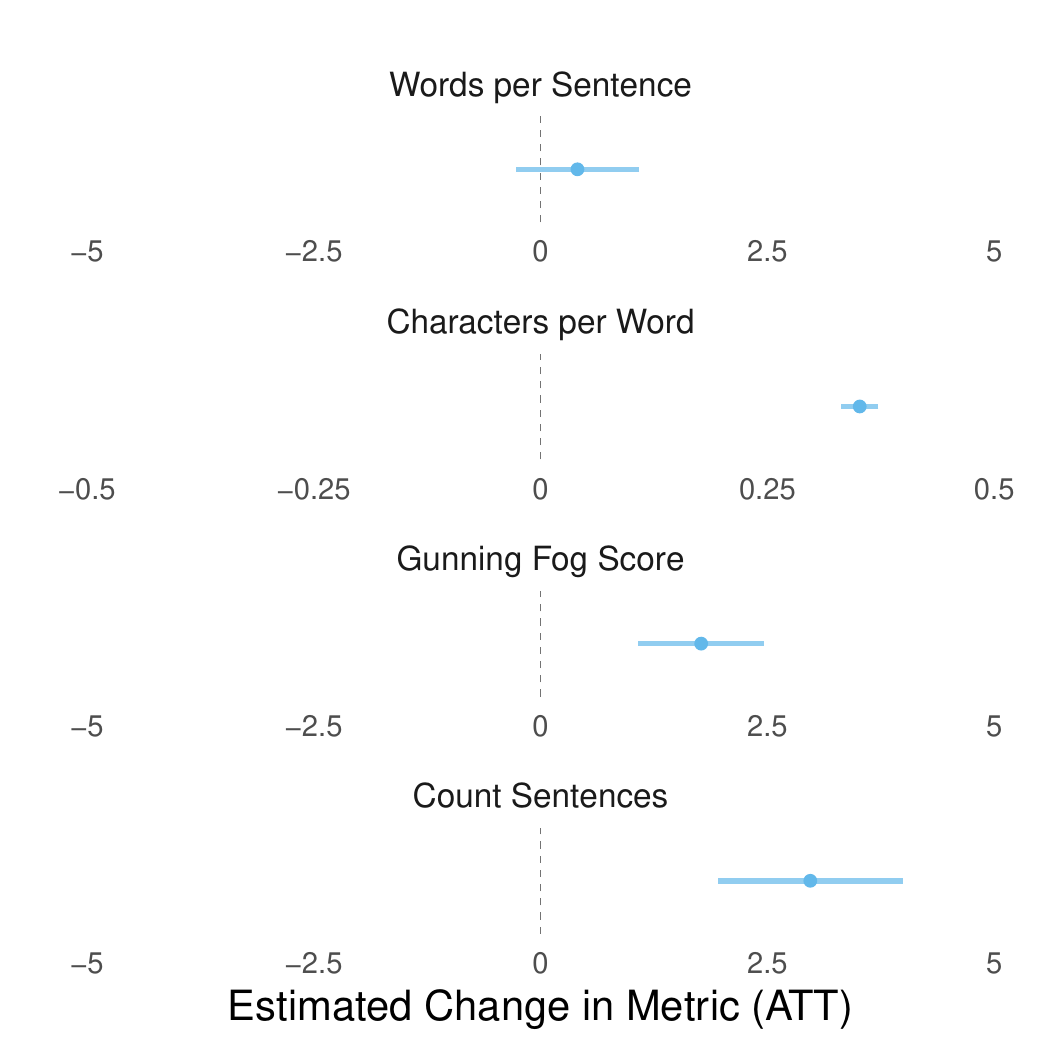}
    \caption{}
\end{subfigure}
\caption[Difference-in-differences for Alternative Lexical Features]{
Panel (a) shows dynamic difference-in-differences and Panel (b) shows static difference-in-differences for additional measures of lexical style.
We report 95\% CI bands and use the same model specification as thresholds in main text. 
Estimates show expected change in median lexical feature for treated countries with access to AI.
Pre-AI period was 65 weeks (N = 237,020 petitions), A/B test period was 26 weeks (N = 82,504 petitions), and the Post-AI period was 11 weeks (N = 42,111 petitions).}
\label{fig:additional_lexical_features}
\end{figure}

\begin{figure}[H]
\centering

\begin{subfigure}[b]{0.48\textwidth}
    \includegraphics[width=\textwidth]{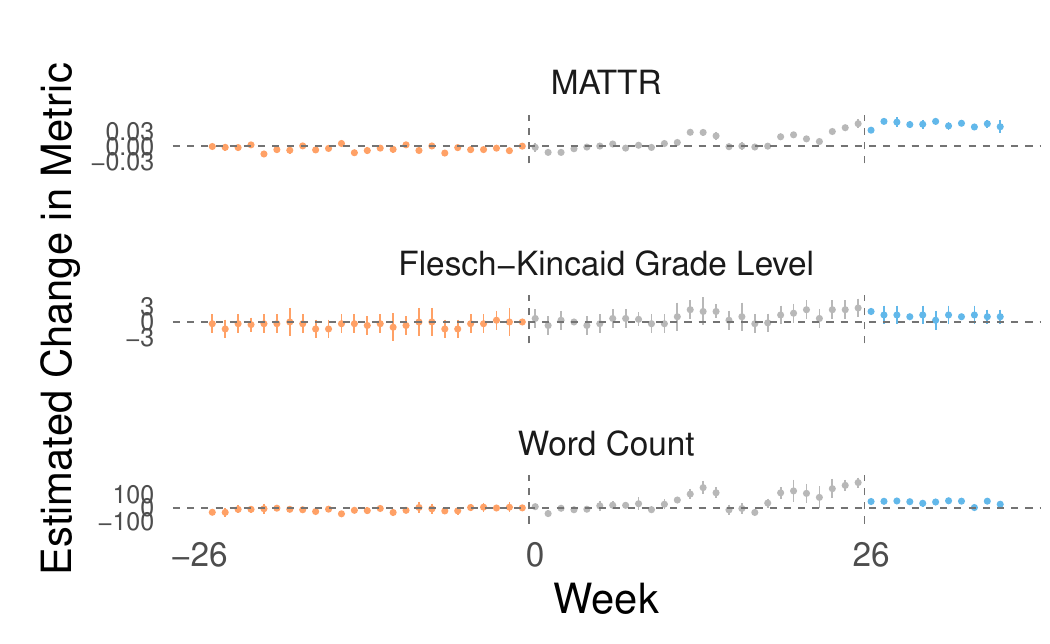}
    \caption{}
\end{subfigure}
\hfill
\begin{subfigure}[b]{0.48\textwidth}
    \includegraphics[width=\textwidth]{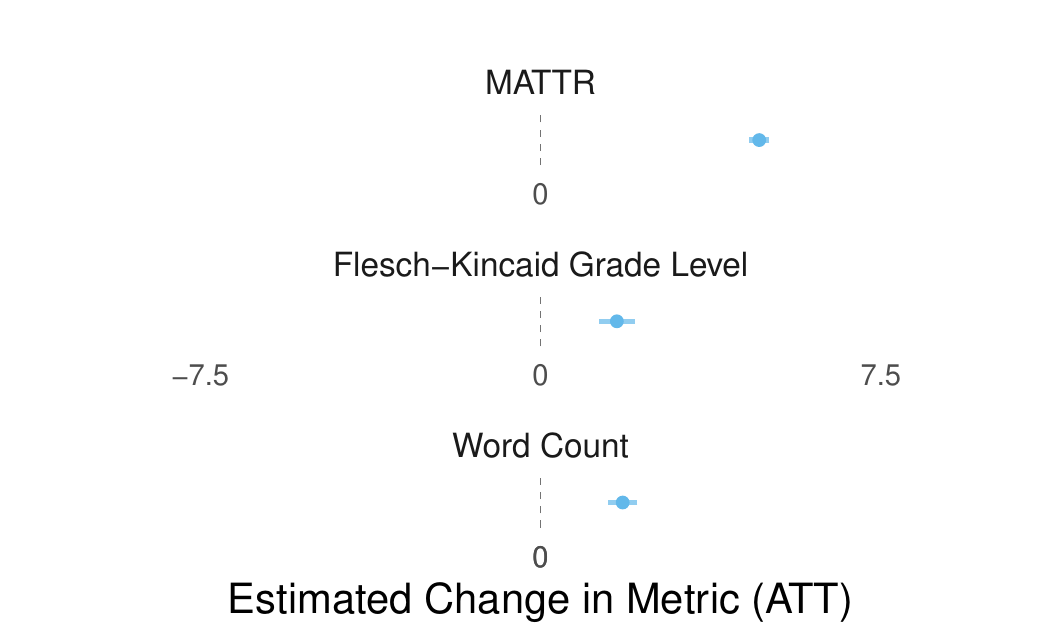}
    \caption{}
\end{subfigure}

\vspace{0.5em}

\begin{subfigure}[b]{0.48\textwidth}
    \includegraphics[width=\textwidth]{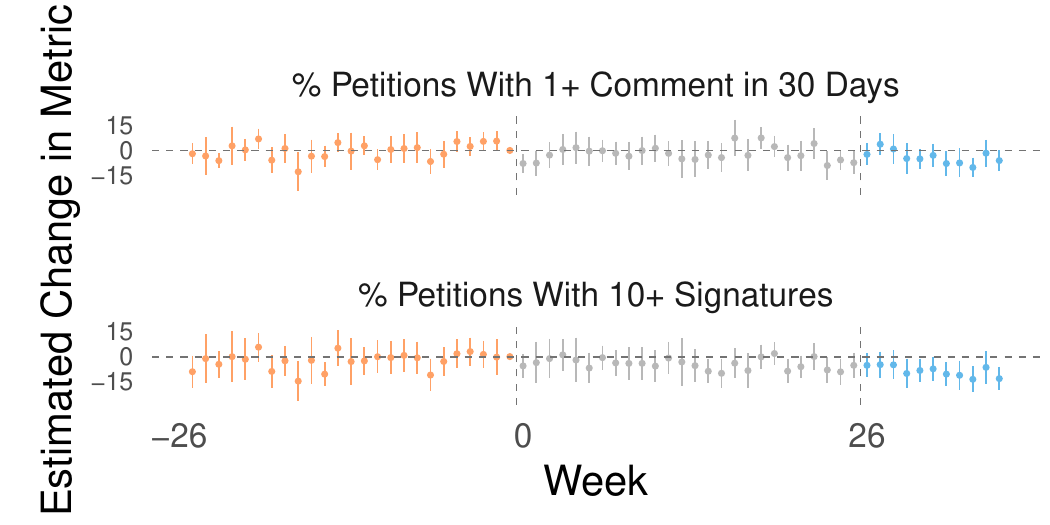}
    \caption{}
\end{subfigure}
\hfill
\begin{subfigure}[b]{0.48\textwidth}
    \includegraphics[width=\textwidth]{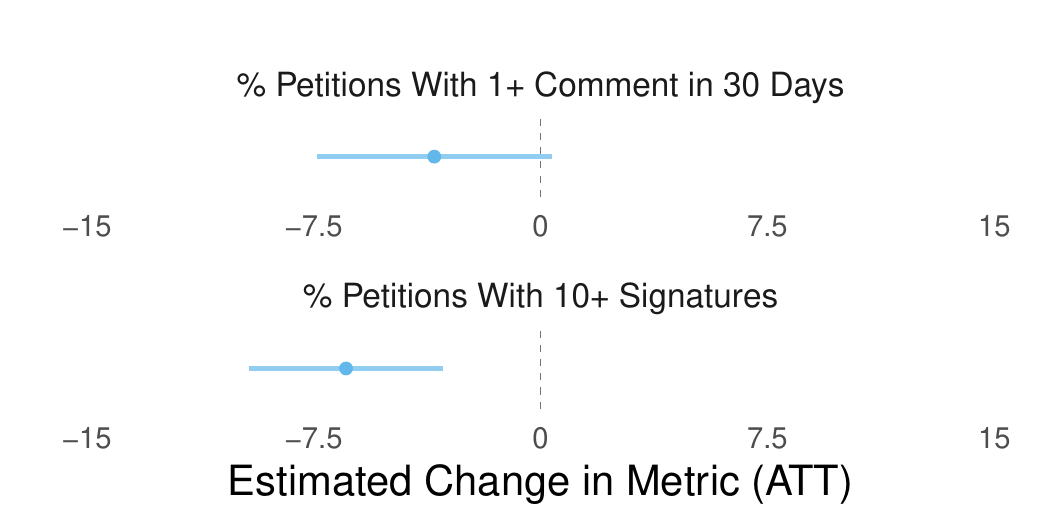}
    \caption{}
\end{subfigure}

\caption[Difference-in-differences for Hypothetical Narrow Pre-Treatment Period]{
Panels (a) and (c) show dynamic difference-in-differences analysis and Panels (b) and (d) show static difference-in-differences estimates for lexical features and outcomes.
Here, we report the same metrics captured in the main text, with a shorter time frame (25 weeks pre-AI, rather than 65 weeks). 
We use the shorter time frame as a robustness check to demonstrate results are not driven by trends in early 2022.
Our results are consistent with the shortened pre-AI period.
This data includes 83,363 petitions posted during the pre-AI period, 82,504 petitions posted during the A/B Test period, and 42,111 petitions posted in the post-AI period.}
\label{fig:narrow_did}
\end{figure}

\begin{figure}[H]
\centering

\begin{subfigure}[b]{0.48\textwidth}
    \includegraphics[width=\textwidth]{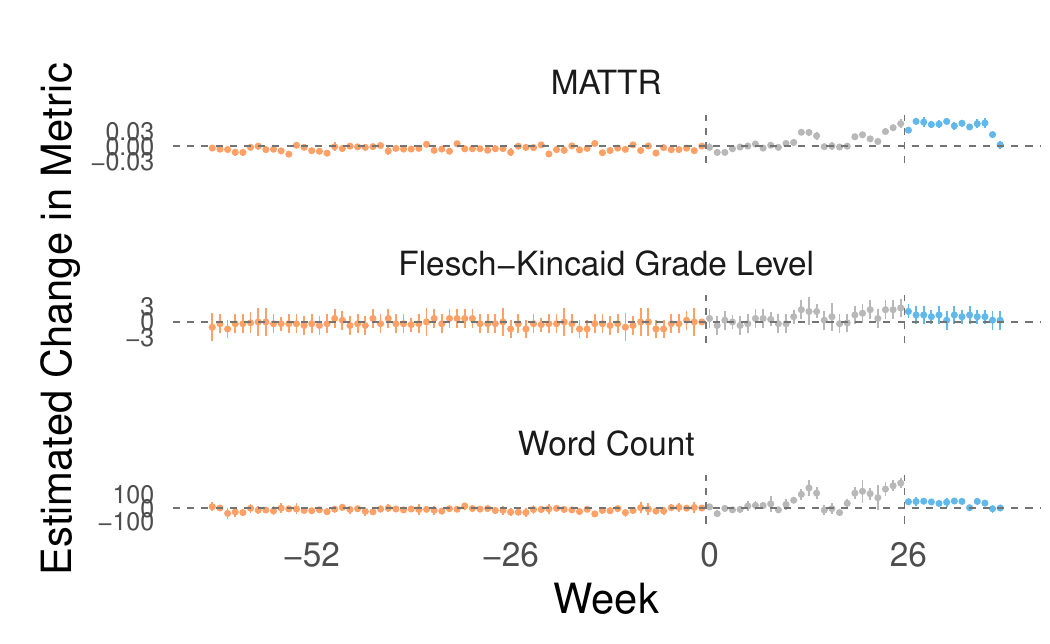}
    \caption{}
\end{subfigure}
\hfill
\begin{subfigure}[b]{0.48\textwidth}
    \includegraphics[width=\textwidth]{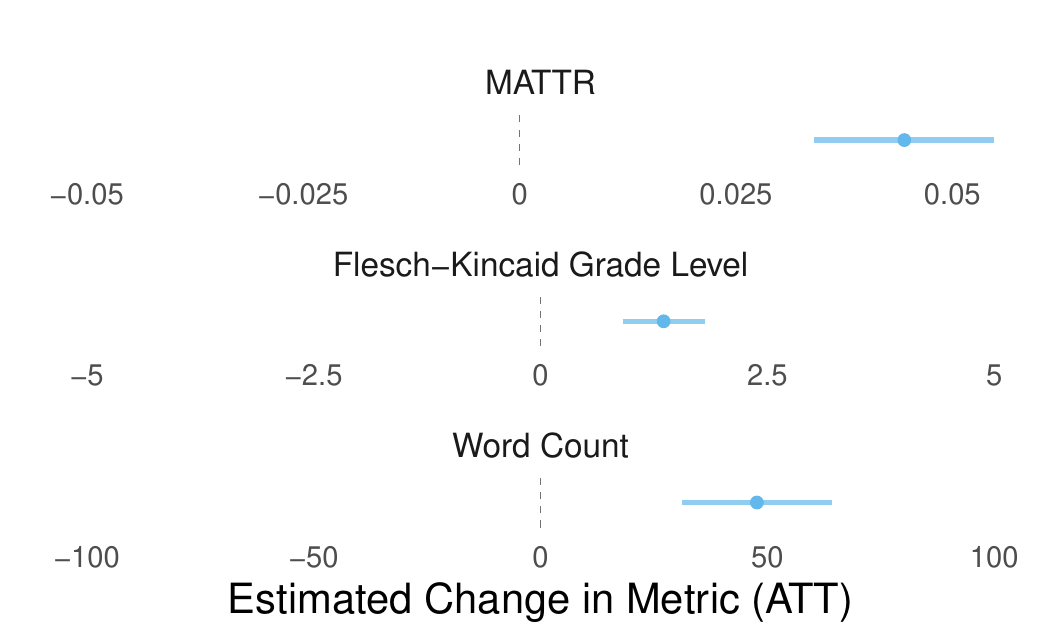}
    \caption{}
\end{subfigure}

\vspace{0.5em}

\begin{subfigure}[b]{0.48\textwidth}
    \includegraphics[width=\textwidth]{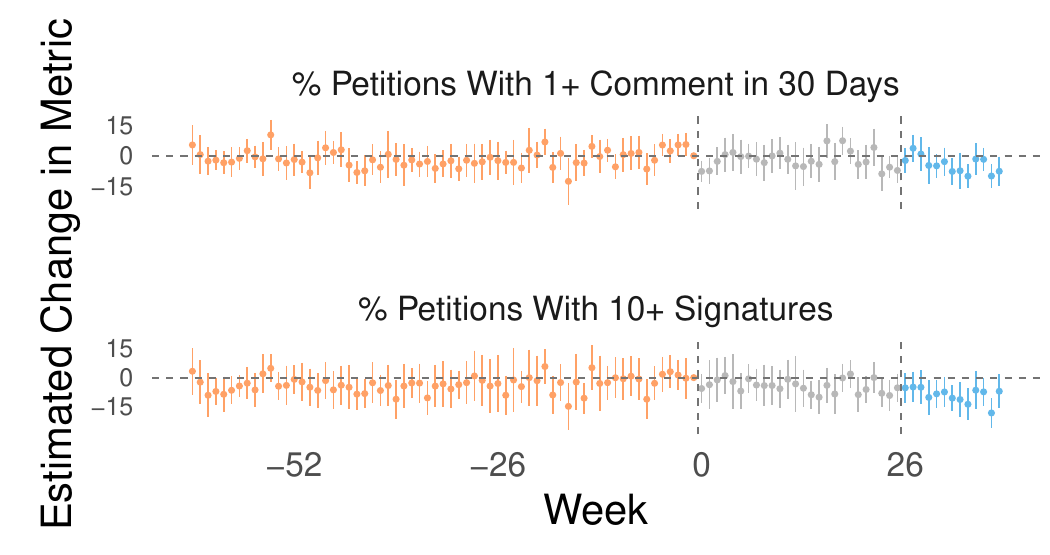}
    \caption{}
\end{subfigure}
\hfill
\begin{subfigure}[b]{0.48\textwidth}
    \includegraphics[width=\textwidth]{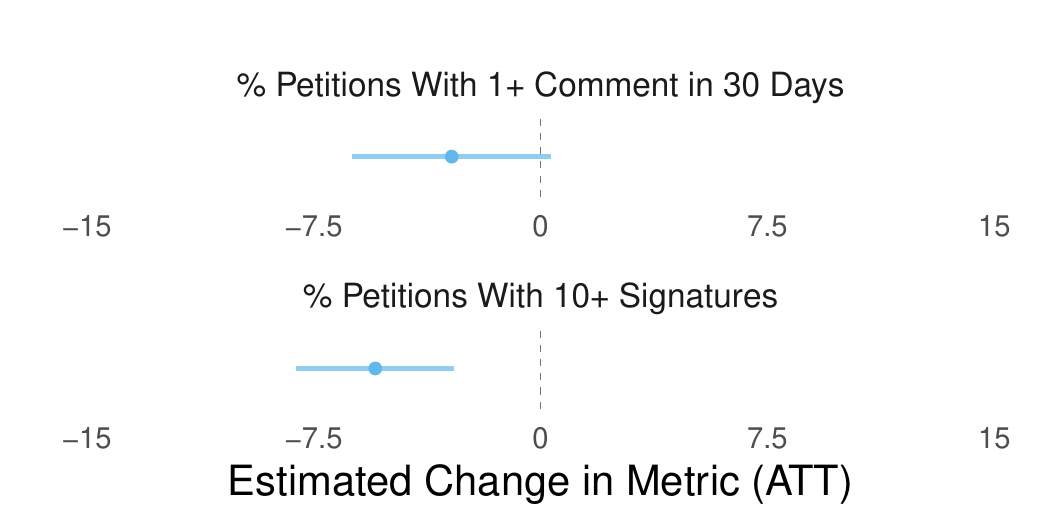}
    \caption{}
\end{subfigure}

\caption[Difference-in-differences for Hypothetical Longer Post-Treatment Period]{Difference-in-differences with the full month of December included as the post-AI period of differential access. 
Analysis includes 237,020 petitions during the Pre-AI period, 82,504 petitions during the A/B test period, and 47,539 petitions during the post-AI period. 
The post-AI period is extended from 11 weeks to 13 weeks (spanning October 2, 2023-December 31, 2023). 
Results are similar to those calculated using the eleven-week period in our main-text; however, the reduced gap between treated and control country style metrics is apparent in the final two weeks, as we expect control users did have access to in-platform AI during this period.}
\label{fig:december_did}
\end{figure}

\begin{figure}[H]
\centering

\begin{subfigure}[b]{0.48\textwidth}
    \includegraphics[width=\textwidth]{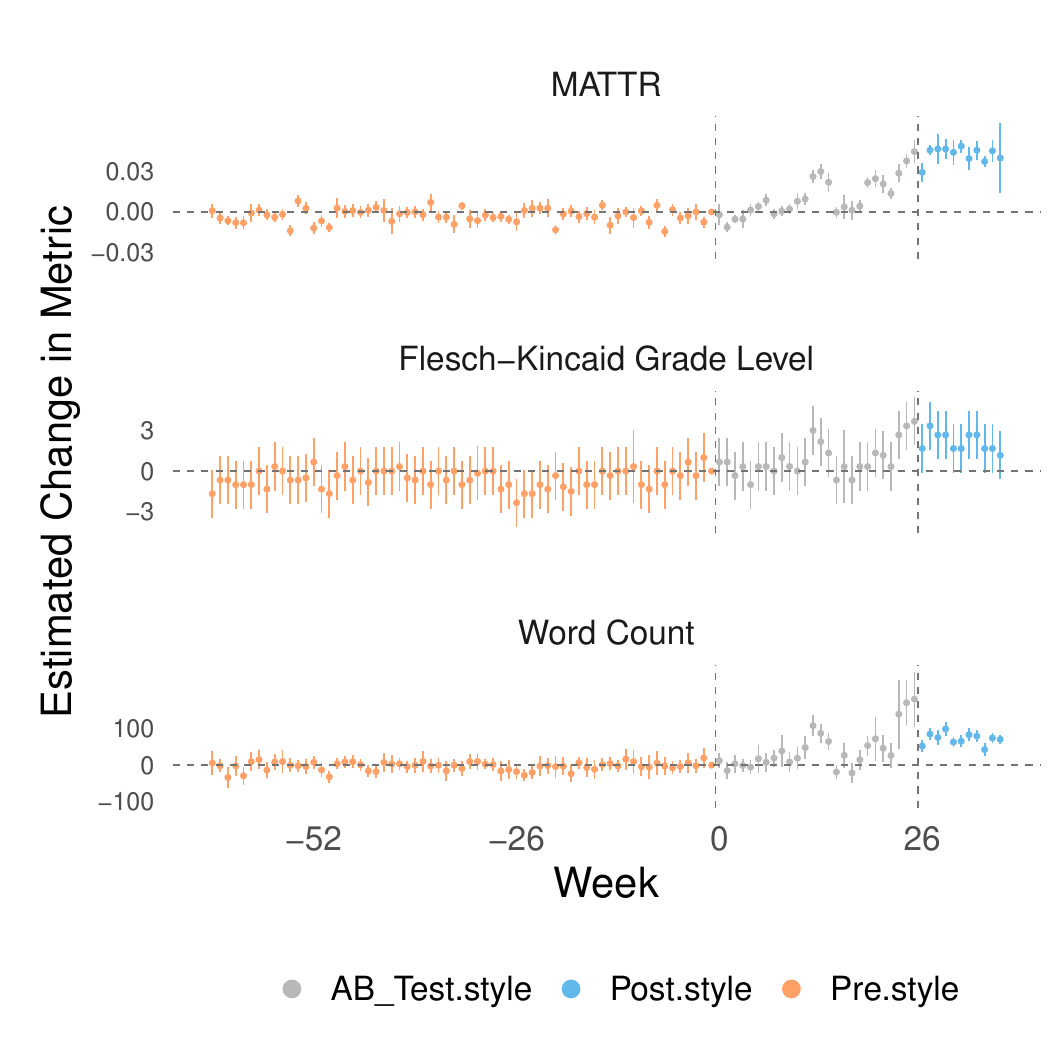}
    \caption{}
\end{subfigure}
\hfill
\begin{subfigure}[b]{0.48\textwidth}
    \includegraphics[width=\textwidth]{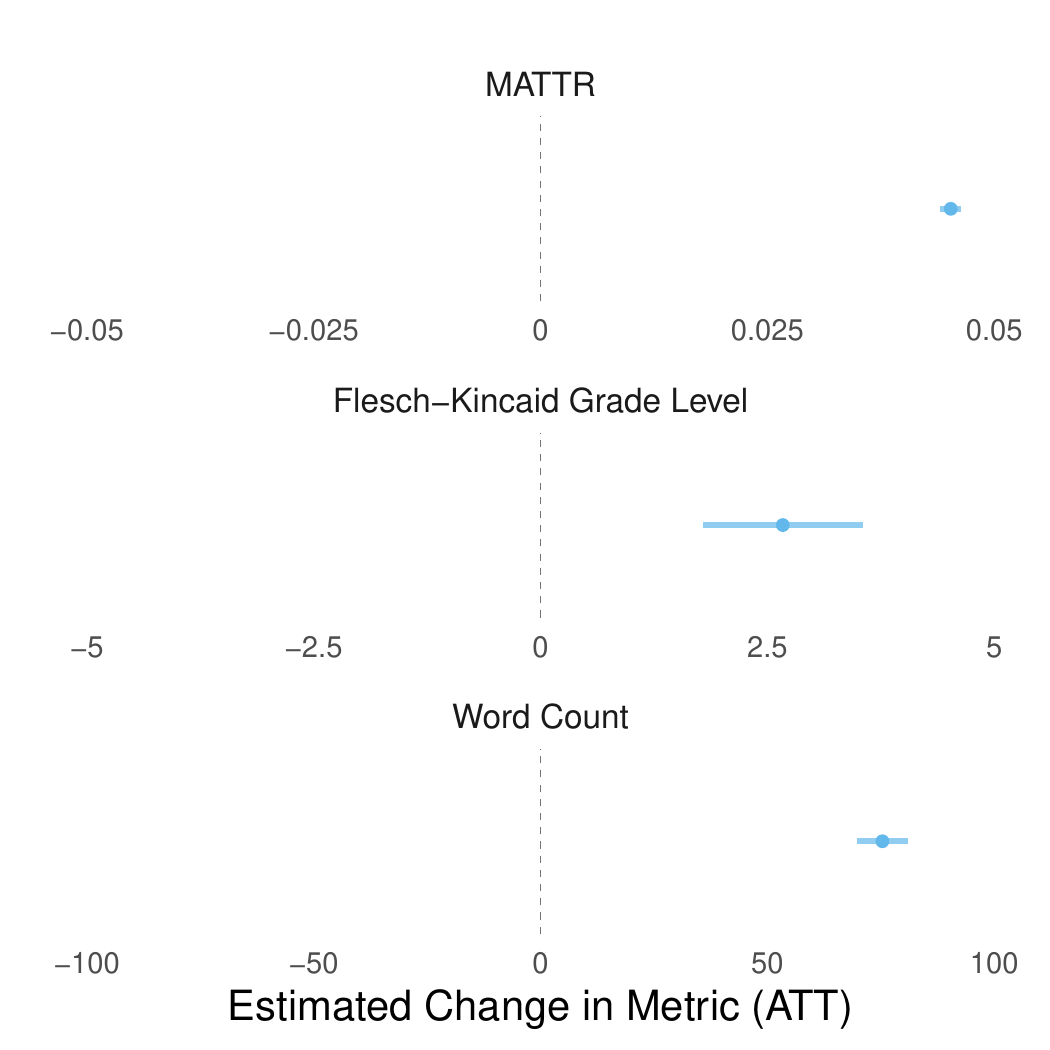}
    \caption{}
\end{subfigure}

\vspace{0.5em}
\begin{subfigure}[b]{0.48\textwidth}
    \includegraphics[width=\textwidth]{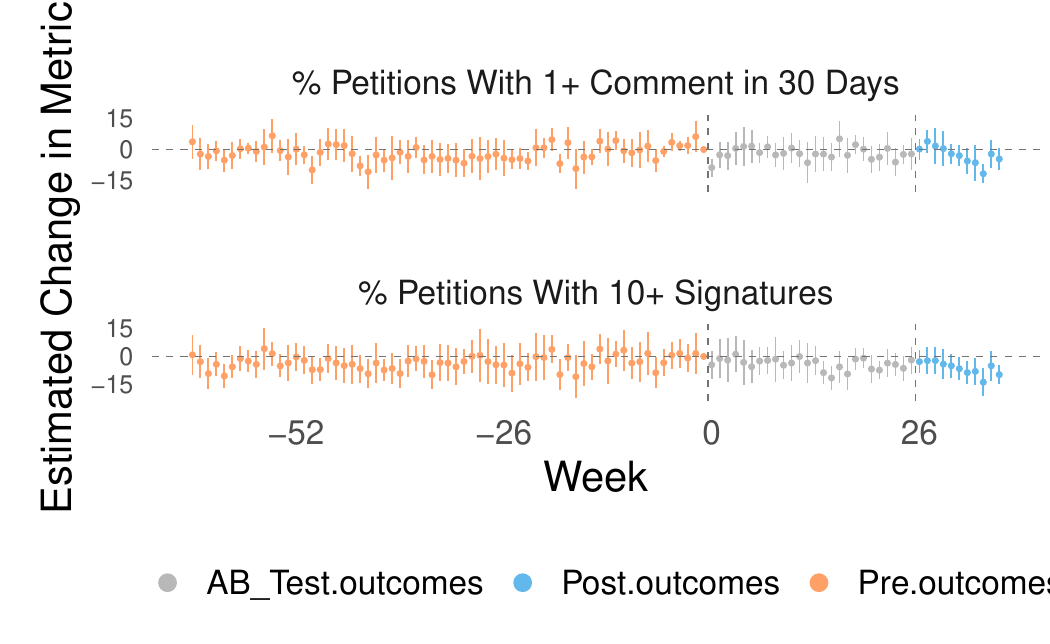}
    \caption{}
\end{subfigure}
\hfill
\begin{subfigure}[b]{0.48\textwidth}
    \includegraphics[width=\textwidth]{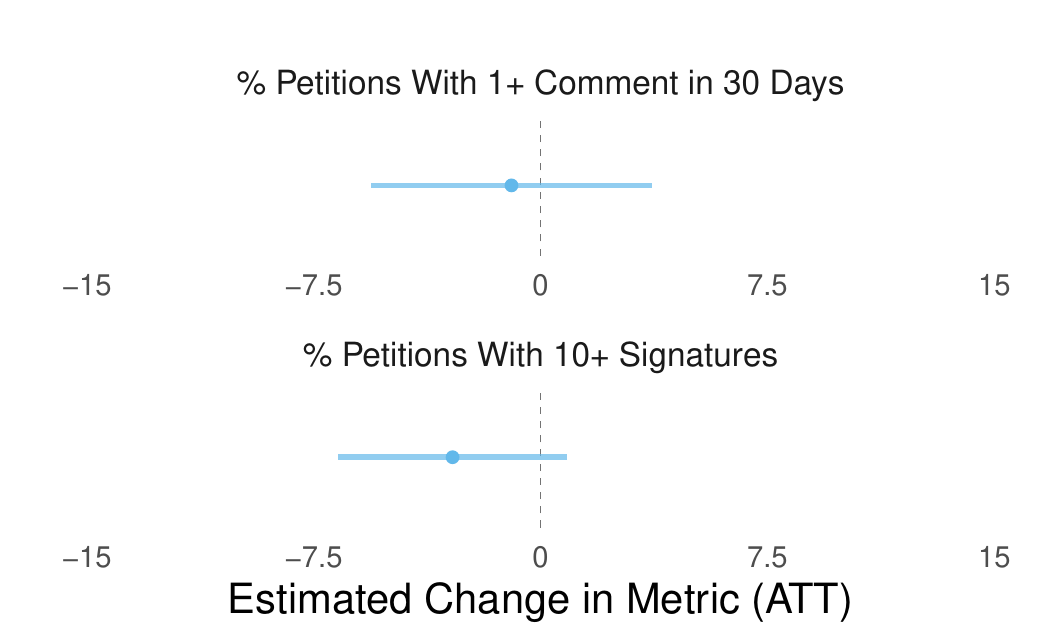}
    \caption{}
\end{subfigure}

\caption[Difference-in-differences with Short Petitions Included]{Dynamic and static difference-in-difference estimates for ATT estimate for lexical features (a,b) and outcomes (c,d) using a full dataset that did not exclude short text (fewer than 5 words; 2 sentences). This analysis includes 322,217 petitions posted during the pre-AI period, 104,843 petitions posted during the A/B test period, and 48,803 petitions posted during the post-AI period.}
\label{fig:short_text_included_did}
\end{figure}

\begin{figure}[H]
\centering
\begin{subfigure}[b]{0.48\textwidth}
    \includegraphics[width=\textwidth]{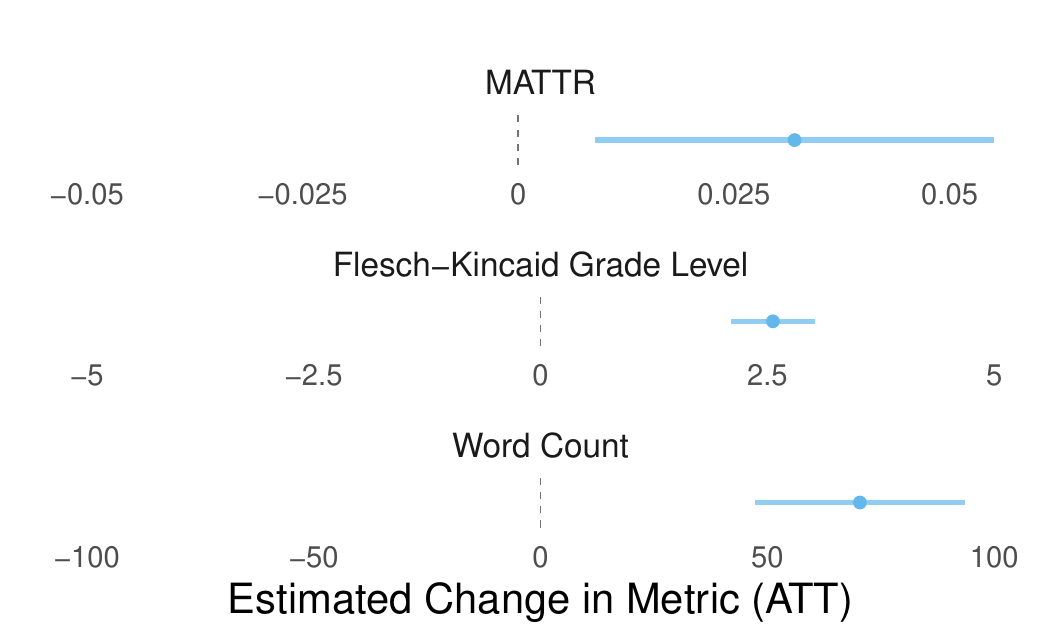}
    \caption{}
\end{subfigure}
\hfill
\begin{subfigure}[b]{0.48\textwidth}
    \includegraphics[width=\textwidth]{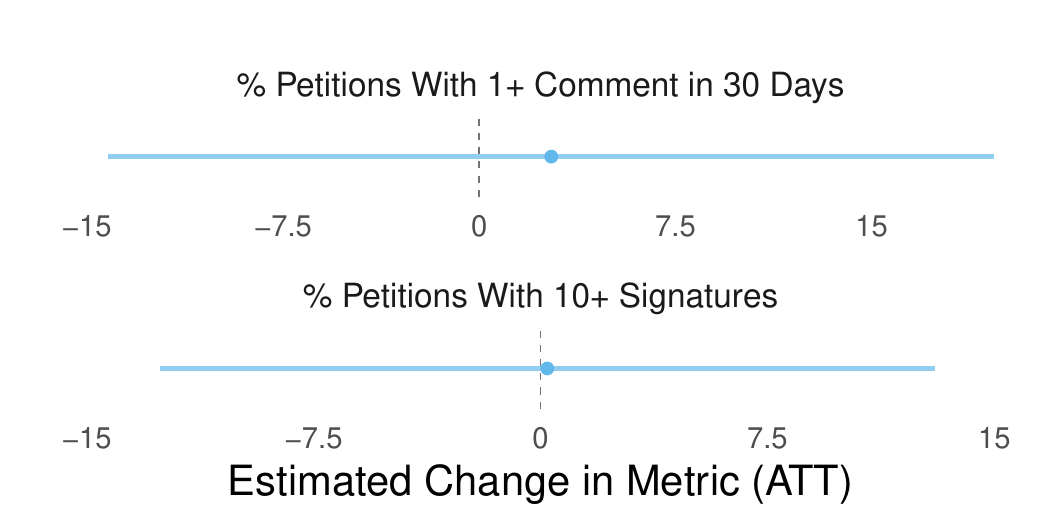}
    \caption{}
\end{subfigure}

\caption[Difference-in-differences with Covariates]{Static difference-in-differences with covariates (the share of petition-writers that had commented previously, the share of petition-writers that were new to the platform, the share of petitions with decision makers labeled, and the share of petitions that had been posted over the weekend). Panel (a) shows average treatment effect on treated units (ATT) for lexical features, (b) shows average treatment effect on treated units (ATT) for outcomes. Estimates are doubly robust. Figure shows 95\% CI.
Pre-AI period was 65 weeks (N = 237,020 petitions), A/B test period was 26 weeks (N = 82,504 petitions), and the Post-AI period was 11 weeks (N = 42,111 petitions).}
\label{fig:full_covar_did}
\end{figure}

\begin{figure}[H]
\centering
\begin{subfigure}[b]{0.48\textwidth}
    \includegraphics[width=\textwidth]{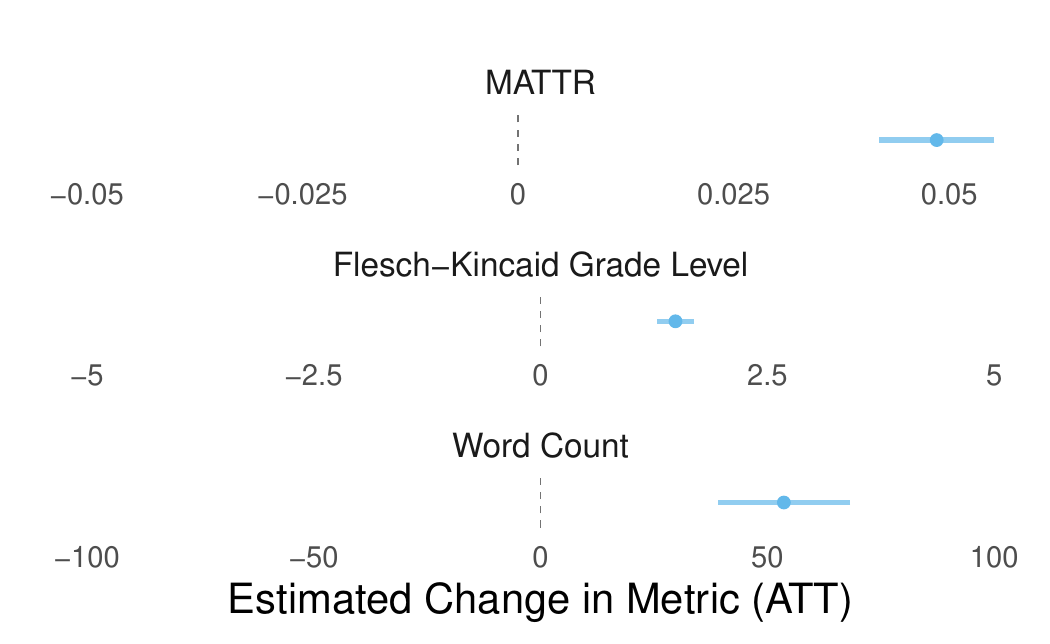}
    \caption{}
\end{subfigure}
\hfill
\begin{subfigure}[b]{0.48\textwidth}
    \includegraphics[width=\textwidth]{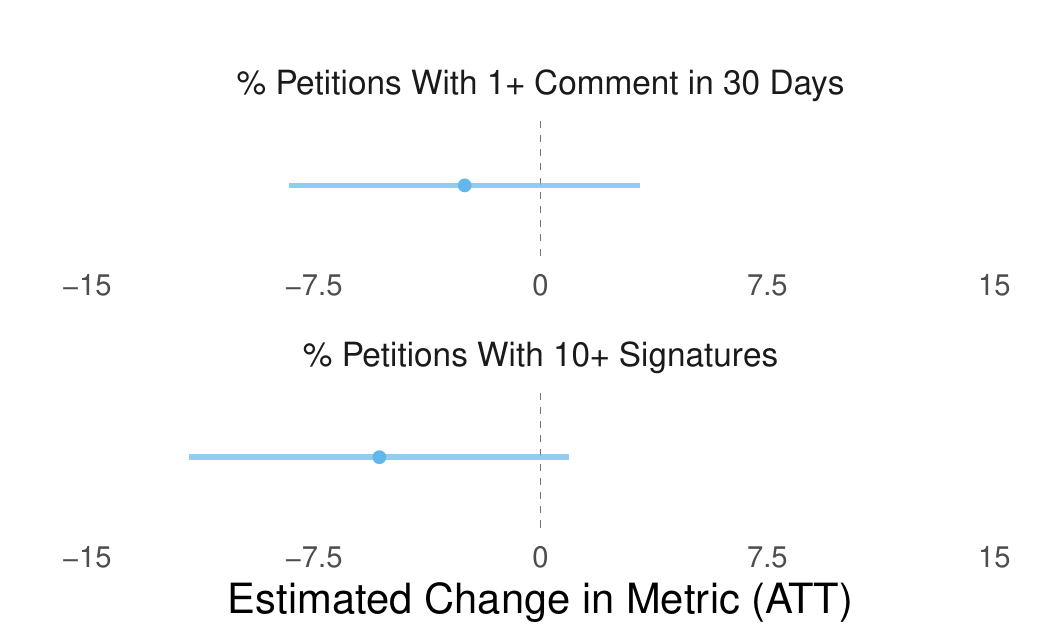}
    \caption{}
\end{subfigure}

\caption[Difference-in-differences with SE Clustered By Country]{Static difference-in-differences with standard errors clustered by country. Panel (a) shows average treatment effect on treated units (ATT) for lexical features, (b) shows average treatment effect on treated units (ATT) for outcomes. The figure shows 95\% CI.
Pre-AI period was 65 weeks (N = 237,020 petitions), A/B test period was 26 weeks (N = 82,504 petitions), and the Post-AI period was 11 weeks (N = 42,111 petitions).}
\label{fig:did_se_country_clustered}
\end{figure}

\begin{figure}[H]
  \centering
  \includegraphics[width=.7\textwidth]{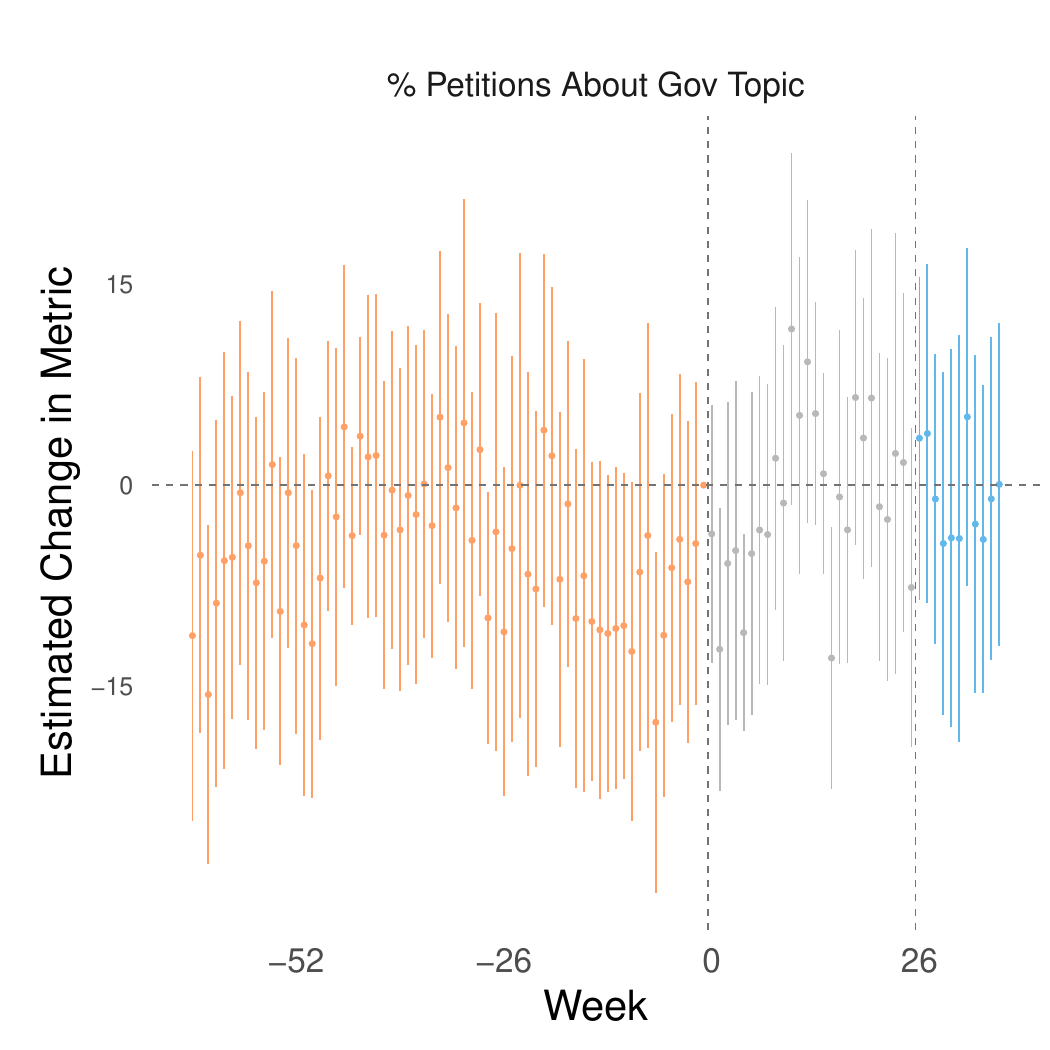}
  \caption[Difference-in-differences for Share of Petitions about Government Topics]{Dynamic difference-in-differences for the percent of petitions that are about a government topic. 
  The y-axis reflects the percent of petitions published that week that we labeled as related to a government or political topic. 
  The figure shows that post-AI weeks do not differ from pre-AI baseline (Week = -1). 
  Pre-AI period was 65 weeks (N = 237,020 petitions), A/B test period was 26 weeks (N = 82,504 petitions), and the Post-AI period was 11 weeks (N = 42,111 petitions).
  The figure shows 95\% CI.}
  \label{fig:gov_event_study}
\end{figure}

\begin{figure}[H]
\centering
\begin{subfigure}[b]{0.48\textwidth}
    \includegraphics[width=\textwidth]{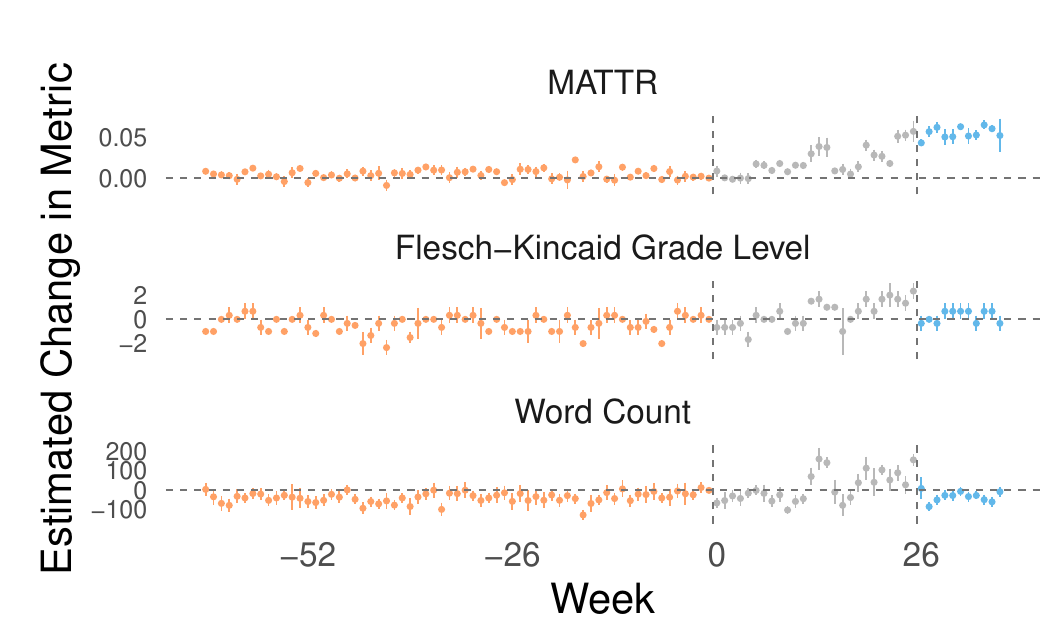}
    \caption{}
\end{subfigure}
\hfill
\begin{subfigure}[b]{0.48\textwidth}
    \includegraphics[width=\textwidth]{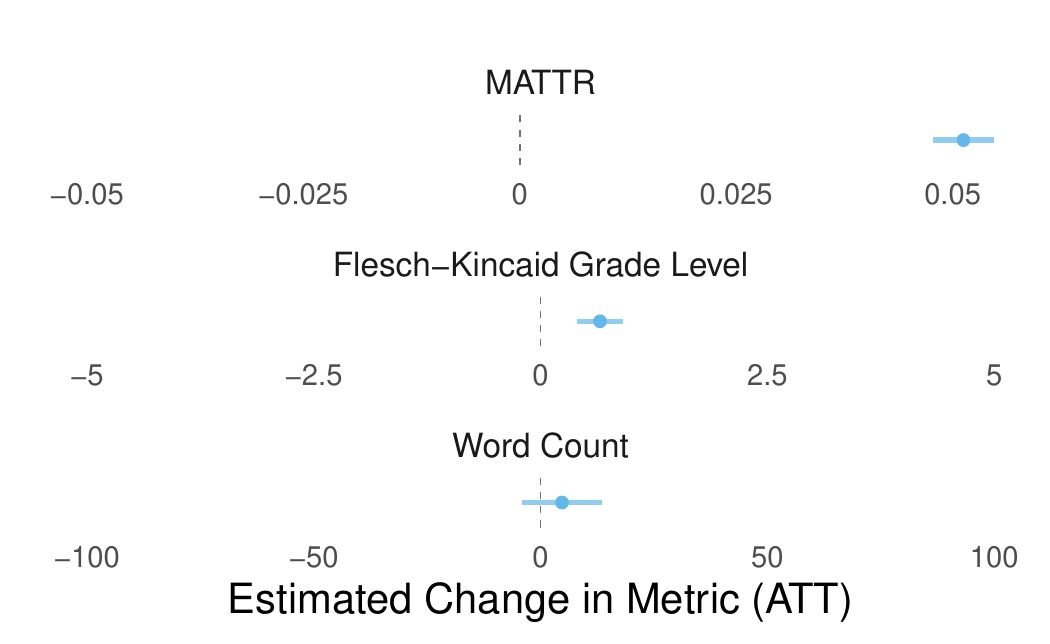}
    \caption{}
\end{subfigure}

\vspace{0.5em}

\begin{subfigure}[b]{0.48\textwidth}
    \includegraphics[width=\textwidth]{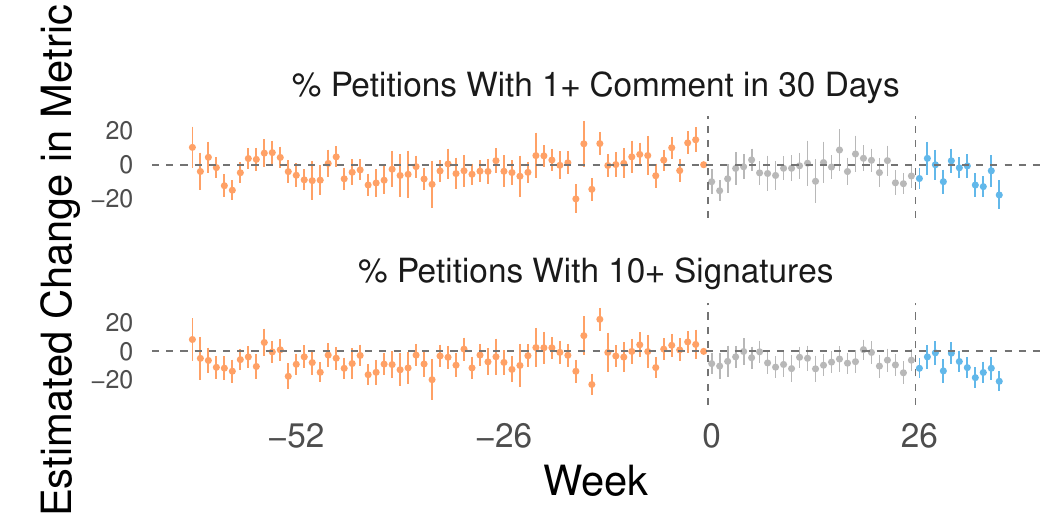}
    \caption{}
\end{subfigure}
\hfill
\begin{subfigure}[b]{0.48\textwidth}
    \includegraphics[width=\textwidth]{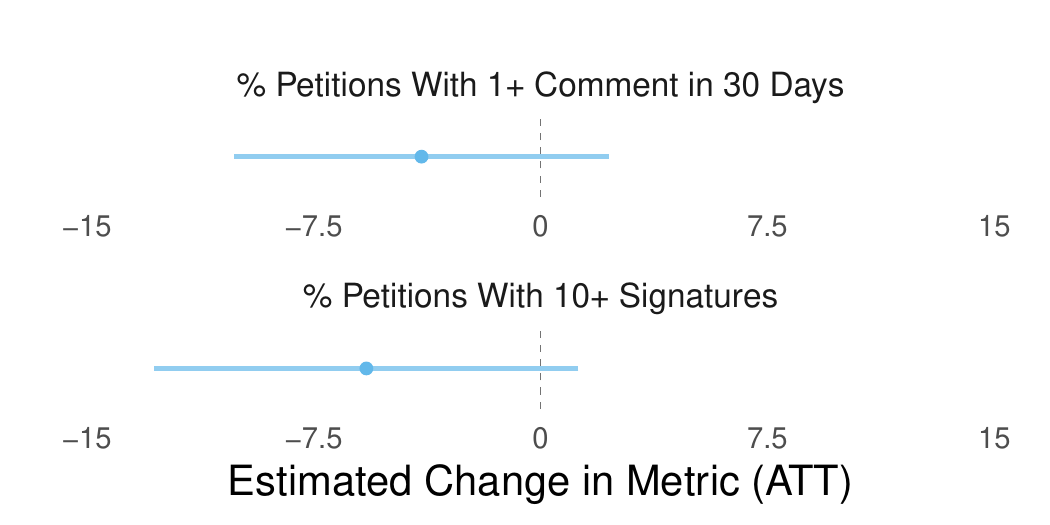}
    \caption{}
\end{subfigure}

\caption[Difference-in-differences for Government Topic Subset]{Difference-in-differences on a subset of petitions labeled as pertaining to a government topic, using 86,638 petitions written pre-AI, 33,957 petitions written during the A/B test period, and 17,051 petitions written post-AI. (a) shows dynamic difference-in-differences of lexical features, (b) static difference-in-differences of lexical features, (c) dynamic difference-in-differences of outcomes, (d) static difference-in-differences of outcomes. The figure shows 95\% point-wise CI.}
\label{fig:did_on_gov_petitions}
\end{figure}

\begin{figure}[H]
\centering
\begin{subfigure}[b]{0.48\textwidth}
    \includegraphics[width=\textwidth]{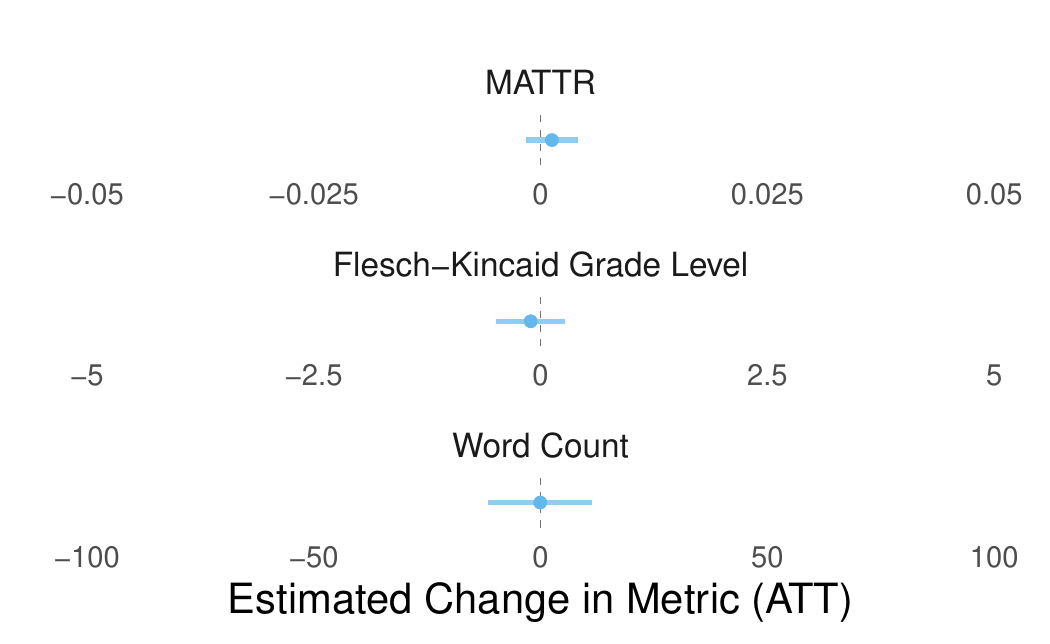}
    \caption{}
\end{subfigure}
\hfill
\begin{subfigure}[b]{0.48\textwidth}
    \includegraphics[width=\textwidth]{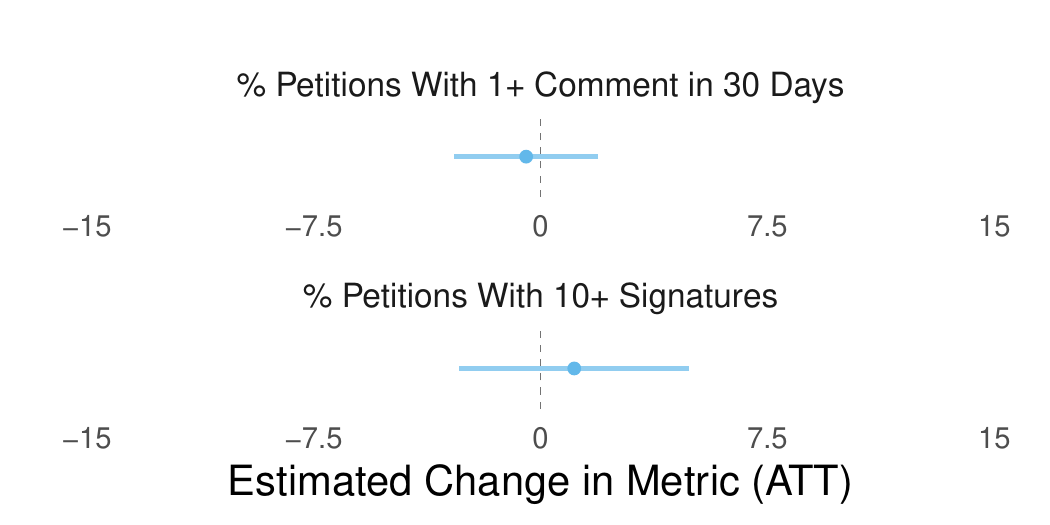}
    \caption{}
\end{subfigure}

\caption[Placebo Difference-in-differences]{Static difference-in-differences estimates using placebo intervention date and A/B test period. Data is from January 3, 2022 to March 31, 2023. The placebo A/B test period is April 1, 2022 -October 1, 2022 and the placebo intervention date is October 1, 2022.
The placebo data includes 13 weeks pre-intervention (59,186 petitions) and 26 weeks post-intervention (87,066 petitions).
Panel (a) shows lexical features, Panel (b) shows outcomes. The figure shows 95\% CI.}
\label{fig:placebo_did}
\end{figure}

\begin{figure}[H]
\centering
\begin{subfigure}[b]{0.7\textwidth}
    \includegraphics[width=\textwidth]{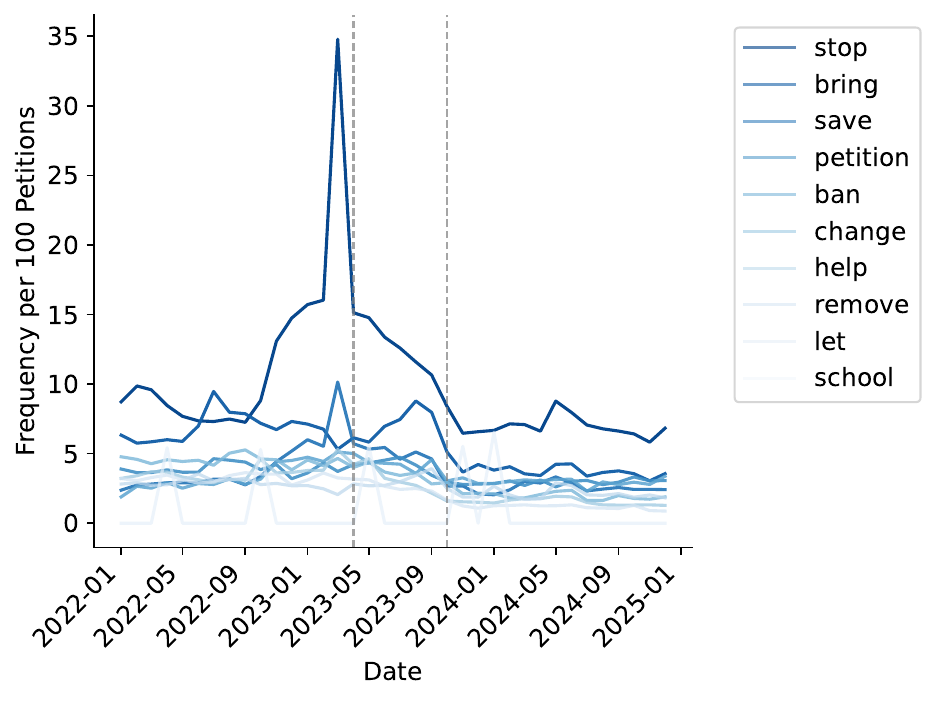}
    \caption{}
    \label{top_words_pre_ai}
\end{subfigure}

\begin{subfigure}[b]{0.7\textwidth}
    \includegraphics[width=\textwidth]{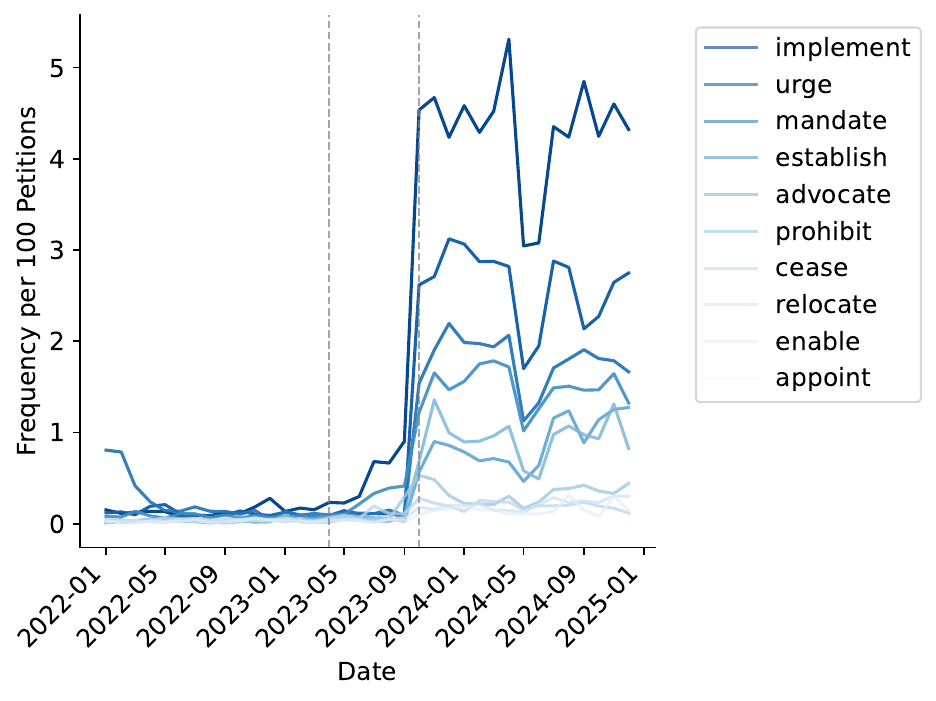}
    \caption{}
    \label{top_words_post_ai}
\end{subfigure}
\caption[Time Series of Petition Title Verb Usage]{
Time series showing most common verbs in petition titles written by users in the United States, Great Britain, and Canada (517,658 petitions).
Both plots show the frequency per 100 words of verbs in petition titles.
Panel (a) shows the 10 verbs that were most common in petition titles before November 2022. We select November 2022 as a threshold to identify terms that were popular before the release of on- or off-platform AI tools (November 2022: ChatGPT release). 
While some verbs spiked with relevant causes (e.g., ``stop'' during the viral \#StoptheWillowProject movement in March, 2023), many of the verbs that were popular in 2022 titles retained consistent use after the introduction of AI.
Panel (b) shows the 10 verbs that had the greatest ratio of post-ChatGPT frequency to pre-ChatGPT frequency.
In both figures, the first dashed line shows the start of A/B testing of the AI feature, the second dashed line shows the start of full AI feature launch. 
The increase in verbs like ``implement'' with access to both general and in-platform AI tools suggests AI tools may change the language used in petitions.
}
\label{fig: Top words}
\end{figure}

\begin{figure}[H]
    \centering
    \includegraphics[width=0.7\textwidth]{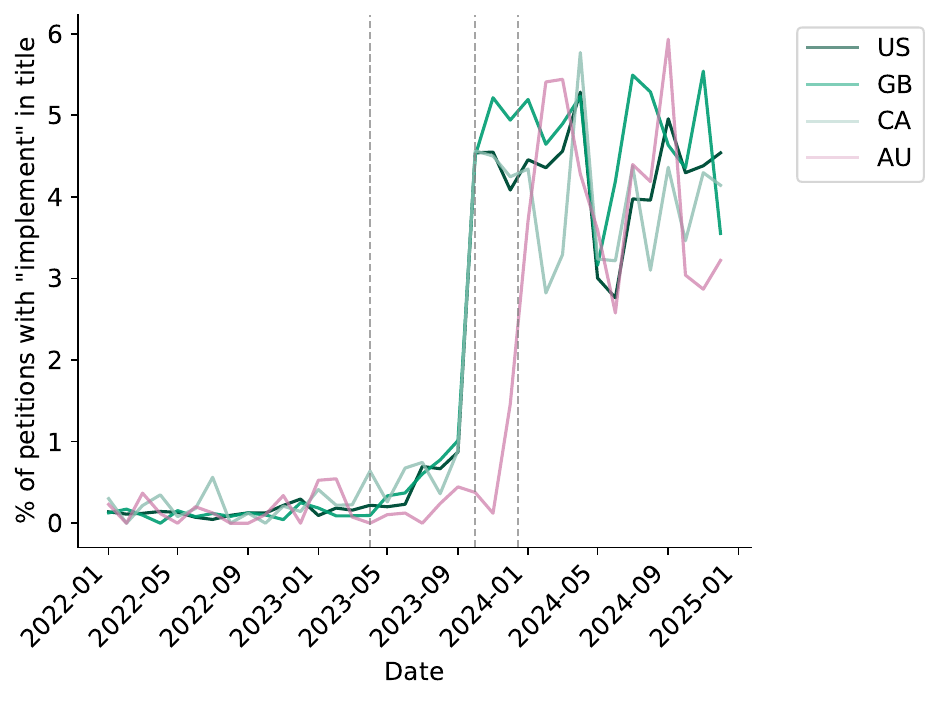}
    \caption[Time Series of ``Implement'' Usage in Petition Titles]{Time series of the percent of petitions per month, by country, that had the word ``implement'' in the title, among petitions written in the United States, Great Britain, Canada, and Australia. The word ``implement'' was the verb with the greatest ratio of post-AI frequency to pre-AI frequency among these countries (549,292 petitions).
    The figure demonstrates that the spike in the verb ``implement'' follows the timeline by which users received access to the in-platform AI tool.}
    \label{fig:implement_by_country}
\end{figure}

\begin{figure}[H]
    \centering
    \begin{subfigure}[b]{0.3\textwidth}
        \centering
        \includegraphics[width=\linewidth]{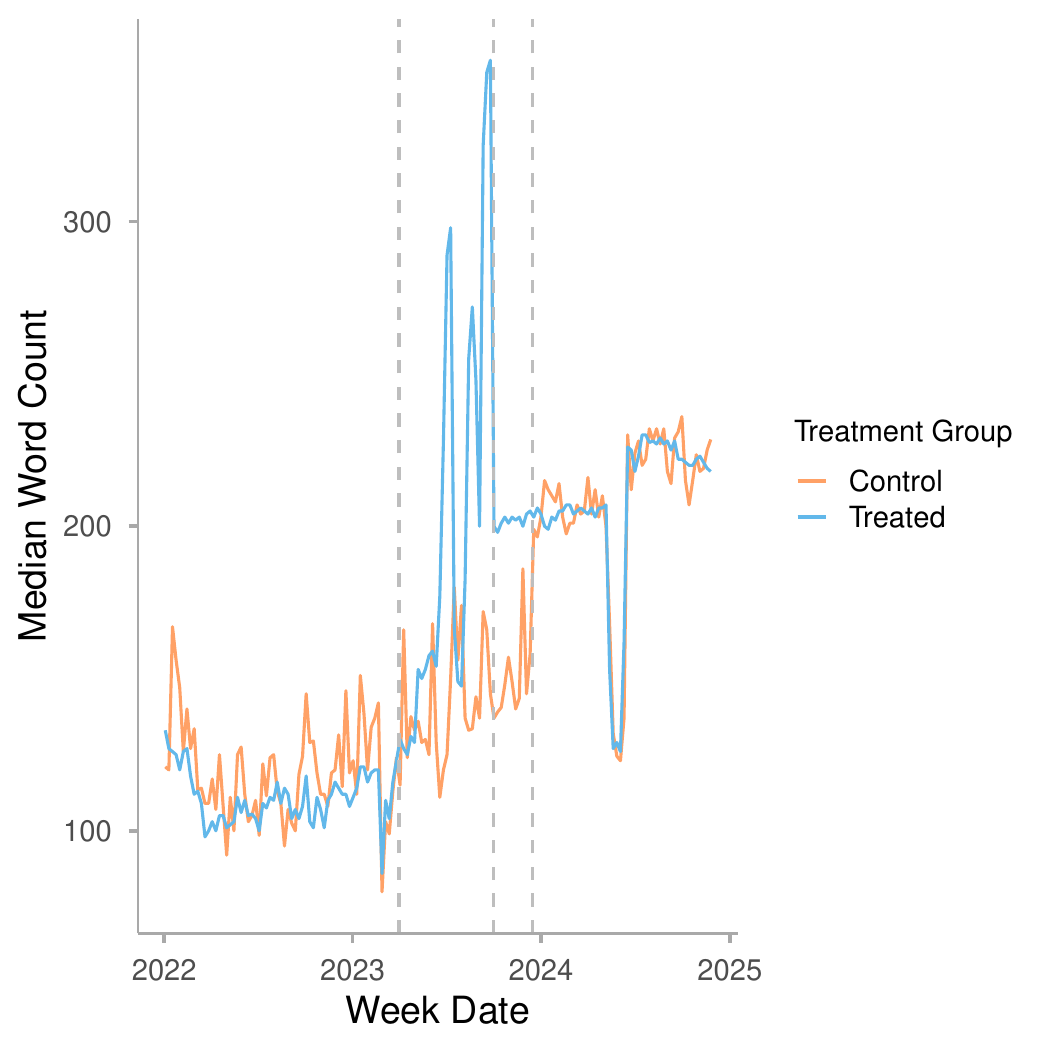}
        \caption{}
        \label{fig:word_count_lineplot}
    \end{subfigure}
    \hfill
    \begin{subfigure}[b]{0.3\textwidth}
        \centering
        \includegraphics[width=\linewidth]{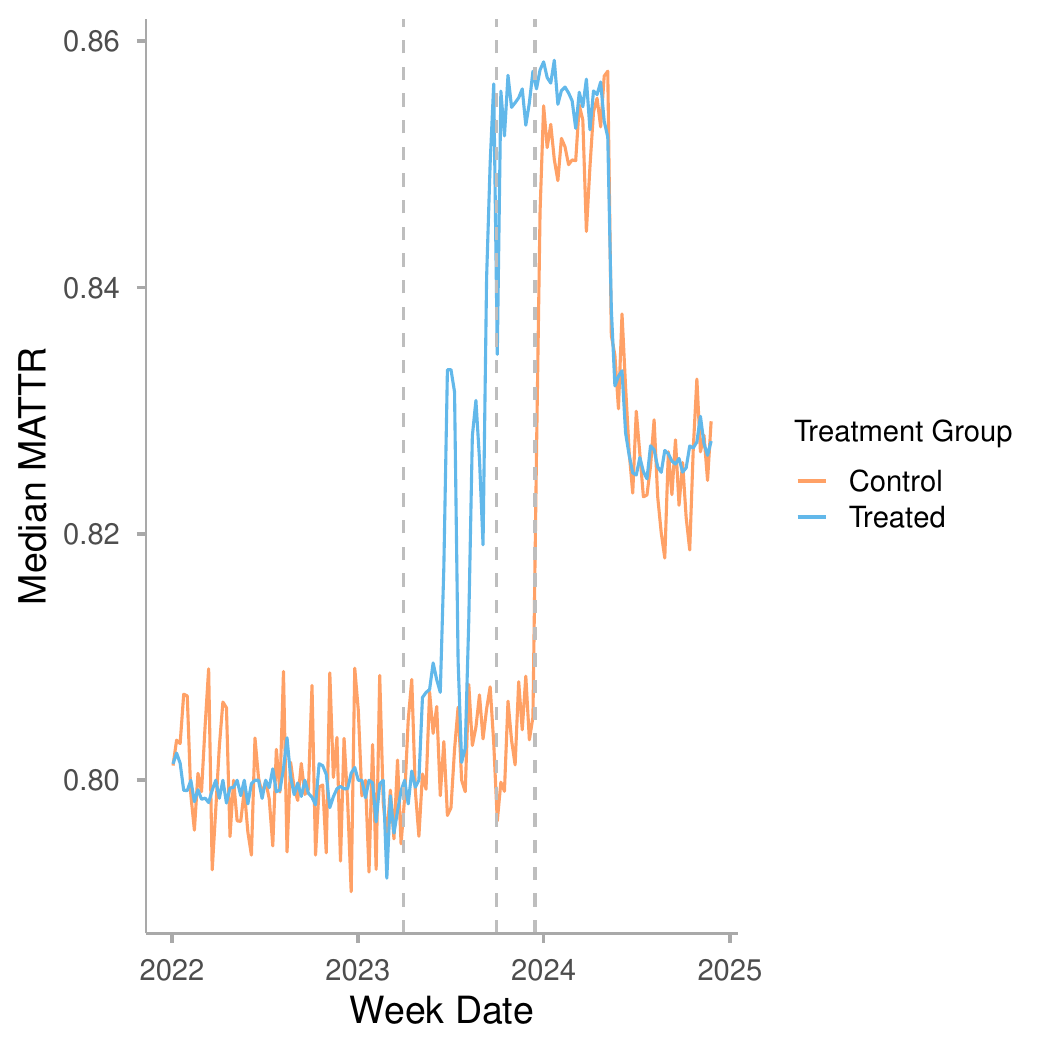}
        \caption{}
        \label{fig:mattr_lineplot}
    \end{subfigure}
    \hfill
    \begin{subfigure}[b]{0.3\textwidth}
        \centering
        \includegraphics[width=\linewidth]{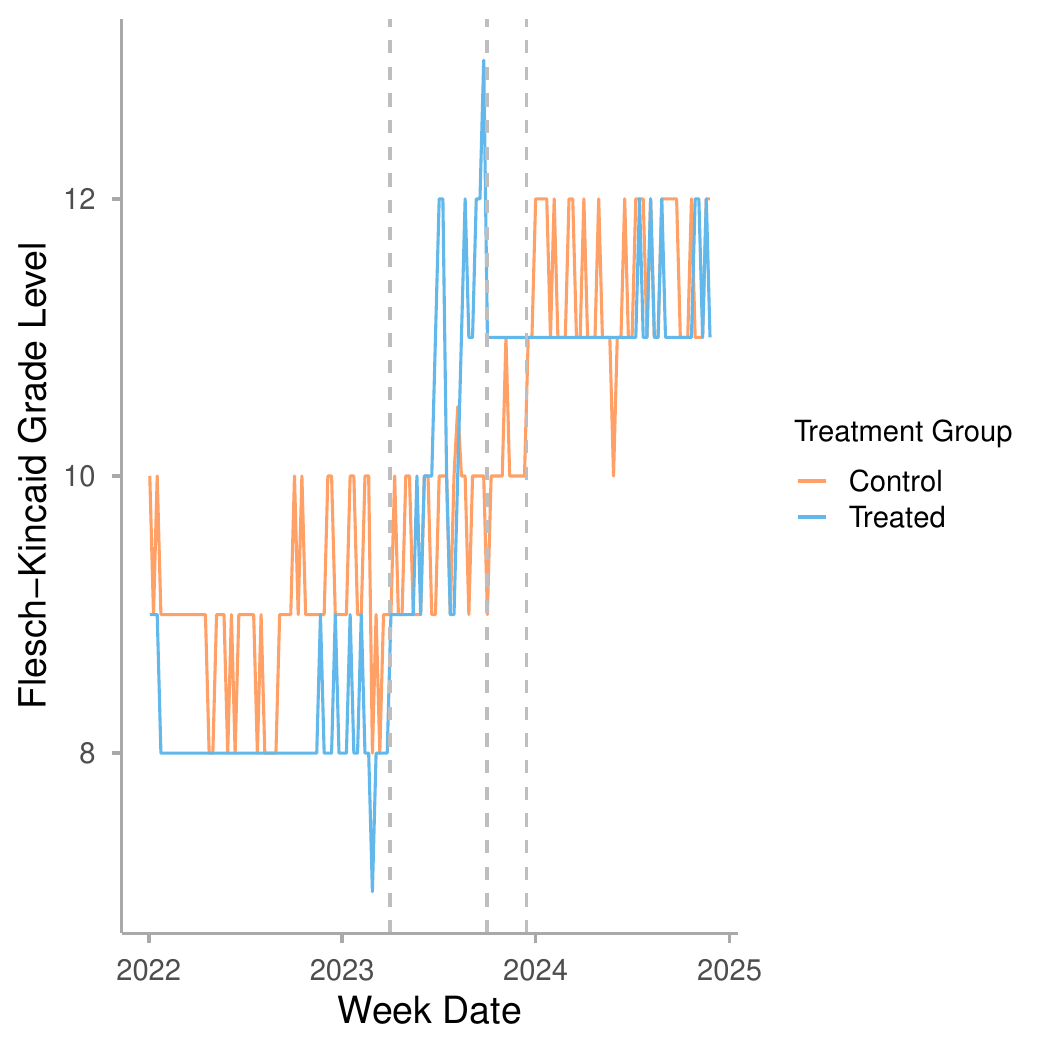}
        \caption{}
        \label{fig:fkgl_lineplot}
    \end{subfigure}

    \vskip\baselineskip
    \begin{subfigure}[b]{0.45\textwidth}
        \centering
        \includegraphics[width=\linewidth]{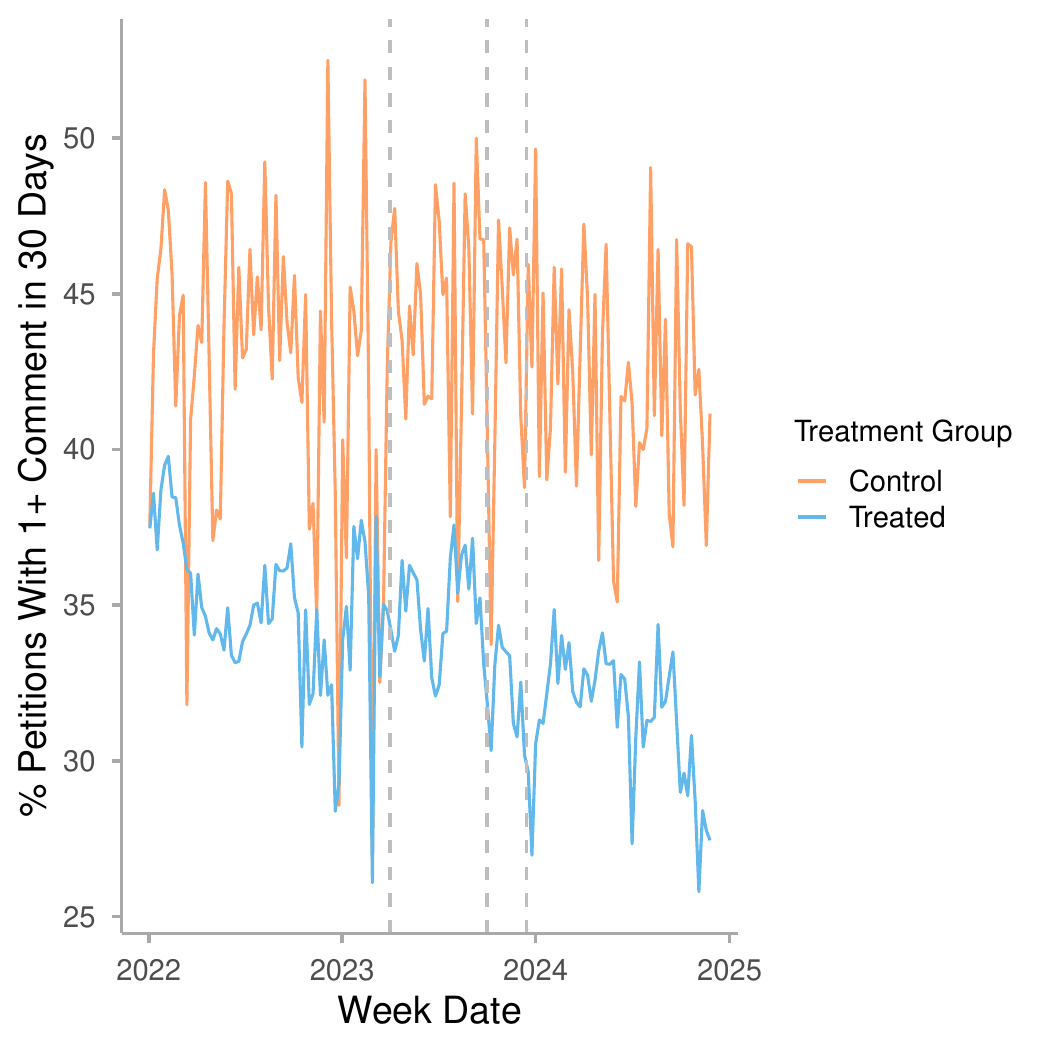}
        \caption{}
        \label{fig:pct_1_comment_lineplot}
    \end{subfigure}
    \hfill
    \begin{subfigure}[b]{0.45\textwidth}
        \centering
        \includegraphics[width=\linewidth]{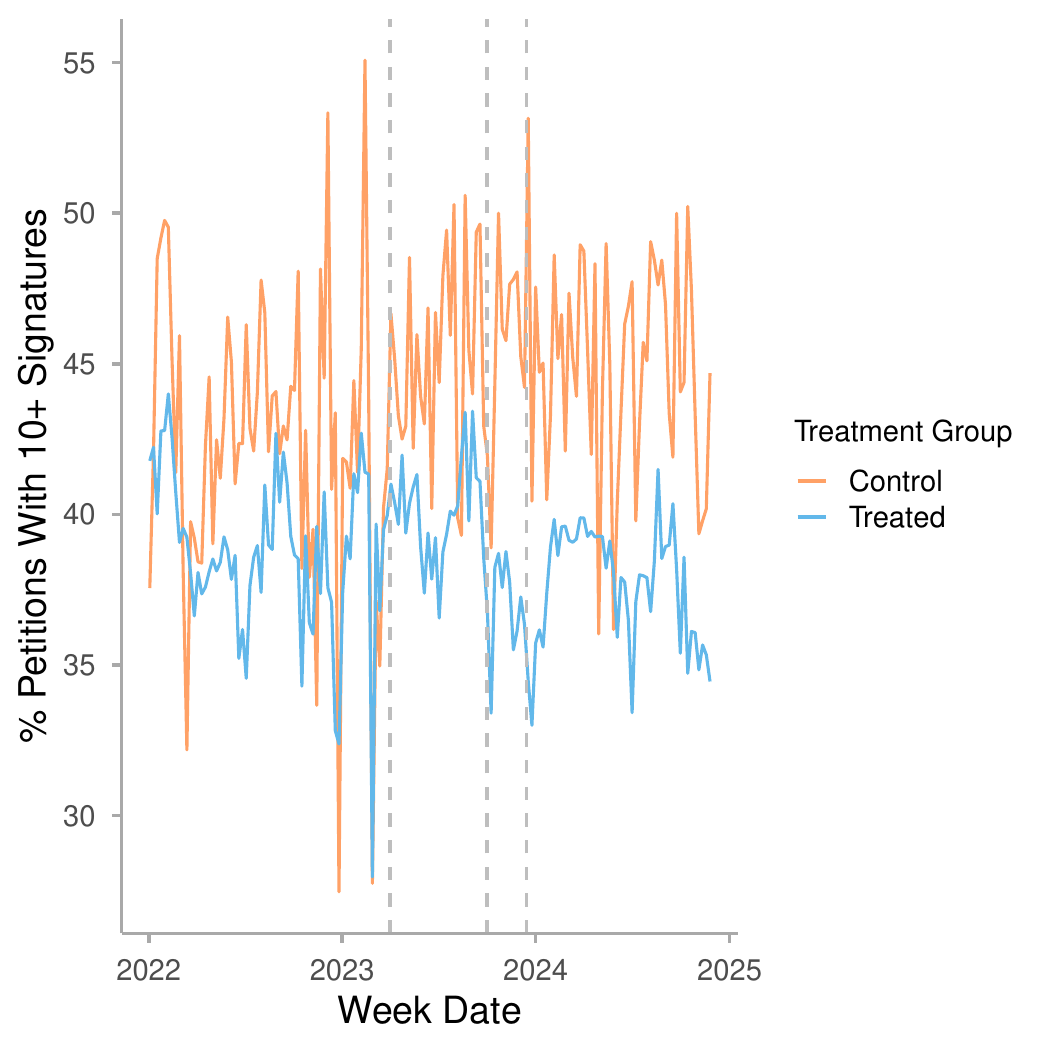}
        \caption{}
        \label{fig:pct_10_sig_lineplot}
    \end{subfigure}

    \caption[Time Series of main Lexical and Outcome Metrics]{Time series of main lexical features and outcome metrics, by week. Treated countries include the United States, Canada, and Great Britain. The control country is Australia. Data is shown for weeks between January 3, 2022 and November 30, 2024 (N = 536,598 petitions).
    We exclude the final month of data collection to have a full month of data for comment and signature outcome metrics.
    Each figure has three dashed vertical lines, from left to right these lines have the intercepts of April 2, 2023; October 2, 2023; December 15, 2023. These lines correspond to the start of A/B test period in the treated countries; the full launch of the AI feature in treated countries; the full launch of the AI feature in Australia. }
    \label{fig:time_series_lexical_features}
\end{figure}

\begin{figure}[H]
\centering
    \begin{subfigure}[b]{0.45\textwidth}
        \includegraphics[width=\textwidth]{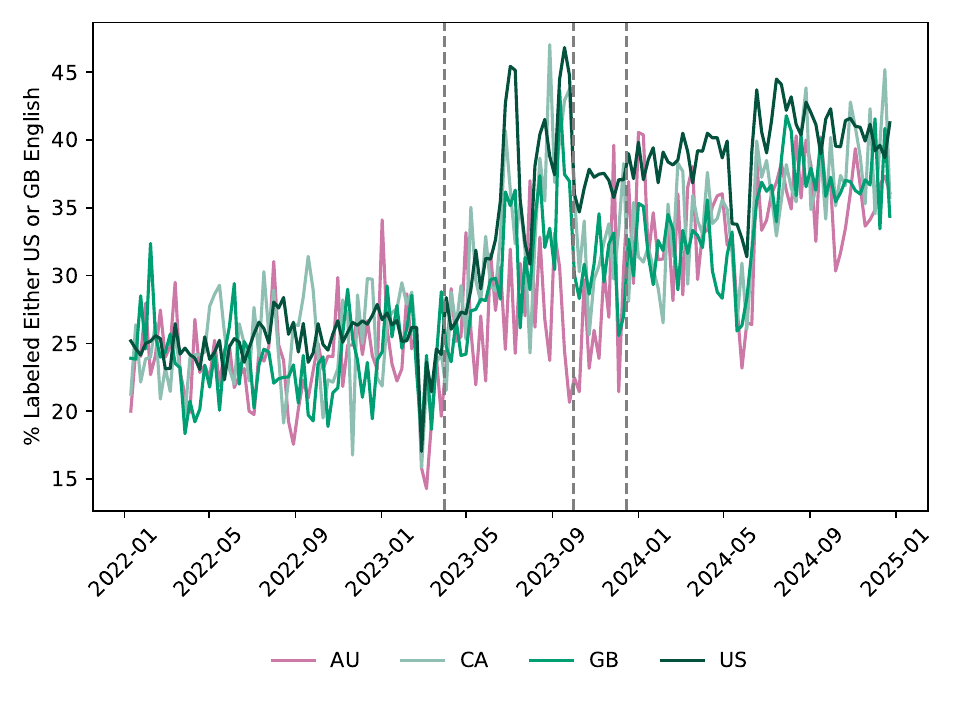}
        \caption{Percent of petitions labeled American or British English}
        \label{subfigure: percent labeled american or british}
    \end{subfigure}

    \vspace{1em} 
    \begin{subfigure}[b]{0.4\textwidth}
        \includegraphics[width=\textwidth]{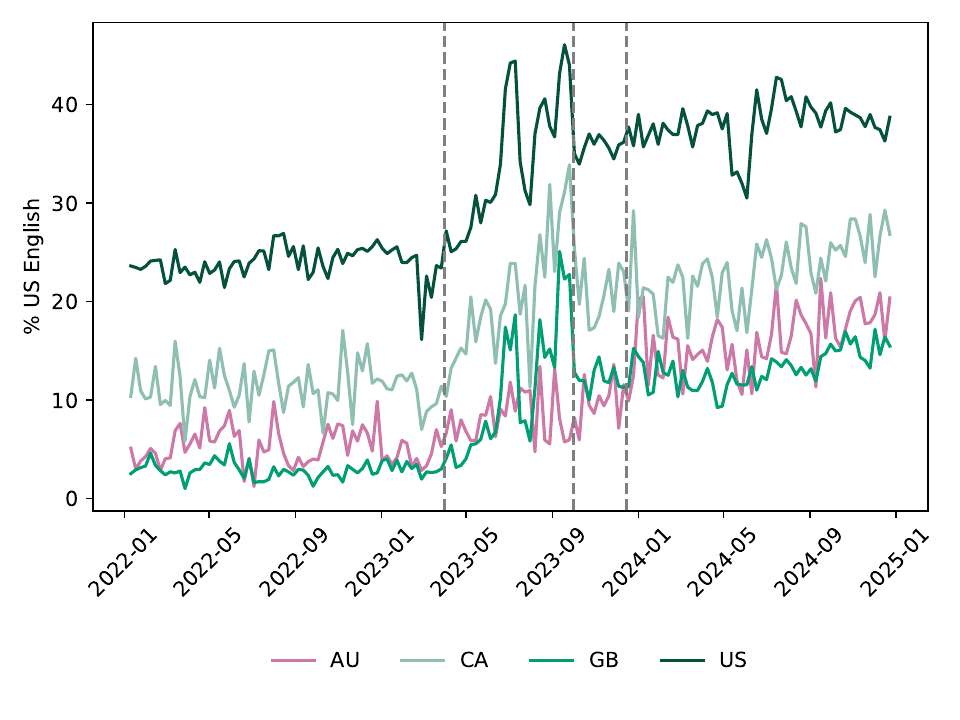}
        \caption{Percent of petitions labeled American English, overall}
        \label{subfigure: percent american english overall}
    \end{subfigure}
    \hfill
    \begin{subfigure}[b]{0.4\textwidth}
        \includegraphics[width=\textwidth]{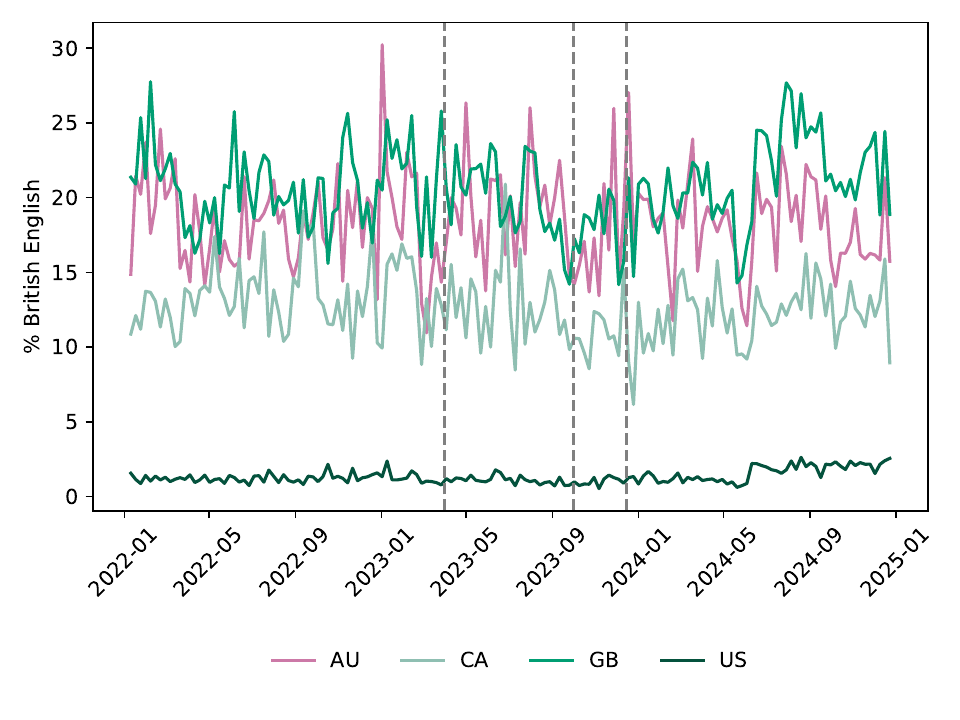}
        \caption{Percent of petitions labeled British English, overall}
        \label{subfigure: percent gb english overall}
    \end{subfigure}
    
    \vspace{1em} 
    \begin{subfigure}[b]{0.4\textwidth}
        \includegraphics[width=\textwidth]{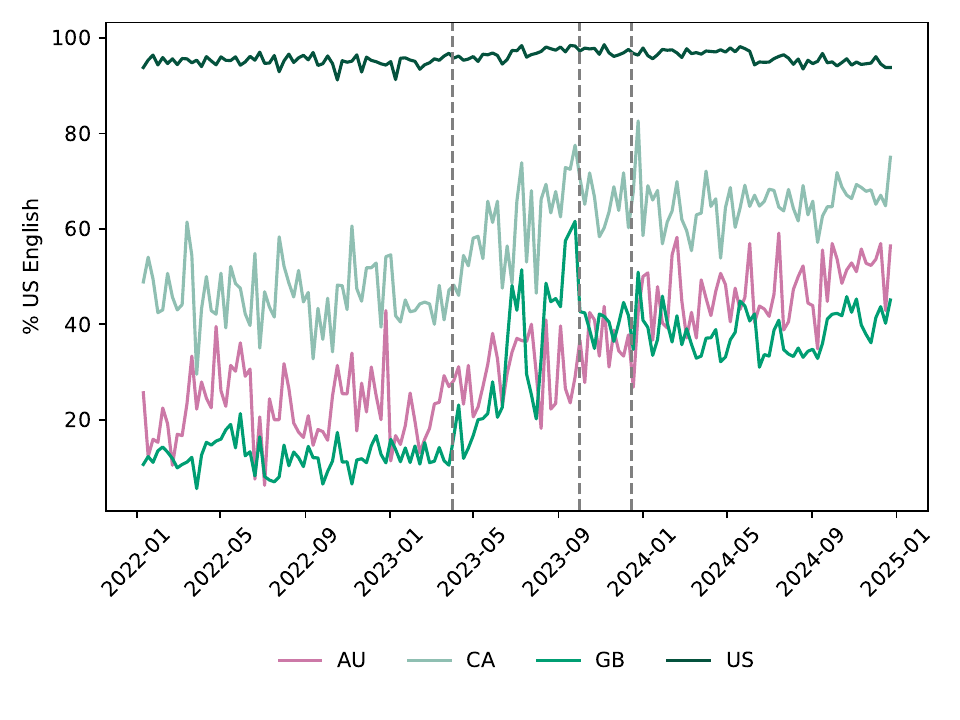}
        \caption{Percent of petitions labeled American English, conditional on receiving a label}
        \label{subfigure: percent american english conditional}
    \end{subfigure}
    \hfill
    \begin{subfigure}[b]{0.4\textwidth}
        \includegraphics[width=\textwidth]{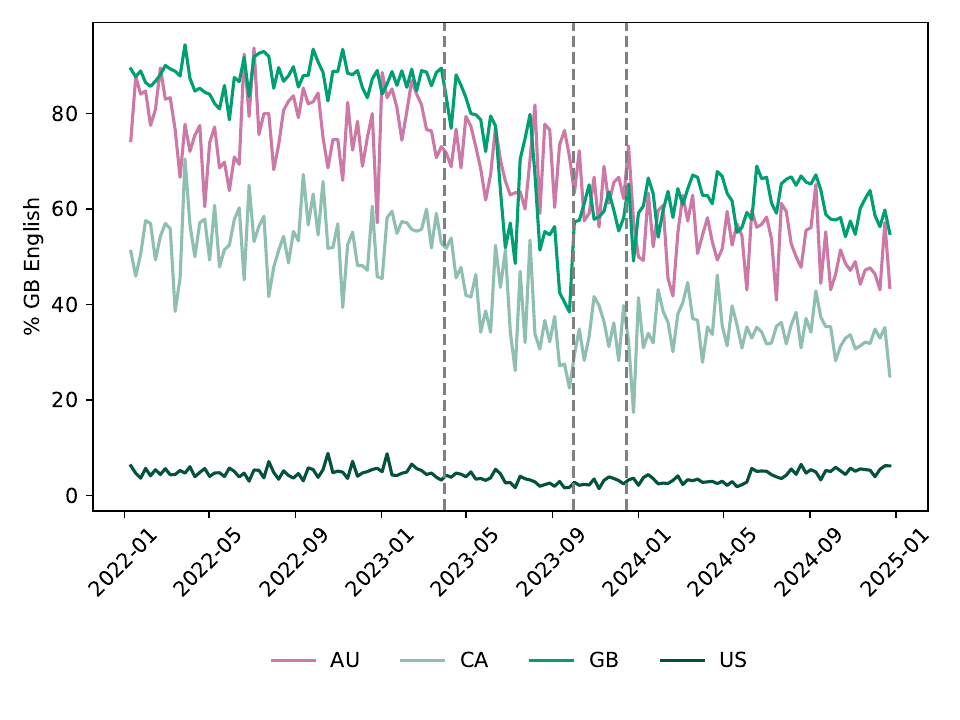}
        \caption{Percent of petitions labeled British English, conditional on receiving a label}
        \label{subfigure: percent gb english conditional}
    \end{subfigure}

\caption[Time Series of English Petitions Classified by English Variety]{Percent of English language petitions by week (x-axis), and country (color saturation), that are classified as (a) either British or American English (b) American English (c) British English, (d) American English, conditional on being assigned a single language variety label, (e) British English, conditional on being assigned a single language variety label. 
549,292 petitions were classified as either British English, American English, Unknown English, or Mix of English varieties.
Of these, 171,875 petitions were assigned a single language variety label (British or American English).
A petition is labeled as either American English or British English using the abclf Python package, which relies on vocabulary and spelling to assign a label.
Overall, the share of petitions that receive a label of either American or British English increases with access to the AI tool: this increase is driven by an increase in petitions labeled as American English.}
\label{fig: english variety}
\end{figure}

\begin{figure}[H]
\centering
\begin{subfigure}[b]{0.48\textwidth}
    \includegraphics[width=\textwidth]{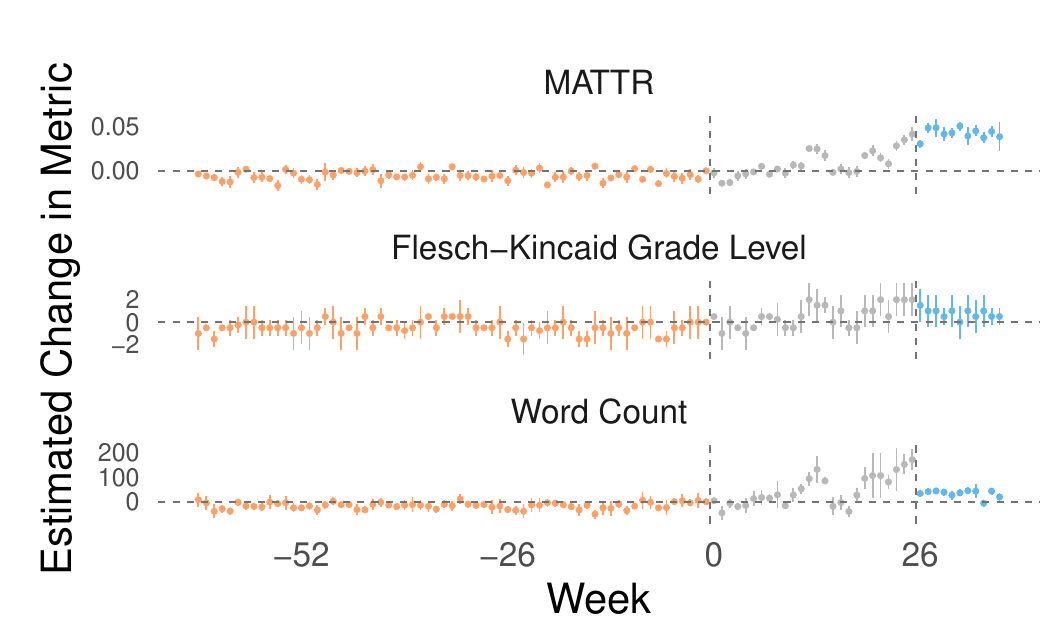}
    \caption{}
\end{subfigure}
\hfill
\begin{subfigure}[b]{0.48\textwidth}
    \includegraphics[width=\textwidth]{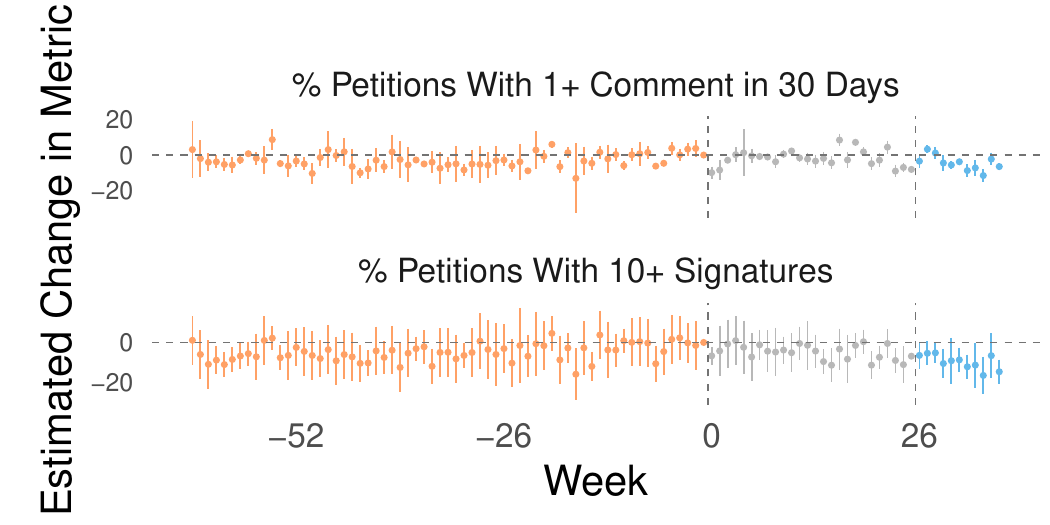}
    \caption{}
\end{subfigure}

\vspace{0.5em}
\begin{subfigure}[b]{0.48\textwidth}
    \includegraphics[width=\textwidth]{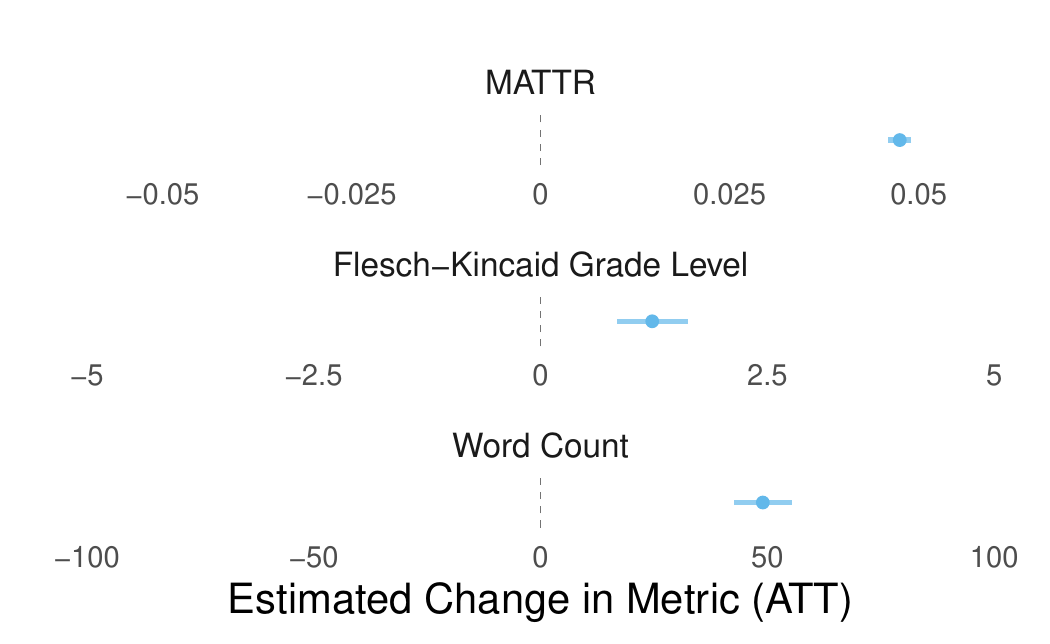}
    \caption{}
\end{subfigure}
\hfill
\begin{subfigure}[b]{0.48\textwidth}
    \includegraphics[width=\textwidth]{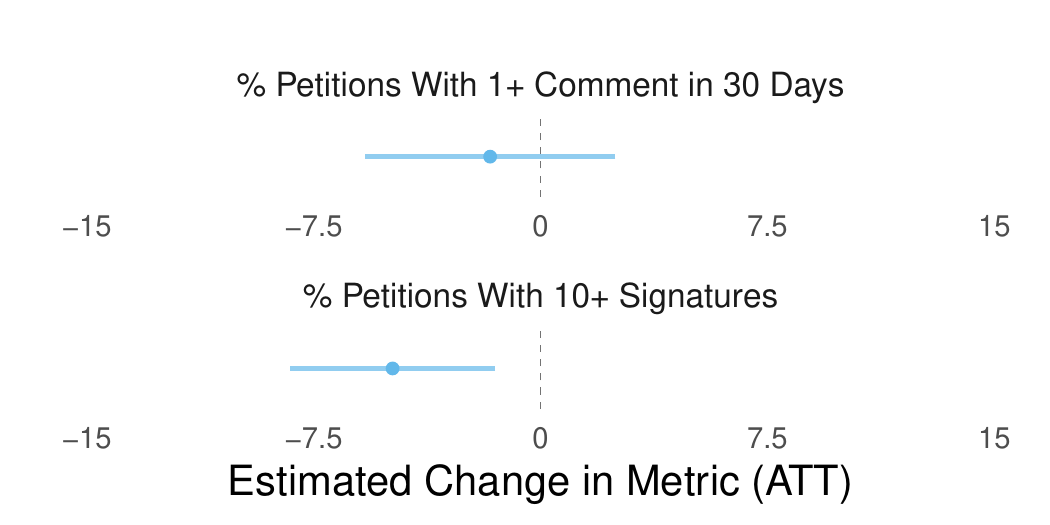}
    \caption{}
\end{subfigure}

\caption[Difference-in-differences with only U.S. as Treated Group]{Static and dynamic difference-in-differences estimates without the United States as a treated group. Treated group includes Great Britain and Canada; control group contains Australia. Analysis includes 68,264 pre-AI petitions, 24,401 petitions written during the A/B period, and 12,479 petitions written post-AI.
(a) Dynamic difference-in-differences with lexical features, (b) dynamic difference-in-differences with outcomes, (c) static difference-in-differences with lexical features, (d) static difference-in-differences with outcomes. 
The figure shows 95\% CI.
Results show significant changes to lexical features and a lack of improved outcomes, which are consistent with difference-in-difference results that include the United States as a treated country.
We show point-wise, rather than simultaneous, 95\% CI bands to account for weeks lacking sufficient variation to estimate week-group ATT.}
\label{fig:no_us_did}
\end{figure}

\begin{figure}[H]
\centering
\begin{subfigure}[b]{0.48\textwidth}
    \includegraphics[width=\textwidth, height=0.3\textheight, keepaspectratio]{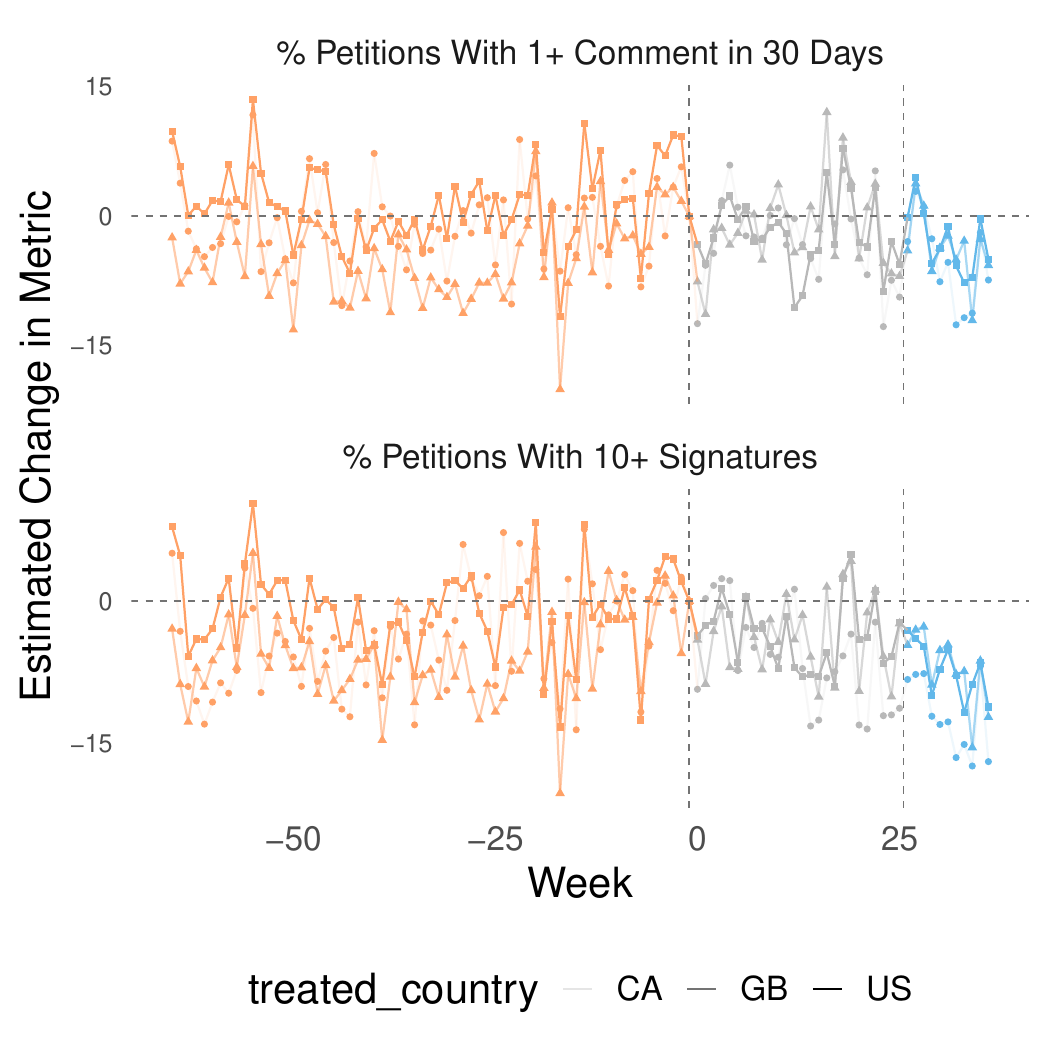}
    \caption{}
\end{subfigure}
\hfill
\begin{subfigure}[b]{0.48\textwidth}
    \includegraphics[width=\textwidth, height=0.3\textheight, keepaspectratio]{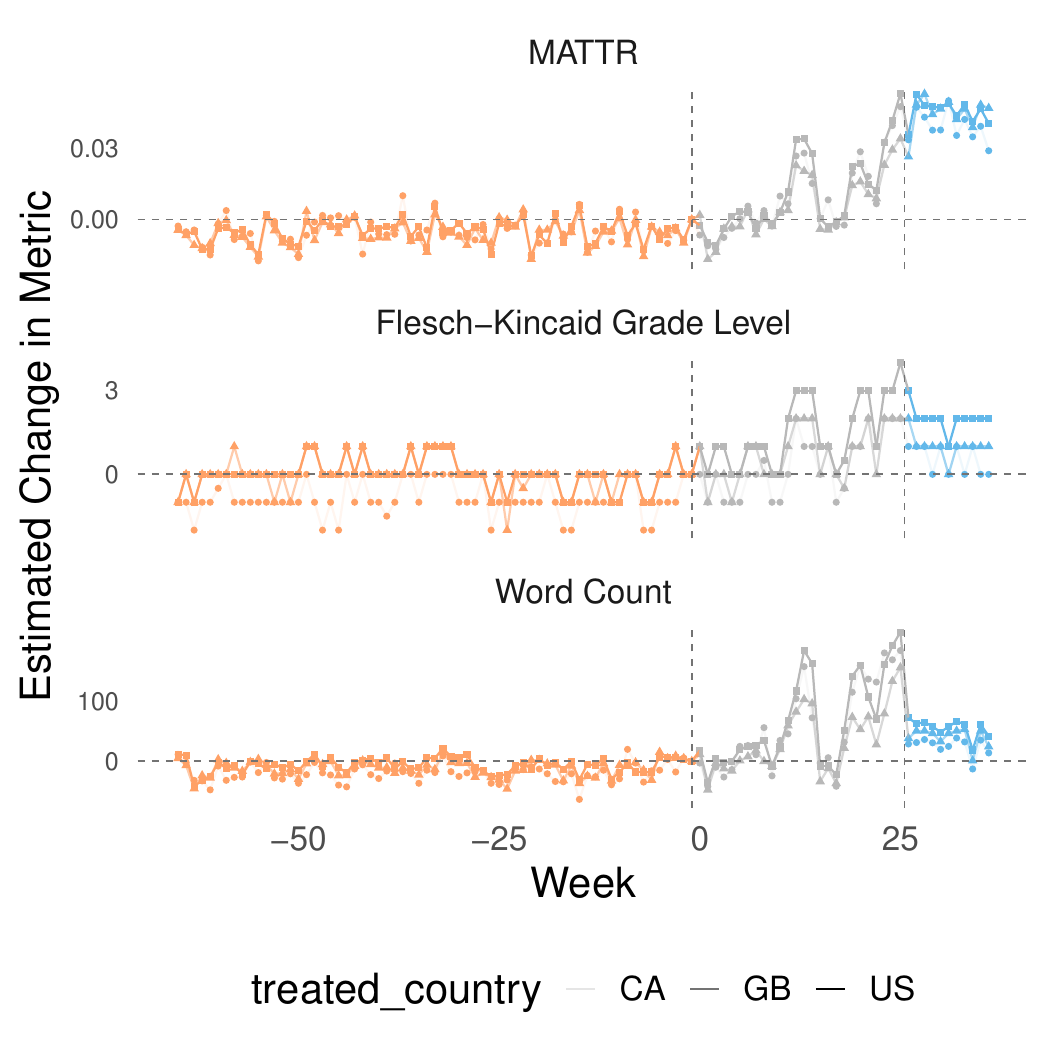}
    \caption{}
\end{subfigure}
\caption[Difference-in-differences with Each Country Compared Alone to Australia]{Dynamic difference-in-difference estimates comparing each treated country (United States: N = 256,491 petitions; Canada: N = 25,250 petitions; Great Britain: N = 59,651 petitions) to Australia (N = 20,243 petitions). Each point reflects the difference between the treated country's weekly metric value and Australia's weekly metric value; relative to the baseline week. The figure demonstrates that shifts in (a) outcomes and (b) lexical features are consistent across countries.}
\label{fig:single_country_comparisons}
\end{figure}

\begin{figure}[H]
\centering
\begin{subfigure}[t]{0.5\textwidth}
    \centering
    \includegraphics[width=\linewidth]{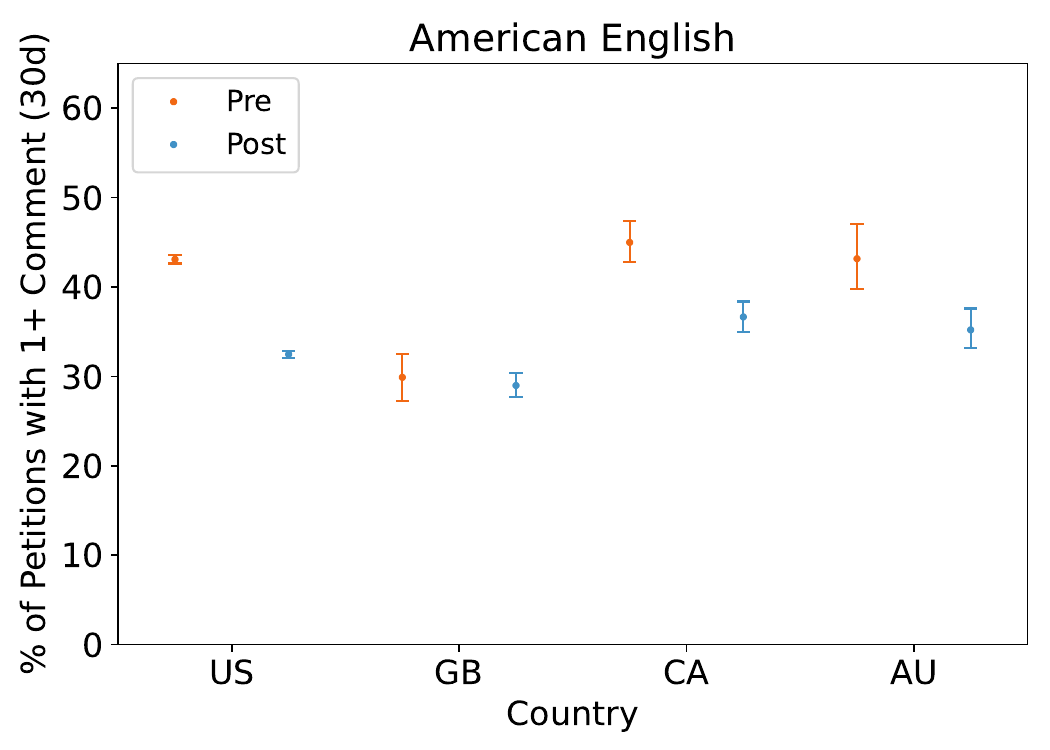}
    \caption{}
    \label{subfig:american english variety}
    \end{subfigure}\hfill
    \begin{subfigure}[t]{0.5\textwidth}
      \centering
      \includegraphics[width=\linewidth]{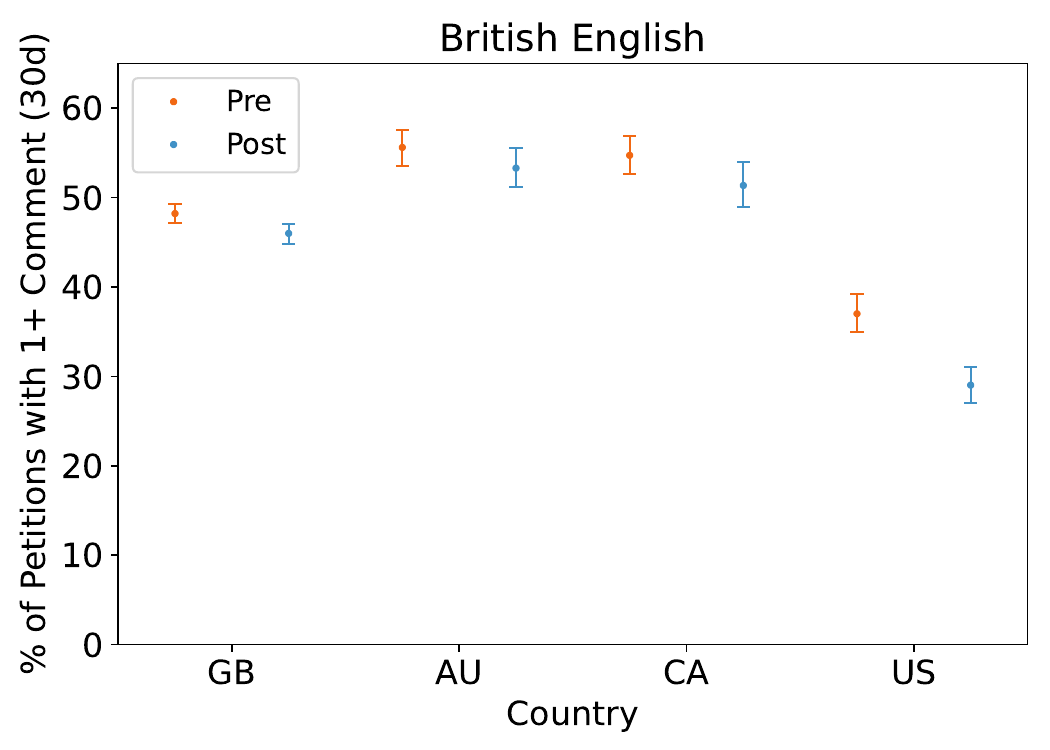}
      \caption{}
      \label{subfig:british english variety}
    \end{subfigure}
    
    \vspace{0.8em}
    
    \begin{subfigure}[t]{0.5\textwidth}
      \centering
      \includegraphics[width=\linewidth]{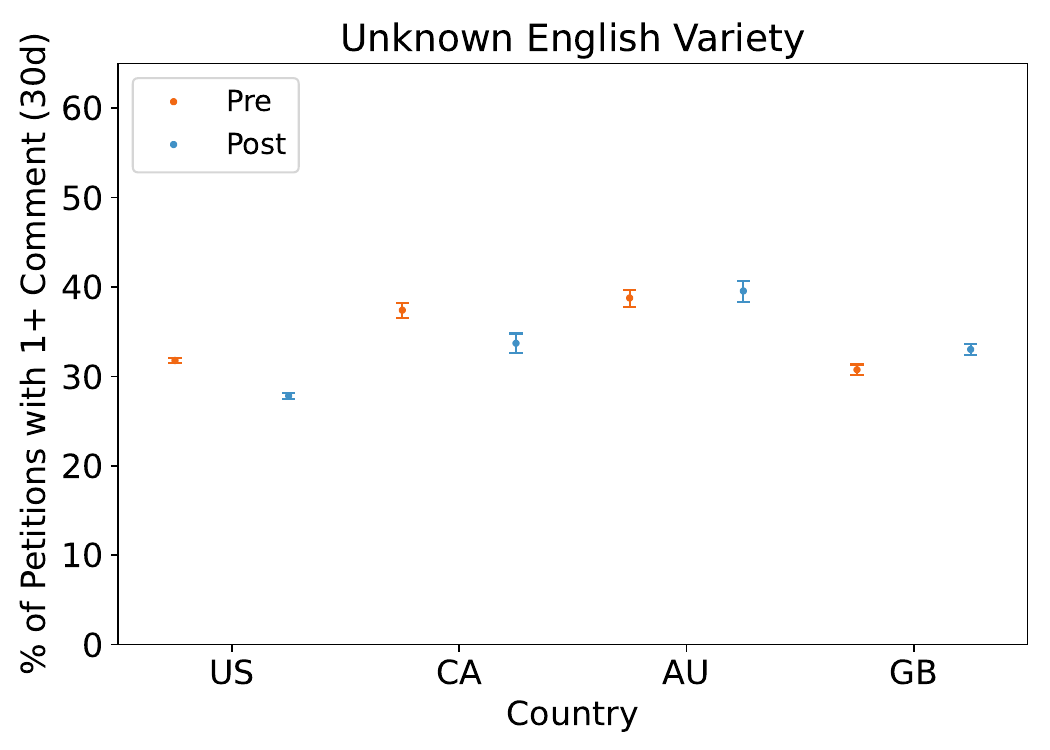}
      \caption{}
      \label{subfig:unknown english variety}
    \end{subfigure}
  \caption[Outcomes by English Variety by Country]{
  The share of petitions that reach the outcome threshold of 1+ comment in 30 days when written in (a) American English, (b) British English, or (c) Unknown English variety. Data is drawn from January 2022 - December 2024, excluding the period of differential access and A/B testing and includes petitions with five or more words and 2 sentences.
  During this period, 100,199 petitions were labeled American English, 27,243 petitions were labeled as British English,
  4,619 labeled a mix of American and British, and 287,188 labeled as Unknown English variety.
  We observe that the share of American English variety petitions that reach 1 comment in 30 days decreases post-AI in the United States, Canada, and Australia, with no observable difference in Great Britain.
  The share of British English petitions that reach 1+ comment in 30 days has the greatest decrease post-AI in the United States, with slight decreases in Canada, Great Britain, and Australia. 
  The share of unknown English variety petitions that reach 1+ comment in 30 days decreases post-AI in the United States, Canada, and Australia, and increases modestly in Great Britain.
  The figure shows 95\% CI.}
  \label{fig:outcomes_english_variety}
\end{figure}

\begin{figure}[H]
  \centering
  \includegraphics[width=0.9\textwidth]{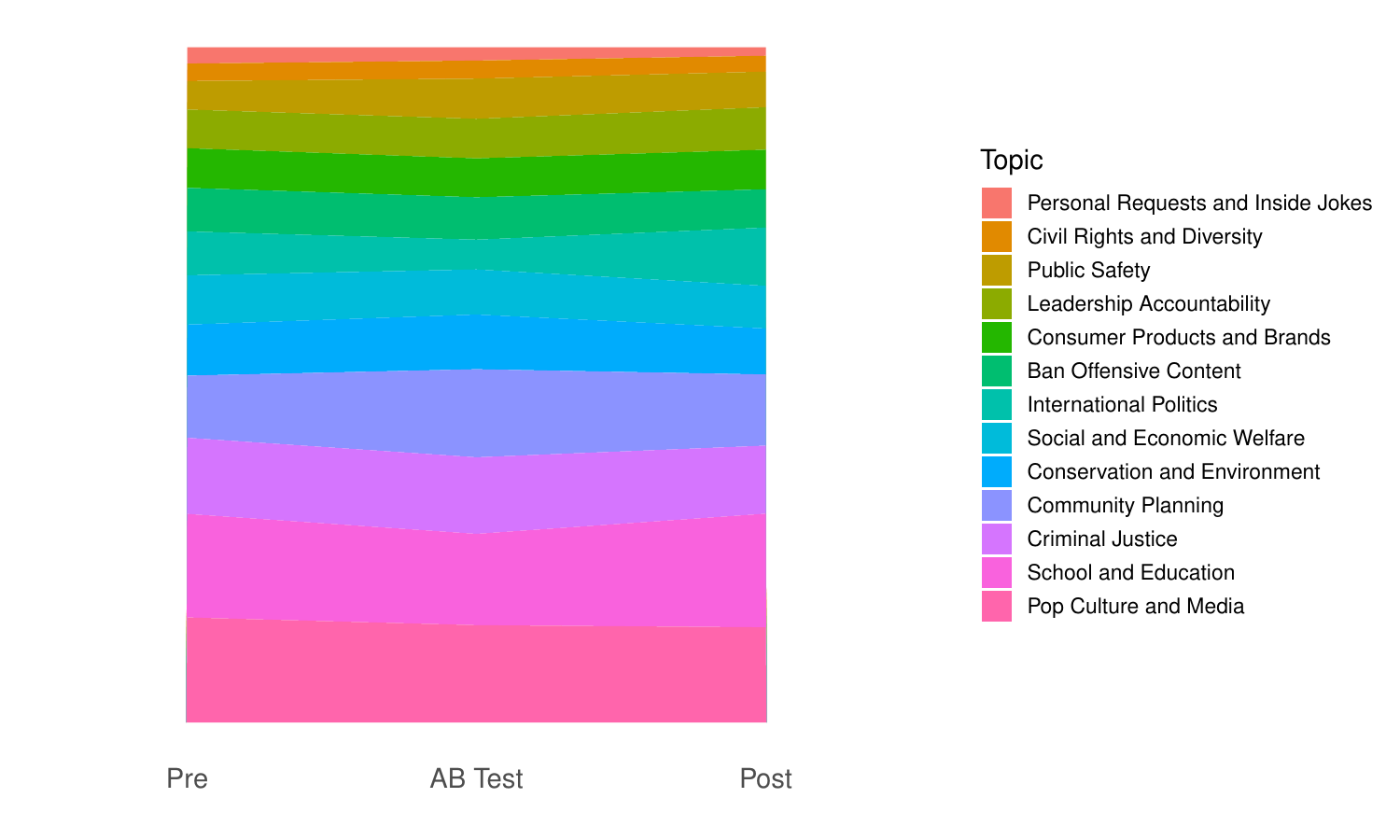}
  \caption[Distribution of Petition Topics]{The distribution of petition topics between January 2022 and December 2023 over pre-AI (N = 237,020 petitions), A/B period (82,504 petitions), and post-AI period (N = 42,111 petitions) in the United States, Great Britain, Australia, and Canada.
  We observe a consistent distribution of petitions by topic across the three periods. 
  A 13x3 chi-square test comparing the distribution of topics over the three periods (pre-AI, A/B period, post-AI) showed period to have a negligible effect on topic distribution ($\chi^2(24)$ = 2950.5, $p<$ 0.0001, Cramer's V=0.064), suggesting little to no relationship between topic and time period, despite statistical significance (likely due to the large sample size).}
  \label{fig:topic distribution}
\end{figure}

\begin{figure}[H]
  \centering
  \includegraphics[width=0.9\textwidth]{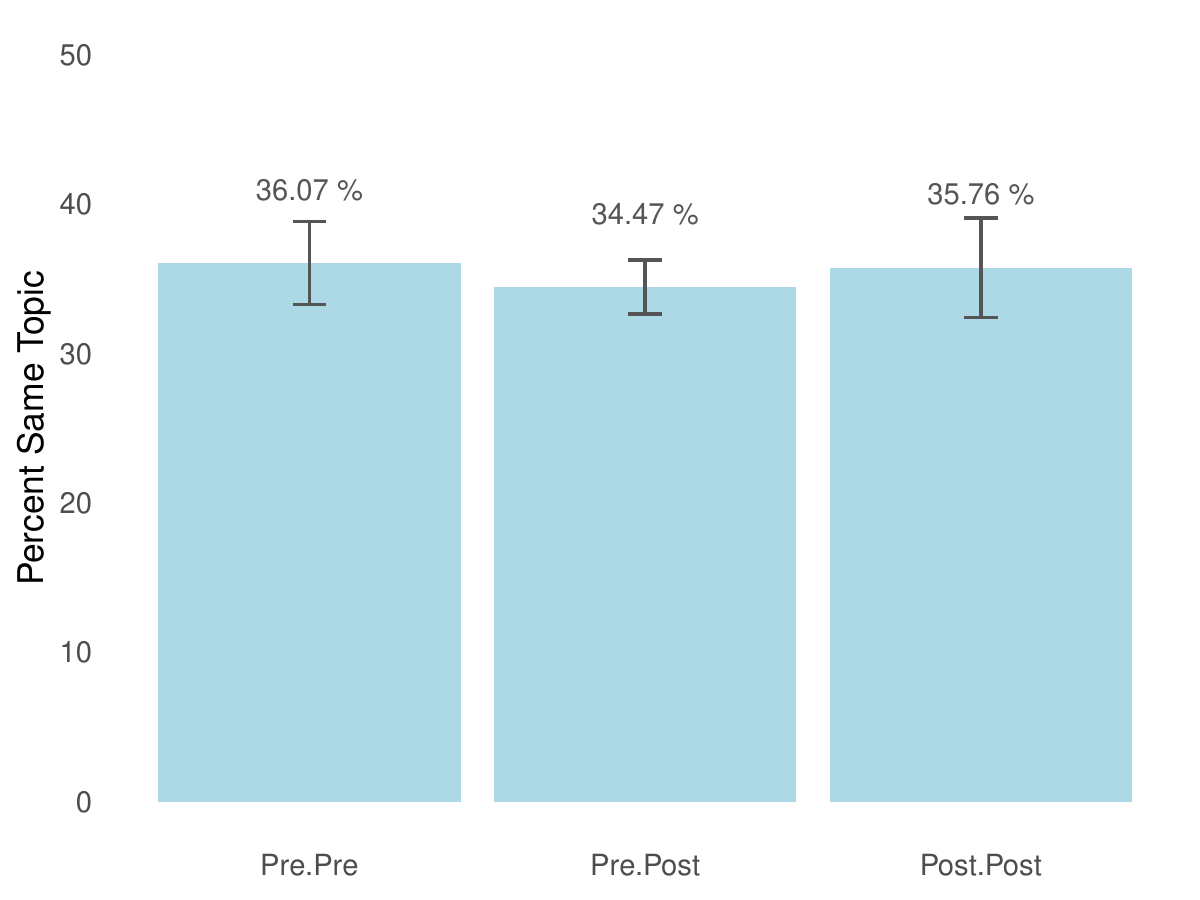}
  \caption[Incident of Same Topic Among Repeat Petition Writers]{
  Among repeat petition-writers included in the main analysis (N=4,611 users), between 34-36\% of users in the Pre/Pre, Pre/Post, and Post/Post cohorts wrote both their first and second petitions about the same topic. The bars represent the proportion of second petitions that had the same topic as the first for the repeat writers in each cohort, with 95\% CI.}
  \label{fig:same topic repeat writers}
\end{figure}

\begin{figure}[H]
  \centering
  \begin{subfigure}[b]{0.48\linewidth}
    \centering
    \includegraphics[width=\linewidth]{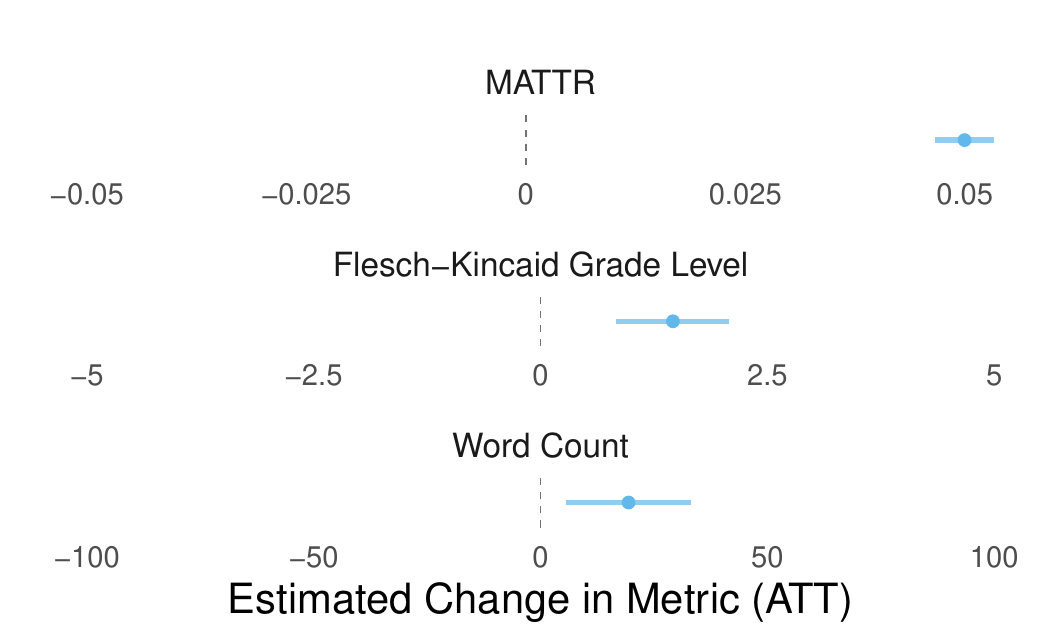}
    \caption{}
    \label{fig:civic engagement style}
  \end{subfigure}\hfill
  \begin{subfigure}[b]{0.48\linewidth}
    \centering
    \includegraphics[width=\linewidth]{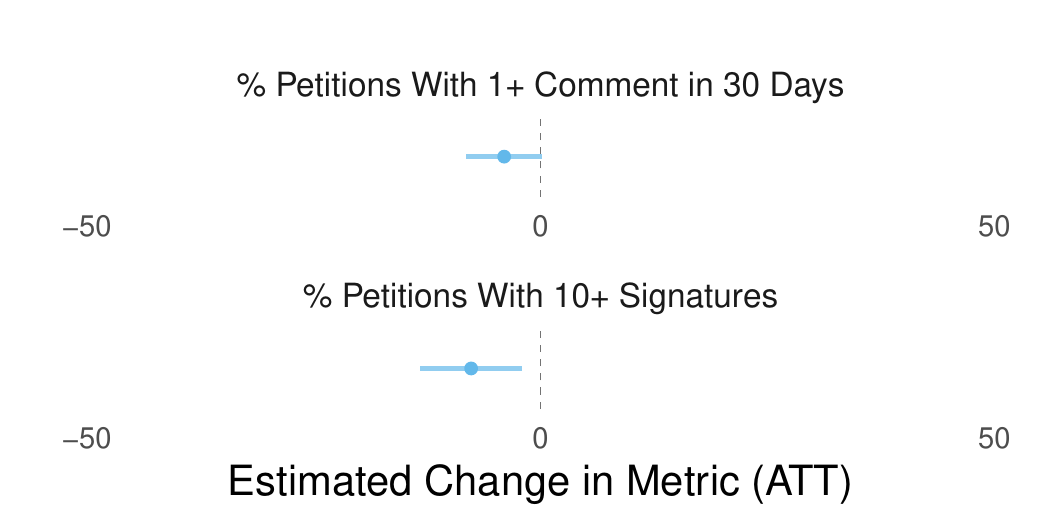}
    \caption{}
    \label{fig:civic engagement outcomes}
  \end{subfigure}
  \caption[Difference-in-differences on Civic Engagement Topic Subset]{
  Static difference-in-differences on a subset of petitions (115,284 petitions pre-AI, 43,061 petitions during the A/B period, 21,055 petitions post-AI) about civic engagement topics (Conservation and Environment, Community Planning, International Politics, Social and Economic Welfare, Civil Rights and Diversity, Criminal Justice, Public Safety). 
  Panel (a) shows civic engagement petitions lexical features shift significantly: petitions have greater lexical diversity, readability, and are longer. 
  Panel (b) shows outcomes for civic engagement petitions do not appear to improve, with significant negative ATT estimates for the percent of petitions that reach minimum comment and signature thresholds.}
  \label{fig:civic engagement}
\end{figure}

\begin{figure}[H]
  \centering
  \begin{subfigure}[b]{0.48\linewidth}
    \centering
    \includegraphics[width=\linewidth]{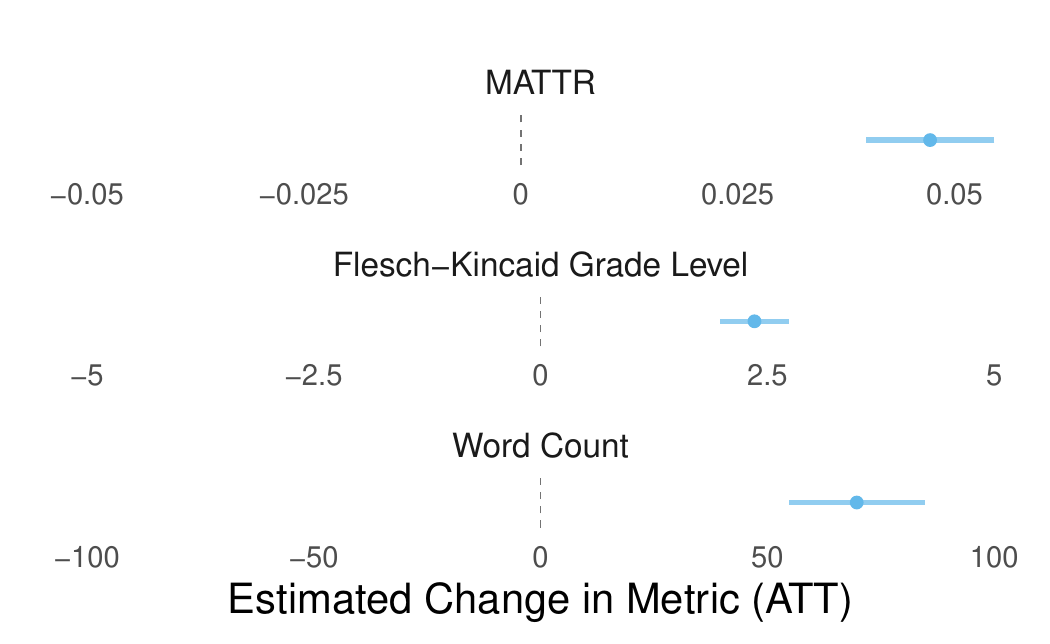}
    \caption{}
    \label{fig:other topic style}
  \end{subfigure}\hfill
  \begin{subfigure}[b]{0.48\linewidth}
    \centering
    \includegraphics[width=\linewidth]{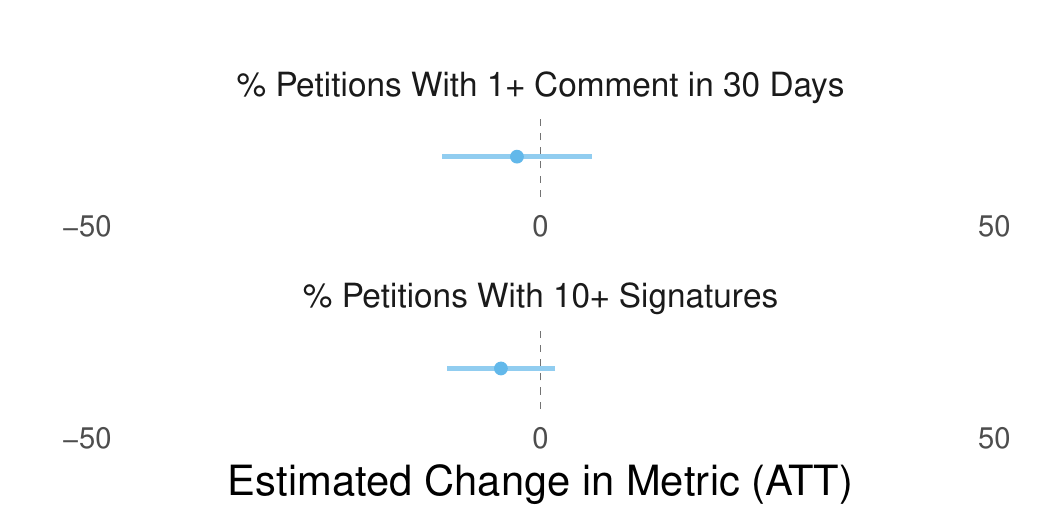}
    \caption{}
    \label{fig:other topic outcomes}
  \end{subfigure}
  \caption[Difference-in-differences on Non-Civic-Engagement Topic Subset]{Static difference-in-differences on subset of petitions about ``other'' topics (121,736 petitions pre-AI, 39,443 petitions during the A/B test period, 21,056 petitions post-AI), which do not pertain to civic engagement (Pop Culture and Media, Consumer Products and Brands, Leadership Accountability, School and Education, Personal Requests and Inside Jokes, Ban Offensive Content). 
  Panel (a) shows lexical features shift significantly: petitions have greater lexical diversity, readability, and are longer. 
  Panel (b) shows outcomes do not appear to improve, with negative ATT estimates for the percent of petitions that reach minimum comment and signature thresholds.}
  \label{fig:other topic}
\end{figure}

\begin{figure}[H]
  \centering
  
  \begin{subfigure}{0.48\linewidth}
    \centering
    \includegraphics[width=\linewidth]{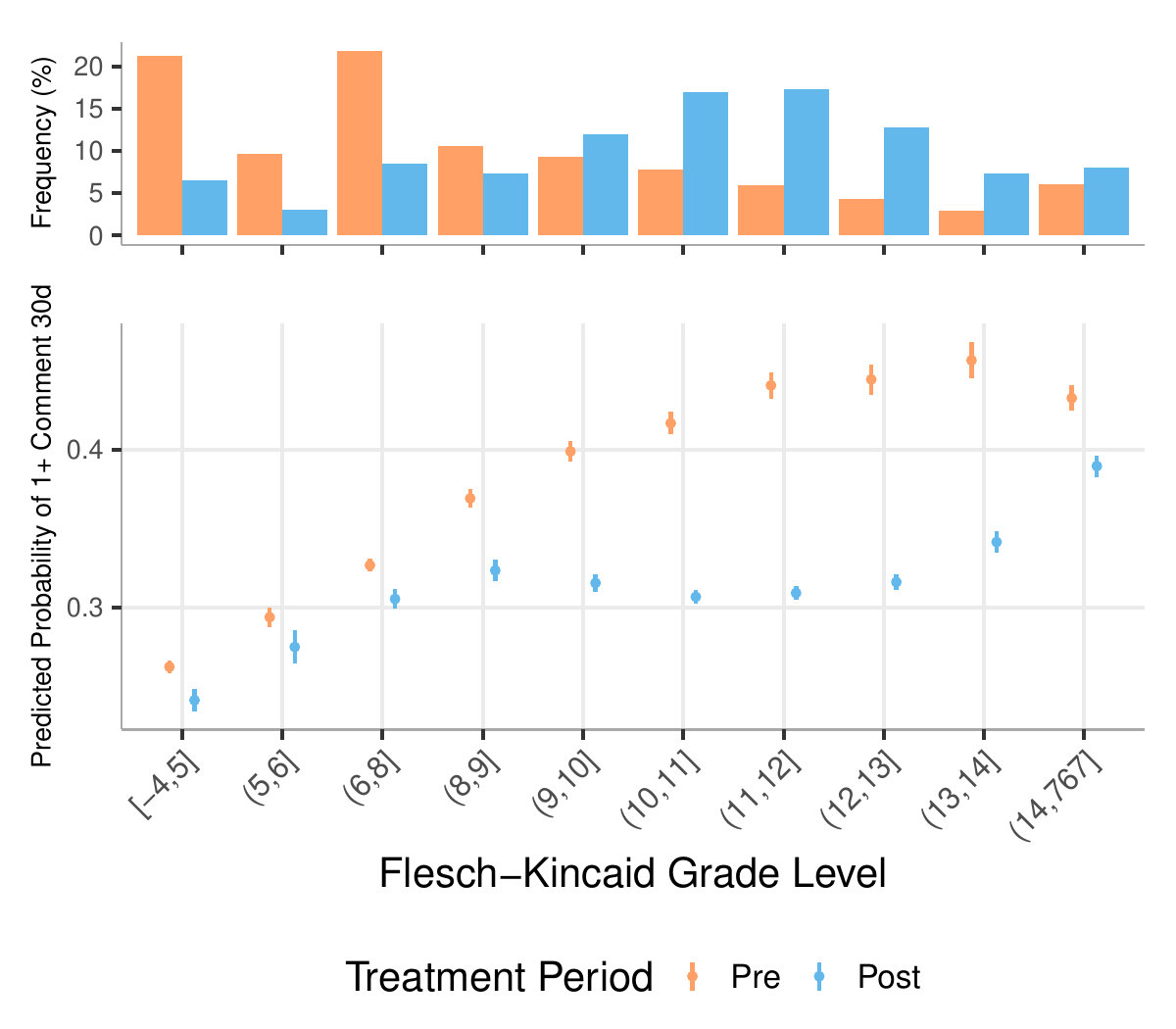}
    \caption{}
  \end{subfigure}
  \hfill
  \begin{subfigure}{0.48\linewidth}
    \centering
    \includegraphics[width=\linewidth]{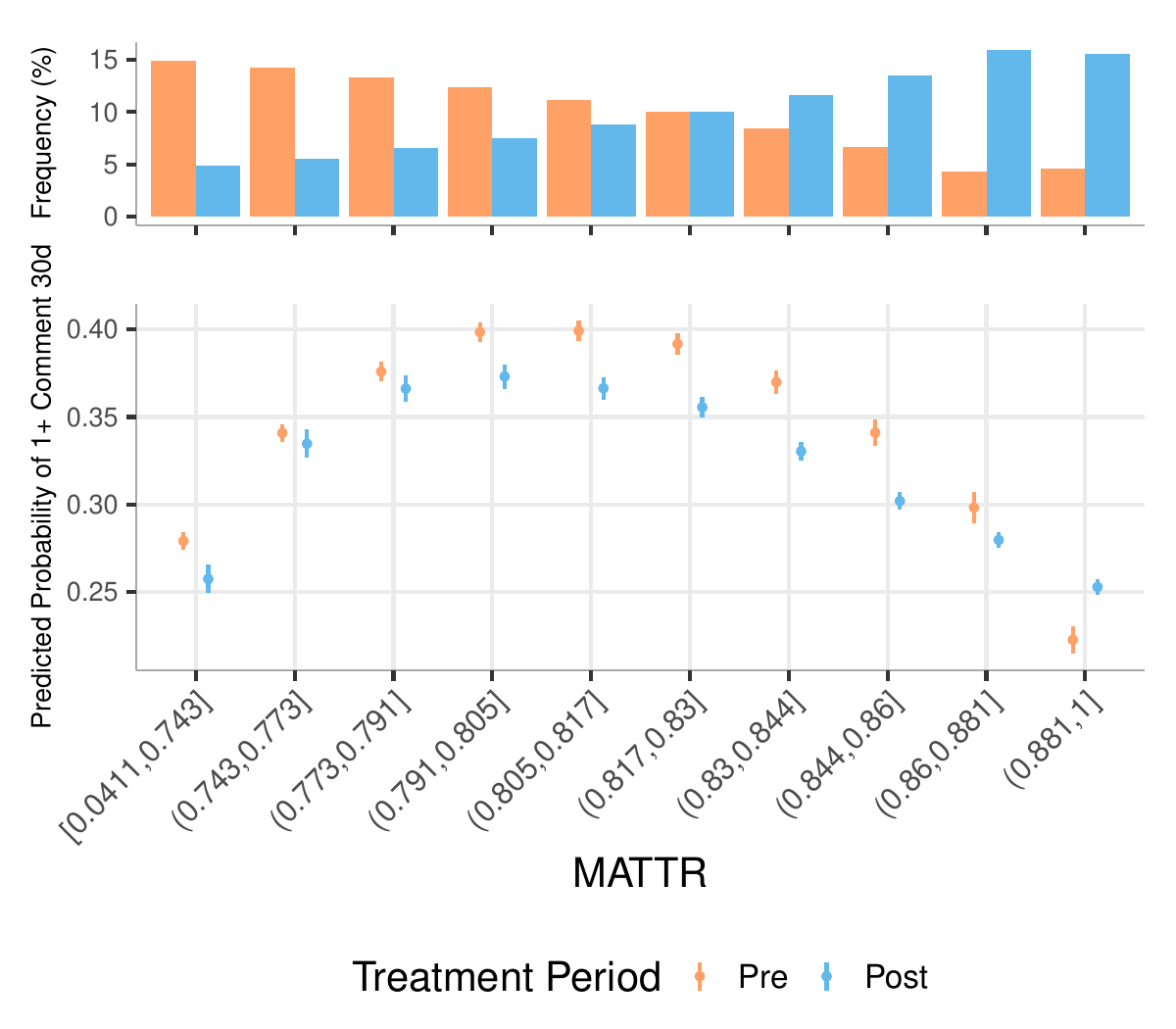}
    \caption{}
  \end{subfigure}
  
  \vspace{0.7em}
  
  \begin{subfigure}{0.6\linewidth}
    \centering
    \includegraphics[width=\linewidth]{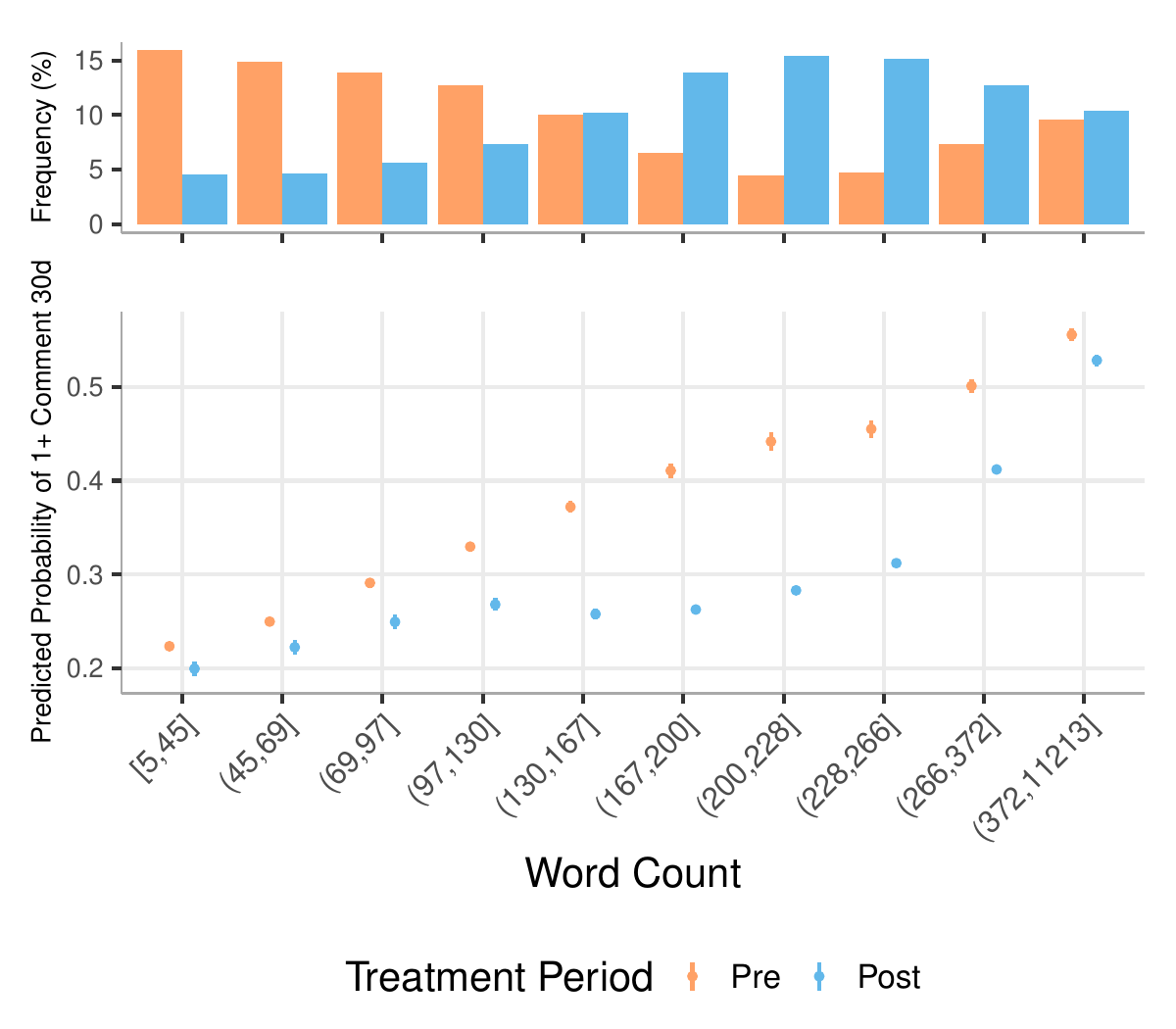}
    \caption{}
  \end{subfigure}

  \caption[Marginal Estimates of Lexical Features Regressed on 1+ Comment Outcome]{
  Marginal estimates for lexical features regressed on outcome metric (1+ comment in 30 days), using petitions written between 2022 and 2024 in the US, GB, and CA (257,590 petitions pre-AI, 244,521 petitions post-AI).
  (a) Flesch-Kincaid grade level, (b) MATTR, and (c) word count are less predictive of petition probability of receiving 1+ comment in 30 days when petitions are written with AI access.
  With AI access, word count deciles in the 130-372 range are more common, but have significantly lower marginal predicted probability of 1+ comment in 30 days, as predicted by a logistic regression model. 
  The figure shows 95\% CIs with HC3 heteroskedasticity-consistent standard errors.
  }
  \label{fig:word_count_predictive}
\end{figure}

\begin{figure}[H]
  \centering
  \begin{subfigure}{0.6\linewidth}
    \centering
    \includegraphics[width=.9\linewidth]{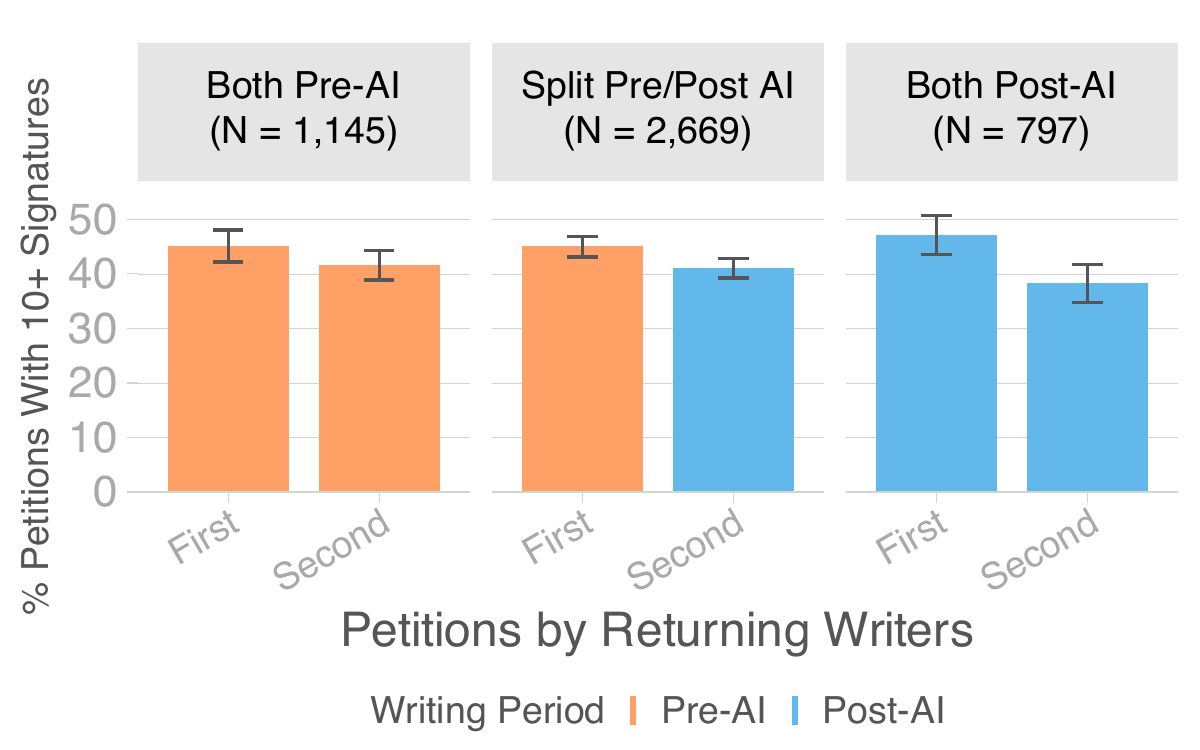}
    \caption{10+ Signature Count}
    \label{subfig:signature_count}
  \end{subfigure}
  
  \begin{subfigure}{0.6\linewidth}
    \centering
    \includegraphics[width=.9\linewidth]{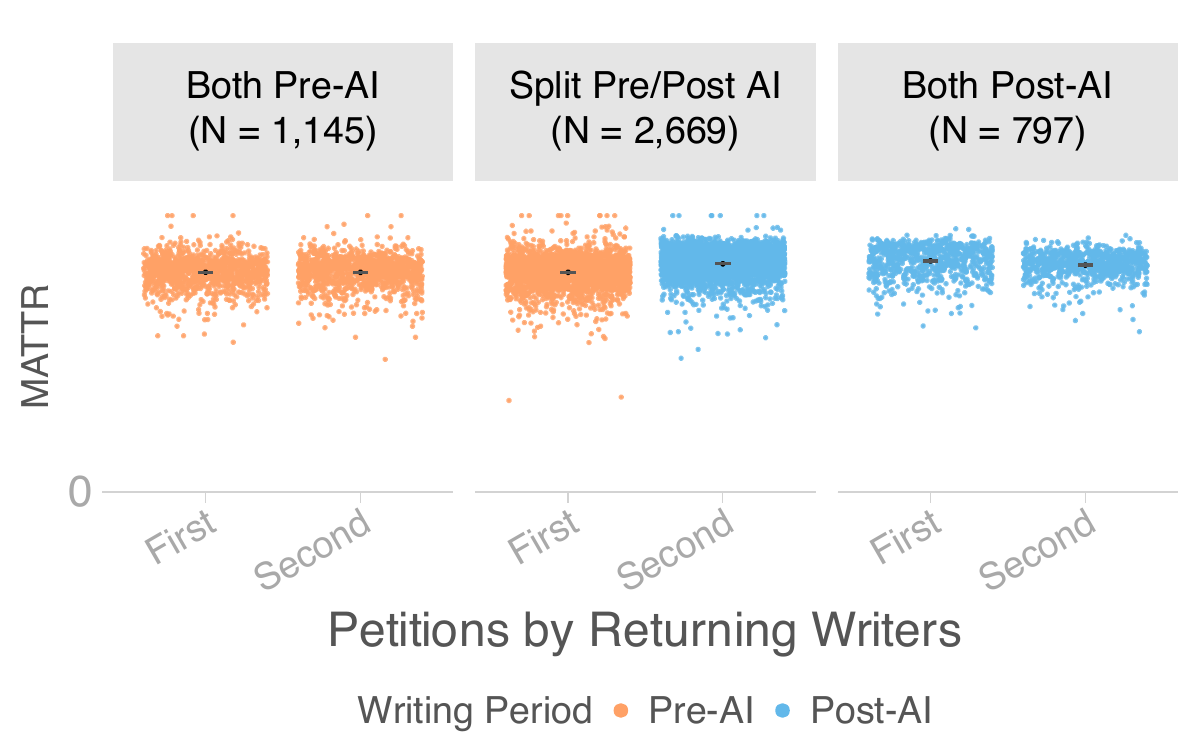}
    \caption{MATTR}
    \label{subfig:mattr_plot}
  \end{subfigure}
  
  \begin{subfigure}{0.6\linewidth}
    \centering
    \includegraphics[width=.9\linewidth]{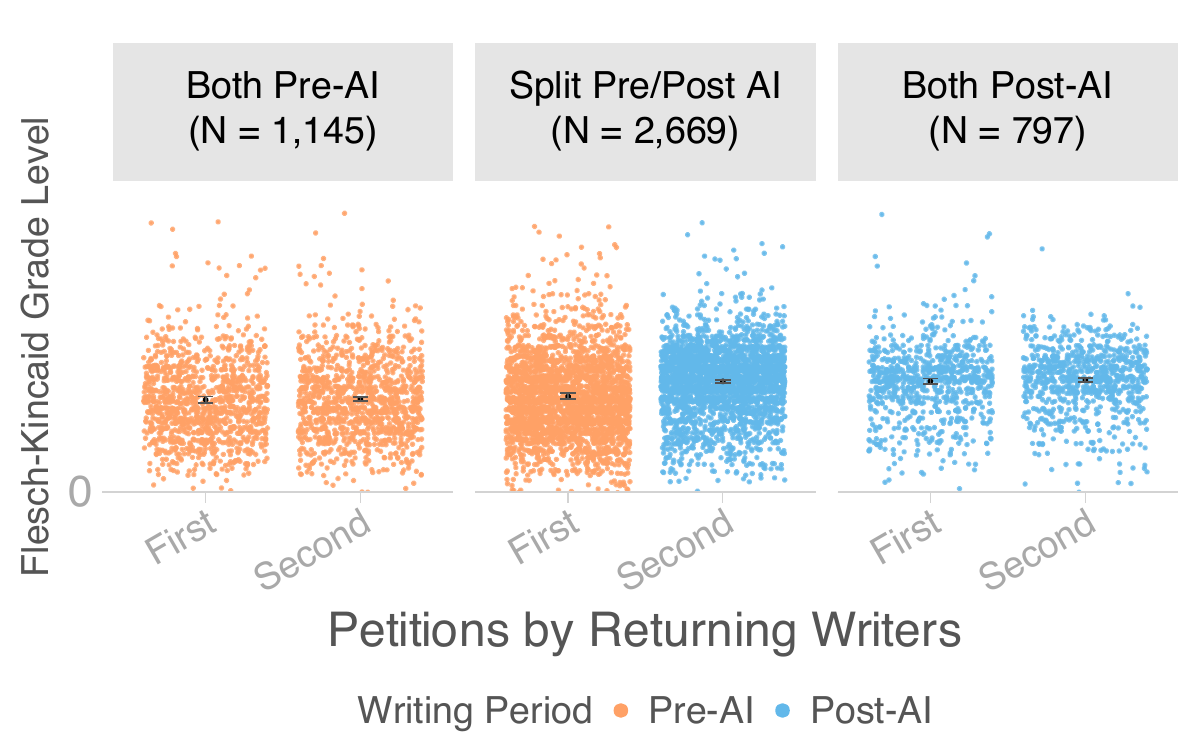}
    \caption{Flesch-Kincaid Score}
    \label{subfig:flesch_kincaid}
  \end{subfigure}
  
  \caption[Repeat Petition Writer Outcomes and Lexical Features]{Average petition outcomes and lexical features for petitions written by repeat-petition-writers (N = 4,611 users). 
  (a) shows the share of petitions in each cohort that reach 10+ signatures,
  (b) shows average MATTR
  (c) shows average Flesch-Kincaid grade level
  Points show petition metric value. Outliers (beyond +/- 3 standard deviation) are excluded from the visualization. The figure shows 95\% CI, bootstrapped with 1,000 iterations.}
  \label{fig:repeat_petition_writers_barplots}
\end{figure}

\begin{figure}[H]
    \centering
    \includegraphics[width=\textwidth]{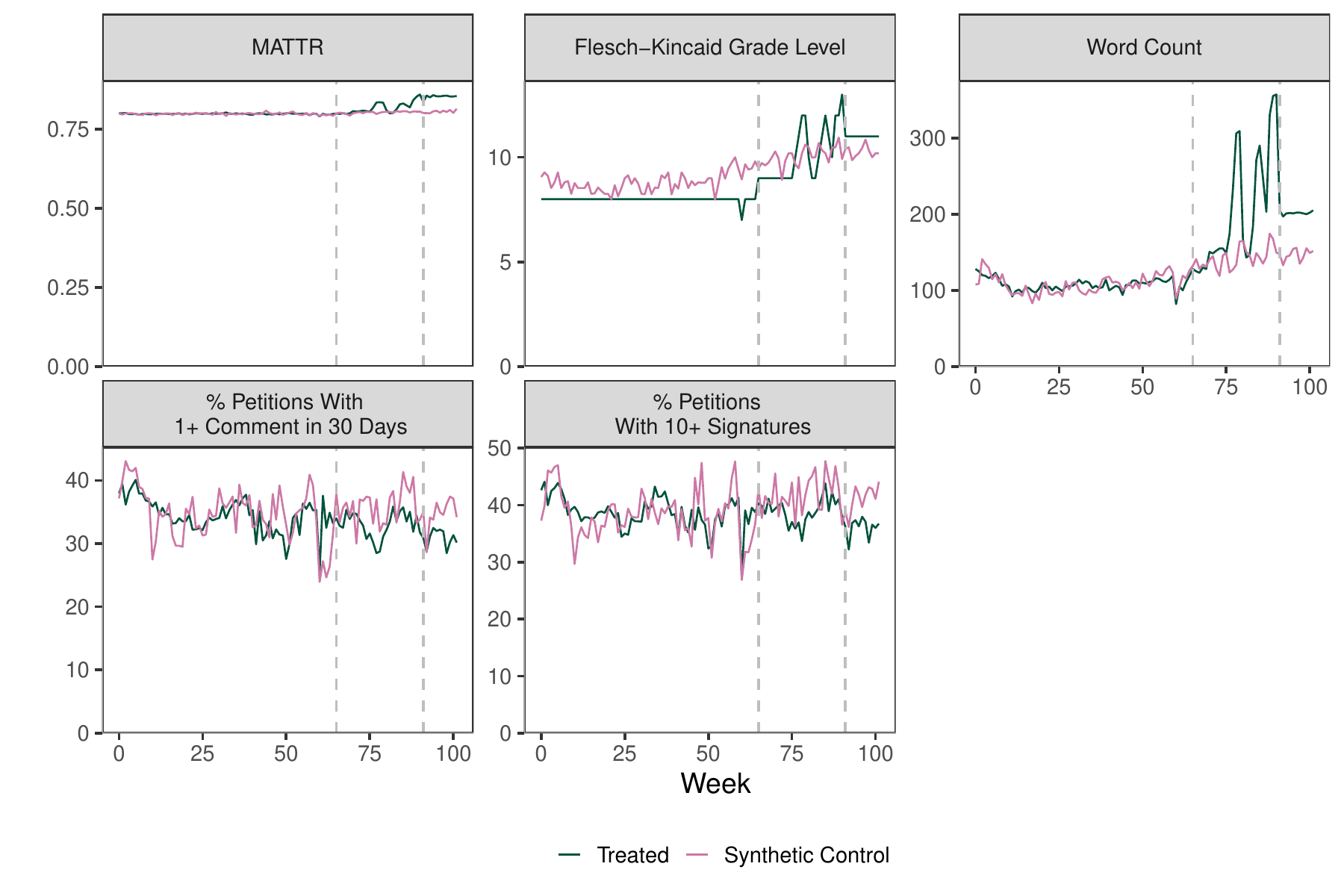}
    \caption[Synthetic Control Time Series: US Treated Unit]{Synthetic control analysis comparing a treated time series against a weighted combination of control time series. 
    This figure compares the United States as a treated unit against a weighted combination of Australia, New Zealand, and India as control units (N = 301,876 petitions).
    The y-axis for each facet is defined in the title of each facet plot.
    Because New Zealand and India fail to demonstrate parallel trends with our treated countries we exclude them from the difference-in-differences analysis.
    However, because the synthetic control method creates a balanced control time series to compare with treated units, we can include these countries in the synthetic control analysis. 
    We find that the estimated effects derived from synthetic controls are directionally consistent and similar in scale to the estimates derived from our difference-in-differences analysis. }
\label{fig:synthetic_control_results_us}
\end{figure}

\begin{figure}[H]
    \centering
    \includegraphics[width=\textwidth]{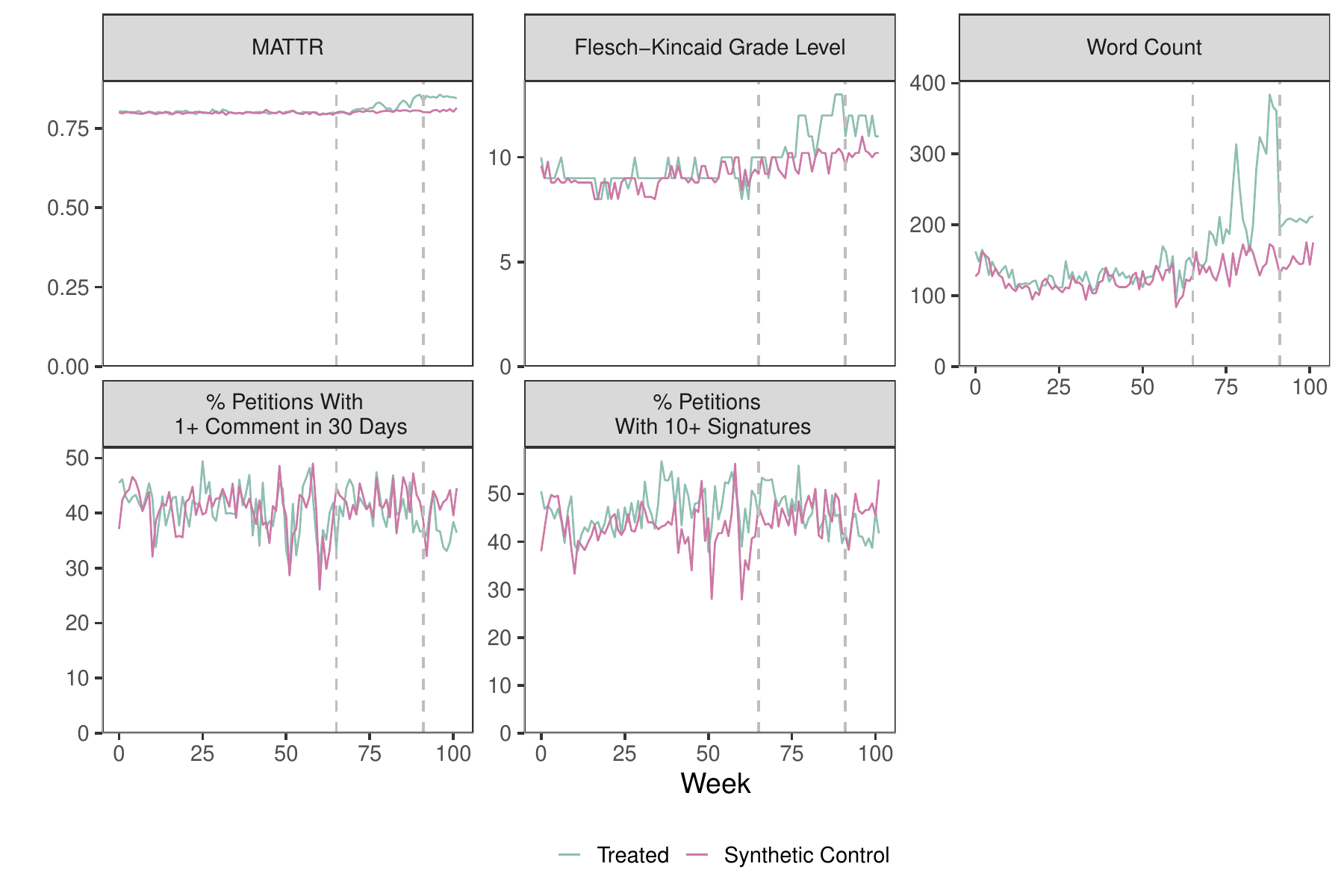}
    \caption[Synthetic Control Time Series: CA Treated Unit]{
    This figure shows a synthetic control analysis comparing Canada as a treated unit against a weighted combination of Australia, New Zealand, and India as control units (N = 70,635 petitions). 
We find that the estimated effects derived from synthetic controls are directionally consistent and similar in scale to the estimates derived from our difference-in-differences analysis. }
\label{fig:synthetic_control_results_ca}
\end{figure}

\begin{figure}[H]
    \centering
    \includegraphics[width=\textwidth]{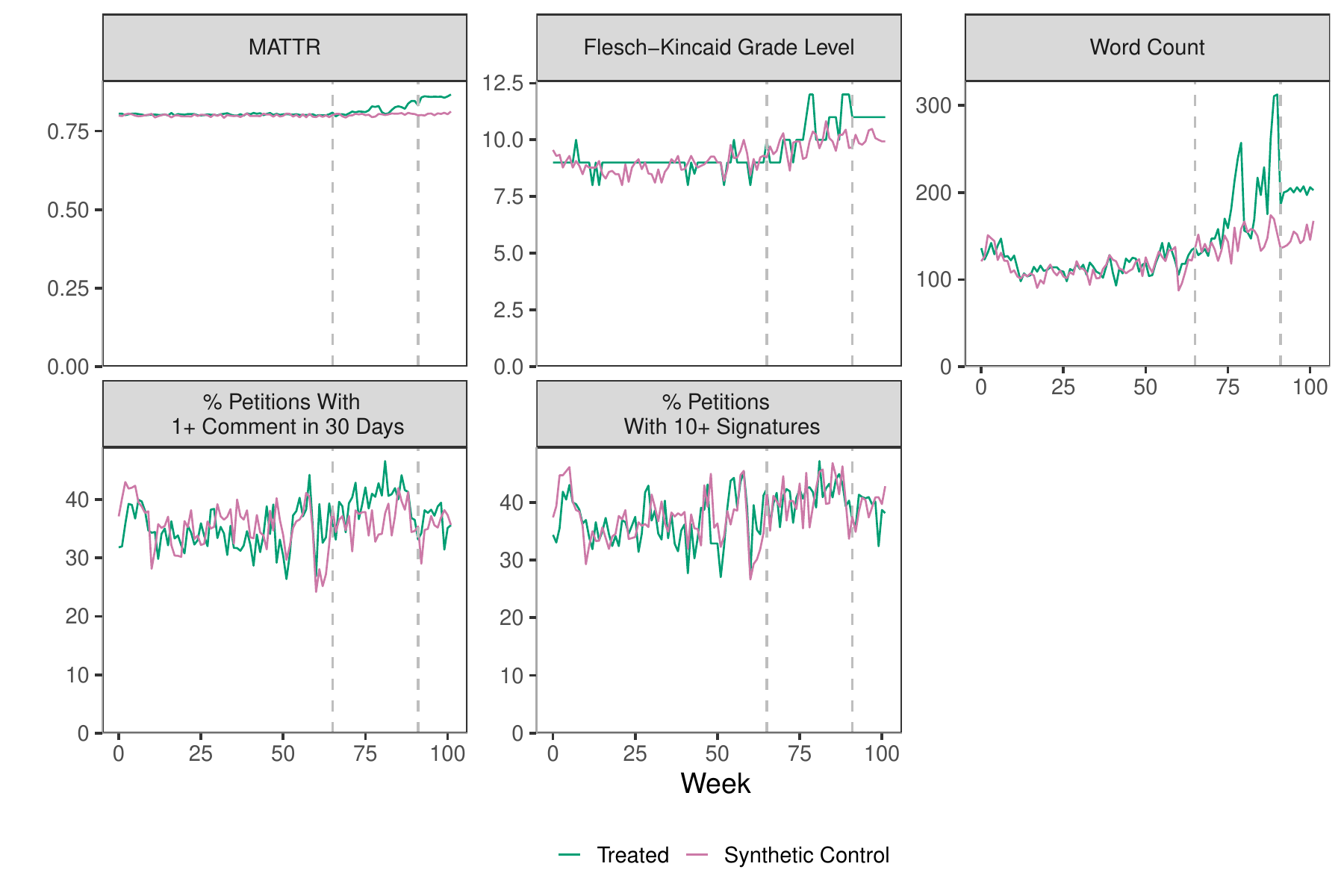}
    \caption[Synthetic Control Time Series: GB Treated Unit]{ 
This figure shows a synthetic control analysis comparing Great Britain as a treated unit against a weighted combination of Australia, New Zealand, and India as control units (N = 105,036 petitions). 
We find that the estimated effects derived from synthetic controls are directionally consistent and similar in scale to the estimates derived from our difference-in-differences analysis. }
\label{fig:synthetic_control_results_gb}
\end{figure}

\begin{figure}[H]
  \centering
  \begin{subfigure}{0.8\linewidth}
    \centering
    \includegraphics[width=\linewidth]{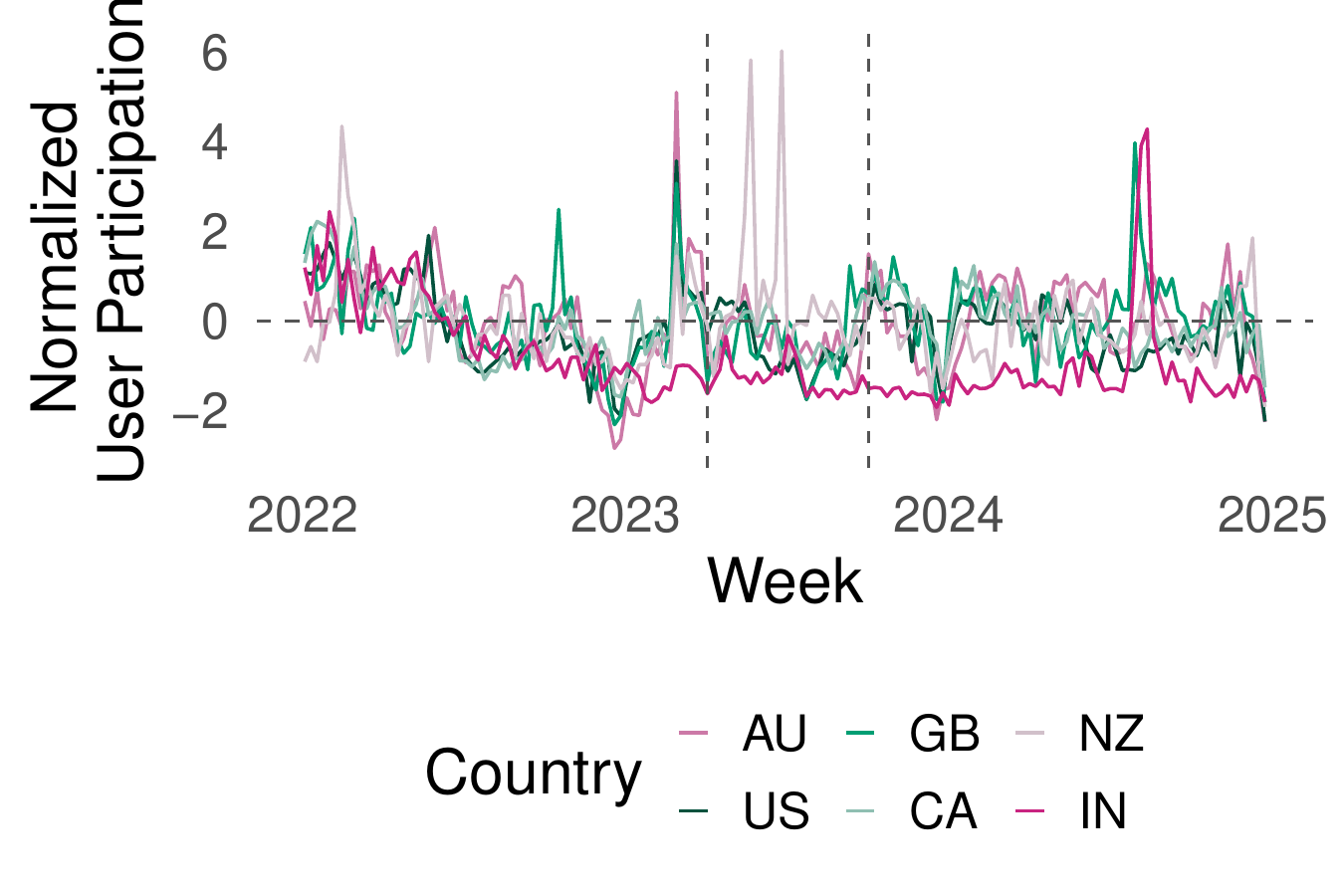}
    \caption{}
    \label{fig:participation_normalized_time_series}
  \end{subfigure}

  \vspace{1em}

  \begin{subfigure}{0.45\linewidth}
    \centering
    \includegraphics[width=\linewidth]{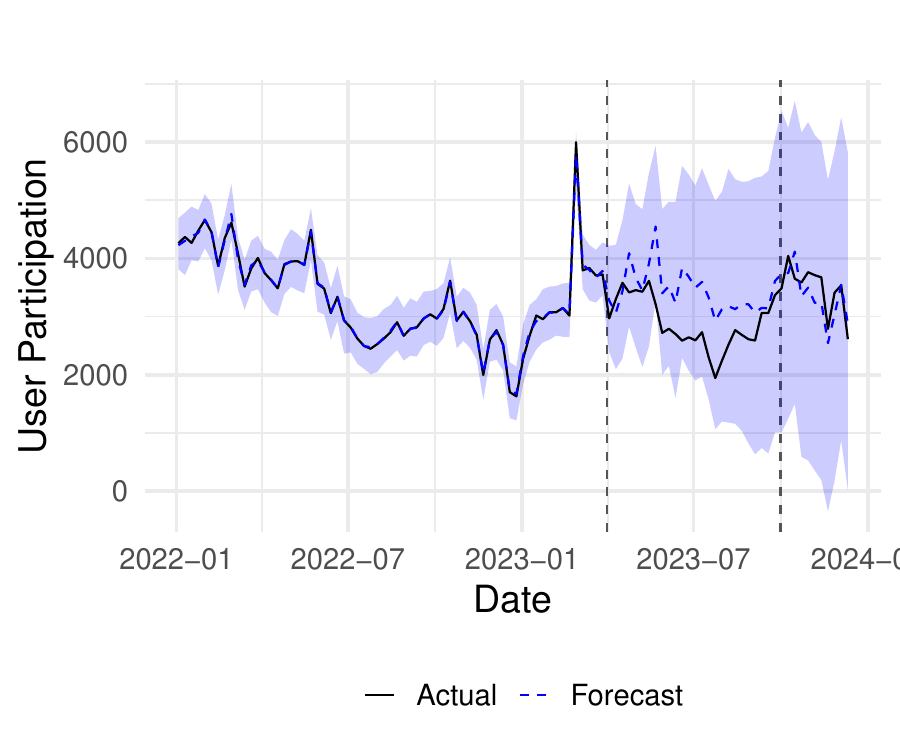}
    \caption{}
    \label{subfig:participation_bsts_time_series}
  \end{subfigure}
  \hfill
  \begin{subfigure}{0.45\linewidth}
    \centering
    \includegraphics[width=\linewidth]{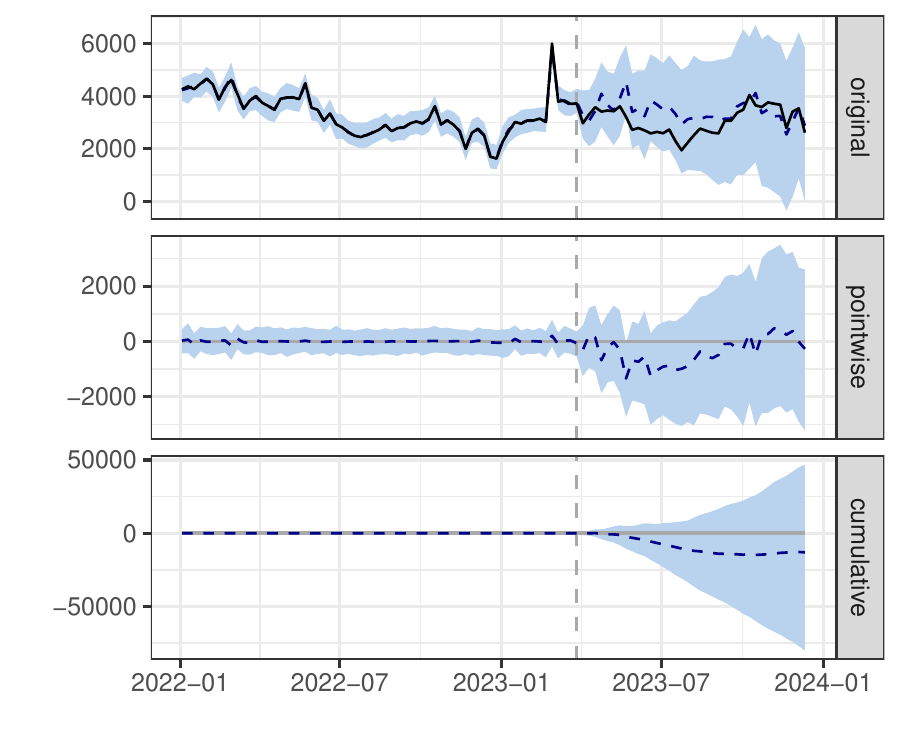}
    \caption{}
    \label{subfig:participation_bsts}
  \end{subfigure}

  \caption[BSTS: Time Series of User Participation]{Time series of platform participation, as measured by normalized (by pre-AI distribution) count of petition-writers among countries that received access in October 2023 (US, GB, CA) and countries that had known delayed release dates (AU, NZ, IN).
  NZ and IN are excluded from main analyses due to their lack of parallel trends with treated countries.
 Panel (a) shows normalized counts of petition-writers by country (AU, IN, NZ, US, GB, CA) through the full data collection period (N = 497,426 users).
  Vertical lines reflect the start of the A/B testing began in the US, GB, and CA and the start of full access in the US, GB, and CA.
  Panels (b) and (c) show our Bayesian Structural Time Series (BSTS) analysis, which we train on pre-AI data to predict platform participation through the period of differential AI access (December 2023), using counts of petition-writers (N = 356,745 users). 
  In the BSTS analysis control countries are AU, IN, NZ and treated countries are the US, GB, and CA.
  This analysis creates a synthetic counterfactual for platform participation in the absence of an intervention. 
  Panel (b) shows our forecast vs the actual user participation. 
  Panel (c) shows 95\% credible intervals for the estimated difference between the synthetic counterfactual and actual user participation. 
  As bands include 0 throughout post-intervention period, we do not find evidence of participation changing due to the shock of in-platform AI access.}
  \label{fig:participation}
\end{figure}

\begin{figure}[H]
  \centering
  \begin{subfigure}{0.8\linewidth}
    \centering
    \includegraphics[width=\linewidth]{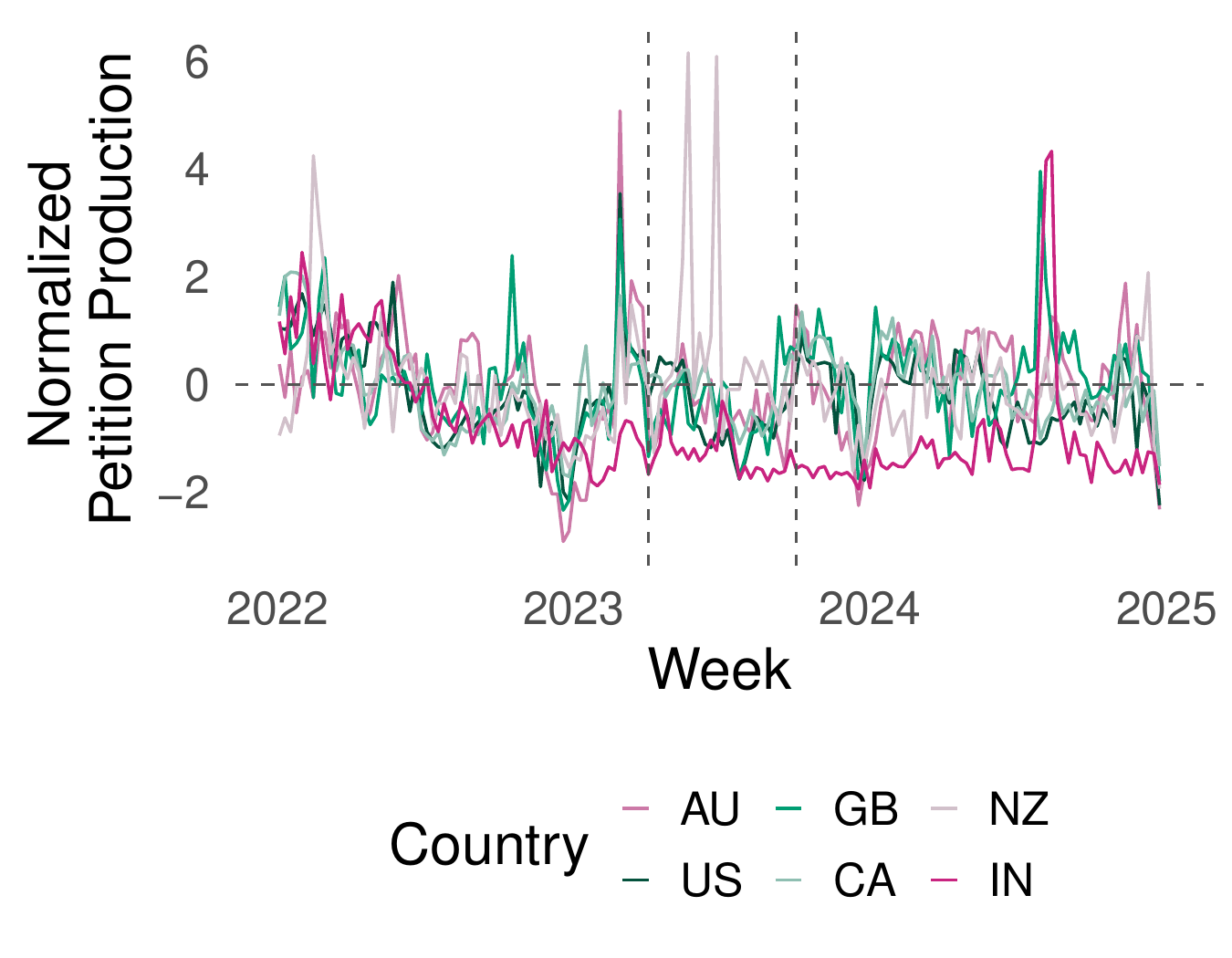}
    \caption{}
  \end{subfigure}

  \vspace{1em}

  \begin{subfigure}{0.45\linewidth}
    \centering
    \includegraphics[width=\linewidth]{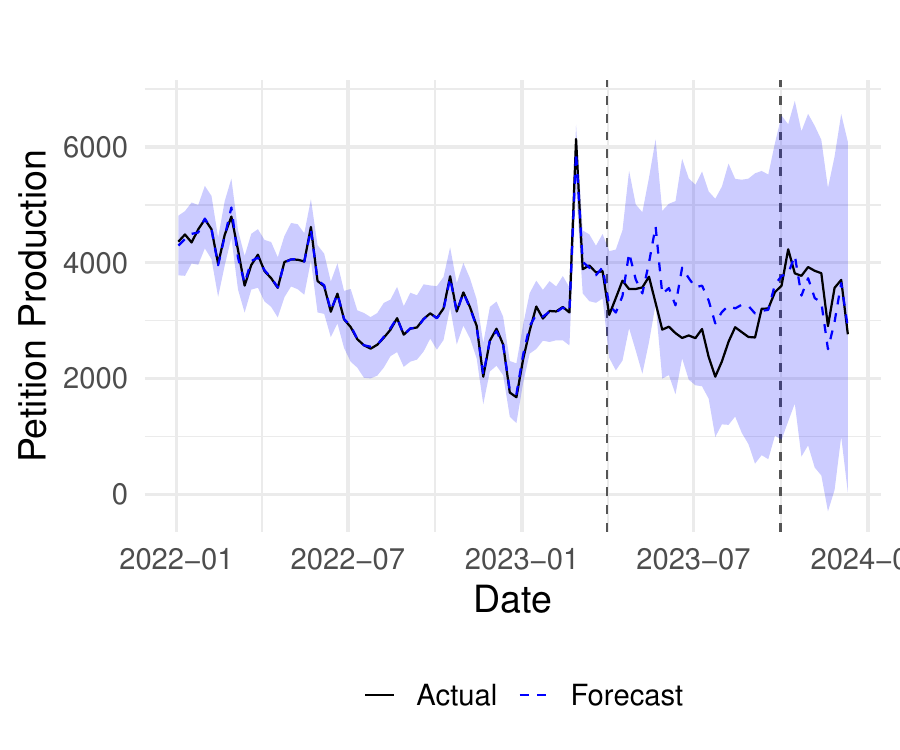}
    \caption{}
    \label{subfig:petition_production_bsts_time_series}
  \end{subfigure}
  \hfill
  \begin{subfigure}{0.45\linewidth}
    \centering
    \includegraphics[width=\linewidth]{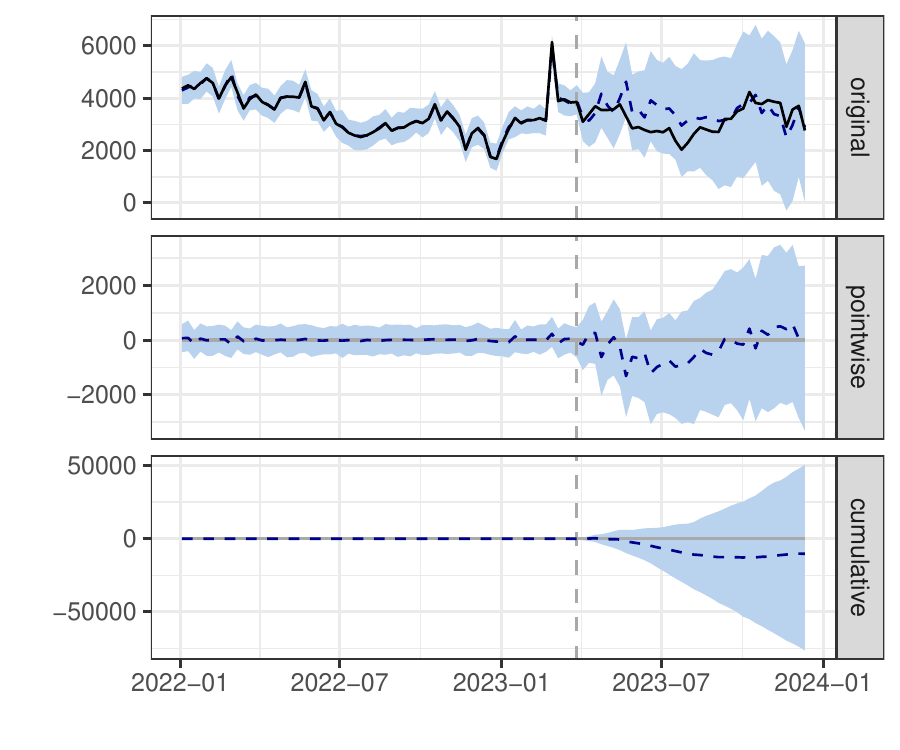}
    \caption{}
  \end{subfigure}

  \caption[BSTS: Time Series of Petition Production]{Time series of normalized (by pre-AI distribution) count of published petitions.
  NZ and IN are excluded from main analyses due to their lack of parallel trends with treated countries.
  Panel (a) shows normalized petition counts by country (AU, IN, NZ, US, GB, CA) across the full period of data collection (N = 548,656 petitions)
  Panels (b) and (c) show our Bayesian Structural Time Series (BSTS) analysis, which we train on pre-AI data to predict petition counts, using published petition counts in that period (N = 386,777 petitions). 
  In the BSTS analysis control countries are AU, IN, NZ and treated countries are the US, GB, and CA.
  This analysis creates a synthetic counterfactual for petition counts in the absence of an intervention. 
  Panel (b) shows our forecast vs the actual petition count. 
  Panel (c) shows 95\% credible intervals for the estimated difference between the synthetic counterfactual and actual petition count. 
  As bands include 0 throughout post-intervention period, we do not find evidence of petition counts changing due to the shock of AI access.}
  \label{fig:participation_production}
\end{figure}

\begin{figure}[H]
    \centering
    \begin{subfigure}[b]{0.45\textwidth}
        \centering
        \includegraphics[width=\textwidth]{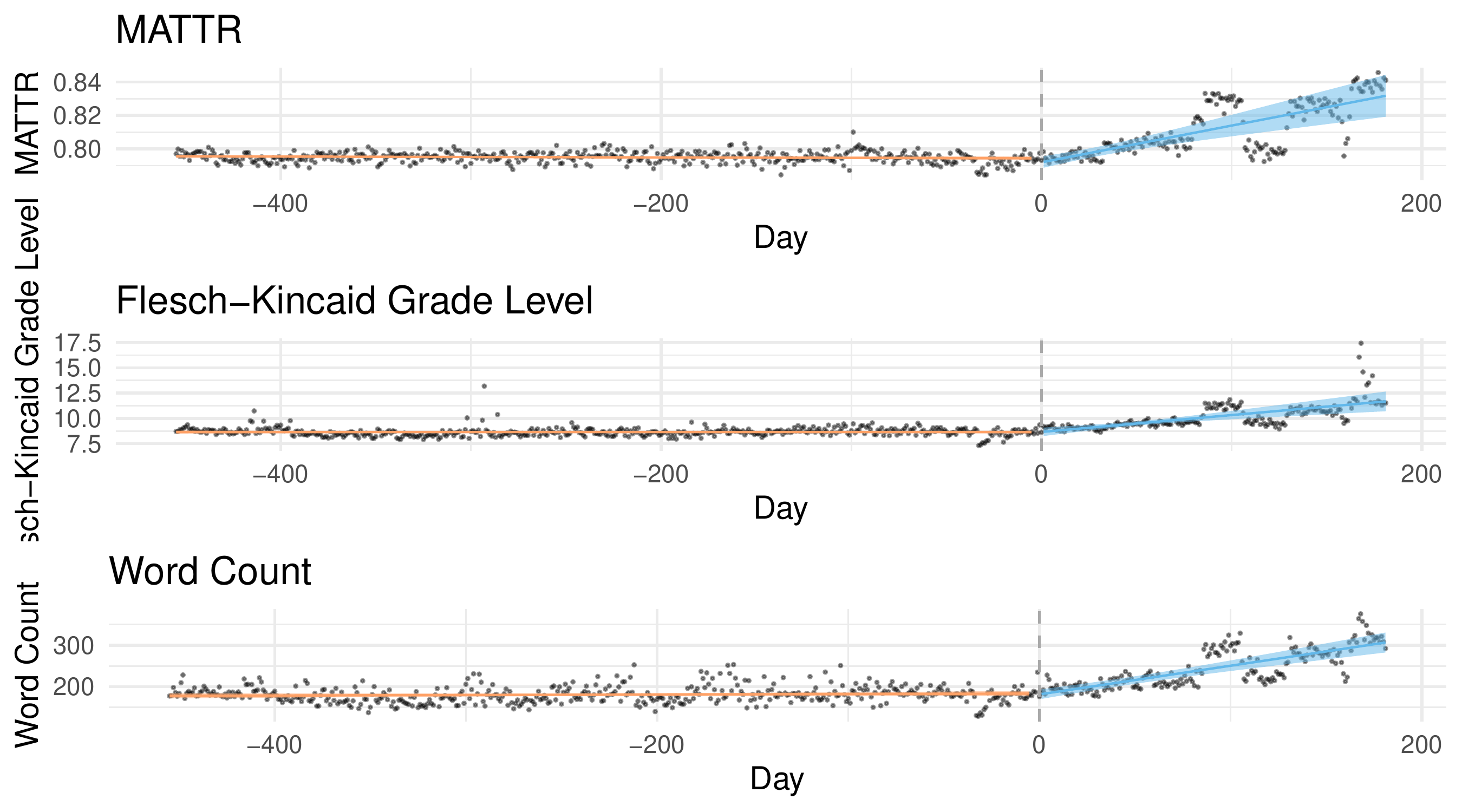}
        \caption{Treated Lexical Features}
        \label{fig:its_treated_style}
    \end{subfigure}
    \hfill
    \begin{subfigure}[b]{0.45\textwidth}
        \centering
        \includegraphics[width=\textwidth]{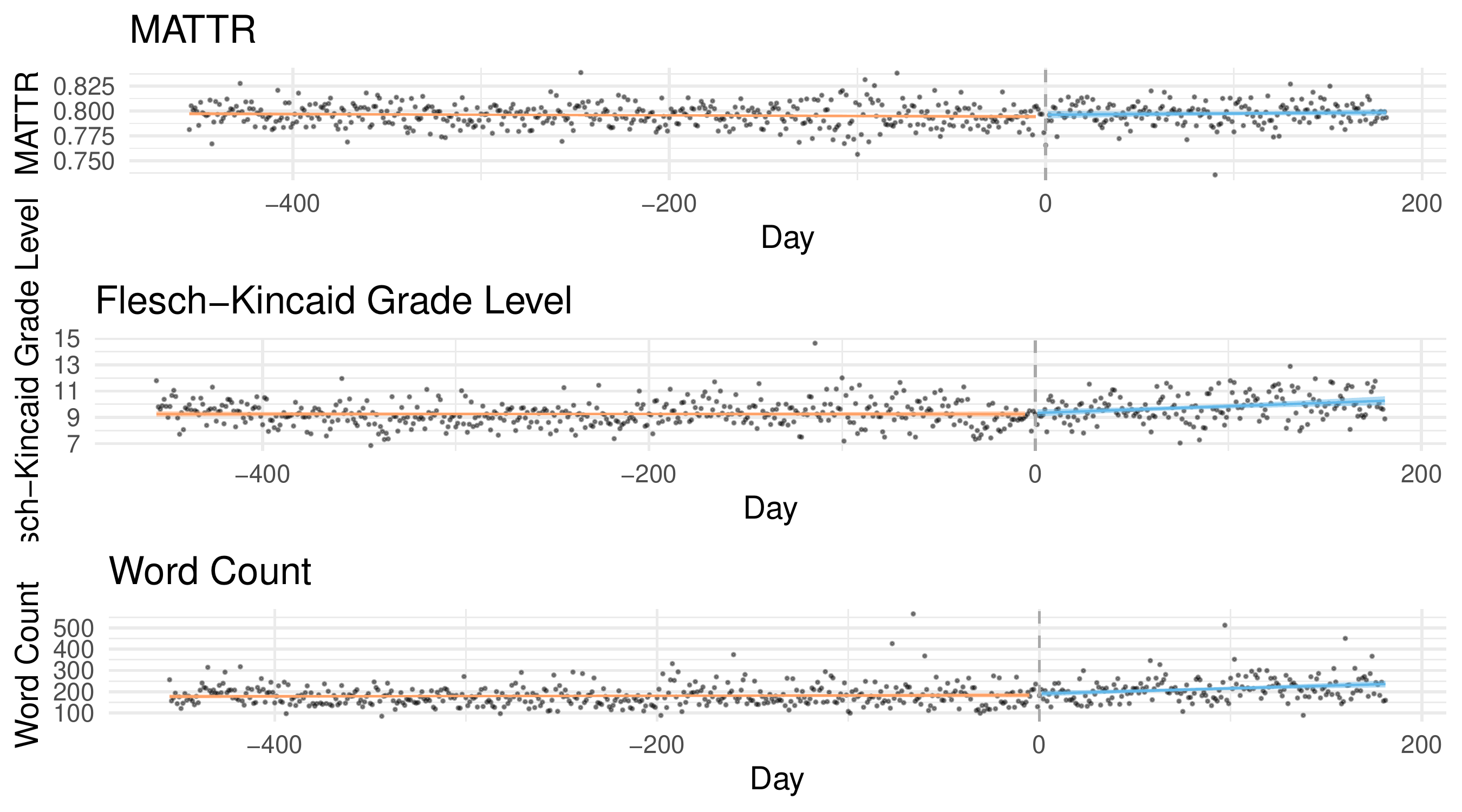}
        \caption{Control Lexical Features}
        \label{fig:its_control_style}
    \end{subfigure}
    
    \vspace{0.5cm} 
    
    \begin{subfigure}[b]{0.45\textwidth}
        \centering
        \includegraphics[width=\textwidth]{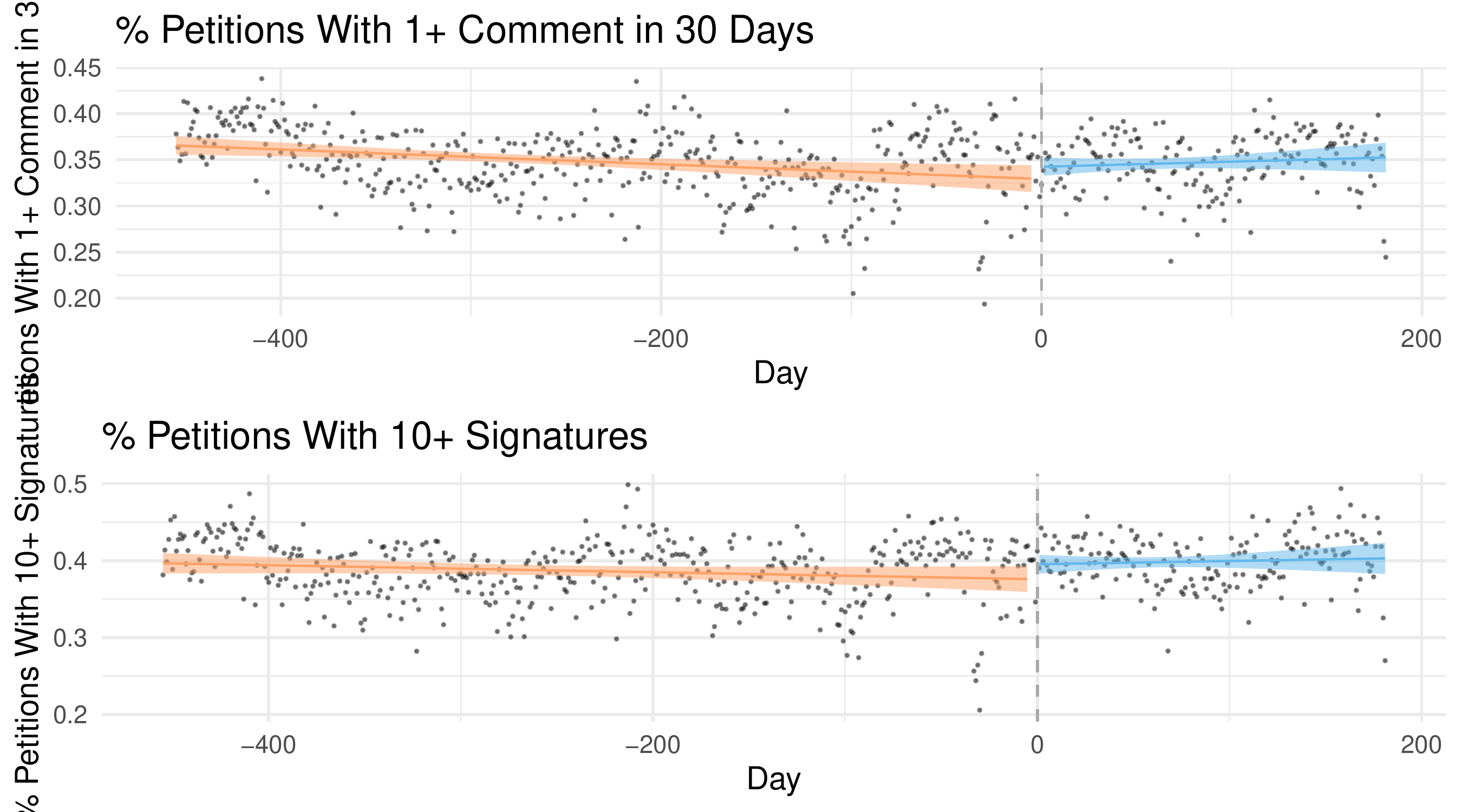}
        \caption{Treated Outcome}
        \label{fig:its_treated_outcomes}
    \end{subfigure}
    \hfill
    \begin{subfigure}[b]{0.45\textwidth}
        \centering
        \includegraphics[width=\textwidth]{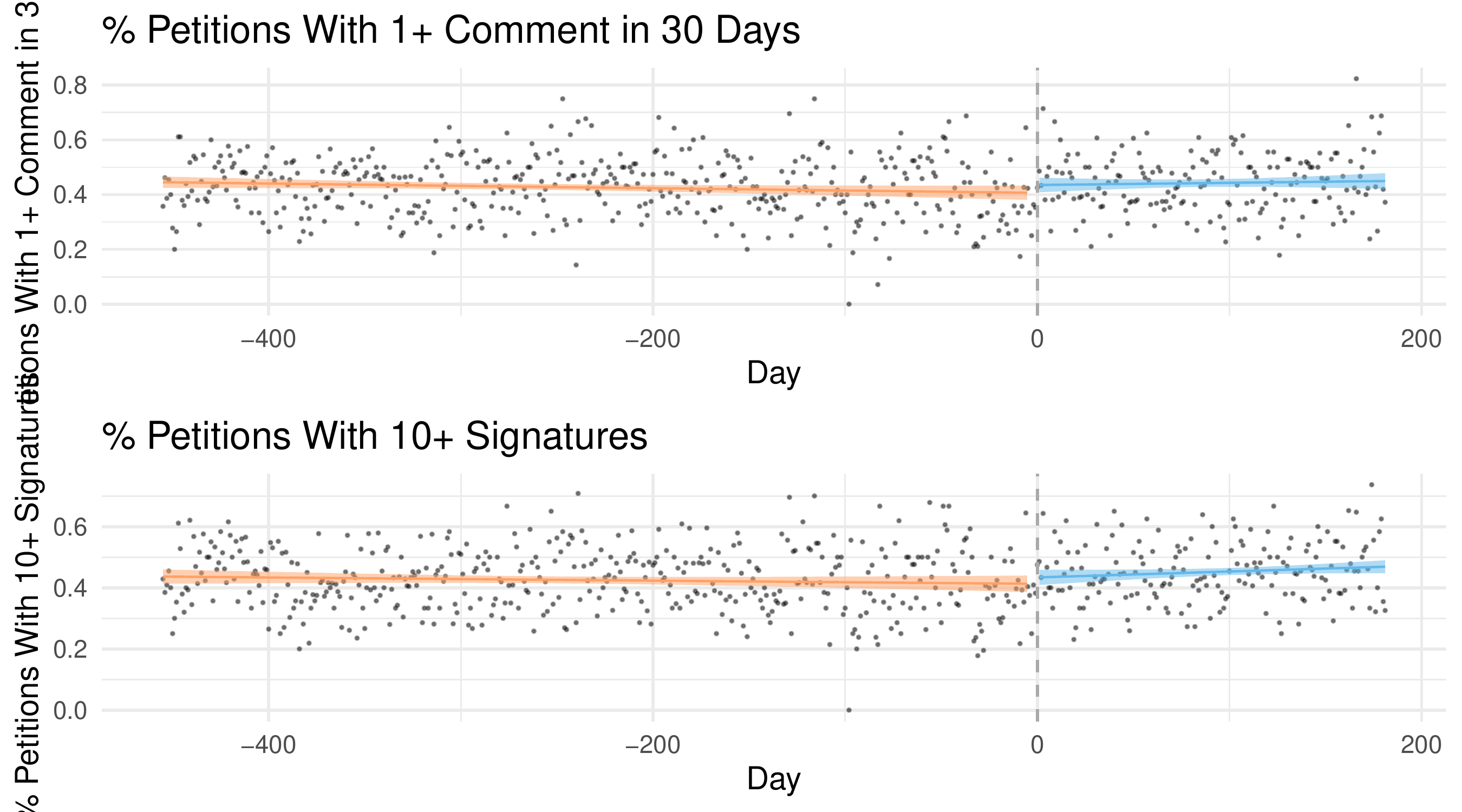}
        \caption{Control Outcome}
        \label{fig:its_control_outcomes}
    \end{subfigure}
    
    \caption[Interrupted Time Series of Lexical and Outcome Metrics]{Interrupted time series (ITS) showing regression discontinuity in time fitted models, where the date of intervention aligns with the start of the A/B testing period April 3, 2023; the start of A/B testing period in treated countries (N=237,020 petitions posted pre-intervention, N=82,504 petitions posted post-intervention, during the A/B test period).
    Here, we measure if there is a substantial change in the slope of regression lines fit to the pre-A/B testing and post-A/B testing periods.
    Panel (a) shows lexical features (top to bottom: MATTR, Flesch-Kincaid grade level, word count) in treated countries (United States, Great Britain, Canada).
    Panel (b) shows lexical features in control country (Australia).
    We observe that while there is a marked increase in the slope of the regression line fit during the A/B testing period for treated countries, and less so in the control countries.
    Panel (c) shows outcomes (top to bottom: \% of petitions that reach 1 comment (30d), \% of petitions that reach 10 signatures) in treated countries.
    (d) Shows outcomes in control countries (Australia).
    We observe little difference in the change in outcomes. 
    In both conditions, the slope of outcomes is slightly increased in the A/B testing period.
    Panels show 95\% CI with Newey-West standard errors.}
    \label{fig:its}
\end{figure}

\begin{table}[!ht]
\centering
\begin{talltblr}[         
caption={Balance table of pre/post covariates for main difference-in-differences, thresholded to the first week of each period, for consistency with difference-in-difference time periods. Pre-AI period is from January 3, 2022 to April 2, 2023. Post-AI period is from October 2, 2023 to December 15, 2023. Difference in means is result of difference in means t-test. Significance reported at $\alpha = 0.05$.},
entry={Balance table}, label={tab:balance_table},
]                     
{                     
colspec={Q[]Q[]Q[]Q[]Q[]Q[]Q[]},
hline{2}={2,4}{solid, black, 0.03em, l=-0.5},
hline{2}={3,5}{solid, black, 0.03em, r=-0.5},
hline{3}={1-7}{solid, black, 0.05em},
hline{1}={1-7}{solid, black, 0.08em},
hline{7}={1-7}{solid, black, 0.08em},
cell{1}{1}={}{halign=c},
cell{1}{2}={c=2}{halign=c},
cell{1}{3}={}{halign=c},
cell{1}{4}={c=2}{halign=c},
cell{1}{5}={}{halign=c},
cell{1}{6}={}{halign=c},
cell{1}{7}={}{halign=c},
cell{2-6}{1}={}{halign=l},
cell{2-6}{2}={}{halign=r},
cell{2-6}{3}={}{halign=r},
cell{2-6}{4}={}{halign=r},
cell{2-6}{5}={}{halign=r},
cell{2-6}{6}={}{halign=r},
cell{2-6}{7}={}{halign=r},
}                     
& Pre (N=237020) &  & Post (N=42111) &  &  &  \\
& Mean & Std. Dev. & Mean & Std. Dev. & Diff. in Means & Std. Error \\
Has Previous Comments & 0.04 & 0.2 & 0.074 & 0.26 & 0.034*** & 0.0013 \\
Has Decision Makers & 0.21 & 0.41 & 0.11 & 0.31 & -0.11*** & 0.0017 \\
New User & 0.93 & 0.26 & 0.88 & 0.32 & -0.044*** & 0.0017 \\
Government Topic & 0.37 & 0.48 & 0.4 & 0.49 & 0.039*** & 0.0026 \\
\end{talltblr}
\end{table}

\begin{table}[!ht]
\centering
\caption{\label{tab:tab:main_metrics_pre}Pre‑treatment (January 3, 2022 - April 2, 2023) descriptive statistics of main text metrics}
\centering
\begin{tabular}[t]{lrrrrrrr}
\toprule
Variable & Unique & MissingPct & Mean & SD & Min & Median & Max\\
\midrule
mattr\_title\_text & 126784 & 0 & 0.79 & 0.06 & 0.04 & 0.8 & 1\\
flesch\_kincaid\_grade\_level & 134 & 0 & 8.65 & 5.57 & -4.00 & 8.0 & 767\\
word\_count\_title\_text & 2262 & 0 & 179.12 & 270.09 & 6.00 & 110.0 & 10778\\
has\_comment\_30d\_1plus & 2 & 0 & 0.35 & 0.48 & 0.00 & 0.0 & 1\\
has\_signature\_count\_10plus & 2 & 0 & 0.39 & 0.49 & 0.00 & 0.0 & 1\\
\bottomrule
\end{tabular}
\end{table}

\begin{table}[!ht]
\centering
\caption{\label{tab:tab:main_metrics_post}Post‑treatment (October 2, 2023 - December 15, 2023) descriptive statistics of main text metrics}
\centering
\begin{tabular}[t]{lrrrrrrr}
\toprule
Variable & Unique & MissingPct & Mean & SD & Min & Median & Max\\
\midrule
mattr\_title\_text & 36360 & 0 & 0.84 & 0.06 & 0.04 & 0.85 & 1\\
flesch\_kincaid\_grade\_level & 63 & 0 & 10.69 & 4.69 & -4.00 & 11.00 & 257\\
word\_count\_title\_text & 1313 & 0 & 228.67 & 225.36 & 7.00 & 201.00 & 10571\\
has\_comment\_30d\_1plus & 2 & 0 & 0.33 & 0.47 & 0.00 & 0.00 & 1\\
has\_signature\_count\_10plus & 2 & 0 & 0.37 & 0.48 & 0.00 & 0.00 & 1\\
\bottomrule
\end{tabular}
\end{table}

\begin{table}[!ht]
\centering
\begin{tabular}{p{4cm} p{10cm}}
\hline
\textbf{Category} & \textbf{Item (1--5 Likert scale)} \\
\hline
{Writing Quality} 
& The text is well-structured and easy to follow. \\
& The paragraphs and sentences are logically organized. \\
& The language is clear and easy to understand. \\
& The text is free of errors in grammar, punctuation, and spelling. \\
\hline
{Persuasiveness} 
& The petition provides reasons for its cause that are believable. \\
& The petition provides reasons for its cause that are convincing. \\
& This petition has the potential to change reader behavior. \\
& This petition has the potential to influence decision-makers. \\
\hline
\end{tabular}
\caption[Labeling Task Measures]{Labeling task measures to evaluate petition writing quality and persuasiveness. Each item is assessed in a 1-5 Likert scale from 1 = strongly disagree to 5 = strongly agree.}
\label{tab:survey_measures}
\end{table}

\begin{table}[!ht]
\centering
\caption[GLM Estimates: Regressing Outcome (1+ Comment 30days) on AI Period and Persuasiveness]{\label{tab:tab:persuasive mod exp comments}GLM estimates (exp. coef.): Logistic regression model regressing outcome (has 1+ comment in 30 days) on persuasiveness score and AI period}
\centering
\begin{tabular}[t]{lrrrr}
\toprule
term & estimate & std.error & statistic & p.value\\
\midrule
(Intercept) & 0.043 & 0.621 & -5.047 & 0.000\\
persuasive\_avg & 2.405 & 0.227 & 3.859 & 0.000\\
treatment\_periodpost\_ai & 3.239 & 0.956 & 1.229 & 0.219\\
persuasive\_avg:treatment\_periodpost\_ai & 0.495 & 0.317 & -2.216 & 0.027\\
\bottomrule
\end{tabular}
\end{table}

\begin{table}[!ht]
\centering
\caption[GLM Estimates: Regressing Outcome (1+ Comment 30days) on AI Period and Writing Quality]{\label{tab:tab:writing quality mod exp comments}GLM estimates (exp. coef.): Logistic regression model regressing comment outcome (1+ comment in 30days) on writing quality and AI period}
\centering
\begin{tabular}[t]{lrrrr}
\toprule
term & estimate & std.error & statistic & p.value\\
\midrule
(Intercept) & 0.129 & 0.622 & -3.288 & 0.001\\
writing\_quality\_avg & 1.516 & 0.214 & 1.947 & 0.052\\
treatment\_periodpost\_ai & 2.934 & 1.023 & 1.052 & 0.293\\
writing\_quality\_avg:treatment\_periodpost\_ai & 0.580 & 0.317 & -1.719 & 0.086\\
\bottomrule
\end{tabular}
\end{table}

\begin{table}[!ht]
\centering
\caption[GLM Estimates: Regressing Outcome (10+ Signature) on AI Period and Persuasiveness]{\label{tab:tab:persuasive mod exp signatures}GLM estimates (exp. coef.): Logistic regression model regressing outcome (has 10+ signatures) on persuasiveness and AI period}
\centering
\begin{tabular}[t]{lrrrr}
\toprule
term & estimate & std.error & statistic & p.value\\
\midrule
(Intercept) & 0.060 & 0.574 & -4.897 & 0.000\\
persuasive\_avg & 2.829 & 0.222 & 4.691 & 0.000\\
treatment\_periodpost\_ai & 2.382 & 0.855 & 1.015 & 0.310\\
persuasive\_avg:treatment\_periodpost\_ai & 0.527 & 0.293 & -2.183 & 0.029\\
\bottomrule
\end{tabular}
\end{table}

\begin{table}[!ht]
\centering
\caption[GLM Estimates: Regressing Outcome (10+ Signature) on AI Period and Writing Quality]{\label{tab:tab:writing quality mod exp signatures}GLM estimates (exp. coef.): Logistic regression model regressing outcome (has 10+ signatures) on writing quality and AI period}
\centering
\begin{tabular}[t]{lrrrr}
\toprule
term & estimate & std.error & statistic & p.value\\
\midrule
(Intercept) & 0.146 & 0.577 & -3.335 & 0.001\\
writing\_quality\_avg & 1.855 & 0.202 & 3.057 & 0.002\\
treatment\_periodpost\_ai & 1.737 & 0.924 & 0.598 & 0.550\\
writing\_quality\_avg:treatment\_periodpost\_ai & 0.659 & 0.287 & -1.452 & 0.146\\
\bottomrule
\end{tabular}
\end{table}

\begin{table}[!ht]
\centering
\caption[Static Difference-in-differences: Percent 1+ Comment (30 Days)]{\label{tab:tab:main_outcomes_1}Static difference-in-differences estimate for outcome metric reported in main text: percentage of petitions with 1+ comment in 30 days. 95\% CIs are shown.}
\centering
\begin{tabular}[t]{lccccc}
\toprule
\multicolumn{1}{c}{ } & \multicolumn{5}{c}{(1)} \\
\cmidrule(l{3pt}r{3pt}){2-6}
\multicolumn{1}{c}{ } & \multicolumn{5}{c}{\% Petitions With 1+ Comment in 30 Days} \\
\cmidrule(l{3pt}r{3pt}){2-6}
\textbf{VARIABLES} & \textbf{Estimate} & \textbf{Std. Error} & \textbf{CIs} & \textbf{t} & \textbf{p}\\
\midrule
Treated Country x Post-AI & -2.508 & 1.826 & {}[-6.360, 1.343] & -1.374 & 0.187\\
Observations & 304 &  &  &  & \\
\bottomrule
\end{tabular}
\end{table}

\begin{table}[!ht]
\centering
\caption[Static Difference-in-differences: Percent with 10+ Signatures]{\label{tab:tab:main_outcomes_2}Static difference-in-differences estimate for outcome metrics reported in main text: percentage of petitions with 10+ signatures in 30 days. 95\% CIs are shown.}
\centering
\begin{tabular}[t]{lccccc}
\toprule
\multicolumn{1}{c}{ } & \multicolumn{5}{c}{(1)} \\
\cmidrule(l{3pt}r{3pt}){2-6}
\multicolumn{1}{c}{ } & \multicolumn{5}{c}{\% Petitions With 10+ Signatures} \\
\cmidrule(l{3pt}r{3pt}){2-6}
\textbf{VARIABLES} & \textbf{Estimate} & \textbf{Std. Error} & \textbf{CIs} & \textbf{t} & \textbf{p}\\
\midrule
Treated Country x Post-AI & -5.329 & 1.363 & {}[-8.205, -2.454] & -3.911 & 0.001\\
Observations & 304 &  &  &  & \\
\bottomrule
\end{tabular}
\end{table}

\begin{table}[!ht]
\centering
\caption[Static Difference-in-differences: MATTR]{\label{tab:tab:main_lexical_1}Static difference-in-differences estimate for lexical feature metric MATTR, as reported in main text. 95\% CIs are shown.}
\centering
\begin{tabular}[t]{lccccc}
\toprule
\multicolumn{1}{c}{ } & \multicolumn{5}{c}{(1)} \\
\cmidrule(l{3pt}r{3pt}){2-6}
\multicolumn{1}{c}{ } & \multicolumn{5}{c}{MATTR} \\
\cmidrule(l{3pt}r{3pt}){2-6}
\textbf{VARIABLES} & \textbf{Estimate} & \textbf{Std. Error} & \textbf{CIs} & \textbf{t} & \textbf{p}\\
\midrule
Treated Country x Post-AI & 0.049 & 0.001 & {}[0.047, 0.050] & 61.482 & 0.000\\
Observations & 304 &  &  &  & \\
\bottomrule
\end{tabular}
\end{table}

\begin{table}[!ht]
\centering
\caption[Static difference-in-differences: Word Count]{\label{tab:tab:main_lexical_2}Static difference-in-differences estimate for lexical feature metric word count, as reported in main text. 95\% CIs are shown.}
\centering
\begin{tabular}[t]{lccccc}
\toprule
\multicolumn{1}{c}{ } & \multicolumn{5}{c}{(1)} \\
\cmidrule(l{3pt}r{3pt}){2-6}
\multicolumn{1}{c}{ } & \multicolumn{5}{c}{Word Count} \\
\cmidrule(l{3pt}r{3pt}){2-6}
\textbf{VARIABLES} & \textbf{Estimate} & \textbf{Std. Error} & \textbf{CIs} & \textbf{t} & \textbf{p}\\
\midrule
Treated Country x Post-AI & 53.628 & 3.105 & {}[47.077, 60.178] & 17.273 & 0.000\\
Observations & 304 &  &  &  & \\
\bottomrule
\end{tabular}
\end{table}

\begin{table}[!ht]
\centering
\caption[Static Difference-in-differences: Flesch-Kincaid Grade Level]{\label{tab:tab:main_lexical_3}Static difference-in-differences estimate for lexical feature metric Flesch-Kincaid grade level, as reported in main text. 95\% CIs are shown.}
\centering
\begin{tabular}[t]{lccccc}
\toprule
\multicolumn{1}{c}{ } & \multicolumn{5}{c}{(1)} \\
\cmidrule(l{3pt}r{3pt}){2-6}
\multicolumn{1}{c}{ } & \multicolumn{5}{c}{Flesch-Kincaid Grade Level} \\
\cmidrule(l{3pt}r{3pt}){2-6}
\textbf{VARIABLES} & \textbf{Estimate} & \textbf{Std. Error} & \textbf{CIs} & \textbf{t} & \textbf{p}\\
\midrule
Treated Country x Post-AI & 1.487 & 0.162 & {}[1.144, 1.830] & 9.157 & 0.000\\
Observations & 304 &  &  &  & \\
\bottomrule
\end{tabular}
\end{table}

\begin{table}[!ht]
\centering
\caption[Difference-in-differences for Mean Pairwise Similarity]{\label{tab:tab:main_similarity}Static difference-in-differences for mean pairwise similarity, as reported in the main text. 95\% CIs are shown.}
\centering
\begin{tabular}[t]{lccccc}
\toprule
\multicolumn{1}{c}{ } & \multicolumn{5}{c}{(1)} \\
\cmidrule(l{3pt}r{3pt}){2-6}
\multicolumn{1}{c}{ } & \multicolumn{5}{c}{Mean Pairwise Similarity} \\
\cmidrule(l{3pt}r{3pt}){2-6}
\textbf{VARIABLES} & \textbf{Estimate} & \textbf{Std. Error} & \textbf{CIs} & \textbf{t} & \textbf{p}\\
\midrule
Treated Country x Post-AI & 0.056 & 0.003 & {}[0.049, 0.063] & 17.825 & 0.000\\
Observations & 304 &  &  &  & \\
\bottomrule
\end{tabular}
\end{table}

\begin{table}[!ht]
\centering
\caption[Static Difference-in-differences with Country-Clustered Errors: Percent with 1+ Comment (30d)]{\label{tab:county_cluster_outcomes_1}Static difference-in-differences estimate for outcome metric: percent of petitions with 1+ comment in 30 days, with standard errors clustered by country. 95\% CIs are shown.}
\centering
\begin{tabular}[t]{lccccc}
\toprule
\multicolumn{1}{c}{ } & \multicolumn{5}{c}{(1)} \\
\cmidrule(l{3pt}r{3pt}){2-6}
\multicolumn{1}{c}{ } & \multicolumn{5}{c}{\% Petitions With 1+ Comment in 30 Days} \\
\cmidrule(l{3pt}r{3pt}){2-6}
\textbf{VARIABLES} & \textbf{Estimate} & \textbf{Std. Error} & \textbf{CIs} & \textbf{t} & \textbf{p}\\
\midrule
Treated Country x Post-AI & -2.508 & 1.822 & {}[-8.308, 3.291] & -1.376 & 0.262\\
Observations & 304 &  &  &  & \\
\bottomrule
\end{tabular}
\end{table}

\begin{table}[!ht]
\centering
\caption[Static Difference-in-differences with Country-Clustered Errors: Percent with 10+ signatures]{\label{tab:county_cluster_outcomes_2}Static difference-in-differences estimate for outcome metric: percent of petitions with 10+ signatures, with standard errors clustered by country. 95\% CIs are shown.}
\centering
\begin{tabular}[t]{lccccc}
\toprule
\multicolumn{1}{c}{ } & \multicolumn{5}{c}{(1)} \\
\cmidrule(l{3pt}r{3pt}){2-6}
\multicolumn{1}{c}{ } & \multicolumn{5}{c}{\% Petitions With 10+ Signatures} \\
\cmidrule(l{3pt}r{3pt}){2-6}
\textbf{VARIABLES} & \textbf{Estimate} & \textbf{Std. Error} & \textbf{CIs} & \textbf{t} & \textbf{p}\\
\midrule
Treated Country x Post-AI & -5.329 & 1.973 & {}[-11.609, 0.950] & -2.701 & 0.074\\
Observations & 304 &  &  &  & \\
\bottomrule
\end{tabular}
\end{table}

\begin{table}[!ht]
\centering
\caption[Static Difference-in-Differences with Country-Clustered Errors: MATTR]{\label{tab:country_cluster_lexical_1}Static difference-in-differences estimate for lexical feature metric MATTR, with standard errors clustered by country. 95\% CIs are shown.}
\centering
\begin{tabular}[t]{lccccc}
\toprule
\multicolumn{1}{c}{ } & \multicolumn{5}{c}{(1)} \\
\cmidrule(l{3pt}r{3pt}){2-6}
\multicolumn{1}{c}{ } & \multicolumn{5}{c}{MATTR} \\
\cmidrule(l{3pt}r{3pt}){2-6}
\textbf{VARIABLES} & \textbf{Estimate} & \textbf{Std. Error} & \textbf{CIs} & \textbf{t} & \textbf{p}\\
\midrule
Treated Country x Post-AI & 0.049 & 0.002 & {}[0.042, 0.055] & 23.150 & 0.000\\
Observations & 304 &  &  &  & \\
\bottomrule
\end{tabular}
\end{table}

\begin{table}[!ht]
\centering
\caption[Static Difference-in-Differences with Country-Clustered Errors: Word Count]{\label{tab:country_cluster_main_lexical_2}Static difference-in-differences estimate for lexical feature metric word count, with standard errors clustered by country. 95\% CIs are shown.}
\centering
\begin{tabular}[t]{lccccc}
\toprule
\multicolumn{1}{c}{ } & \multicolumn{5}{c}{(1)} \\
\cmidrule(l{3pt}r{3pt}){2-6}
\multicolumn{1}{c}{ } & \multicolumn{5}{c}{Word Count} \\
\cmidrule(l{3pt}r{3pt}){2-6}
\textbf{VARIABLES} & \textbf{Estimate} & \textbf{Std. Error} & \textbf{CIs} & \textbf{t} & \textbf{p}\\
\midrule
Treated Country x Post-AI & 53.628 & 4.584 & {}[39.040, 68.215] & 11.699 & 0.001\\
Observations & 304 &  &  &  & \\
\bottomrule
\end{tabular}
\end{table}

\begin{table}[!ht]
\centering
\caption[Static Difference-in-Differences with Country-Clustered Errors: Flesch-Kincaid Grade Level]{\label{tab:country_cluster_main_lexical_3}Static difference-in-differences estimate for lexical feature metric Flesch-Kincaid Grade Level, with standard errors clustered by country. 95\% CIs are shown.}
\centering
\begin{tabular}[t]{lccccc}
\toprule
\multicolumn{1}{c}{ } & \multicolumn{5}{c}{(1)} \\
\cmidrule(l{3pt}r{3pt}){2-6}
\multicolumn{1}{c}{ } & \multicolumn{5}{c}{Flesch-Kincaid Grade Level} \\
\cmidrule(l{3pt}r{3pt}){2-6}
\textbf{VARIABLES} & \textbf{Estimate} & \textbf{Std. Error} & \textbf{CIs} & \textbf{t} & \textbf{p}\\
\midrule
Treated Country x Post-AI & 1.487 & 0.064 & {}[1.284, 1.690] & 23.299 & 0.000\\
Observations & 304 &  &  &  & \\
\bottomrule
\end{tabular}
\end{table}

\begin{table}
\centering
\begin{talltblr}[         
caption={Mixed effects generalized logistic regression model predicting outcome metrics. Metric is a binary value: petitions either meet this threshold or fail to meet the threshold. We include the interaction between petition order and repeat-petition writer cohort, to understand the predictive relationship between user cohort and the outcome metric, relative to the user's experience and past performance.},
note{}={+ p $<$ 0.1, * p $<$ 0.05, ** p $<$ 0.01, *** p $<$ 0.001},
entry={Mixed Effects Model for Outcome Metrics}, label={tab:mixed_effect_outcomes},
]                     
{                     
colspec={Q[]Q[]Q[]},
hline{2}={1-3}{solid, black, 0.05em},
hline{14}={1-3}{solid, black, 0.05em},
hline{1}={1-3}{solid, black, 0.08em},
hline{21}={1-3}{solid, black, 0.08em},
column{2-3}={}{halign=c},
column{1}={}{halign=l},
}                     
& 1 + Comment (30d) & 10+ Signatures \\
Intercept & -0.403*** & -0.291*** \\
& (0.082) & (0.085) \\
Second Petition & -0.193+ & -0.212* \\
& (0.101) & (0.100) \\
Cohort: Pre/Post & -0.136 & -0.004 \\
& (0.098) & (0.101) \\
Cohort: Post/Post & 0.002 & 0.123 \\
& (0.128) & (0.132) \\
Interaction: Second Petition × Cohort (Pre/Post) & -0.090 & -0.033 \\
& (0.121) & (0.120) \\
Interaction: Second Petition × Cohort (Post/Post) & -0.220 & -0.328* \\
& (0.158) & (0.157) \\
Num.Obs. & 9222 & 9222 \\
R2 Marg. & 0.005 & 0.004 \\
R2 Cond. & 0.358 & 0.418 \\
AIC & 11978.9 & 12126.9 \\
BIC & 12028.8 & 12176.8 \\
ICC & 0.4 & 0.4 \\
RMSE & 0.36 & 0.34 \\
\end{talltblr}
\end{table}

\begin{table}
\centering
\begin{talltblr}[         
caption={Mixed effects linear regression model for lexical style metrics. While word count and Flesch-Kincaid scores are log transformed, raw MATTR does not show skew and is therefore not log transformed. We include the interaction between petition order and repeat-petition writer cohort, to understand the predictive relationship between user cohort and the lexical feature metric, relative to the user's experience and past performance. Note 72 observations are removed from the FKGL model due to missingness (negative FKGL, invalid for log transformation).},
note{}={+ p $<$ 0.1, * p $<$ 0.05, ** p $<$ 0.01, *** p $<$ 0.001},
entry={Mixed Effects Model for Lexical Metrics}, label={tab:mixed_effect_model_style},
]                     
{                     
colspec={Q[]Q[]Q[]Q[]},
hline{2}={1-4}{solid, black, 0.05em},
hline{14}={1-4}{solid, black, 0.05em},
hline{1}={1-4}{solid, black, 0.08em},
hline{21}={1-4}{solid, black, 0.08em},
column{2-4}={}{halign=c},
column{1}={}{halign=l},
}                     
& MATTR & Log(Flesch-Kincaid Grade Level) & Log(Word Count) \\
Intercept & 0.795*** & 2.096*** & 4.890*** \\
& (0.002) & (0.013) & (0.025) \\
Second Petition & -0.001 & 0.021 & 0.059* \\
& (0.002) & (0.015) & (0.025) \\
Cohort: Pre/Post & -0.000 & 0.027+ & 0.031 \\
& (0.002) & (0.016) & (0.029) \\
Cohort: Post/Post & 0.041*** & 0.222*** & 0.329*** \\
& (0.002) & (0.020) & (0.038) \\
Interaction: Second Petition × Cohort (Pre/Post) & 0.031*** & 0.176*** & 0.273*** \\
& (0.002) & (0.017) & (0.030) \\
Interaction: Second Petition × Cohort (Post/Post) & -0.014*** & -0.001 & 0.005 \\
& (0.003) & (0.023) & (0.039) \\
Num.Obs. & 9222 & 9150 & 9222 \\
R2 Marg. & 0.087 & 0.052 & 0.038 \\
R2 Cond. & 0.186 & 0.410 & 0.508 \\
AIC & -28101.9 & 10322.3 & 115177.6 \\
BIC & -28044.8 & 10379.3 & 115234.6 \\
ICC & 0.1 & 0.4 & 0.5 \\
RMSE & 0.05 & 0.30 & 0.49 \\
\end{talltblr}
\end{table}

\clearpage

\subsection*{Change.org and the Write with AI Tool}
\addcontentsline{toc}{subsection}{Change.org and the Write with AI Tool}
Change.org launched their in-platform AI tool to select countries, including the United States, Canada, Great Britain, in October 2023. 
Change.org then launched the AI tool to Australia in late December 2023.
We received confirmation of the delayed release in Australia through personal communications with two Change.org employees.
The delay was due to internal logistics, and was not an intentional experiment (as far as the authors are aware). 
While a staggered roll-out is typical practice for tech companies, including Change.org, this delay may also be attributable to the new and quickly-shifting landscape of generative AI legislation.
At the time, users learned about the existence of the AI tool only upon entering the AI-assisted writing workflow after starting a petition~\cite{change_2023}.
During the staggered roll out period that we study in the main text difference-in-differences analysis (October-December 2023), the home page did not advertise the existence of an AI tool. 
Furthermore, users routed directly to petition pages did not see any information about whether or not the petition had been written with the support of the in-platform AI tool. 
However, by 2025, Change.org had updated the platform home page to advertise their AI tool~\cite{change_2025}.
On one Change.org home page, visible on February~14, 2025, there was a text box that advertised the AI tool as providing help for users' productivity and the quality of their petitions~\cite{change_2025}. 
The home page read, ``Create a compelling petition in minutes!'' It is possible that this messaging could increase AI suspicion for petition readers who visit the home page before reading petitions. 
Furthermore, this message suggests that the platform's intent in releasing the AI tool was to enable petition-writers to create high quality petitions with reduced effort. 
By lowering the barrier to petition-creation, we might speculate that the platform sought to increase the volume of petitions produced on the platform or increase petition completion rates.
In 2026, Change.org launched an AI-first home page, in which visitors to the home page could only start a petition with the AI tool~\cite{change_2026}. 
While it is unclear how broadly this home page design was rolled out, it is clear that the platform has increased their commitment to supporting petition writing with AI.

We note that we conducted our study without formal data-sharing collaboration from Change.org. 
Without access to internal data, our focus is limited to platform-level trends rather than the impact of AI on individual petitions or petition-writers.
For example, while we suspect that the AI tool might have increased petition completion rates, we are unable to study this with publicly available data. 
Prior work has discussed the challenges of studying online platforms without internal data access~\cite{lazer2020computationalsocialscience, bak2025moving}. 
We use this opportunity to echo previous calls for open collaboration between platforms and researchers.
With appropriate transparency~\cite{bakcoleman2026industryinfluencehighprofilesocial}, such collaborations could improve public understanding of the trade-offs, harms, and opportunities of in-platform AI-integration.

\subsection*{Data}
\addcontentsline{toc}{subsection}{Data}
We collected 1,517,433 petitions for our dataset, and used a subset of 361,635 for the difference-in-differences analysis in the main text. 
We constructed this subset by including only English language petitions that were posted by users in the United States, Canada, Great Britain, and Australia. 
We derive user language by selecting petitions that have an English language tag on the petition's web page and for which the \texttt{nltk} library recognized at least 50\% of the words as English~\cite{nltk}.
We also exclude petitions that have fewer than 5 words or 2 sentences. 
In Table~\ref{tab:balance_table}, we show the distribution of petition and user attributes for this difference-in-differences data subset before and after the release of AI feature.
We also show the distribution of our primary outcome and lexical feature metrics pre-AI access in Table~\ref{tab:tab:main_metrics_pre}, and the distribution of these metrics post-AI access in Table~\ref{tab:tab:main_metrics_post}.
We also show the distribution of density of our outcome metrics, comments and signatures, on a log scale in SI Appendix Figure~\ref{fig:outcome_hists}.
Note that references to treatment refer to the full launch of the in-platform AI tool.
We define the AI periods using the first complete week after each threshold date (January 1, 2022 for the start of pre-AI period; April 1, 2023 for start of A/B test period; October 1, 2023 for the start of the post-AI period). 
We use the week-level dates for consistency across analyses with our main difference-in-difference analyses, which operate using week-level time series. 
We mapped dates to weeks through the \texttt{pandas} Python package, version 2.2.2.
We use the default week mapping, by which \texttt{pandas} anchors weeks to end on Sunday (so a week runs Monday through Sunday). 
With the weekly aggregation, our pre-AI period includes 65 weeks, from the week of January 3, 2022 to the week of March 27, 2023 (N = 237,020 petitions); the A/B period includes 26 weeks, from the week of April 3, 2023 to the week of September 25, 2023 (N = 82,504 petitions); and the post-AI period of differential access includes 11 weeks, from the week of October 2, 2023 to the week of December 11, 2023 (N = 42,111 petitions). 

\subsection*{Preregistration}
We separately preregistered our main causal analysis and our supplemental petition labeling task.
The preregistration for the causal analysis is at: \url{https://aspredicted.org/xtcd-tvqk.pdf}.
The preregistration for the labeling task is at: \url{https://aspredicted.org/z6g8xe.pdf}.

We did not deviate from the preregistration for our labeling task.
We note three deviations from the preregistration for the causal analysis. 
First, in our preregistration we defined English-language petitions as those with a ``25\% match to the NLTK English corpus''. However, upon inspecting our data, we found that this threshold was far too low, with many non-English petitions passing the threshold. We then increased the minimum threshold for English language detection to 50\%, which is the threshold we used for all analyses in the main and SI text. To allow for consistent interpretation of lexical features, it would have been inappropriate to include non-English petitions despite their happening to fall above the threshold of 25\%. 
Second, we defined our exclusion criteria as ``anomalies, which we define as petitions without any alphanumeric characters and those with a negative number of signatures.'' However, in our main analysis we exclude petitions with fewer than 5 words or 2 sentences. 
We removed these short petitions because they are inappropriate for our analyses of lexical features or text homogeneity, which require at least 2 sentences for reasonable interpretation.
Furthermore, these short petitions are often meaningless. 
For example, Change.org does not allow users to delete their petitions directly, and as a result, some users simply edit their petition text to contain a single word or punctuation in place of full removal.
We replicated our main analysis and observe consistent results when we use the initial exclusion criteria, which includes petitions with any word count or sentence count (SI Appendix Figure~\ref{fig:short_text_included_did}). 
Third, our preregistration also noted that we would use inverse propensity weighting.
Our main analysis does not use weighting due to a lack of reliable pre-treatment covariates (described further under ``Difference-in-Differences Robustness Checks'' below). 
We include a doubly robust difference-in-differences model, which includes inverse propensity weighting, in SI Appendix Figure~\ref{fig:full_covar_did}. 
This analysis supports our main finding that access to in-platform AI changes lexical features but does not improve platform-level petition outcomes.

\subsection*{AI Detection}
\addcontentsline{toc}{subsection}{AI Detection}
We created a Change.org AI tool classifier, and applied it to the sample of English petitions used in the main analysis (N = 361,635).
We assigned this label with an ensemble method that combined an open-source AI-detection model's predictions with two heuristics for in-platform AI usage~\cite{desklib_2025_ai_text_detector}. 
We identified two key heuristics for the Change.org AI in-platform tool.
The first heuristic was word count, as petitions created with the AI feature tended to have at least 100 words.
The second heuristic was the presence of even text breaks, as petitions written with the AI feature tended to have multiple paragraphs that were separated by an even number of text break html tags.
We then used an open-source AI detection model created by Desklib and trained for general use AI detection in education contexts~\cite{desklib_2025_ai_text_detector}. 
As follows, we created three labels: word count over 100 words, an even number of text breaks (i.e., 2,4,6,8,10), and a Desklib AI prediction probability over 0.9. 
We then compiled these labels through a decision tree classifier predicting if text was produced with the in-platform AI feature or not. 

In order to train and test the ensemble classifier we generated 1,125 petitions using the Change.org in-platform AI feature. 
To generate these petitions, we prompted the Change.org AI tool to produce petitions using a random sample of petitions that had been posted in 2022, before general-purpose AI tools were released. 
We did not post any of the petitions that were generated, we only saved the text for use in AI detection. 
We then created a new random sample of 1,125 petitions from the pre-AI period to use for training. 
With this set of 2,250 annotated petitions, we used a random sample of 70\% (n = 1,576) of the annotated petitions for training and the remaining 30\% (n = 674) for testing.
We did not train the model using any of the human-written petitions that we prompted the AI tool with. 
Our ensemble classifier achieved 99\% accuracy, 99\% precision, and 100\% recall on the test set of 674 annotated human and AI petitions. 
SI Appendix Figure~\ref{fig:ai_detection} shows the percentage of petitions that we classify as AI-generated per week, in English-market countries with access to the AI feature (US, CA, GB) and without access to the feature (AU). 
Prior to Australian users receiving access, during the week of December 11, 67.9\% of petitions written in countries with access to the AI feature were labeled as AI-generated, while only 27.7\% of petitions created in Australia were labeled as AI-generated. 
We can also assess false positive rates through AI detection scores prior to the release of general purpose off-platform AI tools: in early 2022 (pre-ChatGPT release in November 2022), our AI detection ensemble method reported 8.8\% of petitions across countries as AI generated. Prior to April 1, 2023, we see an uptick to 11.0\% of petitions flagged as AI generated. As such, we see that our approach identifies at least some off-platform AI assisted petitions, which explains the 27.7\% AI usage rate that we estimate in Australia during the period for which users did not have access to the in-platform tool.
This figure demonstrates high utilization of the AI tool feature; just weeks after its full release most petitions with access to the feature were AI-generated.

\subsection*{Petition Writing Quality and Persuasiveness}
\addcontentsline{toc}{subsection}{Petition Writing Quality and Persuasiveness}
To better understand petition quality before and after the introduction of the AI tool, we labeled petition writing quality and persuasiveness in a sample of petitions across the two time periods. 
Our additional data collection shows that petitions written with access to in-platform AI have higher writing quality and persuasiveness, on average, than petitions written without access to AI. 
Furthermore, writing quality and persuasiveness, while positively predictive of petition outcomes in the pre-AI period, are weaker, or negative, predictors of petition outcomes in the post-AI period. 

We study petition writing quality and persuasiveness through a labeling task in which human labelers assigned writing quality and persuasiveness scores to a random sample of petitions.
We recruited 14 Cornell Tech graduate students to participate in the study as labelers.
We randomly sampled 432 English language petitions from our dataset, such that 216 were from the pre-AI period and 216 were from the post-AI period. 
Post-hoc we applied our AI detection method to predict if our sampled petitions written with AI or not. 
We show in SI Appendix Figure~\ref{fig:survey ai detection} that the ensemble detection model labeled 9.4\% of pre-AI petitions as AI-written and 53.3\% of post-AI petitions as AI written.
We selected a sample size of 432 to achieve power=0.8, \(\alpha=0.05\) to detect an estimated moderate effect size of~0.3. 
We adjusted our sample size calculation to account for the three ratings per petition under a conservative value of \(ICC = 0.6\), and \(ICC_{k=3} = 0.81\).
We sampled from petitions with fewer than 5,000 characters, which accounted for more than 98\% of petitions, to account for Qualtrics word limits. 
To maintain consistency with petitions included in our main analysis, we also excluded petitions with fewer than five words or two sentences. 
We then removed 15 petitions from the sample of 432 that used offensive language or included personal identifying information of petition authors (e.g., email address, online handles).
This left us with a sample of 417 petitions (203 pre-AI, 214 post-AI).
We assigned each rater a random set of 80-90 petitions written both pre- and post- AI.
The Cornell Institutional Review Board (Cornell University's FWA is FWA00004513) deemed the labeling task exempt from formal ethics review.
Human labelers provided informed consent for the task, and were compensated \$20.00 per hour (\$60.00 for a three-hour task).

We asked participants to rate petitions for writing quality and persuasiveness. 
We adapted the measures for writing quality from the measures of email writing quality used by Noy and Zhang~\cite{noyzhang2023science}. 
We share these measures in Table~\ref{tab:survey_measures}.
We drew from prior work on message credibility to inform our measures for persuasiveness~\cite{appelman2016messagecredibility}. 
We adapt them for the purposes of the annotation task, in which our labelers were unfamiliar with the context or communities of sampled petitions and thus could not assess petition authenticity.
We ask about two specific forms of influence that a petition can exert: influencing readers and decision-makers.
We then take the average of the writing quality measures as a writing quality score and the average of the persuasiveness measures as the persuasiveness score.

We find that on average, petition writing quality and persuasiveness is significantly higher in petitions written post-AI, i.e. those who had in-platform AI access, compared to petitions written before AI access. 
SI Appendix Figure~\ref{fig:survey_boxplots} shows the distribution of writing quality and persuasiveness scores among petitions written pre-AI and post-AI.
The average writing quality in the pre-AI petitions was 2.73, while the average writing quality in the post-AI petitions was 3.43. 
A Welch two-sample t-test rejects the null hypothesis of no difference in writing quality between the periods, with a large effect size (\(\mu_{post-pre}=0.70\), t(412.46) = 9.76, \( p<\) 0.001, Cohen's d = 0.96, 95\% CI=[0.75, 1.16]).
The average persuasiveness score for pre-AI petitions was 2.46, while the average persuasiveness score post-AI was 3.12. 
A Welch two-sample t-test rejects the null hypothesis of no difference in persuasiveness scores between periods \(\mu_{post-pre}=0.66\), (t(414.56) = 8.75, \(p<\) 0.001, Cohen's d = 0.86, 95\% CI=[0.65, 1.06]).
In sum, we find that petitions written with access to the AI feature have, on average, higher writing quality and persuasiveness scores.

Our analysis also shows that prior to in-platform AI, higher petition writing quality and persuasiveness scores were positively associated with petition outcomes.
Post-AI, when predicting likelihood of receiving 1+ comment in 30 days, the predictive strength of text persuasiveness weakened, and the predictive relationship between writing quality and petition outcomes reversed.
SI Appendix Figure~\ref{fig:writing_quality_marginal_effects} demonstrates that the predictive relationship between writing quality and outcome (1+ comment in 30 days) observed pre-AI is weaker when petitions are written post-AI. 
We fit a logistic regression model to predict the likelihood of a petition receiving at least 1 comment in 30 days, regressed on the full interaction of average petition writing quality and treatment period. 
We show model coefficients in Table~\ref{tab:tab:writing quality mod exp comments}. 
We observe a similar significant interaction effect between treatment period and persuasiveness score, as shown in Table~\ref{tab:tab:persuasive mod exp comments}.
SI Appendix Figure~\ref{fig:persuasive_marginal_effects} shows that while pre-AI, higher persuasiveness scores were predictive of higher probability of a positive outcome (1+ comment in 30 days), post-AI, persuasiveness was a weaker predictor of outcome. 
We observe consistent results for our other main outcome metric (10+ signatures), as shown in SI Appendix Figure~\ref{fig:survey_marginal_effects_signatures}, with model coefficients in Tables~\ref{tab:tab:persuasive mod exp signatures} and~\ref{tab:tab:writing quality mod exp signatures}.
With in-platform AI, the predictive relationship between text quality (i.e., writing quality, persuasiveness) and outcomes weakened or reversed: the interaction terms indicate that higher writing quality and persuasiveness no longer increase, and in the case of writing quality, may decrease the probability of receiving a comment.
Writing quality and persuasiveness cease to be strong signals of petition outcomes; instead they become weakly, or negatively, associated with outcomes post-AI.

To understand the relationship between lexical features and our measures of petition quality, we consider correlations between variables. 
SI Appendix Figure~\ref{fig:correlation plot lexical features and survey measures}
shows the linear relationship between lexical features (MATTR, log-transformed word count, Flesch-Kincaid grade level) and our measures of text quality (writing quality score, persuasiveness score). 
Pearson correlation coefficients with writing quality range from modest (MATTR) to moderate (log-transformed word count, Flesch-Kincaid grade) positive relationships.
The same pattern holds for correlations with persuasiveness scores.

\subsection*{Sentiment Analysis}
\addcontentsline{toc}{subsection}{Sentiment Analysis}

Our analysis shows that petition sentiment is more positive after the introduction of the in-platform AI tool, as measured as the average ratio of positive/negative/neutral sentiment per sentence.
SI Appendix Figure~\ref{fig:vader} shows the distribution of petition sentiment, as calculated with VADER (Valence Aware Dictionary and sEntiment Reasoner) tool.
VADER is a tool for sentiment analysis that uses lexical and rule-based features to assign sentiment scores to short text~\cite{Hutto_Gilbert_2014}. 
We use the VADER python library to assign sentiment scores to each petition. 
To do this, we first decompose each petition into sentences, as is the recommended approach for longer text. 
VADER provides the ratio of each sentence that is positive, neutral, and negative. 
To retrieve average per-sentence sentiment for each petition, we then take the average of each positive, neutral, and negative score across the sentences within a petition.
SI Appendix Figure~\ref{fig:vader} shows the distribution of average sentence sentiment for each English language petition written pre- and post- AI in the United States, Canada, Great Britain, and Australia. 
When petitions are written with access to in-platform AI, the average share of neutral sentiment in petition sentences decreases, as shown by the distribution shifting to the left in SI Appendix Figure~\ref{fig:neutral}.
Conversely, the average share of positive sentiment in petition sentences increases, as the distribution shifts to the right in SI Appendix Figure~\ref{fig:positive}.
The distribution of negative sentiment does not shift, but rather narrows about the peak of 0.1, as shown in SI Appendix Figure~\ref{fig:negative}, thus demonstrating a concentration of average sentence negativity ratio at 0.1.
Previous literature has documented a positivity bias in text produced by generative AI tools, which supports our finding of a modest increase in positive sentiment~\cite{donmez-etal-2025-ai, markowitz2024}.

\subsection*{Title Concreteness}
\addcontentsline{toc}{subsection}{Title Concreteness}

Our data analysis shows that the language used in petition titles is less concrete 
when petitions are written with access to in-platform AI.
We study the specificity of words used in petition titles through the headline concreteness measure developed by Aubin le Quéré and Matias~\cite{quere_Matias_2025}.
The authors define concreteness as defined the ``degree to which the concept denoted by a word refers to a perceptible entity.''
The measure combines Named Entity Recognition tagging with a psychologically validated scale of English word concreteness. 
We apply the measure to English Change.org petition titles, and display the distribution of title concreteness in SI Appendix Figure~\ref{fig:title_concreteness}.
We apply the measure to Change.org petitions posted from the US, GB, CA, and AU during the pre-AI period (January 2022-March 2023) and post-AI period (January 2024-December 2024). 
The measure crafted by Aubin le Quéré and Matias excludes text that does not have at least one word tagged with their concreteness score.
We observe a small decrease in median petition title concreteness with in-platform AI access, as the median title concreteness decreased from 3.63 to 3.50 (Kruskal-Wallis test H(1)=5341.433, p \( <\) 0.001, \( \eta^2=0.011\)).
We hypothesize that the in-platform AI tool may contribute to a decline in petition title specificity
if users fail to prompt the in-platform AI tool with specific enough information.

\subsection*{Difference-in-Differences}
\addcontentsline{toc}{subsection}{Difference-in-Differences}

In the main text, we report on static and dynamic difference-in-difference estimates. Here, we provide a series of further analyses that evidence the robustness of our findings. 
We use the software provided by Callaway and Sant'Anna to create our static and dynamic difference-in-difference estimations~\cite{callaway_santanna_2025_did}.
We use the R \texttt{did} package (version 2.1.2) to estimate the dynamic and static difference-in-difference analyses. 
For the static difference-in-differences analysis we clustered standard errors at the month level. 
For the dynamic difference-in-differences, we calculated standard errors with a multiplier bootstrap (1,000 iterations) to create a uniform confidence band. 
In the main text, we found that lexical features changed significantly, homogeneity increased, and petition outcomes were either unchanged or worsened by access to the AI tool. 
While we include figures in the main text to visualize these results, we also provide tables for the static difference-in-differences outcomes here. 
We do not include tables for the dynamic difference-in-differences, as each table has as many terms as there are time periods (76). However, code for producing these tables is publicly available on OSF.\footnote{\url{https://osf.io/q3hj5/?view_only=3fc5b89d05eb4741957af431b107cf2a}}
Tables~\ref{tab:tab:main_outcomes_1}-\ref{tab:tab:main_outcomes_2} include the main text outcome variables (percent of petitions with 1+ comment (30d) and percent of petitions with 10+ signatures). 
Tables~\ref{tab:tab:main_lexical_1}-\ref{tab:tab:main_lexical_3} include the static difference-in-differences estimates for lexical features in the main text (MATTR, Flesch-Kincaid Grade Level, and word count).
Table~\ref{tab:tab:main_similarity} includes the static difference-in-differences estimation of petition similarity.

In order to satisfy the assumption that trends in treated and control countries were similar prior to the in-platform AI intervention, we used ordinary least squares regression to fit trend lines to the pre-AI period in each group of countries. SI Appendix Figure~\ref{fig:parallel_trends} shows that the interaction term was insignificant in all cases, with no significant difference in slope for pre-AI trends in treated and control countries.

\subsection*{Difference-in-Differences Robustness Checks}
\addcontentsline{toc}{subsection}{Difference-in-Differences Robustness Checks}
We demonstrate the robustness of our results through several additional checks. 
First, in SI Appendix Figure~\ref{fig:additional_outcomes} we demonstrate that dynamic and static difference-in-differences analysis using additional outcome thresholds reveal similar results and the raw count of comments and signatures.
SI Appendix Figure~\ref{fig:additional_outcomes_thresholds_event_study} and SI Appendix Figure~\ref{fig:additional_outcomes_thresholds_pre_post} show that for a range of comment and signature thresholds, the share of petitions that reach these thresholds is either significantly lower or unchanged for treated countries with AI access. 
In SI Appendix Figure~\ref{fig:additional_outcomes_log_event_study} and SI Appendix Figure~\ref{fig:additional_outcomes_log_pre_post}, we log-transformed raw counts of comments and signatures per week, and then normalized these values by the number of petitions produced in each week. 
We find that this metric, which leverages raw signatures and comments rather than thresholds, reveals the same patterns of unimproved outcomes in the post-AI period.

In addition to estimating the impact of AI on additional outcome measures, we also estimate the impact of AI on additional lexical features. In SI Appendix Figure~\ref{fig:additional_lexical_features} we also conduct dynamic and static difference-in-differences using time series of additional median weekly lexical features.
In particular, we consider words per sentence, characters per word, readability as measured by the Gunning Fog index, and the count of sentences~\cite{gunningfog1969}.
The Gunning Fog index is a measure of readability that is calculated as $0.4\times(\frac{\text{words}}{\text{sentences}}) + 100\times (\frac{\text{complex words}}{\text{words}})$, with complex words defined as select multi-syllabic words.
We find that across these four additional measures, after AI access the median values are significantly increased, as is consistent with the shifts observed in the main text. 

As our pre-AI period is substantially longer than the post-AI period for which we have differential treatment, we repeat the main text difference-in-differences specification using a subset of the pre-AI data, to demonstrate that our estimates are not indicative of any one pre-AI treatment period. 
In SI Appendix Figure~\ref{fig:narrow_did}, we show that dynamic and static estimates for lexical and outcome features that use a shorter time frame for the pre-AI period (25 weeks pre-AI, 11 weeks post-AI) recover the same results as in our main text, where we use our full data. Again, lexical metrics are significantly increased, and in this case, outcome metrics are either unchanged or significantly decreased. This shows that our results are robust to pre-AI time period length. 

We end our main analysis difference-in-differences with December 15th, in order to prevent including weeks in which Australia gained access to the AI feature. To demonstrate that results would hold even for the full month of December included in the period of differential access, we repeat our main analysis with the end date of December 31, 2023. SI Appendix Figure~\ref{fig:december_did} we use the full month of December, rather than curtailing our study with the last full week prior to December 15th. In this study, we show that we recover similar results when using the full month of December for the static difference-in-differences results. However, during the final two weeks of December we observe heterogeneous effects for lexical features, with insignificant weekly estimates of change to lexical features. This is potentially indicative of the time in which Australian users received AI access. 
Outcomes retain the same patterns as in the main text, with significantly reduced share of petitions reaching the signature threshold and unchanged share of petitions reaching the comment threshold. 

In SI Appendix Figure~\ref{fig:short_text_included_did}, we show static and dynamic difference-in-difference estimates had we included short text. In our main report, we drop petitions that have fewer than 2 sentences or 5 words. This is for two reasons: first, such short petitions are unlikely to be legitimate petitions seeking to incite change; second, our lexical metrics require at least two sentences to be coherent. For instance, our readability metric accounts for average number of words per sentence, which would not be meaningful on a one-sentence petition. Likewise, extremely short text is likely to have high lexical diversity by virtue of using few words overall. We show through SI Appendix Figure~\ref{fig:short_text_included_did} that results are consistent with our main text, even when including short petitions. Namely, lexical features increase significantly, and outcomes are unchanged. 

The difference-in-differences model in the main text does not include covariates. 
We use a doubly robust~\cite{callaway_santanna_2025_did} difference-in-difference estimator with both inverse probability weighting and ordinary least-squares regression to control for covariates. 
We include the following covariates: the share of petitions per week and country that were (a) posted on the weekend (b) had decision-makers specified, and the share of users per week and country that (c) were new, and (d) had commented on a petition before. 
We use these covariates as they were available to us from our scrape of the Change.org platform. 
However, it is possible that these covariates are post-treatment (impacted by whether or not a petition was written with AI access).
For example, perhaps the share of petition-writers that had themselves commented on another user's petition is \textit{lower} in the post-AI period, if users are less likely to comment on petitions when they know other authors may use AI. 
For this reason, we interpret these results with caution. 
From this analysis we again recover estimates for change to style and outcomes as we did in the main analysis. 
SI Appendix Figure~\ref{fig:full_covar_did} shows that lexical features increase significantly.
However, the analysis using this model is slightly different with regard to the outcomes than the main analysis: whereas the main text shows that outcomes become worse with access to AI, this doubly robust estimate shows no directional difference in outcomes.
Overall, this analysis supports the claim that access to in-platform AI changes lexical features but does not improve platform-level petition outcomes.

In SI Appendix Figure~\ref{fig:did_se_country_clustered} and Tables~\ref{tab:county_cluster_outcomes_1}-\ref{tab:country_cluster_main_lexical_3}, we present a static difference-in-differences analysis with standard errors clustered by country. 
In our main analysis, we cluster standard errors by month to account for temporal shocks that would affect all countries~\cite{riochanonaAreLargeLanguage2024}.
Another approach to clustering standard errors would be to cluster at the unit of treatment assignment, which would be country in this case~\cite{ROTH20232218}.
However, as we have only four countries used in our main analysis, this would result in too few units for reliable clustering and increase the risk of severe over-rejection~\cite{mackinnonRandomizationInferenceDifferenceindifferences2020}.
As expected, the 95\% confidence intervals calculated with country-clustered standard errors remain significant for lexical features, and are insignificant for outcome features.
These results are consistent with our main finding: access to the in-platform AI tool significantly changed platform-level trends in petition lexical style, without improving outcomes. 

In SI Appendix Figure~\ref{fig:did_on_gov_petitions} we repeat our main text metrics on a subset of petitions that we labeled as being about government and politics topics (we describe methods in the Petition Topic section). While many petitions are posted on Change.org, some pertain to trivial topics or appear to be made without any serious intent to enact change. We therefore sought to understand changes within a subset of petitions that pertain to topics relevant to civic engagement and democratic participation. We found that outcomes and most lexical features were consistent with main text results: lexical features increased significantly while outcomes did not change significantly. The one notable deviation from main text results is that word count is no longer significantly increased when petitions are written with access to AI. Instead, these petitions are unchanged in length. This is likely because government petitions tended to be longer than the average petition in the pre-AI period. As a result, the in-platform AI feature did not make a difference, even if it was highly utilized, for word count. 

In SI Appendix Figure~\ref{fig:placebo_did} we conduct a placebo static difference-in-differences. It is unnecessary to conduct a placebo dynamic difference-in-differences, as we can understand week-to-week changes in average treatment effects by time period in the pre-AI period by simply observing the pre-AI part of the dynamic figures in the main text. 
Our placebo test uses the pre-AI data, with the intervention date of October 1, 2022. This is to ensure that any seasonal effects in the main report would be replicated here. 
To maintain consistency, we also include a placebo A/B test period (which we skip), from April 1 - October 1, 2022. 
We therefore specify the placebo pre-intervention period as January 1, 2022 - March 31, 2022 and the post-intervention period as October 1, 2022 - March 31, 2023. 
We stop the placebo post-period before the actual A/B test intervention.
The placebo test, as expected, does not recover significant estimates in lexical features or outcomes for the placebo static difference-in-differences.

\subsection*{Verb Frequency}
\addcontentsline{toc}{subsection}{Verb Frequency}

We demonstrate the changes to the linguistic features of petitions written with access to the AI tool in SI Appendix Figure~\ref{fig: Top words}. In particular, we show time series of the frequency of select verbs in petition titles. We focus on verbs as these are less specific to the subject or location of the petitions. 
In SI Appendix Figure~\ref{top_words_pre_ai}, we show the 10 words with the highest frequency in the pre-AI period. We observe that the word ``stop'' spikes in frequency in March, 2023: this is because of the viral \#StoptheWillowProject movement, in which many individuals took to Change.org to create petitions about stopping an oil drilling project that would threaten the environment. 
Otherwise, we observe consistent usage of these top 10 words across the time period. 
In SI Appendix Figure~\ref{top_words_post_ai}, we show the 10 words that have the highest post-AI to pre-AI frequency ratio. One notable case is the word ``implement'', which has the greatest jump in pre to post AI frequency, with utilization hovering around 4 titles per 100 petitions after full AI release. 
In the post-AI period we can also observe the influence of changes to the underlying model that powers the AI tool: in mid 2024, the Change.org AI tool was briefly updated to a different large language model, which produced different text. This period falls outside of our main difference-in-differences analysis. 

The influence of the brief model update in 2024 is also apparent in SI Appendix Figure~\ref{fig:time_series_lexical_features}, in which the word count trends dip substantially. 
The time series figures demonstrate that trends prior to AI release were parallel in treated and control countries. Furthermore, after all countries have AI access we observe parallel trends. 
The only period for which lexical features vary in treated and control countries is during the period of differential AI access.

In SI Appendix Figure~\ref{fig:implement_by_country} we demonstrate that this increase in occurrence of the word ``implement'' in petition titles was consistent across countries, with \(<\)1\% of petitions per month using the word pre-AI, and 4-5\% of petitions per month using the tool post-AI access. The United States, Great Britain, and Canada (blue lines) had an earlier spike in occurrence of ``implement'' in October 2023, which was the full roll-out of the AI tool.
Australia then had a spike in word usage around December 2023, which was when Australian users gained access to the tool. In 2024 when all four countries had access to the AI tool we observe similar share of petitions in each country used the term ``implement'' in the titles.

\subsection*{Petition Language Variant}
\addcontentsline{toc}{subsection}{Petition Language Variant}

The current edition of the Change.org AI tool does not appear personalized by country (as of time of writing: January 2026).
While the AI tool produces petitions in the language of the user's input text, our exploratory interactions with the tool suggested that if the user inputs text in English varieties or dialects other than American English, the AI tool may yet produce a petition in American English.
SI Appendix Figure~\ref{subfigure: percent american english overall} supports this claim, showing that the percent of petitions classified as American English, using a simple spelling and vocabulary based classifier, increases in Canada, Great Britain, and Australia when petitions are written with access to the AI tool.
We use the \texttt{abclf} Python package to classify petition English variety. The package uses spelling and vocabulary to classify English variety (American vs. British). 
The \texttt{abclf} package classifies text as either British English, American English, unknown, or as a mix of British and American English. 
The classifier checks if words are present in the VarCon dictionary, and if not, assigns unknown.
The classifier assigns a text to a variant if it has more than twice as many identified words as the other variant. 
Otherwise it classifies the text as a mix~\cite{abclf101}.
As shown in SI Appendix Figure~\ref{fig: english variety}, access to the AI tool increases the share of petitions that are assigned either the American English or British English label. 
This is likely because many petitions do not use any spelling or vocabulary that is specific to a single English variety, as identified by \texttt{abclf}. 
With access to the AI tool, however, petitions had longer and more complicated vocabularies, which were more likely to include words that trigger a vocabulary-based classifier (e.g. ``organize''  or ``organise''). 
SI Appendix Figure~\ref{subfigure: percent gb english conditional} shows that conditional on receiving a label at all (British or American English), the percent of petitions labeled British English decreased while in SI Appendix Figure~\ref{subfigure: percent american english conditional}, the percent of petitions that use American English increased. As SI Appendix Figure~\ref{subfigure: percent american english overall} shows, the \textit{relative} increase of American English petitions seems to be driven by an increase in petitions labeled American English rather than a decrease in petitions labeled British English. 
SI Appendix Figure~\ref{subfigure: percent gb english overall} shows that the overall share of petitions labeled British English remains consistent.

The increase in percent of petitions labeled as American English could amount to a quality of service harm~\cite{shelby2023sociotechnical},
as users may find that the petition produced with the tool does not reflect their intended dialect, language, and grammar. 
However, for the purposes of our study, we do not expect that the tool's predilection for American English responses to factor into the results we observe. 
While users in our main control country, Australia, may find the tool to be misaligned with their Australian English, it is also the case that two of our treatment countries (Canada and Great Britain) would experience analogous misalignment. 
As shown in SI Appendix Figure~\ref{subfigure: percent american english overall}, the percent of petitions labeled as American English has a parallel increase in all countries.
While this classification is likely an underestimate considering the low detection rate of American English petitions among petition produced in the US, the trends in English varieties suggest that English variety does not explain the patterns that we observe with AI access. 
In SI Appendix Figure~\ref{fig:no_us_did} we conducted a difference-in-differences analysis where we exclude the United States as a treated country, to compare changes in lexical features and outcomes between countries where the local variety of English is non-American. 
We find results are consistent with those that include the US as a treated country (as in the main text). 

SI Appendix Figure~\ref{fig:single_country_comparisons} further demonstrates that it is unlikely that quality of service harms by English language variety drive our results.
 Here, we show the trends reported in the main text are consistent with the single-country comparisons between Australia to Great Britain; Australia to Canada; Australia to the United States. 
Since Australian and British English similarly diverge from American English, if it were the case that country-level varieties of English affected the outcomes reported in our main analysis, we would expect to see differences in countries' pairwise comparisons. 
While it may be the case that the AI tool fails to generate petitions in dialects and varieties of English other than Standard American English (SAE), we do not anticipate that our causal estimates were driven by this aspect of the AI tool. 

We also compare petition outcomes by English variety, before and after the introduction of AI to the platform. SI Appendix Figure~\ref{fig:outcomes_english_variety} shows  that there is no single shift in all countries' outcome metrics for petitions written in American English. 
In the United States, Canada, and Australia, the share of American English petitions that reach 1+ comment in 30 days is lower when petitions are written with access to AI. 
On the other hand, we see a much more minor difference in the share of British English petitions that reach 1+ comment before and after AI access. 
If there were a quality-of-service harm introduced by the AI tool in countries for which the English variety is non-American, we would expect to see a consistent decrease in outcomes for American English petitions written post-AI in Canada, Great Britain, and Australia. Though we do see decreases in Canada and Australia, there is no notable difference in Great Britain. 
Furthermore, there is a similar sharp difference in pre- and post- AI outcomes in the United States for American English petitions, which we would not attribute to a quality-of-service harm from national language variety.

\subsection*{Petition Topic}
\addcontentsline{toc}{subsection}{Petition Topic}

Change.org petitions cover a wide array of topics, from education to corporate policies, to pop culture and more.
The petition topic distribution did not meaningfully shift between the periods we study in the main analysis difference-in-differences. 
Further, the difference-in-differences results replicate even when focusing the analysis on specific topics that are related to civic action and politics.

To analyze topics, we followed the topic extraction pipeline used by Costello et al., who took open-ended participant text, summarized the text, created embeddings for summarizations, and then clustered the embeddings to define topics~\cite{costello2024}. 
We created one-sentence petition summaries with the OpenAI Chat Completions API with the model gpt-4o-mini. 
Our prompt read: ``You are summarizing Change.org petitions. Summarize the following petition, which describes the author’s claim or request, in a single sentence. Do not say that it is a petition, and do not add any normative judgment. Merely accurately describe the content in a way the author who wrote the petition would concur with. Frame it as an assertion. If the petition content is already short, no need to change it very much. If it is quite long and detailed, be sure to capture the core, high-level points, and key words relevant to the topic. Do not focus on the evidence provided for the petition, but merely focus on the basic assertion. Return ONLY strict JSON with keys: petition\_id, title, claim.'' 
To control input size and cost, we truncated petition text to a maximum of 500 words prior to summarization, while always including the petition title to preserve context. Only 4.37\% of petitions were long enough to be truncated.

We then generated embeddings for each summarization with the OpenAI model text-embedding-3-small, which produces 1536-dimensional embeddings. 
During verification, we identified that approximately 0.01\% of English language petitions (115 of 964,751 English language petitions) had missing/empty embeddings. Manual inspection revealed that these petitions had empty summaries, typically originating from petitions with placeholder titles or non-substantive text (e.g., ``.'', ``ignore'', ``null'', `` ''). These likely stem from the nature of the Change.org platform, which does not allow users to delete their petitions. 
Instead, users might simply edit the text to contain non-substantive placeholder text, like the examples above.

We clustered embeddings with MiniBatchKMeans with K=20, which approximates standard K-means by repeatedly updating cluster centroids based on small random batches of observations. 
We used both the inertia (within-cluster sum of squares) and the silhouette score to determine the value of K. 
We used this method due to the size of our dataset (964,636 English language petitions) and the high dimensionality of embeddings. 
Following clustering, clusters were interpreted by manually inspecting random samples of petition claims within each cluster.
We assigned each cluster a descriptive topic label based on its dominant theme, with labels determined independently by two annotators, then combined. 
In a final consolidation step, we grouped clusters reflecting closely related subject matter under broader thematic categories. For example, clusters related to schooling were consolidated under ``School and Education,'' entertainment-related clusters under ``Pop Culture and Media,'' environmental and animal-related clusters under ``Conservation and Environment.'' 
This manual review and consolidation process mirrors the qualitative validation emphasized by Costello et al., ensuring that unsupervised clusters correspond to substantially meaningful topics~\cite{costello2024}.

SI Appendix Figure~\ref{fig:topic distribution} shows the distribution of topics for 361,635 petitions in the main analysis during the pre-AI, A/B test, and post AI periods.
The distribution of topics across the three periods remains consistent, with the most frequent topics being Pop Culture and Media, School and Education, Criminal Justice, and Community Planning. 
A 13x3 chi-squared test comparing the distribution of topic by period (pre-AI, A/B test, post-AI) had a negligible effect size, as shown by Cramer's V \(<\) 0.1 ($\chi^2(24)$ = 2950.5, $p<$ 0.0001, Cramer's V=0.064),
suggesting little to no relationship between topic and time period.
Relatedly, SI Appendix Figure~\ref{fig:same topic repeat writers}
shows a consistent share (about one-third) of repeat-petition writers return to the platform to write about the same topic. This trend is consistent across cohorts (Pre/Pre, Pre/Post, Post/Post).

We then focused our analysis on a set of topics related to civic engagement (Conservation and Environment, Community Planning, International Politics, Social and Economic Welfare, Civil Rights and Diversity, Criminal Justice, Public Safety). 
To verify the topic selection, the first author (IC) sampled 100 petitions randomly (50 civic engagement topics and 50 other topics) and manually classified them as related to civic engagement, or not. The manual ratings had 82\% agreement with the label based on the topic assignment via clustering (i.e., civic engagement topic or other). 
We used this civic engagement label to repeat the static difference-in-differences analysis.
In SI Appendix Figure~\ref{fig:civic engagement}, we show that the trends reported in the main analysis, by which lexical features change significantly and outcomes do not show evidence of improvement, are consistent when limited to the set of civic engagement petitions. 
We also repeat the static difference-in-differences analysis with other petition topics (Pop Culture and Media, Consumer Products and Brands, Leadership Accountability, School and Education, Personal Requests and Inside Jokes, Ban Offensive Content), and observe consistent trends in SI Appendix Figure~\ref{fig:other topic}.

In addition to labeling petitions through unsupervised clustering, we sought to understand trends around the specific topic of political petitions, as classified through a large language model.
Previous literature in petitioning has often focused on petitions related to politics and government, as these have the potential for the greatest impact. 
In order to understand if the share of petitions produced on Change.org that related to government topics changed after the introduction of AI, we labeled petitions as pertaining to a government or non-government topic. 
We used zero‑shot classification with the OpenAI gpt-4o-mini model. 
Each petition’s text (truncated to 8,000 tokens) is sent to the model with a system prompt plus a category‑specific user prompt, and the model responds with just “A” (in cluster) or “B” (not in cluster). The decoding is deterministic (temperature=0, no fine‑tuning), and we run the requests in parallel with up to 50 threads.
We then conducted a dynamic difference-in-differences analysis to understand how the share of petitions pertaining to government topics changed after the release of the in-platform AI feature.
As shown in SI Appendix Figure~\ref{fig:gov_event_study}, there was no change in trend around the share of petitions pertaining to government topics. Instead, the share of petitions that concerned government topics in the post-AI period (blue, right of week 26) was consistently insignificantly different from the baseline week (Week -1) prior to A/B testing of the AI feature. 

The system prompt is as follows: 
``You are a highly efficient information detection and extraction engine, specialized in analyzing natural language data.
You value accuracy: when the user asks you to extract certain information from given text data, you will try your best
to adhere to what is directly mentioned in the text and the extraction criteria. You value efficiency: your response
will be concise and follow formatting conventions specified in the extraction prompt.''

The user prompt is as follows:
``The following text is a petition. Is the petition topic Government and Public Safety? 
This includes petitions that are about matters related to local government, government and politics, or public safety. 
Output a letter only.
A) Yes, government and public safety topic
B) No, other topic
Text: {text}
Answer:''

We then validated responses through a manually labeled set of 153 petitions (2 raters, cohen's kappa = 0.583). The LLM-annotated results used 104 of the labeled petitions to achieve 86.5\% accuracy, 82.3\% precision, 89.3\% recall on this set.

\subsection*{Predictive Strength of Lexical Features}
\addcontentsline{toc}{subsection}{Predictive Strength of Lexical Features}

As we show AI, significantly changes lexical features, and it is expected that the distribution of each feature would be different in the pre-AI and post-AI periods. 
We show this through the histograms captured in each panel of SI Appendix Figure~\ref{fig:word_count_predictive}.
For instance, the distribution of petition word count shifts from the 5-45 word decile to the 200-228 word decile.
The post-AI word count mode corresponds to the word count that the AI feature tends to produce, which is the 100-250 word range. 
SI Appendix Figure~\ref{fig:word_count_predictive} also shows the marginal means of word count deciles in the pre-AI and post-AI periods. 
The marginal mean of each decile corresponds to the predicted probability of a positive outcome, as measured by the presence of at least one comment within 30 days of petition post.
We calculated CIs with HC3 heteroskedasticity-consistent
standard errors~\cite{longervin2000}.
Before access to AI, higher word count deciles corresponded to higher average predicted probability of a positive outcome. 
Thus, with the shift of petition word count to the 100-300 word range, we would expect to see petitions that have average mean predicted probability of positive outcome in the 0.3-0.4 range. 
However, with access to the feature, the predictive strength of word count plateaus once it enters the AI-range: lengthier petitions are no longer more likely to receive a positive outcome than their shorter counterparts. 
In fact, petitions in the 97-130 word range have higher, though insignificantly so, average marginal means than petitions in the 167-200 word range. 
We observe similar patterns for the Flesch-Kincaid grade level and the MATTR.

\subsection*{Repeat Petition Writers}
\addcontentsline{toc}{subsection}{Repeat Petition Writers}
We studied a group of 4,611 repeat petition writers that wrote multiple petitions between January 2022 and November 2024. 
We exclude the final month of 2024 to ensure all petitions had at least one month before time of data collection, since primary outcome metric is indication of receiving a comment within 30 days. We drop observations that have missing log-transformed Flesch-Kincaid grade levels, which are those petitions for which the Flesch-Kincaid grade level is negative. Furthermore, we only consider repeat petition-writers whose second petitions were written at least 6 months after their first petitions, and exclude the A/B testing period. 

We conduct an exploratory descriptive analysis of repeat petition writers. The main text includes the distribution of word count and the share of petitions that reach 1 comment in 30 days across the three cohorts of petition writers. In Figure~\ref{fig:repeat_petition_writers_barplots} we show the distribution of MATTR, Flesch-Kincaid grade level, and the share of petitions that reach 10 signatures. We find consistent results to the main text: in the case of lexical measures, the average MATTR and Flesch-Kincaid grade level is higher in users' second petitions when users write their first petition without AI access and their second petition with AI access (Pre/Post cohort). In the case of the 10+ signature outcome, users' second petitions have reduced outcomes when their first petition was written without AI access and their second petition was written with AI access (Pre/Post cohort). 

In our main analysis, we design a mixed effects model for inference into the predictive strength of user cohort among repeat petition writers. 
We show tables with the model summary for outcomes in Table~\ref{tab:mixed_effect_outcomes}, and for lexical features in Table~\ref{tab:mixed_effect_model_style}.
In both cases we use the \texttt{lme4} R package to fit our models~\cite{lme4}: we use the BOBYQA optimizer with 200,000 maximum iterations to fit the model for outcomes.
In Table~\ref{tab:mixed_effect_outcomes}, the first column shows the coefficients and standard deviation in parentheses for each term of the model trained to predict if a petition received at least 1 comment in 30 days. Likewise, the second column shows the coefficients and standard error for each term of the model trained to predict if the petition received at least 10 signatures.
Both models were generalized logistic regression models with random effects for user ID. 
In this model, the intercept reflects the first petitions written in the Pre/Pre cohort.
In Table~\ref{tab:mixed_effect_model_style}, we show the coefficients and standard error for linear regression model trained to predict petition MATTR, Flesch-Kincaid grade level, and word count. We log transform Flesch-Kincaid and word count, but not MATTR, as it is normally distributed without transformation. Similarly, the intercept reflects the first petitions written in the Pre/Pre cohort.
We then use pairwise contrasts with a Bonferroni adjustment to compare users' first and second petitions, across the three cohorts.

\subsection*{Synthetic Control}
\addcontentsline{toc}{subsection}{Synthetic Control}

Synthetic control is an approach that allows us to compare trends in treated countries to trends in control countries without a strict requirement of parallel trends. 
To confirm the findings we report in the main text, we conduct a synthetic control analysis in which we use the \texttt{synth} package version 1.1.8 in R to compare a single treated unit against a synthetic control~\cite{synth}. 
With the synthetic control, we can leverage data from countries that did not qualify for the difference-in-differences analysis in the main text. 
In particular, we use trends from English petitions written in Australia, India, and New Zealand to create our synthetic control.

The \texttt{synth} package identifies country-level weights that minimize the pre-treatment gap between treated and control countries with the Broyden–Fletcher–Goldfarb–Shanno algorithm.
This synthetic control is created by using a weighted combination of control units to approximate the characteristics of the treated units. 
We then use the actual time series for each control country, weighted by the optimized weights, to produce a synthetic counterfactual. 
In order to create the weighted synthetic controls we provide the same set of covariates as are included in the full-covariate difference-in-differences robustness check. These are as follows: the share of petitions per week and country that were (a) posted on the weekend (b) had decision makers specified, and the share of users per week and country that (c) were new, and (d) had commented on a petition before.
We also include the lagged outcome (lagged at 4, 8, and 12 weeks) for matching across groups~\cite{cunningham_mixtape}.
We can then compare the time series in the treated countries to the time series in the synthetic counterfactual, which provides us with an estimate of the causal effect of the intervention (in-platform AI access).
SI Appendix Figures~\ref{fig:synthetic_control_results_us},~\ref{fig:synthetic_control_results_ca},~\ref{fig:synthetic_control_results_gb} show the time series of the actual treated countries against the synthetic control estimates, which are consistent with our difference-in-differences analysis.
In particular, the word count, lexical diversity, and readability of petitions written in treated countries are higher than in the synthetic control counterfactual trend during the period of differential access.
Outcome metrics (i.e., the share of petitions that reach 1+ comment in 30 days or 10+ signatures overall) in treated countries are lower than the synthetic control counterfactual during the period of differential access.

Through the synthetic control analysis, we can see that after full AI access, the actual treated petition lexical features (MATTR, Grade Level, Word Count) are greater than the synthetic control estimates. Likewise, the actual treated outcome metrics are lower than the synthetic control outcome metrics. 
These trends are consistent with the patterns we observed in our main differences-in-differences analysis.

\subsection*{Bayesian Structural Time Series}
\addcontentsline{toc}{subsection}{Bayesian Structural Time Series}

Another form of synthetic control is the Bayesian Structural Time Series (BSTS) model, which is leveraged through the \texttt{CausalImpact} R software ~\cite{brodersen2015} to create a synthetic control that accounts for both trends in control countries as well as patterns of seasonality, and general trends. 
We choose to use the BSTS approach to predict participation, because unlike outcomes and lexical features, trends in participation appear to have a strong seasonal component. We use participation in control countries (same as earlier synthetic control: Australia, India, New Zealand) to predict synthetic control outcomes. 
We fit a BSTS model with two components: a semi-local linear trend, and a seasonal component. 
We fit the BSTS model with the \texttt{bsts} R package using package-default priors throughout (including a spike-and-slab prior on the control-series regression coefficient) and 1,000 MCMC iterations.
In SI Appendix Figure~\ref{fig:participation} we show three sub-figures that describe platform participation on Change.org after AI access. 
In SI Appendix Figure~\ref{fig:participation_normalized_time_series}, we show that weekly count of users that produce petitions (normalized by pre-AI average and standard deviation) does not appear to shift from the 2022 mean.
In Figure~\ref{subfig:participation_bsts_time_series}, we show raw user participation counts in a solid line for the period before AI access, during the A/B period, and after the full launch. 
Figure~\ref{subfig:participation_bsts} further elucidates this result, through the actual vs. predicted time series (top), point-wise estimate, for which the effect is estimated for every week, (middle), and cumulative estimate, for which the overall impact over the post-AI time period is shown (bottom). 
We find that user participation fell well within a 95\% credible interval of predictions. The estimated effect of the intervention, according to the BSTS model, is insignificant (\(\Delta_{treated-control}\) = -354 users (s.d. = 901), $p$ = 0.343, 95\% credible interval [-2167, 1263]).

We repeat the BSTS analysis process on petition production, rather than user participation, to see if results are consistent. Indeed, Figure~\ref{fig:participation_production} demonstrates that the actual trends in petition production do not diverge from predicted trends estimated using pre-AI seasonal trends and production from control countries. This is not surprising as only 6.38\% of users wrote multiple petitions, and as such we would expect to see similar trends in user participation and petition production over time.

\subsection*{Interrupted Time Series}
\addcontentsline{toc}{subsection}{Interrupted Time Series}
In our preregistration we planned to use an interrupted time series analysis through a regression discontinuity in time method to support our analysis. 
Figure~\ref{fig:its} shows the output of this analysis, in which we plot changes to the slope of fit lines in lexical features and outcomes, before and after the start of A/B testing. 
In this case, the intervention is the start of A/B testing rather than the full launch. 
We use the start of A/B testing as the intervention date for two reasons.
First, this date enables us to have a clean threshold for the discontinuity, rather than a 6-month gap when comparing pre-AI to post-AI. 
Second, because we do not know how the A/B testing period differed from the full-launch, we would not be able to interpret changes between the A/B testing and the post-AI period: shifts to level or slope of the trend line could be a result of expanding access or shifting features during experimentation.
However, comparing pre-AI period to A/B testing period we can hypothesize that we would expect to see a change in slope if A/B testing occurred incrementally and changed lexical features and outcomes. 
In Figure~\ref{fig:its_treated_style}, we show that the slope of the line fit to predict lexical features in treated countries (top to bottom: MATTR, Flesch-Kincaid grade level, word count) increased after the introduction of A/B testing for the in-platform AI feature. Conversely, Figure~\ref{fig:its_control_style} shows that the slope of the line fit to predict lexical features in control countries, meaning those that were not included in the A/B test, did not have the same increase. 
Figure~\ref{fig:its_treated_outcomes} and Figure~\ref{fig:its_control_outcomes} show that the slope of the line fit to predict outcomes seems to increase post-A/B testing, for both treated and control countries. 
These results are consistent with those found with our difference-in-differences analysis and synthetic control methods studying the full-launch.

\putbib[si_ref]
\end{bibunit}


\begin{thebibliography}{59}
\providecommand{\natexlab}[1]{#1}
\providecommand{\url}[1]{\texttt{#1}}
\expandafter\ifx\csname urlstyle\endcsname\relax
  \providecommand{\doi}[1]{doi: #1}\else
  \providecommand{\doi}{doi: \begingroup \urlstyle{rm}\Url}\fi

\bibitem[Noy and Zhang(2023)]{Noy2023-sf}
Shakked Noy and Whitney Zhang.
\newblock Experimental evidence on the productivity effects of generative artificial intelligence.
\newblock \emph{Science}, 381\penalty0 (6654):\penalty0 187--192, 2023.
\newblock \doi{10.1126/science.adh2586}.
\newblock URL \url{https://www.science.org/doi/abs/10.1126/science.adh2586}.

\bibitem[Doshi and Hauser(2024)]{doshiHauser2024}
Anil~R. Doshi and Oliver~P. Hauser.
\newblock Generative {AI} enhances individual creativity but reduces the collective diversity of novel content.
\newblock \emph{Science Advances}, 10\penalty0 (28):\penalty0 eadn5290, 2024.
\newblock \doi{10.1126/sciadv.adn5290}.
\newblock URL \url{https://www.science.org/doi/abs/10.1126/sciadv.adn5290}.

\bibitem[Matz et~al.(2024)Matz, Teeny, Vaid, Peters, Harari, and Cerf]{matz2024potential}
Sandra~C Matz, Jacob~D Teeny, Sumer~S Vaid, Heinrich Peters, Gabriella~M Harari, and Moran Cerf.
\newblock The potential of generative {AI} for personalized persuasion at scale.
\newblock \emph{Scientific Reports}, 14\penalty0 (1):\penalty0 4692, 2024.

\bibitem[Argyle et~al.(2025)Argyle, Busby, Gubler, Lyman, Olcott, Pond, and Wingate]{argyle2025testing}
Lisa~P Argyle, Ethan~C Busby, Joshua~R Gubler, Alex Lyman, Justin Olcott, Jackson Pond, and David Wingate.
\newblock Testing theories of political persuasion using {AI}.
\newblock \emph{Proceedings of the National Academy of Sciences}, 122\penalty0 (18):\penalty0 e2412815122, 2025.

\bibitem[Kumar et~al.(2025)Kumar, Vincentius, Jordan, and Anderson]{kumar2025human}
Harsh Kumar, Jonathan Vincentius, Ewan Jordan, and Ashton Anderson.
\newblock Human creativity in the age of {LLMs}: Randomized experiments on divergent and convergent thinking.
\newblock In \emph{Proceedings of the 2025 CHI Conference on Human Factors in Computing Systems}, pages 1--18, 2025.

\bibitem[Westmoreland(2024)]{experienceAmazonSellersCan2024}
Mary~Beth Westmoreland.
\newblock Amazon sellers can now automatically improve product listings with our new {{Gen AI}} tool, September 2024.

\bibitem[Meta(2025)]{WriteAIFacebook}
Meta.
\newblock Write with {{AI}} on {{Facebook}} {\textbar} {{Facebook Help Center}}, 2025.
\newblock URL \url{https://www.facebook.com/help/7745061278859290?cms_platform=iphone-app}.

\bibitem[Vaccaro et~al.(2024)Vaccaro, Almaatouq, and Malone]{vaccaroWhenCombinationsHumans2024}
Michelle Vaccaro, Abdullah Almaatouq, and Thomas Malone.
\newblock When combinations of humans and {{AI}} are useful: {{A}} systematic review and meta-analysis.
\newblock \emph{Nature Human Behaviour}, 8\penalty0 (12):\penalty0 2293--2303, December 2024.
\newblock ISSN 2397-3374.
\newblock \doi{10.1038/s41562-024-02024-1}.

\bibitem[Wu et~al.(2025)Wu, Black, and Chandrasekaran]{wu2024generativemonoculturelargelanguage}
Fan Wu, Emily Black, and Varun Chandrasekaran.
\newblock Generative monoculture in large language models.
\newblock In \emph{International Conference on Learning Representations}, volume 2025, pages 33068--33107, 2025.
\newblock URL \url{https://proceedings.iclr.cc/paper_files/paper/2025/file/5178b2f2d7c44aa390c0777dc77b3f0c-Paper-Conference.pdf}.

\bibitem[Hao et~al.(2026)Hao, Xu, Li, and Evans]{haoArtificialIntelligenceTools2026}
Qianyue Hao, Fengli Xu, Yong Li, and James Evans.
\newblock Artificial intelligence tools expand scientists' impact but contract science's focus.
\newblock \emph{Nature}, 649\penalty0 (8099):\penalty0 1237--1243, January 2026.
\newblock ISSN 1476-4687.
\newblock \doi{10.1038/s41586-025-09922-y}.

\bibitem[Agarwal et~al.(2025)Agarwal, Naaman, and Vashistha]{agarwal2025ai}
Dhruv Agarwal, Mor Naaman, and Aditya Vashistha.
\newblock {AI} suggestions homogenize writing toward western styles and diminish cultural nuances.
\newblock In \emph{Proceedings of the 2025 CHI Conference on Human Factors in Computing Systems}, pages 1--21, 2025.

\bibitem[Anderson et~al.(2024)Anderson, Shah, and Kreminski]{anderson2024homogenization}
Barrett~R Anderson, Jash~Hemant Shah, and Max Kreminski.
\newblock Homogenization effects of large language models on human creative ideation.
\newblock In \emph{Proceedings of the 16th Conference on Creativity \& Cognition}, C\&C '24, page 413–425, New York, NY, USA, 2024. Association for Computing Machinery.
\newblock ISBN 9798400704857.
\newblock \doi{10.1145/3635636.3656204}.
\newblock URL \url{https://doi.org/10.1145/3635636.3656204}.

\bibitem[Brynjolfsson et~al.(2025)Brynjolfsson, Li, and Raymond]{brynjolfsson2025generativeAIatwork}
Erik Brynjolfsson, Danielle Li, and Lindsey Raymond.
\newblock Generative {{AI}} at work.
\newblock \emph{The Quarterly Journal of Economics}, 140\penalty0 (2):\penalty0 889--942, 02 2025.
\newblock ISSN 0033-5533.
\newblock \doi{10.1093/qje/qjae044}.
\newblock URL \url{https://doi.org/10.1093/qje/qjae044}.

\bibitem[Callaway and Sant’Anna(2021)]{callaway2021difference}
Brantly Callaway and Pedro~H.C. Sant’Anna.
\newblock Difference-in-differences with multiple time periods.
\newblock \emph{Journal of Econometrics}, 225\penalty0 (2):\penalty0 200--230, 2021.
\newblock ISSN 0304-4076.
\newblock \doi{https://doi.org/10.1016/j.jeconom.2020.12.001}.
\newblock URL \url{https://www.sciencedirect.com/science/article/pii/S0304407620303948}.
\newblock Themed Issue: Treatment Effect 1.

\bibitem[Mesgar and Strube(2018)]{mesgar2018neural}
Mohsen Mesgar and Michael Strube.
\newblock A neural local coherence model for text quality assessment.
\newblock In \emph{Proceedings of the 2018 conference on empirical methods in natural language processing}, pages 4328--4339, 2018.

\bibitem[Bizzoni et~al.(2023)Bizzoni, Moreira, Dwenger, Lassen, Thomsen, and Nielbo]{bizzoni2023good}
Yuri Bizzoni, Pascale Moreira, Nicole Dwenger, Ida Lassen, Mads Thomsen, and Kristoffer Nielbo.
\newblock Good reads and easy novels: Readability and literary quality in a corpus of {US}-published fiction.
\newblock In \emph{Proceedings of the 24th Nordic Conference on Computational Linguistics (NoDaLiDa)}, pages 42--51, 2023.

\bibitem[Cowgill et~al.(2026)Cowgill, Hern\'{a}ndez-Lagos, and Wright]{cowgill2026}
Bo~Cowgill, Pablo Hern\'{a}ndez-Lagos, and Nataliya~Langburd Wright.
\newblock Does {{AI}} cheapen talk? theory and evidence from global entrepreneurship and hiring.
\newblock \emph{Management Science}, 0\penalty0 (0), 2026.
\newblock \doi{10.1287/mnsc.2024.07027}.
\newblock URL \url{https://doi.org/10.1287/mnsc.2024.07027}.

\bibitem[Padmakumar and He(2024)]{padmakumar2023does}
Vishakh Padmakumar and He~He.
\newblock Does writing with language models reduce content diversity?
\newblock In B.~Kim, Y.~Yue, S.~Chaudhuri, K.~Fragkiadaki, M.~Khan, and Y.~Sun, editors, \emph{International Conference on Learning Representations}, volume 2024, pages 642--669, 2024.
\newblock URL \url{https://proceedings.iclr.cc/paper_files/paper/2024/file/02dec8877fb7c6aa9a79f81661baca7c-Paper-Conference.pdf}.

\bibitem[Brodersen et~al.(2015)Brodersen, Gallusser, Koehler, Remy, and Scott]{brodersen2015causalimpact}
Kay~H. Brodersen, Fabian Gallusser, Jim Koehler, Nicolas Remy, and Steven~L. Scott.
\newblock {Inferring causal impact using Bayesian structural time-series models}.
\newblock \emph{The Annals of Applied Statistics}, 9\penalty0 (1):\penalty0 247 -- 274, 2015.
\newblock \doi{10.1214/14-AOAS788}.
\newblock URL \url{https://doi.org/10.1214/14-AOAS788}.

\bibitem[Maxwell(1980)]{scottpairwise1980}
Scott~E. Maxwell.
\newblock Pairwise multiple comparisons in repeated measures designs.
\newblock \emph{Journal of Educational Statistics}, 5\penalty0 (3):\penalty0 269--287, 1980.
\newblock \doi{10.3102/10769986005003269}.
\newblock URL \url{https://doi.org/10.3102/10769986005003269}.

\bibitem[Dell'Acqua et~al.(2023)Dell'Acqua, McFowland~III, Mollick, {Lifshitz-Assaf}, Kellogg, Rajendran, Krayer, Candelon, and Lakhani]{dellacquaNavigatingJaggedTechnological2023}
Fabrizio Dell'Acqua, Edward McFowland~III, Ethan~R. Mollick, Hila {Lifshitz-Assaf}, Katherine Kellogg, Saran Rajendran, Lisa Krayer, Fran{\c c}ois Candelon, and Karim~R. Lakhani.
\newblock Navigating the {{Jagged Technological Frontier}}: {{Field Experimental Evidence}} of the {{Effects}} of {{AI}} on {{Knowledge Worker Productivity}} and {{Quality}}, September 2023.
\newblock SSRN [Preprint] \url{https://doi.org/10.2139/ssrn.4573321} (accessed 25 July 2025).

\bibitem[Hagen et~al.(2016)Hagen, Harrison, Uzuner, May, Fake, and Katragadda]{HAGEN2016783}
Loni Hagen, Teresa~M. Harrison, {\"O}zlem Uzuner, William May, Tim Fake, and Satya Katragadda.
\newblock E-petition popularity: {{Do}} linguistic and semantic factors matter?
\newblock \emph{Government Information Quarterly}, 33\penalty0 (4):\penalty0 783--795, 2016.
\newblock ISSN 0740-624X.
\newblock \doi{10.1016/j.giq.2016.07.006}.

\bibitem[Chen et~al.(2019)Chen, Deng, Kwak, Elnoshokaty, and Wu]{chen2019multi}
Yan Chen, Shuyuan Deng, Dong-Heon Kwak, Ahmed Elnoshokaty, and Jiao Wu.
\newblock A multi-appeal model of persuasion for online petition success: A linguistic cue-based approach.
\newblock \emph{Journal of the Association for Information Systems}, 20\penalty0 (2):\penalty0 3, 2019.

\bibitem[Schoenegger et~al.(2025)Schoenegger, Salvi, Liu, Nan, Debnath, Fasolo, Leivada, Recchia, Günther, Zarifhonarvar, Kwon, Islam, Dehnert, Lee, Reinecke, Kamper, Kobaş, Sandford, Kgomo, Hewitt, Kapoor, Oktar, Kucuk, Feng, Jones, Gainsburg, Olschewski, Heinzelmann, Cruz, Tappin, Ma, Park, Onyonka, Hjorth, Slattery, Zeng, Finke, Grossmann, Salatiello, and Karger]{schoenegger2025large}
Philipp Schoenegger, Francesco Salvi, Jiacheng Liu, Xiaoli Nan, Ramit Debnath, Barbara Fasolo, Evelina Leivada, Gabriel Recchia, Fritz Günther, Ali Zarifhonarvar, Joe Kwon, Zahoor~Ul Islam, Marco Dehnert, Daryl Y.~H. Lee, Madeline~G. Reinecke, David~G. Kamper, Mert Kobaş, Adam Sandford, Jonas Kgomo, Luke Hewitt, Shreya Kapoor, Kerem Oktar, Eyup~Engin Kucuk, Bo~Feng, Cameron~R. Jones, Izzy Gainsburg, Sebastian Olschewski, Nora Heinzelmann, Francisco Cruz, Ben~M. Tappin, Tao Ma, Peter~S. Park, Rayan Onyonka, Arthur Hjorth, Peter Slattery, Qingcheng Zeng, Lennart Finke, Igor Grossmann, Alessandro Salatiello, and Ezra Karger.
\newblock Large language models are more persuasive than incentivized human persuaders, 2025.
\newblock URL \url{https://arxiv.org/abs/2505.09662}.
\newblock (accessed 25 July 2025).

\bibitem[Wojtowicz and DeDeo(2025)]{wojtowicz2025undermining}
Zachary Wojtowicz and Simon DeDeo.
\newblock Undermining mental proof: How ai can make cooperation harder by making thinking easier.
\newblock \emph{Proceedings of the AAAI Conference on Artificial Intelligence}, 39\penalty0 (2):\penalty0 1592--1600, 4 2025.
\newblock \doi{10.1609/aaai.v39i2.32151}.
\newblock URL \url{https://ojs.aaai.org/index.php/AAAI/article/view/32151}.

\bibitem[Feuerriegel et~al.(2023)Feuerriegel, DiResta, Goldstein, Kumar, Lorenz-Spreen, Tomz, and Pr{\"o}llochs]{feuerriegel2023research}
Stefan Feuerriegel, Ren{\'e}e DiResta, Josh~A Goldstein, Srijan Kumar, Philipp Lorenz-Spreen, Michael Tomz, and Nicolas Pr{\"o}llochs.
\newblock Research can help to tackle {AI}-generated disinformation.
\newblock \emph{Nature Human Behaviour}, 7\penalty0 (11):\penalty0 1818--1821, 2023.

\bibitem[Floridi and Chiriatti(2020)]{floridi2020gpt}
Luciano Floridi and Massimo Chiriatti.
\newblock {GPT-3}: Its nature, scope, limits, and consequences.
\newblock \emph{Minds and Machines}, 30:\penalty0 681--694, 2020.

\bibitem[M{\o}ller et~al.(2025)M{\o}ller, Romero, Jurgens, and Aiello]{moller2025impact}
Anders~Giovanni M{\o}ller, Daniel~M. Romero, David Jurgens, and Luca~Maria Aiello.
\newblock The impact of generative {AI} on social media: An experimental study.
\newblock arXiv [Preprint], 2025.
\newblock URL \url{https://arxiv.org/abs/2506.14295}.
\newblock (accessed 25 July 2025).

\bibitem[Tullis(2025)]{tullis2025sifting}
Jane Tullis.
\newblock Sifting through the slop: How generative {AI} created a market for lemons for text-based works.
\newblock \emph{SSRN [Preprint]}, 2025.
\newblock \doi{10.2139/ssrn.5266660}.
\newblock Available at: \url{https://dx.doi.org/10.2139/ssrn.5266660} (accessed 25 July 2025).

\bibitem[Jakesch et~al.(2023)Jakesch, Hancock, and Naaman]{jakesch2023human}
Maurice Jakesch, Jeffrey~T Hancock, and Mor Naaman.
\newblock Human heuristics for {AI}-generated language are flawed.
\newblock \emph{Proceedings of the National Academy of Sciences}, 120\penalty0 (11):\penalty0 e2208839120, 2023.

\bibitem[Reif et~al.(2025)Reif, Larrick, and Soll]{reif2025socialpenalty}
Jessica~A. Reif, Richard~P. Larrick, and Jack~B. Soll.
\newblock Evidence of a social evaluation penalty for using {AI}.
\newblock \emph{Proceedings of the National Academy of Sciences}, 122\penalty0 (19):\penalty0 e2426766122, 2025.
\newblock \doi{10.1073/pnas.2426766122}.
\newblock URL \url{https://www.pnas.org/doi/abs/10.1073/pnas.2426766122}.

\bibitem[Kadoma et~al.(2025)Kadoma, Metaxa, and Naaman]{kadoma2025generative}
Kowe Kadoma, Dana\'{e} Metaxa, and Mor Naaman.
\newblock Generative {AI} and perceptual harms: Who's suspected of using llms?
\newblock In \emph{Proceedings of the 2025 CHI Conference on Human Factors in Computing Systems}, CHI '25, New York, NY, USA, 2025. Association for Computing Machinery.
\newblock ISBN 9798400713941.
\newblock \doi{10.1145/3706598.3713897}.
\newblock URL \url{https://doi.org/10.1145/3706598.3713897}.

\bibitem[Jakesch et~al.(2019)Jakesch, French, Ma, Hancock, and Naaman]{jakesch2019ai}
Maurice Jakesch, Megan French, Xiao Ma, Jeffrey~T Hancock, and Mor Naaman.
\newblock {AI}-mediated communication: How the perception that profile text was written by {AI} affects trustworthiness.
\newblock In \emph{Proceedings of the 2019 CHI conference on human factors in computing systems}, pages 1--13, 2019.

\bibitem[Hohenstein et~al.(2023)Hohenstein, Kizilcec, DiFranzo, Aghajari, Mieczkowski, Levy, Naaman, Hancock, and Jung]{hohenstein2023artificial}
Jess Hohenstein, Rene~F Kizilcec, Dominic DiFranzo, Zhila Aghajari, Hannah Mieczkowski, Karen Levy, Mor Naaman, Jeffrey Hancock, and Malte~F Jung.
\newblock Artificial intelligence in communication impacts language and social relationships.
\newblock \emph{Scientific Reports}, 13\penalty0 (1):\penalty0 5487, 2023.

\bibitem[Duan et~al.(2024)Duan, Zhou, Scalia, Yin, Weng, Zhang, Freeman, McNeese, Gorman, and Tolston]{duan2024understanding}
Wen Duan, Shiwen Zhou, Matthew~J Scalia, Xiaoyun Yin, Nan Weng, Ruihao Zhang, Guo Freeman, Nathan McNeese, Jamie Gorman, and Michael Tolston.
\newblock Understanding the evolvement of trust over time within human-{AI} teams.
\newblock \emph{Proceedings of the ACM on Human-Computer Interaction}, 8\penalty0 (CSCW2):\penalty0 1--31, 2024.

\bibitem[Matatov et~al.(2024)Matatov, Qu{\'e}r{\'e}, Amir, and Naaman]{matatov2024examining}
Hana Matatov, Marianne Aubin~Le Qu{\'e}r{\'e}, Ofra Amir, and Mor Naaman.
\newblock Examining the prevalence and dynamics of {AI}-generated media in art subreddits.
\newblock \emph{arXiv preprint arXiv:2410.07302}, 2024.

\bibitem[Draxler et~al.(2024)Draxler, Werner, Lehmann, Hoppe, Schmidt, Buschek, and Welsch]{draxler2024ai}
Fiona Draxler, Anna Werner, Florian Lehmann, Matthias Hoppe, Albrecht Schmidt, Daniel Buschek, and Robin Welsch.
\newblock The {AI} ghostwriter effect: When users do not perceive ownership of ai-generated text but self-declare as authors.
\newblock \emph{ACM Trans. Comput.-Hum. Interact.}, 31\penalty0 (2), February 2024.
\newblock ISSN 1073-0516.
\newblock \doi{10.1145/3637875}.
\newblock URL \url{https://doi.org/10.1145/3637875}.

\bibitem[Kadoma et~al.(2024)Kadoma, Aubin Le~Quere, Fu, Munsch, Metaxa, and Naaman]{kadoma2024inclusion}
Kowe Kadoma, Marianne Aubin Le~Quere, Xiyu~Jenny Fu, Christin Munsch, Dana\"{e} Metaxa, and Mor Naaman.
\newblock The role of inclusion, control, and ownership in workplace {AI}-mediated communication.
\newblock In \emph{Proceedings of the 2024 CHI Conference on Human Factors in Computing Systems}, CHI '24, New York, NY, USA, 2024. Association for Computing Machinery.
\newblock ISBN 9798400703300.
\newblock \doi{10.1145/3613904.3642650}.
\newblock URL \url{https://doi.org/10.1145/3613904.3642650}.

\bibitem[Xu et~al.(2024)Xu, Cheng, and Kuzminykh]{xu2024what}
Yuxin Xu, Mengqiu Cheng, and Anastasia Kuzminykh.
\newblock What makes it mine? exploring psychological ownership over human-ai co-creations.
\newblock In \emph{Proceedings of the 50th Graphics Interface Conference}, GI '24, New York, NY, USA, 2024. Association for Computing Machinery.
\newblock ISBN 9798400718281.
\newblock \doi{10.1145/3670947.3670974}.
\newblock URL \url{https://doi.org/10.1145/3670947.3670974}.

\bibitem[He et~al.(2025)He, Houde, and Weisz]{he2025which}
Jessica He, Stephanie Houde, and Justin~D. Weisz.
\newblock Which contributions deserve credit? perceptions of attribution in human-{AI} co-creation.
\newblock In \emph{Proceedings of the 2025 CHI Conference on Human Factors in Computing Systems}, CHI '25, New York, NY, USA, 2025. Association for Computing Machinery.
\newblock ISBN 9798400713941.
\newblock \doi{10.1145/3706598.3713522}.
\newblock URL \url{https://doi.org/10.1145/3706598.3713522}.

\bibitem[Pierce et~al.(2003)Pierce, Kostova, and Dirks]{pierce2003state}
Jon~L Pierce, Tatiana Kostova, and Kurt~T Dirks.
\newblock The state of psychological ownership: Integrating and extending a century of research.
\newblock \emph{Review of general psychology}, 7\penalty0 (1):\penalty0 84--107, 2003.

\bibitem[Zhao et~al.(2013)Zhao, Salehi, Naranjit, Alwaalan, Voida, and Cosley]{zhao2013faces}
Xuan Zhao, Niloufar Salehi, Sasha Naranjit, Sara Alwaalan, Stephen Voida, and Dan Cosley.
\newblock The many faces of facebook: experiencing social media as performance, exhibition, and personal archive.
\newblock In \emph{Proceedings of the SIGCHI Conference on Human Factors in Computing Systems}, CHI '13, page 1–10, New York, NY, USA, 2013. Association for Computing Machinery.
\newblock ISBN 9781450318990.
\newblock \doi{10.1145/2470654.2470656}.
\newblock URL \url{https://doi.org/10.1145/2470654.2470656}.

\bibitem[Prosser et~al.(2011)Prosser, Panagiotopoulos, Sams, Elliman, and Fitzgerald]{prosser2011DoSocialNetworking}
Alexander Prosser, Panagiotis Panagiotopoulos, Steven Sams, Tony Elliman, and Guy Fitzgerald.
\newblock Do social networking groups support online petitions?
\newblock \emph{Transforming Government: People, Process and Policy}, 5\penalty0 (1):\penalty0 20--31, 03 2011.
\newblock ISSN 1750-6166.
\newblock \doi{10.1108/17506161111114626}.
\newblock URL \url{https://doi.org/10.1108/17506161111114626}.

\bibitem[Proskurnia et~al.(2017)Proskurnia, Grabowicz, Kobayashi, Castillo, Cudr\'{e}-Mauroux, and Aberer]{proskurnia2017PredictingtheSuccess}
Julia Proskurnia, Przemyslaw Grabowicz, Ryota Kobayashi, Carlos Castillo, Philippe Cudr\'{e}-Mauroux, and Karl Aberer.
\newblock Predicting the success of online petitions leveraging multidimensional time-series.
\newblock In \emph{Proceedings of the 26th International Conference on World Wide Web}, WWW '17, page 755–764, Republic and Canton of Geneva, CHE, 2017. International World Wide Web Conferences Steering Committee.
\newblock ISBN 9781450349130.
\newblock \doi{10.1145/3038912.3052705}.
\newblock URL \url{https://doi.org/10.1145/3038912.3052705}.

\bibitem[Dumas et~al.(2015)Dumas, LaManna, Harrison, Ravi, Kotfila, Gervais, Hagen, and Chen]{dumas2015ExaminingPoliticalMobilization}
Catherine~L Dumas, Daniel LaManna, Teresa~M Harrison, SS~Ravi, Christopher Kotfila, Norman Gervais, Loni Hagen, and Feng Chen.
\newblock Examining political mobilization of online communities through e-petitioning behavior in we the people.
\newblock \emph{Big Data \& Society}, 2\penalty0 (2):\penalty0 2053951715598170, 2015.
\newblock \doi{10.1177/2053951715598170}.
\newblock URL \url{https://doi.org/10.1177/2053951715598170}.

\bibitem[Carlson(2019)]{Carlson2019PleaseSignHere}
Erin~Brock Carlson.
\newblock Please sign here (and share it to your facebook and twitter feeds): Online petitions and inventing for circulation.
\newblock \emph{Computers and Composition}, 52:\penalty0 175--194, 2019.
\newblock ISSN 8755-4615.
\newblock \doi{https://doi.org/10.1016/j.compcom.2019.01.003}.
\newblock URL \url{https://www.sciencedirect.com/science/article/pii/S8755461517300737}.

\bibitem[Harrison et~al.(2022)Harrison, Dumas, DePaula, Fake, May, Atrey, Lee, Rishi, and Ravi]{Harrison2022ExploringEpetitioning}
Teresa~M. Harrison, Catherine Dumas, Nic DePaula, Tim Fake, Will May, Akanksha Atrey, Jooyeon Lee, Lokesh Rishi, and S.S. Ravi.
\newblock Exploring e-petitioning and media: The case of \#bringbackourgirls.
\newblock \emph{Government Information Quarterly}, 39\penalty0 (1):\penalty0 101569, 2022.
\newblock ISSN 0740-624X.
\newblock \doi{https://doi.org/10.1016/j.giq.2021.101569}.
\newblock URL \url{https://www.sciencedirect.com/science/article/pii/S0740624X21000058}.

\bibitem[Bu{\c{c}}inca et~al.(2021)Bu{\c{c}}inca, Malaya, and Gajos]{buccinca2021trust}
Zana Bu{\c{c}}inca, Maja~Barbara Malaya, and Krzysztof~Z Gajos.
\newblock To trust or to think: cognitive forcing functions can reduce overreliance on {AI} in {AI}-assisted decision-making.
\newblock \emph{Proceedings of the ACM on Human-computer Interaction}, 5\penalty0 (CSCW1):\penalty0 1--21, 2021.

\bibitem[Lloyd et~al.(2025)Lloyd, Reagle, and Naaman]{lloyd2023there}
Travis Lloyd, Joseph Reagle, and Mor Naaman.
\newblock `{{There}} has to be a lot that we're missing': Moderating {{AI}}-generated content on reddit.
\newblock \emph{Proc. ACM Hum.-Comput. Interact.}, 9\penalty0 (7), October 2025.
\newblock \doi{10.1145/3757445}.
\newblock URL \url{https://doi.org/10.1145/3757445}.

\bibitem[Bak-Coleman et~al.(2025)Bak-Coleman, Lewandowsky, Lorenz-Spreen, Narayanan, Orben, and Oswald]{bak2025moving}
Joseph~B. Bak-Coleman, Stephan Lewandowsky, Philipp Lorenz-Spreen, Arvind Narayanan, Amy Orben, and Lisa Oswald.
\newblock Moving towards informative and actionable social media research, 2025.
\newblock URL \url{https://arxiv.org/abs/2505.09254}.

\bibitem[Lee and Lehdonvirta(2022)]{lee2022NewDigitalSafetyNet}
Sumin Lee and Vili Lehdonvirta.
\newblock New digital safety net or just more ‘friendfunding’? institutional analysis of medical crowdfunding in the united states.
\newblock \emph{Information, Communication \& Society}, 25\penalty0 (8):\penalty0 1151--1175, 2022.
\newblock \doi{10.1080/1369118X.2020.1850838}.
\newblock URL \url{https://doi.org/10.1080/1369118X.2020.1850838}.

\bibitem[Borst et~al.(2018)Borst, Moser, and Ferguson]{borst2018FromFriendfunding}
Irma Borst, Christine Moser, and Julie Ferguson.
\newblock From friendfunding to crowdfunding: Relevance of relationships, social media, and platform activities to crowdfunding performance.
\newblock \emph{New Media \& Society}, 20\penalty0 (4):\penalty0 1396--1414, 2018.
\newblock \doi{10.1177/1461444817694599}.
\newblock URL \url{https://doi.org/10.1177/1461444817694599}.
\newblock PMID: 30581357.

\bibitem[Zenker and Kyle(2021)]{ZENKER2021100505}
Fred Zenker and Kristopher Kyle.
\newblock Investigating minimum text lengths for lexical diversity indices.
\newblock \emph{Assessing Writing}, 47:\penalty0 100505, 2021.
\newblock ISSN 1075-2935.
\newblock \doi{https://doi.org/10.1016/j.asw.2020.100505}.
\newblock URL \url{https://www.sciencedirect.com/science/article/pii/S1075293520300660}.

\bibitem[Kincaid et~al.(1975)Kincaid, Fishburne, P., L., and S.]{kincaid1975derivation}
Kincaid, Fishburne, P., L., and S.
\newblock \emph{Derivation of New Readability Formulas (Automated Readability Index, Fog Count and Flesch Reading Ease Formula) for Navy Enlisted Personnel}.
\newblock Naval Technical Training Command Millington TN Research Branch, Feb 1975.
\newblock \doi{10.21236/ada006655}.

\bibitem[Tanprasert and Kauchak(2021)]{tanprasert2021flesch}
Teerapaun Tanprasert and David Kauchak.
\newblock {{Flesch-Kincaid}} is not a text simplification evaluation metric.
\newblock In \emph{Proceedings of the 1st Workshop on Natural Language Generation, Evaluation, and Metrics (GEM 2021)}, pages 1--14, 2021.

\bibitem[Jindal and MacDermid(2017)]{jindal2017assessing}
Pranay Jindal and Joy~C MacDermid.
\newblock Assessing reading levels of health information: uses and limitations of {{Flesch}} formula.
\newblock \emph{Education for Health}, 30\penalty0 (1):\penalty0 84--88, 2017.

\bibitem[del {Rio-Chanona} et~al.(2024)del {Rio-Chanona}, Laurentsyeva, and Wachs]{rio-chanonaAreLargeLanguage2024}
Maria del {Rio-Chanona}, Nadzeya Laurentsyeva, and Johannes Wachs.
\newblock Are {{Large Language Models}} a {{Threat}} to {{Digital Public Goods}}? {{Evidence}} from {{Activity}} on {{Stack Overflow}}.
\newblock \emph{PNAS Nexus}, 3\penalty0 (9):\penalty0 pgae400, September 2024.
\newblock ISSN 2752-6542.
\newblock \doi{10.1093/pnasnexus/pgae400}.

\bibitem[Sant’Anna and Zhao(2020)]{SANTANNA2020101}
Pedro~H.C. Sant’Anna and Jun Zhao.
\newblock Doubly robust difference-in-differences estimators.
\newblock \emph{Journal of Econometrics}, 219\penalty0 (1):\penalty0 101--122, 2020.
\newblock ISSN 0304-4076.
\newblock \doi{https://doi.org/10.1016/j.jeconom.2020.06.003}.
\newblock URL \url{https://www.sciencedirect.com/science/article/pii/S0304407620301901}.

\bibitem[Callaway and {{Sant'Anna}}(2021)]{callaway_santanna_2025_did}
Brantly Callaway and Pedro~H.C. {{Sant'Anna}}.
\newblock did: Difference in differences, 2021.
\newblock URL \url{https://bcallaway11.github.io/did/}.
\newblock R package version 2.1.2.

\end{thebibliography}


\begin{thebibliography}{25}
\providecommand{\natexlab}[1]{#1}
\providecommand{\url}[1]{\texttt{#1}}
\expandafter\ifx\csname urlstyle\endcsname\relax
  \providecommand{\doi}[1]{doi: #1}\else
  \providecommand{\doi}{doi: \begingroup \urlstyle{rm}\Url}\fi

\bibitem[{Change.org}(2023)]{change_2023}
{Change.org}.
\newblock Change.org {Home}.
\newblock \emph{Change.org}. \url{https://www.change.org}, 2023.
\newblock Accessed June 5, 2026. Archived at \emph{Internet Archive}, \url{https://web.archive.org/web/20231101012225/https://www.change.org/}.

\bibitem[{Change.org}(2025)]{change_2025}
{Change.org}.
\newblock Change.org {Home}.
\newblock \emph{Change.org}. \url{https://www.change.org}, 2025.
\newblock Accessed February 14, 2025.

\bibitem[{Change.org}(2026)]{change_2026}
{Change.org}.
\newblock Change.org {Home}.
\newblock \emph{Change.org}. \url{https://www.change.org}, 2026.
\newblock Accessed January 26, 2026.

\bibitem[Lazer et~al.(2020)Lazer, Pentland, Watts, Aral, Athey, Contractor, Freelon, Gonzalez-Bailon, King, Margetts, Nelson, Salganik, Strohmaier, Vespignani, and Wagner]{lazer2020computationalsocialscience}
David M.~J. Lazer, Alex Pentland, Duncan~J. Watts, Sinan Aral, Susan Athey, Noshir Contractor, Deen Freelon, Sandra Gonzalez-Bailon, Gary King, Helen Margetts, Alondra Nelson, Matthew~J. Salganik, Markus Strohmaier, Alessandro Vespignani, and Claudia Wagner.
\newblock Computational social science: Obstacles and opportunities.
\newblock \emph{Science}, 369\penalty0 (6507):\penalty0 1060--1062, 2020.
\newblock \doi{10.1126/science.aaz8170}.
\newblock URL \url{https://www.science.org/doi/abs/10.1126/science.aaz8170}.

\bibitem[Bak-Coleman et~al.(2026)Bak-Coleman, West, O'Connor, and Bergstrom]{bakcoleman2026industryinfluencehighprofilesocial}
Joseph Bak-Coleman, Jevin West, Cailin O'Connor, and Carl~T. Bergstrom.
\newblock Industry influence in high-profile social media research, 2026.
\newblock URL \url{https://arxiv.org/abs/2601.11507}.

\bibitem[Bird and Loper(2004)]{nltk}
Steven Bird and Edward Loper.
\newblock Natural language toolkit (nltk), 2004.
\newblock URL \url{https://www.nltk.org/}.

\bibitem[Desklib(2025)]{desklib_2025_ai_text_detector}
Desklib.
\newblock {Desklib AI Text Detector}, 2025.
\newblock URL \url{https://huggingface.co/desklib/ai-text-detector-v1.01}.

\bibitem[Noy and Zhang(2023)]{noyzhang2023science}
Shakked Noy and Whitney Zhang.
\newblock Experimental evidence on the productivity effects of generative artificial intelligence.
\newblock \emph{Science}, 381\penalty0 (6654):\penalty0 187--192, 2023.
\newblock \doi{10.1126/science.adh2586}.
\newblock URL \url{https://www.science.org/doi/abs/10.1126/science.adh2586}.

\bibitem[Appelman and Sundar(2016)]{appelman2016messagecredibility}
Alyssa Appelman and S.~Shyam Sundar.
\newblock Measuring message credibility: Construction and validation of an exclusive scale.
\newblock \emph{Journalism \& Mass Communication Quarterly}, 93\penalty0 (1):\penalty0 59--79, 2016.
\newblock \doi{10.1177/1077699015606057}.
\newblock URL \url{https://doi.org/10.1177/1077699015606057}.

\bibitem[Hutto and Gilbert(2014)]{Hutto_Gilbert_2014}
C.~Hutto and Eric Gilbert.
\newblock {{VADER}}: A parsimonious rule-based model for sentiment analysis of social media text.
\newblock \emph{Proceedings of the International AAAI Conference on Web and Social Media}, 8\penalty0 (1):\penalty0 216--225, May 2014.
\newblock \doi{10.1609/icwsm.v8i1.14550}.

\bibitem[D{\"o}nmez et~al.(2025)D{\"o}nmez, Maurer, Lapesa, and Falenska]{donmez-etal-2025-ai}
Esra D{\"o}nmez, Maximilian Maurer, Gabriella Lapesa, and Agnieszka Falenska.
\newblock {AI} argues differently: Distinct argumentative and linguistic patterns of {LLM}s in persuasive contexts.
\newblock In Christos Christodoulopoulos, Tanmoy Chakraborty, Carolyn Rose, and Violet Peng, editors, \emph{Proceedings of the 2025 Conference on Empirical Methods in Natural Language Processing}, pages 34583--34614, Suzhou, China, November 2025. Association for Computational Linguistics.
\newblock ISBN 979-8-89176-332-6.
\newblock \doi{10.18653/v1/2025.emnlp-main.1755}.
\newblock URL \url{https://aclanthology.org/2025.emnlp-main.1755/}.

\bibitem[Markowitz et~al.(2024)Markowitz, Hancock, and Bailenson]{markowitz2024}
David~M. Markowitz, Jeffrey~T. Hancock, and Jeremy~N. Bailenson.
\newblock Linguistic markers of inherently false ai communication and intentionally false human communication: Evidence from hotel reviews.
\newblock \emph{Journal of Language and Social Psychology}, 43\penalty0 (1):\penalty0 63--82, 2024.
\newblock \doi{10.1177/0261927X231200201}.
\newblock URL \url{https://doi.org/10.1177/0261927X231200201}.

\bibitem[Aubin Le~Quéré and Matias(2025)]{quere_Matias_2025}
Marianne Aubin Le~Quéré and J.~Nathan Matias.
\newblock When curiosity gaps backfire: effects of headline concreteness on information selection decisions.
\newblock \emph{Scientific Reports}, 15\penalty0 (1):\penalty0 994, January 2025.
\newblock ISSN 2045-2322.
\newblock \doi{10.1038/s41598-024-81575-9}.
\newblock URL \url{https://www.nature.com/articles/s41598-024-81575-9}.

\bibitem[Gunning(1969)]{gunningfog1969}
Robert Gunning.
\newblock The fog index after twenty years.
\newblock \emph{Journal of Business Communication}, 6\penalty0 (2):\penalty0 3--13, 1969.
\newblock \doi{10.1177/002194366900600202}.
\newblock URL \url{https://doi.org/10.1177/002194366900600202}.

\bibitem[del {Rio-Chanona} et~al.(2024)del {Rio-Chanona}, Laurentsyeva, and Wachs]{riochanonaAreLargeLanguage2024}
Maria del {Rio-Chanona}, Nadzeya Laurentsyeva, and Johannes Wachs.
\newblock Are {{Large Language Models}} a {{Threat}} to {{Digital Public Goods}}? {{Evidence}} from {{Activity}} on {{Stack Overflow}}.
\newblock \emph{PNAS Nexus}, 3\penalty0 (9):\penalty0 pgae400, September 2024.
\newblock ISSN 2752-6542.
\newblock \doi{10.1093/pnasnexus/pgae400}.

\bibitem[Roth et~al.(2023)Roth, Sant'Anna, Bilinski, and Poe]{ROTH20232218}
Jonathan Roth, Pedro~H.C. Sant'Anna, Alyssa Bilinski, and John Poe.
\newblock What's trending in difference-in-differences? {{A}} synthesis of the recent econometrics literature.
\newblock \emph{Journal of Econometrics}, 235\penalty0 (2):\penalty0 2218--2244, 2023.
\newblock ISSN 0304-4076.
\newblock \doi{10.1016/j.jeconom.2023.03.008}.

\bibitem[MacKinnon and Webb(2020)]{mackinnonRandomizationInferenceDifferenceindifferences2020}
James~G. MacKinnon and Matthew~D. Webb.
\newblock Randomization inference for difference-in-differences with few treated clusters.
\newblock \emph{Journal of Econometrics}, 218\penalty0 (2):\penalty0 435--450, October 2020.
\newblock ISSN 0304-4076.
\newblock \doi{10.1016/j.jeconom.2020.04.024}.

\bibitem[Malyala(2025)]{abclf101}
Sreekanth Malyala.
\newblock abclf, version 1.0.1.
\newblock \url{https://pypi.org/project/abclf/1.0.1/}, 2025.
\newblock Python package.

\bibitem[Shelby et~al.(2023)Shelby, Rismani, Henne, Moon, Rostamzadeh, Nicholas, Yilla-Akbari, Gallegos, Smart, Garcia, and Virk]{shelby2023sociotechnical}
Renee Shelby, Shalaleh Rismani, Kathryn Henne, AJung Moon, Negar Rostamzadeh, Paul Nicholas, N'Mah Yilla-Akbari, Jess Gallegos, Andrew Smart, Emilio Garcia, and Gurleen Virk.
\newblock Sociotechnical harms of algorithmic systems: Scoping a taxonomy for harm reduction.
\newblock In \emph{Proceedings of the 2023 AAAI/ACM Conference on AI, Ethics, and Society}, AIES '23, page 723–741, New York, NY, USA, 2023. Association for Computing Machinery.
\newblock ISBN 9798400702310.
\newblock \doi{10.1145/3600211.3604673}.
\newblock URL \url{https://doi.org/10.1145/3600211.3604673}.

\bibitem[Costello et~al.(2024)Costello, Pennycook, and Rand]{costello2024}
Thomas~H. Costello, Gordon Pennycook, and David~G. Rand.
\newblock Durably reducing conspiracy beliefs through dialogues with ai.
\newblock \emph{Science}, 385\penalty0 (6714):\penalty0 eadq1814, 2024.
\newblock \doi{10.1126/science.adq1814}.
\newblock URL \url{https://www.science.org/doi/abs/10.1126/science.adq1814}.

\bibitem[Long and Ervin(2000)]{longervin2000}
J.~Scott Long and Laurie~H. Ervin.
\newblock Using heteroscedasticity consistent standard errors in the linear regression model.
\newblock \emph{The American Statistician}, 54\penalty0 (3):\penalty0 217--224, 2000.
\newblock \doi{10.1080/00031305.2000.10474549}.
\newblock URL \url{https://www.tandfonline.com/doi/abs/10.1080/00031305.2000.10474549}.

\bibitem[Bates et~al.(2015)Bates, M{\"a}chler, Bolker, and Walker]{lme4}
Douglas Bates, Martin M{\"a}chler, Ben Bolker, and Steve Walker.
\newblock Fitting linear mixed-effects models using {lme4}, 2015.

\bibitem[{Jens Hainmueller aut cre} et~al.(2011){Jens Hainmueller aut cre}, {Alexis Diamond aut}, and {Alberto Abadie aut}]{synth}
{Jens Hainmueller aut cre}, {Alexis Diamond aut}, and {Alberto Abadie aut}.
\newblock Synth: An r package for synthetic control methods in comparative case studies, 2011.
\newblock URL \url{https://www.jstatsoft.org/v42/i13/}.

\bibitem[Cunningham(2021)]{cunningham_mixtape}
Scott Cunningham.
\newblock \emph{Causal Inference: The Mixtape}.
\newblock Yale University Press, 2021.
\newblock ISBN 9780300251685.
\newblock URL \url{http://www.jstor.org/stable/j.ctv1c29t27}.

\bibitem[Brodersen et~al.(2015)Brodersen, Gallusser, Koehler, Remy, and Scott]{brodersen2015}
Kay~H. Brodersen, Fabian Gallusser, Jim Koehler, Nicolas Remy, and Steven~L. Scott.
\newblock {Inferring causal impact using Bayesian structural time-series models}.
\newblock \emph{The Annals of Applied Statistics}, 9\penalty0 (1):\penalty0 247 -- 274, 2015.
\newblock \doi{10.1214/14-AOAS788}.
\newblock URL \url{https://doi.org/10.1214/14-AOAS788}.

\end{thebibliography}
\end{document}